 \documentclass[final,5p,times,twocolumn,authoryear]{elsarticle}

\usepackage{amssymb}
\usepackage{lipsum}

\usepackage{xfrac}
\usepackage{xcolor}
\usepackage{amsmath,amssymb,amsfonts}
\usepackage{latexsym}
\usepackage{epstopdf} 
\usepackage{wasysym}
\usepackage{float}
\usepackage{array,multirow}
\usepackage{orcidlink}
\usepackage{longtable}

\journal{High Energy Astrophysics}

\begin{document}

\begin{frontmatter}



\title{Helium Accumulation and Thermonuclear Instabilities on Accreting White Dwarfs: From Recurring Helium Novae to Type Ia Supernovae}


\author[first,second]{Yael Hillman\corref{cor1}\orcidlink{0000-0002-0023-0485}}
\affiliation[first]{organization={Department of Physics Technion - Israel Institute of Technology},
            city={Haifa},
            postcode={3200003}, 
            country={Israel}}
\affiliation[second]{organization={Department of Physics, Azrieli College of Engineering},
            addressline={Ramat Beit Hakerem, POB 3566}, 
            city={Jerusalem},
            postcode={91035}, 
            country={Israel}}
\author[first]{Amir Michaelis\orcidlink{0000-0002-1361-9115}}
\author[first]{Hagai B. Perets\orcidlink{0000-0002-5004-199X}}

\cortext[cor1]{Corresponding author. Email: yaelhl@jce.ac.il}

\begin{abstract}
We investigate helium accumulation on carbon-oxygen (CO) white dwarfs (WDs), exploring a broad parameter space of initial WD masses ($0.65$--$1.0M_{\odot}$) and helium accretion rates ($10^{-10}$--$10^{-4}M_{\odot}\text{yr}^{-1}$). 
Our simulations, which were allowed to run for up to the order of a Gyr, reveal distinct regimes determined by the given accretion rate: at higher rates ($\gtrsim10^{-5}M_\odot\rm yr^{-1}$), the mass is repelled by radiation pressure without accretion; intermediate rates ($\sim10^{-8}$--$10^{-5}M_{\odot}\text{yr}^{-1}$) produce periodically recurring helium nova eruptions, enabling gradual WD mass growth; and lower rates ($\lesssim 10^{-8}M_{\odot}\text{yr}^{-1}$) facilitate prolonged, uninterrupted helium accumulation, eventually triggering a thermonuclear runaway (TNR) which for some cases is at sub-Chandrasekhar masses, indicative of a type Ia supernova (SNe) ignition, i.e. providing a potential single-degenerate channel for sub-Chandra SNe. Our models indicate that the WD mass and the helium accumulation rate critically determine the ignition mass and TNR energetics. We identify compositional and thermal signatures characteristic of each regime, highlighting observational diagnostics relevant to helium-rich transients. We discuss these theoretical results in the context of the observed helium nova V445 Puppis, emphasizing helium accretion's pivotal role in shaping diverse thermonuclear phenomena.
\end{abstract}




\end{frontmatter}




%
%
%
%
%
%
%
%
%
%
%
%
%
%
\section{Introduction} \label{sec:intro}
A white dwarf (WD) in a binary system, in principal, can accrete mass from its companion until the companion is entirely eroded \cite[e.g.,][]{Hillman2020,Hillman2021b,Hillman2021a,Vathachira2024}. The occurrence of this mass transfer, the rate at which the mass is transferred and the consequences of the mass being transferred, such as the amount of accreted and ejected mass, ejected velocity etc., \cite[]{Paczynski1978,Prialnik1982,Starrfield2009,Hillman2019,Aydi2020} all depend on system parameters such as the mass of the WD ($M_{\rm WD}$) \cite[]{Prikov1995,Yaron2005,Starrfield2012,Starrfield2012a}, the composition of the accreted matter \cite[]{Faulkner1972,Kovetz1985,Starrfield1986,Strope2010,Mason2020,Starrfield2020,Hillman2022a}, the binary separation ($a$) and the orbital period ($P_{\rm orb}$) of the system \cite[]{Kenyon1983,Ritter1988,Knigge2011,Abate2013,Hillman2020ASR,Hillman2021b,Hillman2022b}, as well as the nature of the companion. The companion could be a red dwarf (RD) \cite[e.g.,][]{ Shara1980,Shara1981,Prikov1995,Yaron2005,Hachisu2010,Shara2018}, a giant (either on the red giant branch (RGB) or the asymptotic giant branch (AGB))\cite[]{Mikolajewska1992,Mikolajewska2008,Mikolajewska2010,Hillman2021a,Vathachira2024} or it may be another, less massive, possibly helium WD (or an AM CVn system) \cite[]{Bildsten2007,Guillochon2010,Perets2010,Pakmor2012,Pakmor2021,Wong2023,Zenati2023}. If the donor is a RD, an RGB or an AGB, mass transfer will inevitably lead to nova eruptions as the result of hydrogen-rich matter being accumulated on the surface of the WD and ignited under degenerate conditions leading to a runaway fusion process (a thermonuclear runaway, TNR) \cite[]{Kraft1964,Starrfield1972,Prialnik1978,Shara1981,Starrfield2008}. The end result of a nova eruption depends mostly on the mass accretion rate ($\dot{M}_{\rm acc}$) \cite[]{Prikov1995,Yaron2005,Hillman2020ASR,Hillman2020,Hillman2021b}. It may result in the accreted shell being entirely ejected while taking with it some underlying WD core matter, or some fused hydrogen may be left behind in the form of a thin helium envelope that may slowly build up from one nova eruption to the next. Lower accretion rates lead to the former and higher accretion rates result in the latter \cite[]{Prikov1995,Yaron2005,Starrfield2012a,Hillman2015,Hillman2019}, while in the latter case, the accumulation of helium on the surface of a WD will eventually (after hundreds or thousands of hydrogen novae) lead to a \textit{helium} nova in which the helium will fuse into carbon and oxygen, and as in a hydrogen nova, either be entirely or partially ejected \cite[]{Hillman2016}. Moreover, just as in hydrogen novae, after the eruption, the WD will relax and eventually resume accretion, and in principal, after an additional long series of hydrogen novae, a helium novae may occur again. 

For the case of an AGB donor, the transferred mass may be more enriched in helium and heavy metals than a RD donor, and this can effect the timescales of the resulting nova eruptions as well as other characteristics, such as the amount of mass accreted ($m_{\rm acc}$) and ejected ($m_{\rm ej}$) \cite[]{Hillman2022a}. This may essentially speed-up the helium accumulation process. Moreover, if the companion (donor) were to be a WD, i.e., the system were to be a double WD system (DWD), the donor WD could be of similar, slightly less mass, or substantially less massive to the extent that it may be a \textit{helium} WD (or a hot helium donor as in AM CVn systems), in which case, the accreted mass will be mostly helium, and void of hydrogen. 

Various simulations of hydrogen-rich accretion that led to helium eruptions after hundreds or thousands of novae have been carried out, leading to net mass gain or loss depending on the input parameters (e.g., WD mass and accretion rate)  \cite[e.g,][and references within]{Idan2013,Newsham2013,Hillman2016}. 
Direct accretion of helium-rich matter has been simulated over the course of a few cycles for WD masses of $1.0-1.2M_\odot$ resulting in the amount of ejected mass being inversely proportionate to the rate that it is accreted \cite[]{Jose1993,Cassatella2005}. Calculations of helium accumulation onto a massive, $1.3M_\odot$ WD indicate helium flashes with a net mass gain \cite[]{Kato1999,Kato2004}.  
\cite{Piersanti2014} performed calculations of helium accretion onto WDs for a range of models, while each model was followed over a few cycles. 
They reported mild flashes for high accretion rates, dynamical flashes for lower accretion rates, and a detonation regime for cases with even lower accretion rates.

\cite{Hillman2016} have shown that during a series of $\sim$\textit{hundreds} of helium novae on WDs with masses $\geq1.0M_\odot$ at accretion rates of $\gtrsim10^{-7}M_\odot\rm yr^{-1}$, the resulting powerful, helium eruptions slowly heat the WD core, lowering the degeneracy and eventually leading to less (or non) ejective helium novae, thus potentially allowing the WD to grow towards the Chandrasekhar mass. For a lower accretion rate ($\dot{M}_{\rm acc}$), they obtained a rapid temperature rise for sub-Chandrasekhar WDs, possibly implying the detonation of a weak thermonuclear SN \cite[as described by][]{Bildsten2007}. 

The only helium nova known to date is V445 Puppis, which erupted in late 2000. It has been reported to be absent of hydrogen lines, and thus the donor is assumed to have a helium-rich envelope \cite[e.g.,][and references within]{Wagner2001a,Ashok2003,Lynch2004,Iijima2008,Banerjee2023}. Based on the high luminosity of the system prior to its eruption, \cite{Woudt2009} deduced that the donor should be of roughly $\sim1.25M_\odot$ and fusing helium in a shell. 
Estimates of the WD mass range from 0.8 to 1.3 $M_\odot$ \cite[e.g.,][and reference within]{Piersanti2014}.

\cite{Schaefer2025} deduced via orbital period changes that the amount of mass ejected during the 2000 eruption is of order $\gg10^{-3}M_\odot$ and explains that this is more than the estimated accreted mass, assuming the WD to be very massive \cite[]{Kato2008b}, thus concludes that the WD in V445 Puppis can not grow towards the Chandrasekhar mass.

The accretion of helium on the surface of WDs also has potential implications catalyzing the detonation of WDs in type Ia supernovae (SNe) explosions and/or the production of faint peculiar themornuclear SNe. For example, a pre-existing helium layer on the surface of two interacting WDs is a prerequisite for the "dynamically driven double-degenerate double-detonation" (D6) scenario for type Ia SNe \cite[]{Guillochon2010, Pakmor2013, Sato2015, Shen2018} (based on earlier double detonation scenarios not involving two WDs \cite[]{Iben1985,Iben1987}). In this model, a close binary system consisting of two white dwarfs (one typically $>0.85\, M_\odot$) undergoes a gravitational wave inspiral. The primary is assumed to already have a layer of helium on its surface. As the WDs come close to each other, mass-transfer from the companion ensues, and the primary accretes an additional small amount of helium from the outer layers of its companion WD. A helium detonation is triggered in the helium layer of the primary WD, leading to a (second) carbon-detonation in the core and the production of a type Ia supernova explosion. The amount of pre-existing helium on both WDs is a critical issue in this scenario. A too-large mass on the primary would lead to a stronger helium detonation, producing too many iron elements in the outer layer, which will change the early SN spectrum, making it inconsistent with observed normal type Ia SNe. Such cases may give rise to peculiar types of SNe. A large helium shell on a low-mass WD might only result in the helium-shell detonation and the production of a faint, likely Ca-rich SN, \citep{Nomoto84,Woo+86,Bildsten2007,Perets2010,Zenati2023}. Too little of a helium layer on a massive primary might not produce a significant detonation and therefore not lead to a second detonation at all. Any possible helium accretion process that could affect the existence and extent of a helium layer is, therefore, a key issue to the modeling of type Ia SNe and other faint SNe.

This work deals with accreting carbon-oxygen (CO) WDs, focusing on the accumulation of helium, regardless of the source. I.e., the helium is added to the WD's surface at given, constant, external accretion rates ($\dot{M}_{\rm acc}$) and the development is followed in order to investigate the consequences of altering input parameters, namely, the accretion rate ($\dot{M}_{\rm acc}$) and the initial WD mass ($M_{\rm WD,i}$). Our primary goal is to determine the amount of helium that can accumulate quiescently on the surface of a WD before initiating a TNR.

In the next section we describe our models and computational method, followed by \S\ref{sec:results} where we present our findings. We discuss our results in \S\ref{sec:discussion}, relate our results to the currently only known helium nova --- V445 Puppis --- in \S\ref{sec:V445pup} and present our conclusions in \S\ref{sec:conclusions}.

\section{Method and models} \label{sec:method}
We use a hydrodynamic Lagrangian nova evolution code to simulate multiple consecutive nova eruptions resulting from the accretion of helium. This code was originally designed to produce nova eruptions as the result of accretion of Solar-like material and is described in detail in \cite{Prikov1995}, \cite{Yaron2005}, \cite{Epelstain2007}, and \cite{Hillman2015}. We replace the hydrogen with helium as the accreted material (as explained in \cite{Hillman2016}) in order to suit our goal of determining the amount of helium a WD can retain before triggering a TNR. We do not impose a separate, super-Eddington wind prescription during the eruption phase.

We carried out a series of simulations of consecutive helium nova eruptions, on the surface of carbon-oxygen (in equal parts) WD models with initial masses of $0.65$, $0.70$, $0.80$ and $1.0M_\odot$ at an initial core temperature ($T_{\rm c}$) of 30MK. We chose two models with an initial $T_{\rm c}$ of 45MK to assess the robustness of the results and the extent of the influence of the initial $T_{\rm c}$ and found it to have a minor shift on the 30MK results. We explored the entire feasible range of helium accretion rates ($10^{-10}-10^{-4}M_\odot \rm yr^{-1}$). Each initial $M_{\rm WD}$ model was given a constant accretion rate ($\dot{M}_{\rm acc}$) and allowed to run uninterrupted. 

We follow our models' evolutions over the order of $\sim10^3-10^4$ cycles of accretion and eruption (depending on the model), which amounts to the order of $\sim10^5-10^9$ years (also depending on the model). We record relevant parameters throughout the simulations, such as, the accreted mass per cycle ($m_{\rm acc}$\footnote{Which is 98$\%$ helium, the remaining 2$\%$ being Solar Z.}), the ejected mass per cycle ($m_{\rm ej}$), the core temperature ($T_{\rm c}$), the maximum temperature ($T_{\rm max}$) per eruption, temperature profiles, the WD mass ($M_{\rm WD}$) and radius ($R_{\rm WD}$), the composition of the WD, and the bolometric, nuclear and neutrino luminosities ($L_{\rm bol}$, $L_{\rm nuc}$ and $L_{\rm neut}$ respectively). 

Table \ref{tab:mdls} specifies the models used in this work, and differentiates between models that resulted in periodic helium nova eruptions, and models that experienced uninterrupted prolonged helium accretion --- implying possible type Ia supernova (SNIa) progenitors (as elaborated on in \S \ref{sec:nonerup}). The models marked as non-accretive imply an Eddington accretion rate, indicating the upper limit of the allowed helium accretion rate. We elaborate on the different outcome regimes in the following section.

\begin{table}
	\begin{center}
\small
\begin{tabular}{|c|c|c|c|c|c|}
\hline
{Model}&{$M_{\rm WD,i}$}&{$T_{\rm c}$}&{$\dot{M}_{\rm acc}$}&{Accretion}&{Eruption}\\
{$\#$}&{$[M_\odot]$}&{[$10^7$K]}&{$[M_\odot \rm yr^{-1}]$}&{Y/N}&{HN/SN/UT}\\
			\hline\hline
{1}&{0.65}&{3.0}&{$5e{-6}$}&{N}&{---}\\

{2}&{0.65}&{3.0}&{$3e{-6}$}&{N}&{---}\\

{3}&{0.65}&{3.0}&{$2e{-6}$}&{Y}&{HN}\\

{4}&{0.65}&{3.0}&{$1e{-6}$}&{Y}&{HN}\\

{5}&{0.65}&{3.0}&{$5e{-7}$}&{Y}&{HN}\\

{6}&{0.65}&{3.0}&{$3e{-7}$}&{Y}&{HN}\\

{7}&{0.65}&{3.0}&{$1e{-7}$}&{Y}&{HN}\\

{8}&{0.65}&{3.0}&{$5e{-8}$}&{Y}&{UT}\\

{9}&{0.65}&{3.0}&{$3e{-8}$}&{Y}&{UT}\\

{10}&{0.65}&{3.0}&{$2e{-8}$}&{Y}&{SN}\\

{11}&{0.65}&{3.0}&{$1e{-8}$}&{Y}&{SN}\\

\hline

{12}&{0.7}&{3.0}&{$5e{-5}$}&{N}&{---}\\

{13}&{0.7 }&{3.0}&{$4e{-5}$}&{N}&{---}\\

{14}&{0.7}&{3.0}&{$3e{-5}$}&{Y}&{HN}\\

{15}&{0.7}&{3.0}&{$1e{-5}$}&{Y}&{HN}\\

{16}&{0.7}&{3.0}&{$1e{-6}$}&{Y}&{HN}\\

{17}&{0.7}&{3.0}&{$5e{-7}$}&{Y}&{HN}\\

{18}&{0.7}&{3.0}&{$3e{-7}$}&{Y}&{HN}\\

{19}&{0.7}&{3.0}&{$1e{-7}$}&{Y}&{HN}\\

{20}&{0.7}&{3.0}&{$5e{-8}$}&{Y}&{UT}\\

{21}&{0.7}&{3.0}&{$3e{-8}$}&{Y}&{UT}\\

{22}&{0.7}&{3.0}&{$2e{-8}$}&{Y}&{SN}\\

{23}&{0.7}&{3.0}&{$1e{-8}$}&{Y}&{SN}\\

{24}&{0.7}&{3.0}&{$1e{-10}$}&{Y}&{SN}\\
\hline

{25}&{0.8}&{3.0}&{$1e{-4}$}&{N}&{---}\\

{26}&{0.8}&{3.0}&{$5e{-5}$}&{N}&{---}\\

{27}&{0.8}&{3.0}&{$4e{-5}$}&{N}&{---}\\

{28}&{0.8 }&{3.0}&{$3e{-5}$}&{Y}&{HN}\\

{29}&{0.8}&{3.0}&{$1e{-5}$}&{Y}&{HN}\\

{30}&{0.8}&{4.5}&{$1e{-5}$}&{Y}&{HN}\\

{31}&{0.8}&{3.0}&{$5e{-6}$}&{Y}&{HN}\\

{32}&{0.8}&{3.0}&{$1e{-6}$}&{Y}&{HN}\\

{33}&{0.8}&{3.0}&{$1e{-7}$}&{Y}&{HN}\\

{34}&{0.8}&{4.5}&{$1e{-7}$}&{Y}&{HN}\\

{35}&{0.8}&{3.0}&{$5e{-8}$}&{Y}&{HN}\\

{36}&{0.8}&{3.0}&{$4e{-8}$}&{Y}&{HN}\\

{37}&{0.8}&{3.0}&{$3e{-8}$}&{Y}&{SN}\\

{38}&{0.8}&{3.0}&{$1e{-8}$}&{Y}&{SN}\\

{39}&{0.8}&{3.0}&{$1e{-9}$}&{Y}&{SN}\\

{40}&{0.8}&{3.0}&{$1e{-10}$}&{Y}&{SN}\\
\hline

{41}&{1.0}&{3.0}&{$5e{-5}$}&{N}&{---}\\

{42}&{1.0}&{3.0}&{$4e{-5}$}&{N}&{---}\\

{43}&{1.0}&{3.0}&{$3e{-5}$}&{Y}&{HN}\\

{44}&{1.0}&{3.0}&{$1e{-5}$}&{Y}&{HN}\\

{45}&{1.0}&{3.0}&{$1e{-6}$}&{Y}&{HN}\\

{46}&{1.0}&{3.0}&{$1e{-7}$}&{Y}&{HN}\\

{47}&{1.0}&{3.0}&{$5e{-8}$}&{Y}&{HN}\\

{48}&{1.0}&{3.0}&{$4e{-8}$}&{Y}&{SN}\\

{49}&{1.0}&{3.0}&{$3e{-8}$}&{Y}&{SN}\\

{50}&{1.0}&{3.0}&{$1e{-8}$}&{Y}&{SN}\\

{51}&{1.0}&{3.0}&{$1e{-9}$}&{Y}&{SN}\\
\hline

  \end{tabular}
\caption{Model parameters. Columns from left to right: Model $\#$; initial WD mass; initial core temperature; helium rich accretion rate; occurrence of accretion, yes (Y) or no (N); occurrence of recurring helium novae (HN), prolonged accretion leading to supernova signatures (SN) or models that accreted mass for a prolonged epoch, terminated without successfully ejecting mass, but did not show clear signs of SNIa ignition, which we define as undetermined transients (UT).}
\label{tab:mdls}
	\end{center}
\end{table}

\section{Results} \label{sec:results}

We find the regime of helium accretion rates that produce recurring helium nova eruptions to range roughly from a few times $10^{-5}M_\odot\rm yr^{-1}$ to a few times $10^{-8}M_\odot\rm yr^{-1}$ depending on the WD mass. Above this range, no mass is accreted onto the WD due to Eddington accretion. Below this range we obtain \textit{uninterrupted accretion}.
While the bolometric luminosity of the WD can exceed the Eddington limit during helium flashes, potentially contributing to radiatively driven mass loss, we define the upper limit of our helium nova regime as the highest accretion rate that permits mass accumulation. We thus define an approximate lower limit to the helium accretion rate under which the accumulating helium does not lead to helium novae, but rather slowly builds up to a thick shell on the WD's surface until eventually igniting and becoming violently unstable.

Following here (\S \ref{sec:heliumnova}), we present our results of the helium-nova producing models, and we elaborate on the uninterrupted-accretion results in \S \ref{sec:nonerup}. 

\subsection{Periodic helium novae}\label{sec:heliumnova}

\begin{figure}
	\begin{center}
{\includegraphics[trim={1.9cm 0.7cm 6.8cm 0.05cm}, clip, width=0.99\columnwidth]{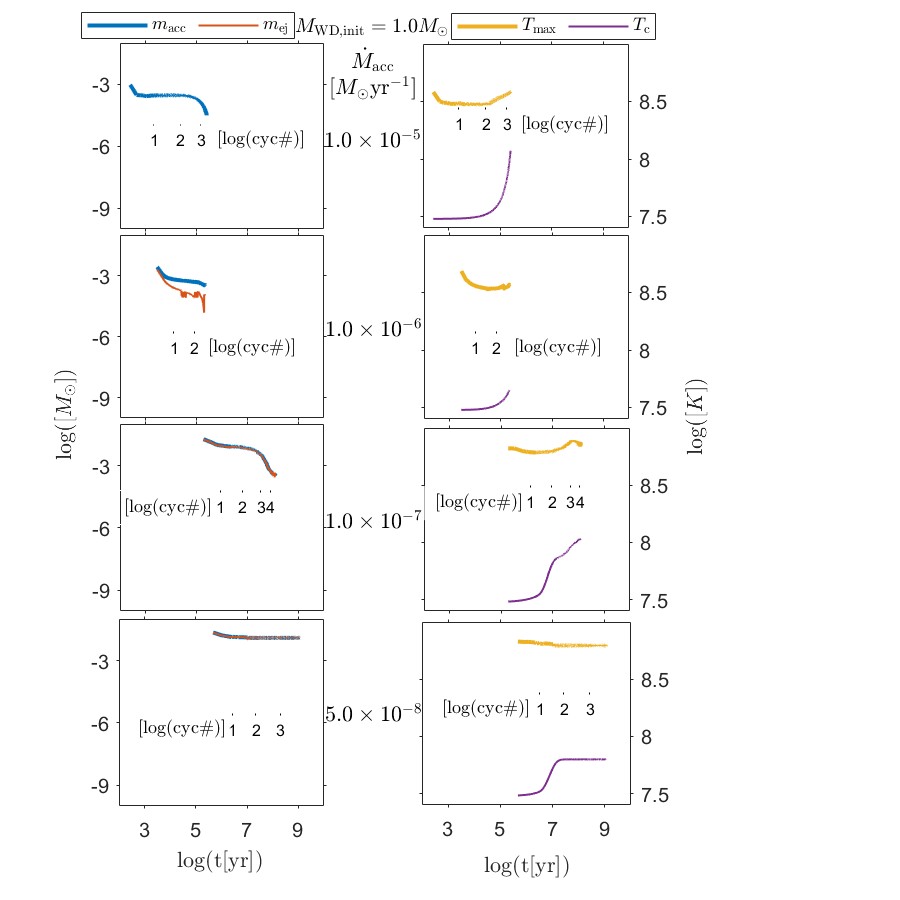}}
\caption{Accreted and ejected mass  ($m_{\rm acc}$ and $m_{\rm ej}$ respectively, left) and core and maximum temperatures ($T_{\rm c}$ and $T_{\rm max}$ respectively, right) per eruption vs. evolutionary time. For visualization, we added the cycle number on logarithmic scales in each plot. This figure shows four of our periodic helium-nova-producing models (HN type) with initial WD masses of $1.0M_\odot$. Refer to \S \ref{sec:A_adtnl_figs}, Figures \ref{fig:macc_mej_065_A}$-$\ref{fig:macc_mej_100_A} for a more detailed presentation, including the data of these models, as well as for the initial WD mass models of $0.65$, $0.7$ and $0.8M_\odot$, and to \S \ref{sec:B_adtnl_tbl}, Table \ref{tab:for_fig_1_A}  for detailed data for all the HN-type models.}\label{fig:macc_mej}
	\end{center}
\end{figure}

\begin{figure}
	\begin{center}
{\includegraphics[trim={0.8cm 0.4cm 1.2cm 0.8cm}, clip, width=0.99\columnwidth]{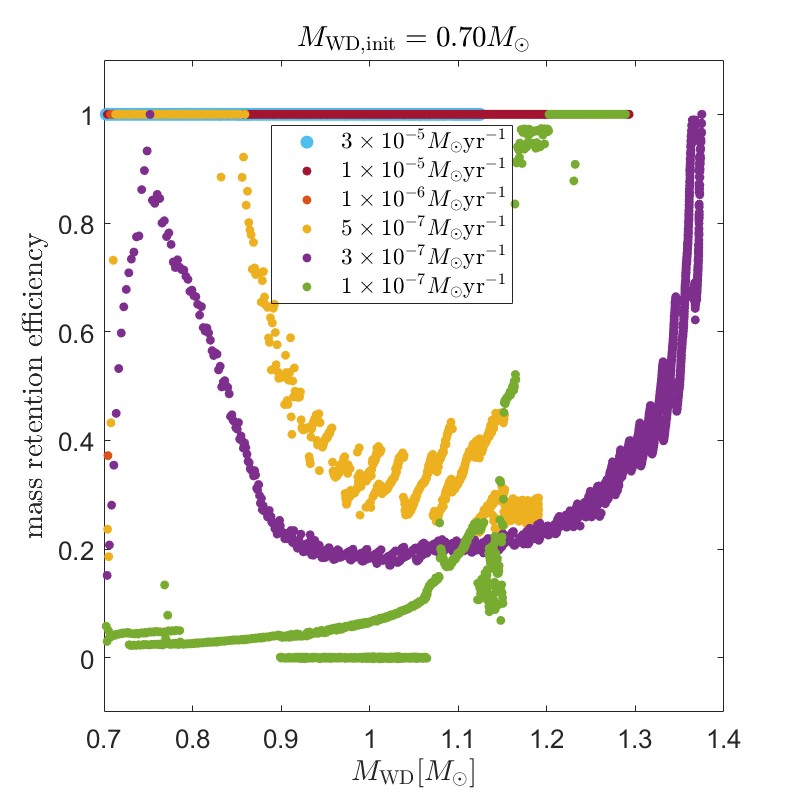}}
\caption{Retention efficiency vs. WD mass for our HN type models with initial WD masses of $0.70M_\odot$. Refer to \S \ref{sec:A_adtnl_figs}, Figure \ref{fig:retention_A} for the initial WD mass models of $0.65$, $0.80$ and $1.0M_\odot$.}\label{fig:retention}
	\end{center}
\end{figure}

\begin{figure*}
	\begin{center}
{\includegraphics[trim={0.0cm 0.4cm 0.0cm 2.4cm}, clip, width=1.9\columnwidth]{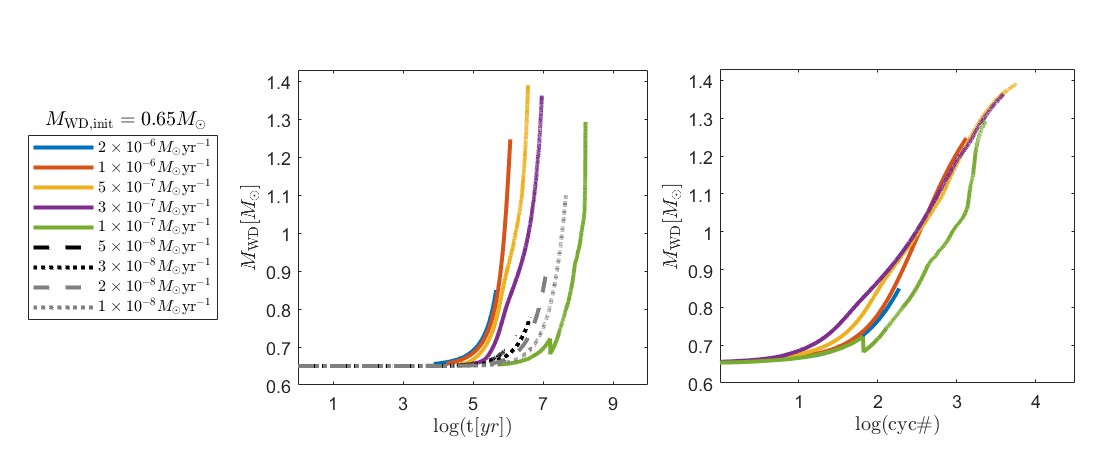}}\\
{\includegraphics[trim={0.0cm 0.4cm 0.0cm 2.4cm}, clip, width=1.9\columnwidth]{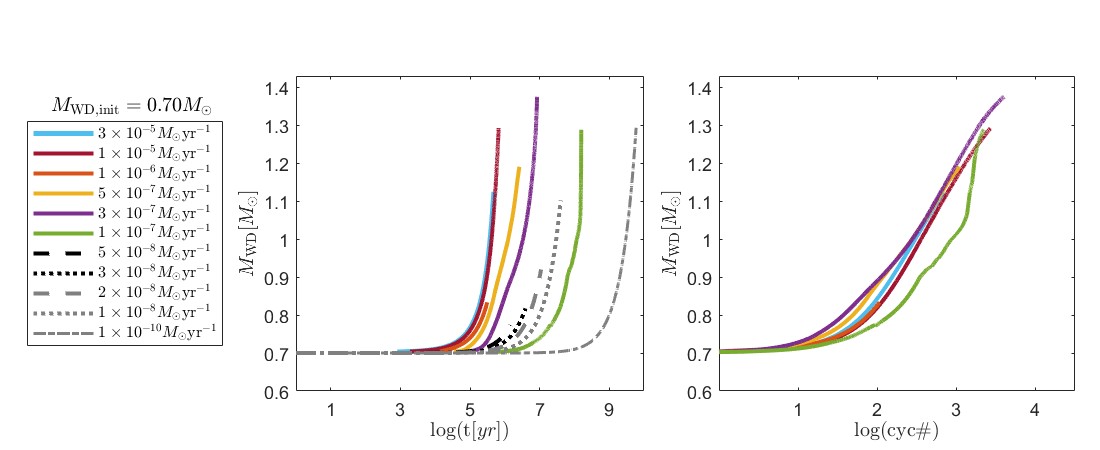}}\\
{\includegraphics[trim={0.0cm 0.3cm 0.0cm 2.4cm}, clip, width=1.9\columnwidth]{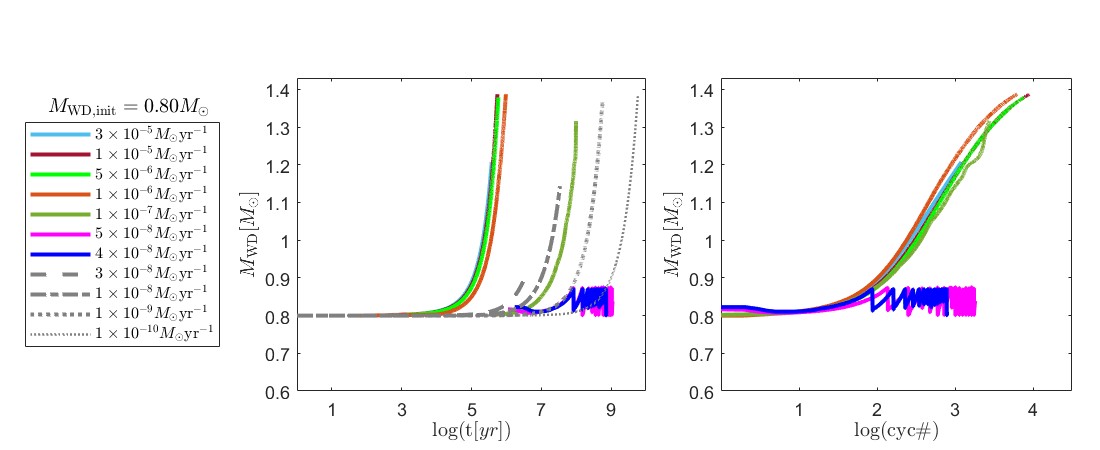}}\\
{\includegraphics[trim={0.0cm 0.4cm 0.0cm 2.3cm}, clip, width=1.9\columnwidth]{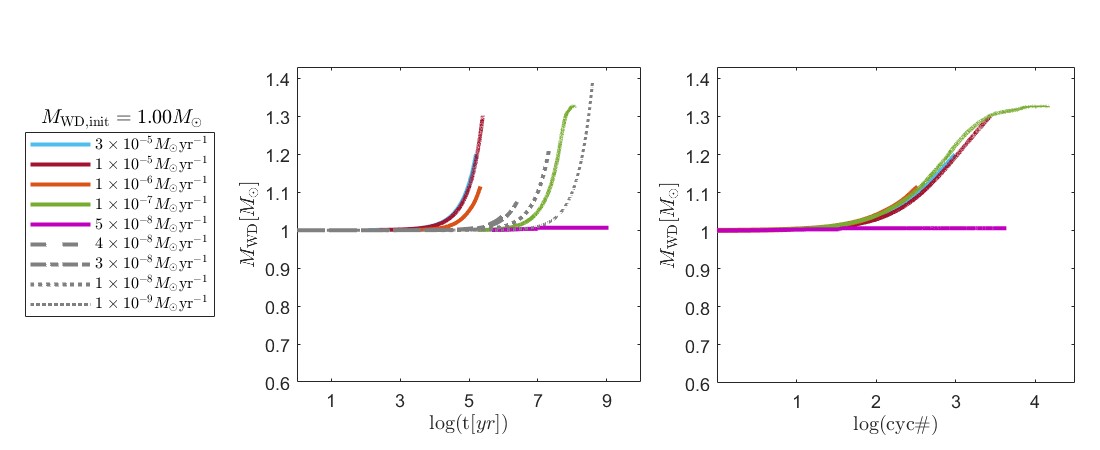}}\\
\caption{WD mass ($M_{\rm WD}$) vs. time (left) and vs. cycle number. Solid color lines represent helium-nova-producing models (HN type); dashed/dotted black lines represent prolonged accretion models w/o signs of SNIa ignition (UT type) and dashed/dotted gray lines represent prolonged accretion models with distinct signs of SNIa ignition (SN type).}\label{fig:Mwd_t_all}
	\end{center}
\end{figure*}

For the regime of recurring helium novae, we show in Figure \ref{fig:macc_mej} the evolution of the accreted and ejected masses ($m_{\rm acc}$ and $m_{\rm ej}$ respectively) as well as the maximum and core temperatures ($T_{\rm max}$ and $T_{\rm c}$ respectively) over time and per eruption, for our helium nova producing models (HN) with initial WD masses of $0.7M_\odot$, demonstrating the similarities and dissimilarities of the behavior of our helium novae with typical hydrogen novae behavior. For instance, while hydrogen novae may result in $m_{\rm ej}$ that is either more than, equal to or less than $m_{\rm acc}$, we find that our helium novae exhibit an $m_{\rm ej}$ that is consistently $\sim$less than $m_{\rm acc}$ --- for the very high accretion rates we obtain $m_{\rm ej}=0$, i.e., non-ejective helium novae, while the difference between $m_{\rm ej}$ and $m_{\rm acc}$ becomes less substantial as $\dot{M}_{\rm acc}$ decreases. 
This behavior is consistent with hydrogen nova models in the sense that for a given $M_{\rm WD}$, more mass is ejected for lower accretion rates, however, here, the low rates do not produce nova eruptions. The resulting mass retention efficiencies $({m_{\rm acc}-m_{\rm ej}})/m_{\rm acc}$, shown in Figure \ref{fig:retention}, reflect competing effects: more massive WDs reach higher pressures at the base of the accreted layer, favoring more energetic eruptions and thus lower retention efficiencies, whereas hotter WDs (with higher core temperatures) have lower degeneracy, leading to less violent ignition and therefore higher retention efficiencies.
This overall trend is consistent with hydrogen nova modeling \cite[e.g.,][]{Yaron2005,Vathachira2024}, although here we find the mass retention efficiency to be positive for the entire HN regime whereas for hydrogen nova it is positive only for the high end of accretion rates. We note that while the trends are similar, the range of accretion rates here is entirely different --- of order $10-100$ times higher --- than for hydrogen novae because the conditions required to ignite helium are different than those needed for hydrogen ignition; specifically, it must be denser and hotter. Thus, more mass is needed to produce these conditions. We find our other three groups of initial WD mass models to behave in a similar manner and show their results in \S \ref{sec:A_adtnl_figs} in Figures \ref{fig:macc_mej_065_A}$-$\ref{fig:retention_A}.  
For a few models, with the lowest accretion rate that still produces eruptions, we obtain $m_{\rm ej}\approx m_{\rm acc}$. These are models with initial $0.8$ and $1.0M_\odot$ WDs at accretion rates of $\sim5\times 10^{-8}M_\odot\rm yr^{-1}$. We allowed these model to run for about a Gyr of evolution and found that not only do these models exhibit $\sim$constant accreted and ejected masses throughout evolution, the WDs also reach constant values of $T_{\rm max}$ and $T_{\rm c}$ while for all the other models in this regime the temperatures show a general trend of moderately increasing while the accreted and ejected masses decrease.
The three models that exhibited $m_{\rm ej}\approx m_{\rm acc}$ throughout evolution\footnote{models \#35, 36 and 47.}, are clearly evident in Figure \ref{fig:Mwd_t_all}, by being the only three curves that do not show the WD mass to increase. 
Accretion rates lower than these shifts the conditions to the uninterrupted accretion regime; thus, the key finding here is that \textit{the WD in helium-nova producing systems grows}.

\begin{figure}
	\begin{center}
{\includegraphics[trim={0.0cm 0.cm 0.cm 0.cm}, clip, width=0.99\columnwidth]{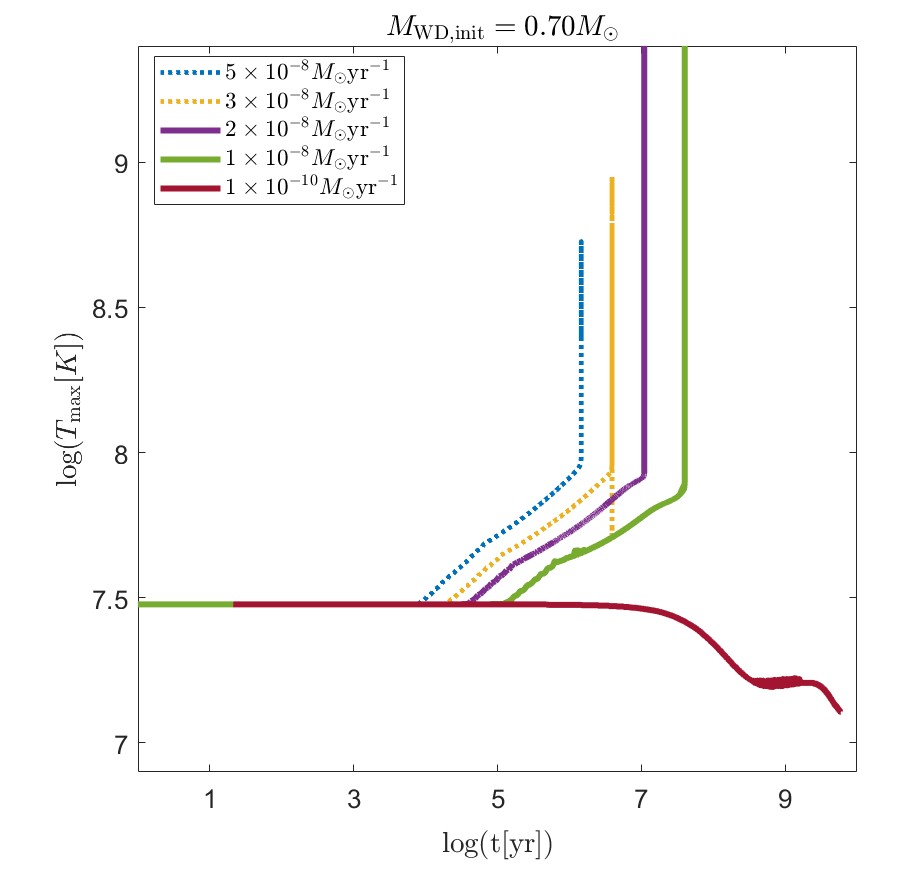}}
\caption{For our $0.70M_\odot$ models: Maximum temperature ($T_{\rm max}$) evolution vs. time for models in our uninterrupted accretion regime. Solid lines represent models that lead to signatures of SNIa ignition (SN type) and dotted lines lead to those that did not show a clear signs of SNIa ignition (UT type). Refer to \S \ref{sec:A_adtnl_figs} Figure \ref{fig:App_T_t_SNIa} for the $0.65$, $0.80$ and $1.0M_\odot$ models.}\label{fig:T_t_SNIa_070}
	\end{center}
\end{figure}

We show this mass growth in Figure \ref{fig:Mwd_t_all} as a function of time and per eruption for our eruptive models, exhibiting a general secular WD mass growth (with the exception of the three models mentioned earlier). Interestingly, the WD mass growth vs. cycle plot reveals that all the high accretion rate models ($>$ a few times $10^{-7}M_\odot\rm yr^{-1}$) converge. This implies that for a given WD mass, the net amount of mass accreted between eruptions (i.e., $m_{\rm acc}-m_{\rm ej}$) is $\sim$constant, regardless of the accretion rate, which is demonstrated in Figures \ref{fig:retention} and \ref{fig:retention_A}. The lower nova-producing rates for each WD mass diverge from the rest, implying a less efficient mass accumulation and in the case of three systems mentioned earlier, $m_{\rm acc}\approx m_{\rm ej}$ meaning that the WD mass remains $\sim$constant. 
Notably, when shifting to even lower accretion rates --- i.e., into the uninterrupted accretion regime --- significant mass accumulation resumes because the accretion is so slow, that even though mass is being piled on to the surface of the WD and compressed, the temperature does not increase efficiently at first, and in some cases even decreases (Figure \ref{fig:T_t_SNIa_070}). This also explains why, within the uninterrupted accretion regime, lower accretion rates achieve a higher WD mass before showing signs of SNIa ignition. This trend continues the temperature trend we found for our eruptive models (Figure \ref{fig:macc_mej}) that showed heating for all the models for which the WD mass consistently increased.

\begin{figure*}
	\begin{center}
{\includegraphics[trim={0.5cm 0.5cm 0.5cm 0.5cm}, clip, width=0.99\columnwidth]{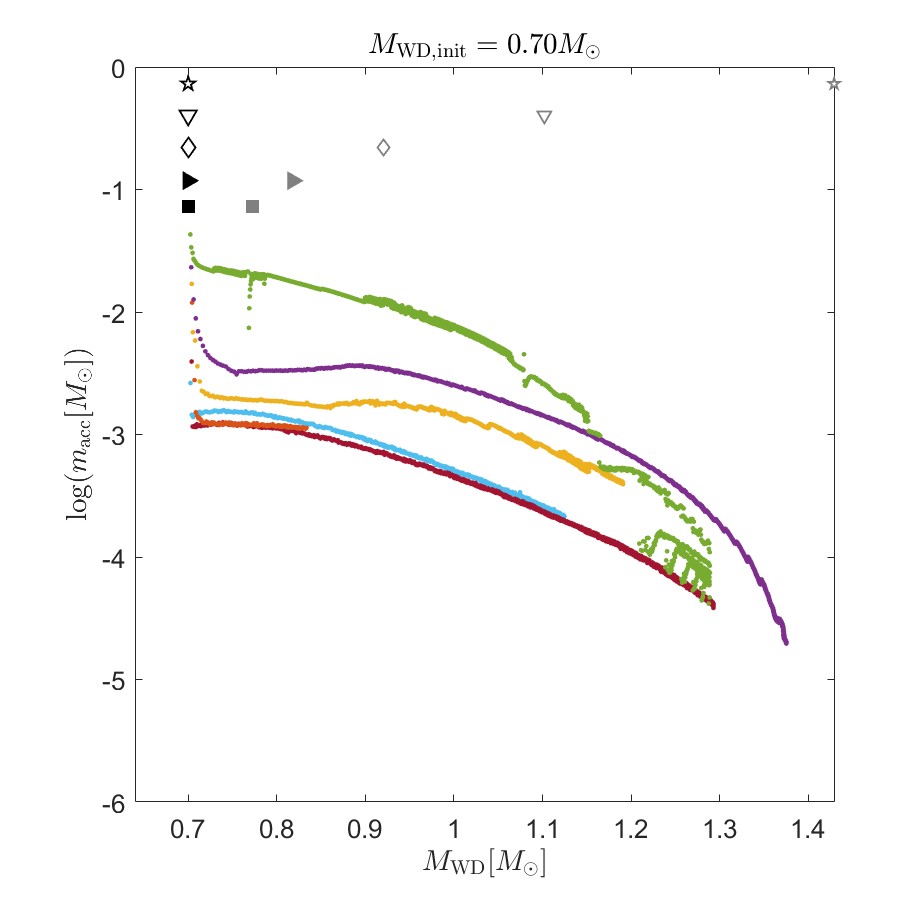}}
{\includegraphics[trim={0.0cm 1.5cm 1.0cm 0.0cm}, clip, width=0.99\columnwidth]{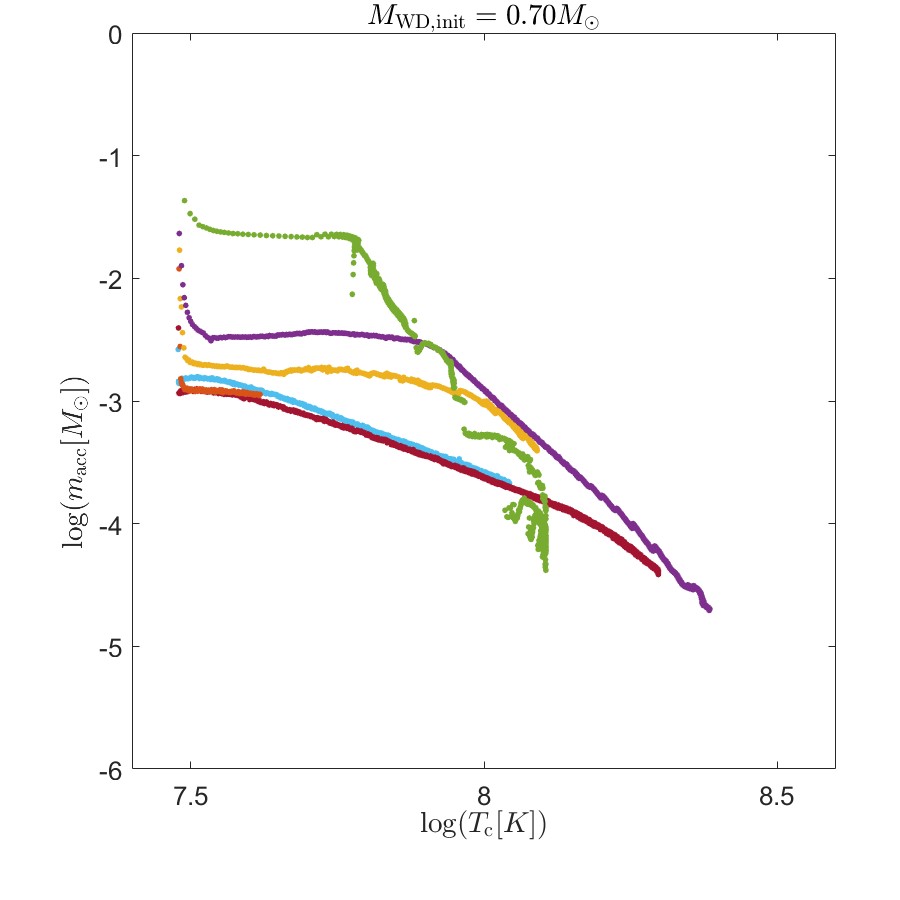}}\\
{\includegraphics[trim={0.0cm 0.5cm 1.0cm 0.5cm}, clip, width=0.99\columnwidth]{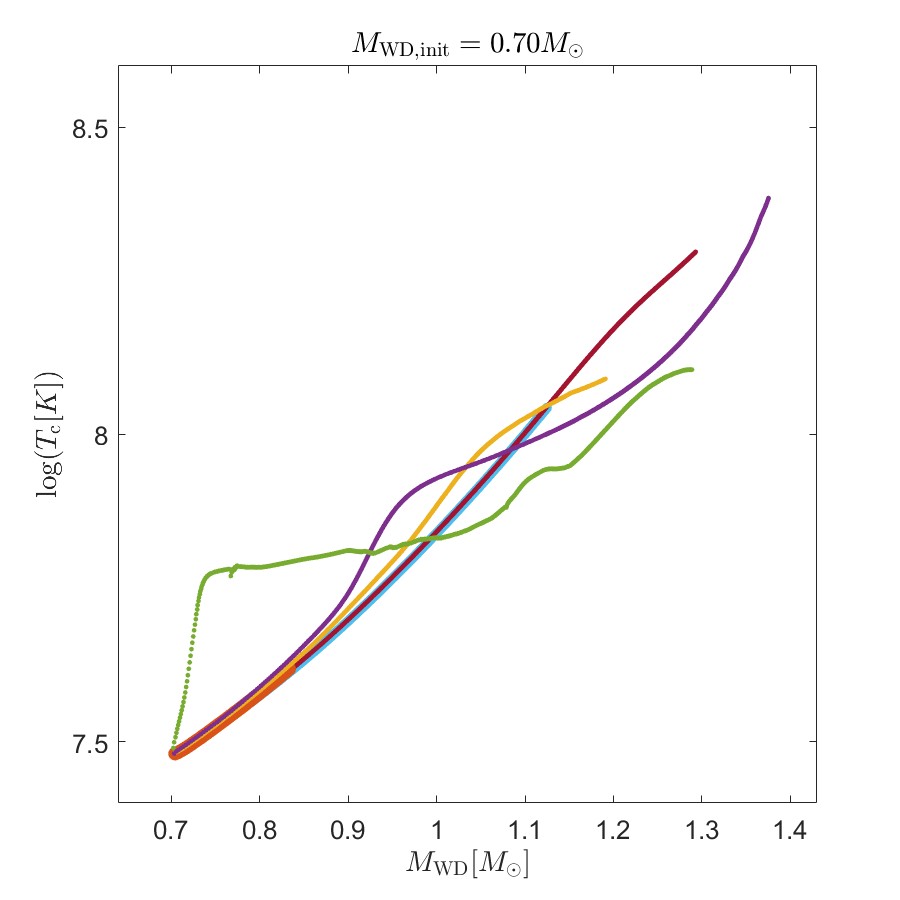}}
{\includegraphics[trim={-1.1cm -3.2cm -3.7cm 0.0cm}, clip, width=0.99\columnwidth]{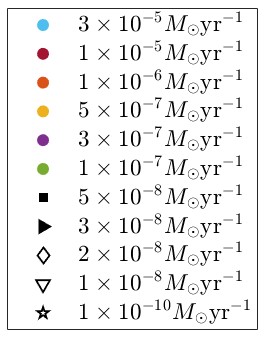}}\\

\caption{Accreted mass ($m_{\rm acc}$) vs. WD mass ($M_{\rm WD}$) (top left); accreted mass ($m_{\rm acc}$) vs. core temperature ($T_{\rm c}$) (top right); and core temperature ($T_{\rm c}$)  vs. WD mass ($M_{\rm WD}$) (bottom) for the $0.7M_\odot$ helium-nova-producing models (HN type) (colors). Additionally marked in top left panel: accreted mass for uninterrupted accretion models that did not show clear signs of SNIa ignition (UT type) (full markers); and accreted mass for models that led to signatures of SNIa ignition (SN type) (outlined markers); the black markers are located at the initial WD mass ($0.7M_\odot$) for convenience, and the gray markers are at the final WD mass. Refer to \S \ref{sec:A_adtnl_figs} Figures \ref{fig:macc_MWD_TC_065_A}$-$\ref{fig:macc_MWD_TC_100_A} for initial WD mass models of $0.65$, $0.8$ and $1.0M_\odot$.
}\label{fig:macc_MWD_Tc_070}
	\end{center}
\end{figure*}

In Figure \ref{fig:macc_MWD_Tc_070}
we present the correlations between the evolutions of the amount of accreted mass ($m_{\rm acc}$), the WD mass ($M_{\rm WD}$) and the core temperature ($T_{\rm c}$) for our models with an initial WD mass of $0.7M_\odot$, demonstrating that the WD mass and temperature are correlated with each other while anti-correlated with the amount of accreted mass. We show in \S \ref{sec:A_adtnl_figs} in Figures \ref{fig:macc_MWD_TC_065_A}$-$\ref{fig:macc_MWD_TC_100_A}  these correlations for the other three initial WD mass models. \cite{Schwab2017} have simulated quiescent accretion of carbon and oxygen (CO) onto oxygen-neon-magnesium (ONeMg) WDs and found that the core temperature secularly cools to a minimum temperature which is higher for higher accretion rates. Our results here resemble this trend, but for a different physical reason --- the mass being added to their WDs releases gravitational energy that heats the core, quiescently slowing the natural cooling process, while in our models, since the accretion is of mostly helium (rather than CO), there are very hot eruptions. So while the high accretion rate may have a minor influence in slowing down the quiescent cooling process, the periodic helium eruptions have a persistent effect of heating the entire core. \cite{Epelstain2007} simulated several thousand consecutive hydrogen nova cycles for two initial models, each with a different WD mass and accretion rate (accreting material of Solar composition), and found the core temperature and WD mass to consistently decrease, 
and \cite{Hillman2021b} carried out long-term self-consistent simulations over several Gyrs for a selection of models of WDs accreting Solar composition material from RD donors in Roch-lobe overflow (RLOF), also to find the core temperature and WD mass to consistently decrease. Both these works are in stark contrast with our helium nova results in Figure \ref{fig:macc_MWD_Tc_070}, where the WD mass increases \textit{significantly}, and with it the core temperature, while since the accretion rate has a strong influence on the eruption frequency, it inadvertently has an effect on the core temperature evolution as well. These correlations also support our earlier deduction that, for a given $M_{\rm WD}$, while the accretion rate is sufficiently high, the amount of accreted mass is independent of the accretion rate. However, for the lower end of nova-producing accretion rates, the amount of accreted mass becomes correlated with the accretion rate, i.e., lower accretion rates allow more mass to be accreted before igniting a TNR that leads to a nova eruption.  Figure \ref{fig:macc_MWD_Tc_070} demonstrates how the increase in core temperature depends on the accretion rate. For a given WD mass, lower accretion rates lead to a slower rise of the core temperature because compressional heating proceeds on a longer timescale than in higher accretion rate models and the inward diffusion of heat deposited in the outer layers is delayed. As discussed by \cite{Piersanti2003} for CO-accreting models, at sufficiently high accretion rates the core evolution is dominated by homologous compression, whereas at lower accretion rates the core temperature can remain nearly constant (or even decrease slightly) until thermal diffusion reaches the center, after which the core temperature again evolves mainly under the effect of compression. This occurs because the core temperature is set by the balance between compressional heating from the added mass and the WD’s thermal cooling timescale, so that, at lower accretion rates, the slower mass loading results in weaker compressional heating and a correspondingly slower temperature rise. (see bottom panel of Figure \ref{fig:macc_MWD_Tc_070}).

\begin{figure}
	\begin{center}
{\includegraphics[trim={1.0cm 0.0cm 1.5cm 0.0cm}, clip, width=0.62\columnwidth]{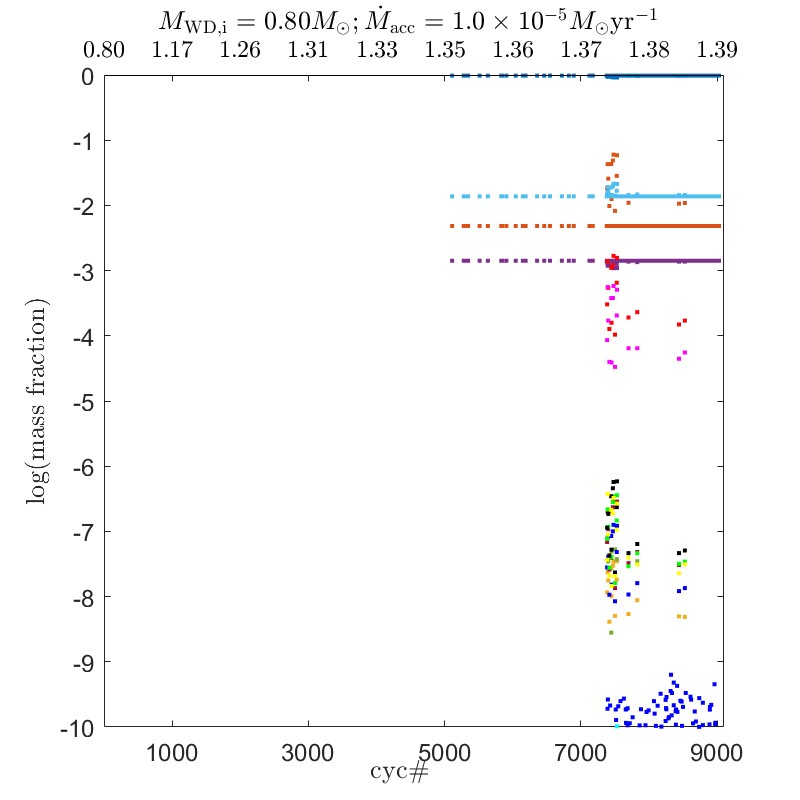}}
{\includegraphics[trim={22.5cm 5.6cm 0.7cm 4.0cm}, clip, width=0.31\columnwidth]{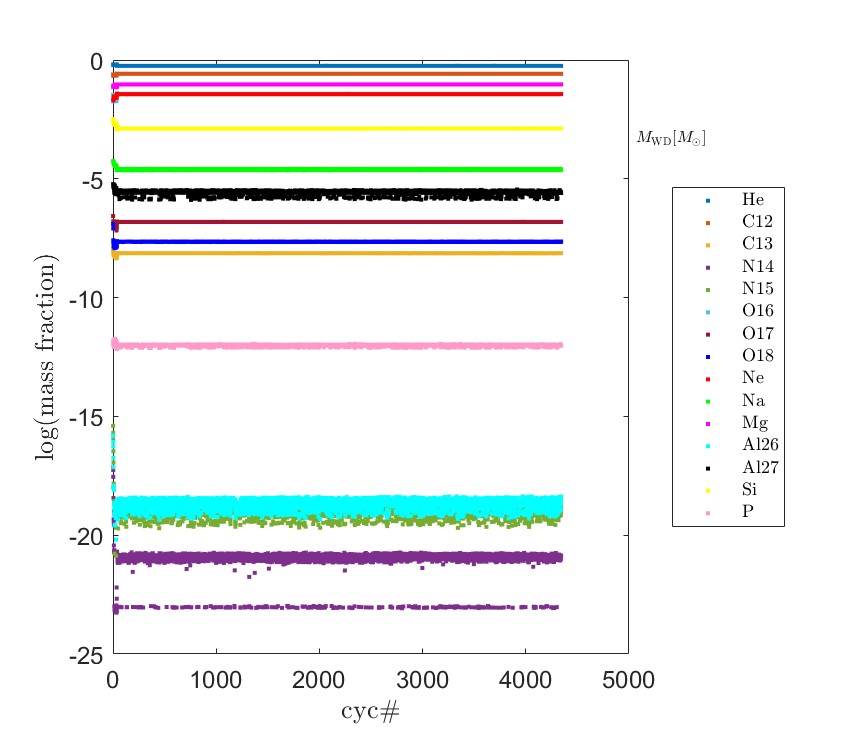}}\\
{\includegraphics[trim={1.0cm 0.0cm 1.5cm 0.0cm}, clip, width=0.62\columnwidth]{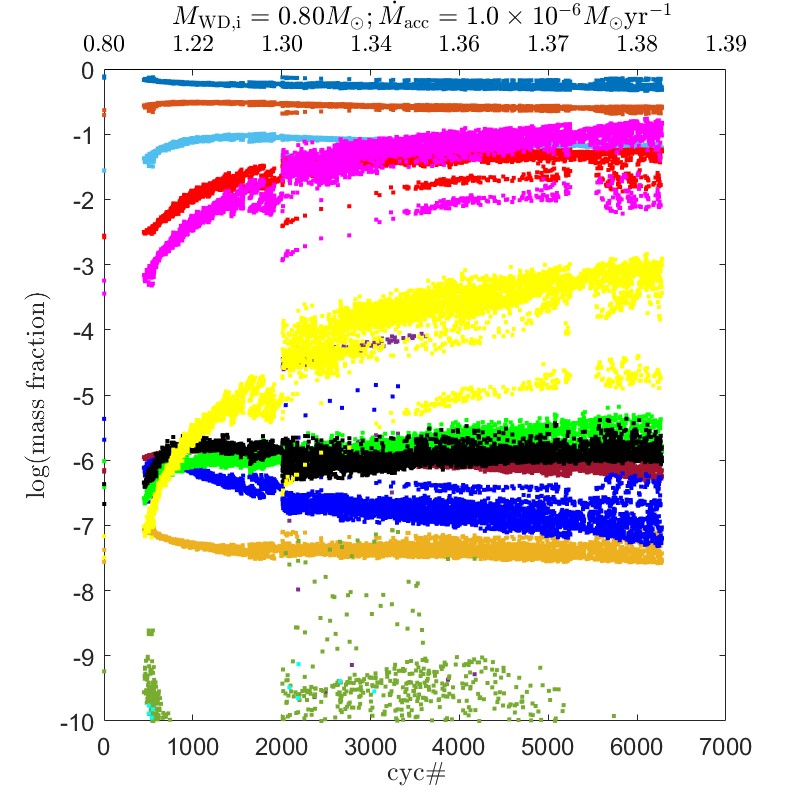}}
{\includegraphics[trim={22.5cm 5.6cm 0.7cm 4.0cm}, clip, width=0.31\columnwidth]{comp_for_legend.jpg}}\\
{\includegraphics[trim={1.0cm 0.0cm 1.5cm 0.0cm}, clip, width=0.62\columnwidth]{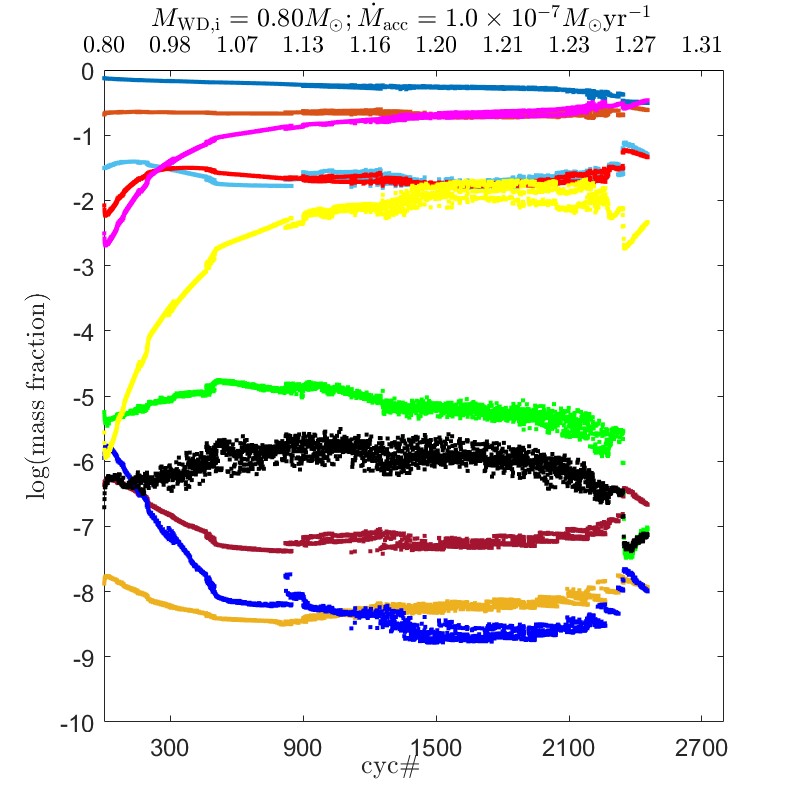}}
{\includegraphics[trim={22.5cm 5.6cm 0.7cm 4.0cm}, clip, width=0.31\columnwidth]{comp_for_legend.jpg}}
\caption{Composition of ejected material per cycle for a sample of ejective periodic helium nova models (HN type): Initial WD mass models of $0.8M_\odot$ accreting at rates of $10^{-5}$, $10^{-6}$ and $10^{-7}M_\odot\rm yr^{-1}$ (models $\#29$, $32$, and $33$ respectively). For the rest of the HN type models for this WD mass as well as all the HN type models of $0.65$, $0.7$ and $1.0M_\odot$ initial WD masses refer to \S \ref{sec:A_adtnl_figs} Figures \ref{fig:comp_nova_065}$-$\ref{fig:comp_nova_100} respectively.}\label{fig:comp_nova}
	\end{center}
\end{figure}

\begin{figure}
	\begin{center}
{\includegraphics[trim={1.0cm 0.0cm 1.5cm 0.0cm}, clip, width=0.62\columnwidth]{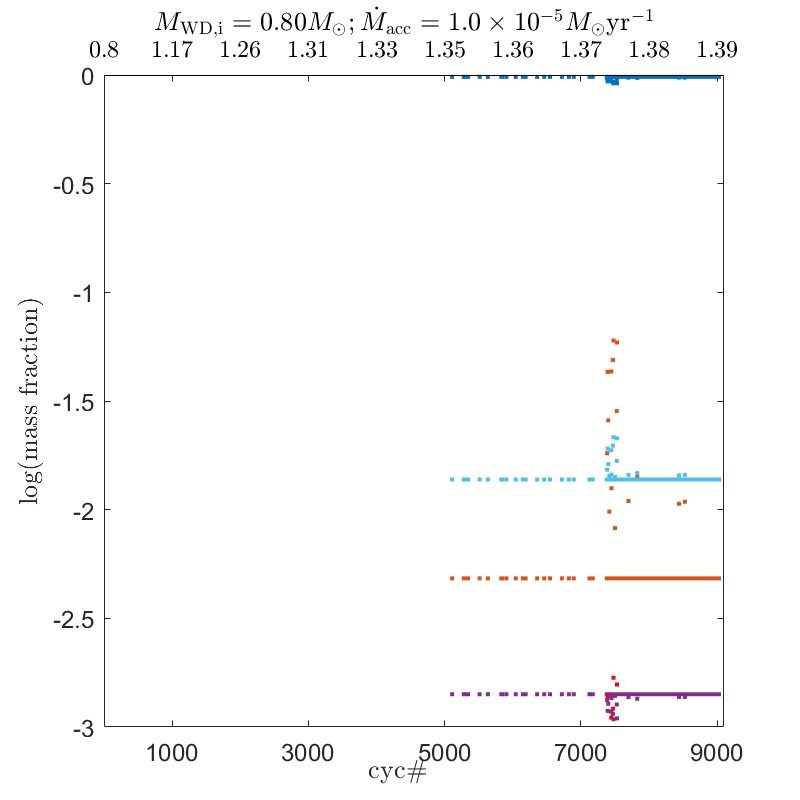}}
{\includegraphics[trim={22.5cm 5.6cm 0.7cm 4.0cm}, clip, width=0.31\columnwidth]{comp_for_legend.jpg}}\\
{\includegraphics[trim={1.0cm 0.0cm 1.5cm 0.0cm}, clip, width=0.62\columnwidth]{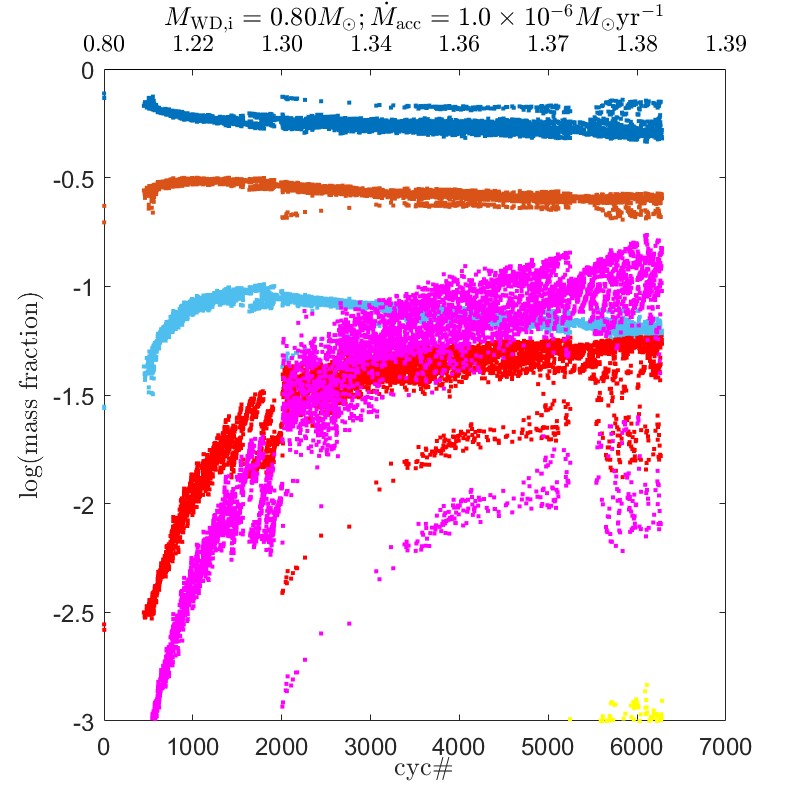}}
{\includegraphics[trim={22.5cm 5.6cm 0.7cm 4.0cm}, clip, width=0.31\columnwidth]{comp_for_legend.jpg}}\\
{\includegraphics[trim={1.0cm 0.0cm 1.5cm 0.0cm}, clip, width=0.62\columnwidth]{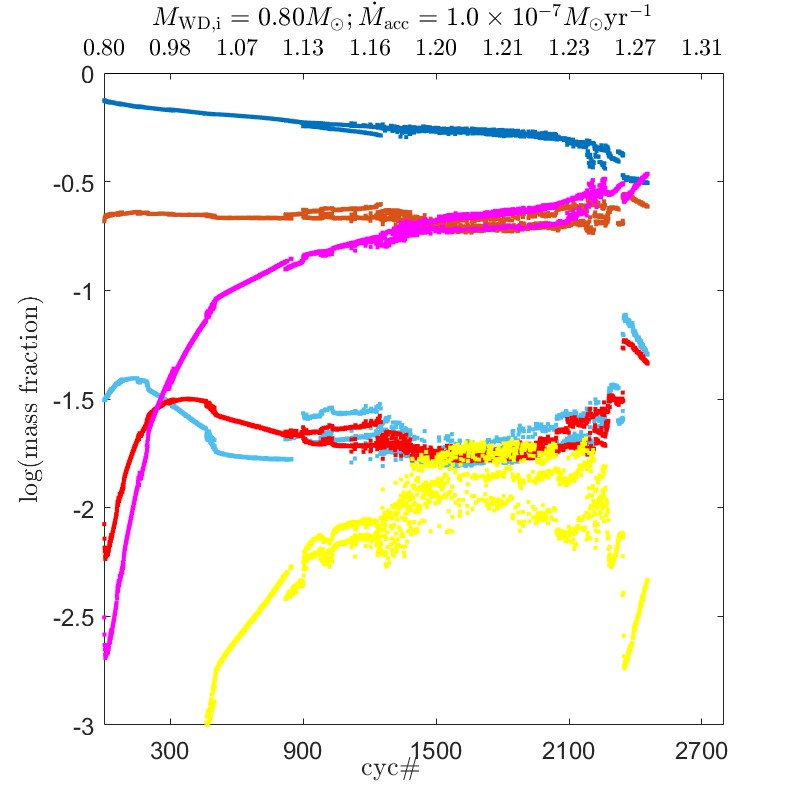}}
{\includegraphics[trim={22.5cm 5.6cm 0.7cm 4.0cm}, clip, width=0.31\columnwidth]{comp_for_legend.jpg}}
\caption{A close-up of Figure \ref{fig:comp_nova}, for mass fractions $\ge0.1\%$. Additional models, as described in Figure \ref{fig:comp_nova} are shown in \S \ref{sec:A_adtnl_figs} Figures \ref{fig:comp_novazoom_065}$-$\ref{fig:comp_novazoom_100}.\label{fig:comp_novazoom}}
	\end{center}
\end{figure}

Nova modeling has established that the time between eruptions has a direct influence on the ejecta composition \cite[e.g.,][]{Kovetz1997,Yaron2005,Starrfield2020,Starrfield2021,Hillman2021b, Hillman2022a,Hillman2022b,Starrfield2024} because a longer quiescent interval (i.e., a lower accretion rate) allows more time for diffusive mixing of the accreted hydrogen with the underlying outer layers of the WD's core. This mixing shifts the ignition to a deeper point, so the overlying envelope then requires more energy to lift, leading to higher peak temperatures, more extended burning, and stronger enrichment of the ejected mass. We find a similar trend for our helium nova models, i.e., more enriched ejecta at lower accretion rates, but the underlying cause differs: although some diffusion occurs, it is not the key driver of the helium nova outcome, since the CNO abundance does not strongly affect helium ignition. Instead, the very long accumulation time at low accretion rates allows a substantially more massive helium envelope to build up, which requires more energy to lift, driving higher peak temperatures, more extensive burning, and consequently more enriched ejecta. Figure \ref{fig:comp_nova}
shows, for a sample our helium nova (HN) models with initial $0.8M_\odot$ WDs, the ejecta composition including all the elements that were present in a non-negligible mass fraction\footnote{Mass fractions $\ge10^{-10}$} and Figure \ref{fig:comp_novazoom} shows the substantial composition components, i.e., mass fractions of $\ge0.1\%$. We show the rest of HN type models for this initial WD mass, as well as all the HN type for the $0.65$, $0.7$ and $1.0M_\odot$ initial WD models and their close-ups in \S \ref{sec:A_adtnl_figs} Figures \ref{fig:comp_nova_065}$-$\ref{fig:comp_novazoom_100}. The very high end of accretion rates within this regime produced non-ejective periodic novae, which is consistent with hydrogen novae evolution theory \cite[e.g.,][]{Yaron2005,Hillman2019} and similar to the mild helium flashes described by \cite{Piersanti2014}. These models\footnote{Non-ejective, periodic helium-nova-producing models: $\#3, 14-16, 28, 43-44$.} have accretion rates of roughly $10^{-6}$ to a few times $10^{-5}M_\odot\rm yr^{-1}$.   
Figures \ref{fig:comp_nova} and \ref{fig:comp_novazoom} show less heavy elements in the ejecta for higher accretion rates, while for the highest rates that produced novae, we see almost entirely only helium, carbon, oxygen, and nitrogen, compliant with the abundances of the accreted material. This expresses how these high accretion rates do not allow substantial mixing, thus, the mass that is ejected is not enriched. As the accretion rate decreases, and as the evolution progresses and the WD mass increases, we see trace amounts of elements such as carbon-13, oxygen-17, oxygen-18, sodium and aluminum, and substantial amounts of magnesium and neon which are higher for lower accretion rates and higher WD masses, while the few border-line models for which the WD mass did not increase, the composition remained $\sim$constant. Additionally, we see the production of silicon for evolved models with lower accretion rates, reaching up to a few percent. 

\begin{figure*}
	\begin{center}
{\includegraphics[trim={0.0cm 0.0cm 0.0cm 0.0cm}, clip, width=0.67\columnwidth]{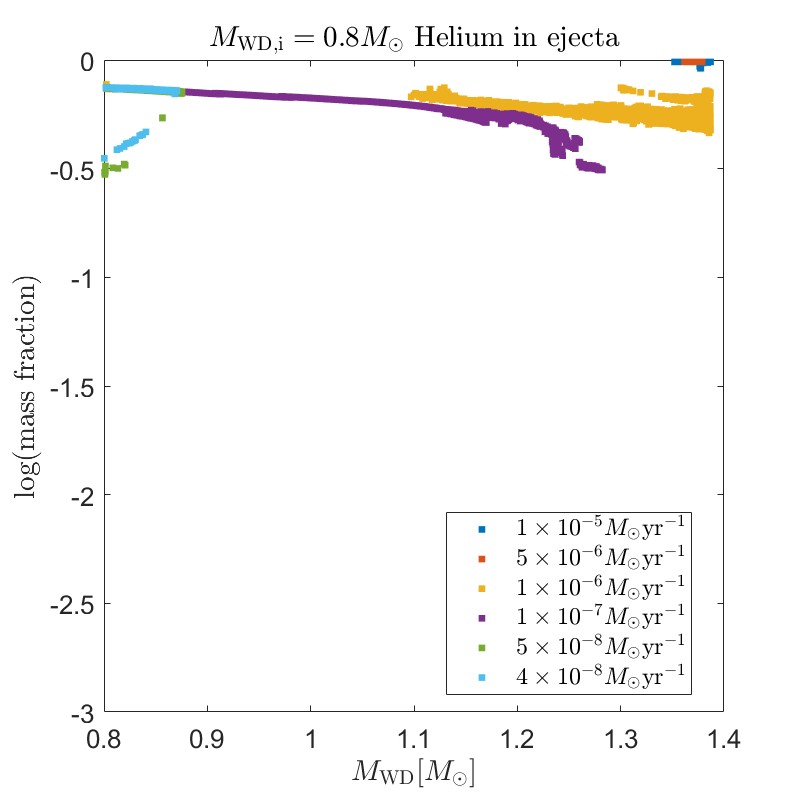}}
{\includegraphics[trim={0.0cm 0.0cm 0.0cm 0.0cm}, clip, width=0.67\columnwidth]{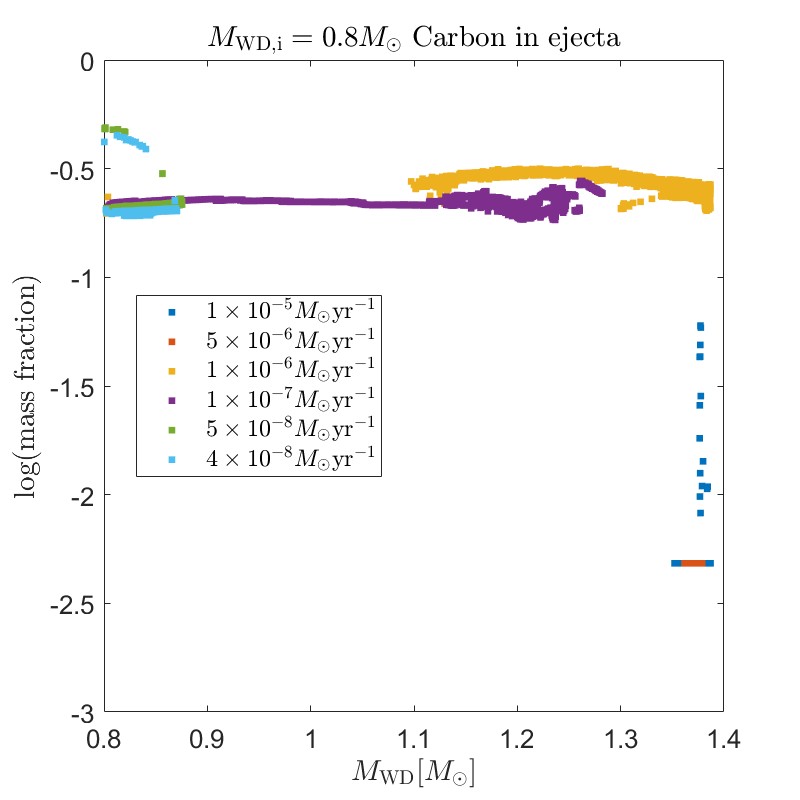}}
{\includegraphics[trim={0.0cm 0.0cm 0.0cm 0.0cm}, clip, width=0.67\columnwidth]{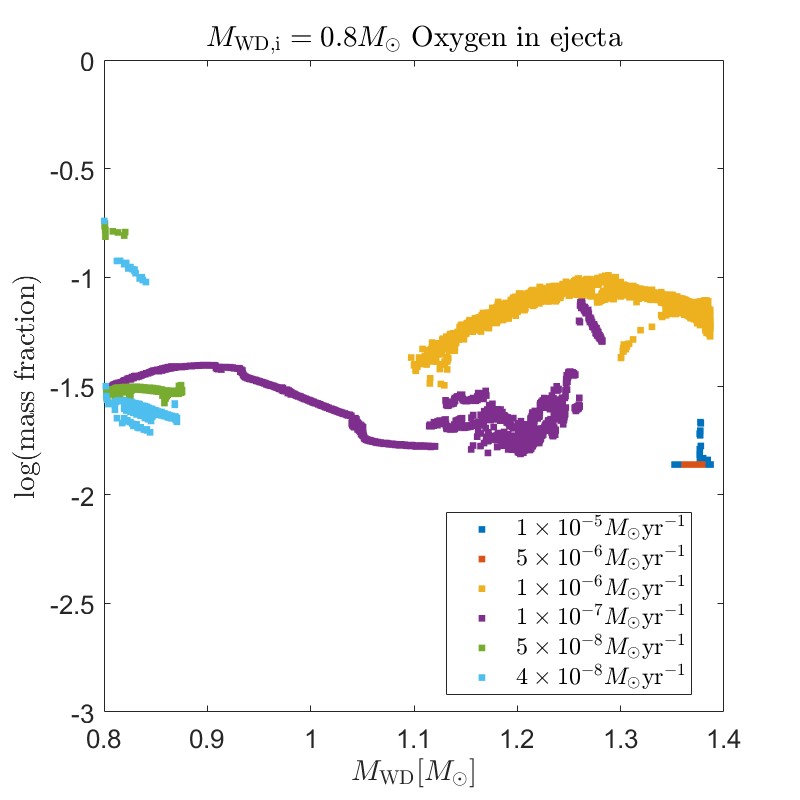}}\\
{\includegraphics[trim={0.0cm 0.0cm 0.0cm 0.0cm}, clip, width=0.67\columnwidth]{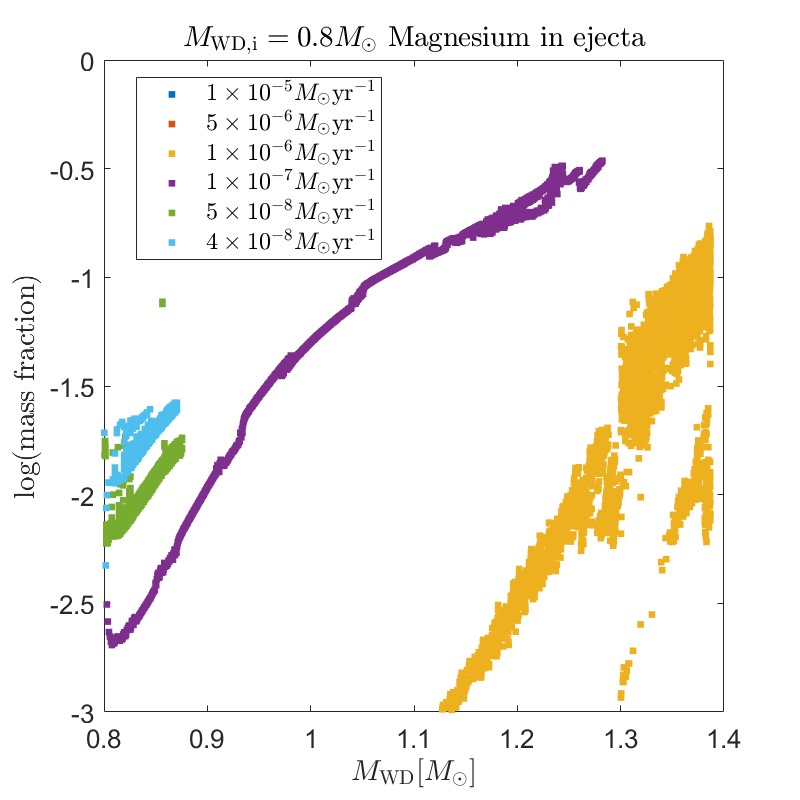}}
{\includegraphics[trim={0.0cm 0.0cm 0.0cm 0.0cm}, clip, width=0.67\columnwidth]{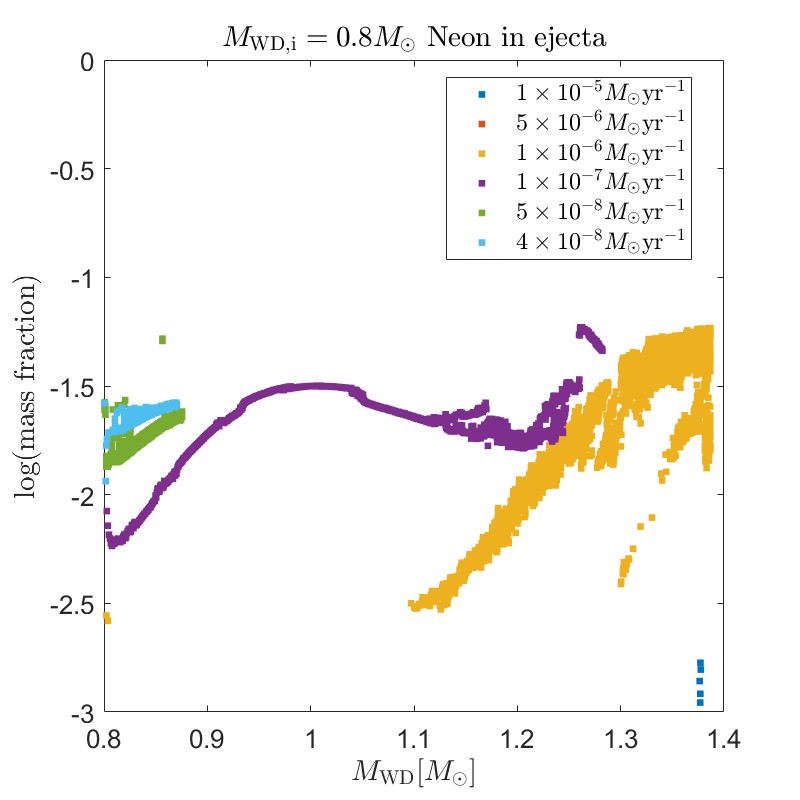}}
{\includegraphics[trim={0.0cm 0.0cm 0.0cm 0.0cm}, clip, width=0.67\columnwidth]{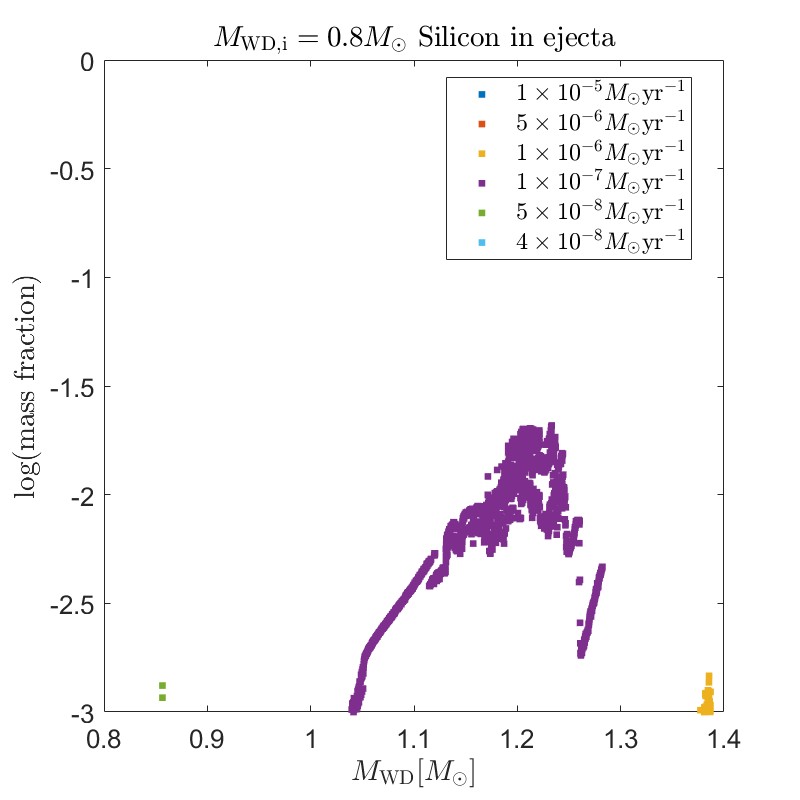}}
\caption{Composition of substantial elements ejecta over all eruptive HN type models with $M_{\rm WD,i}=0.8M_\odot$.}\label{fig:comp_080_specific}
	\end{center}
\end{figure*}

To further understand the correlations between ejecta enrichment, accretion rate and WD mass, we chose one of our initial WD masses ($M_{\rm WD,i}=0.8M_\odot$) and show in Figure \ref{fig:comp_080_specific} the evolution of five significant abundances for all eruptive HN type models. The figure demonstrates how the helium, carbon and oxygen abundances are weakly dependent on the accretion rate, while the abundance of magnesium is higher for higher WD masses and lower accretion rates, and it is negligible for the highest accretion rates. The silicon shows a non-negligible abundance only for an intermediate accretion rate (within the eruptive HN-type models) --- we do not see silicon for our $\dot{M}_{\rm acc}>10^{-6}M_\odot\rm yr^{-1}$ models; for our $\dot{M}_{\rm acc}=10^{-6}M_\odot\rm yr^{-1}$ model we see trace amounts of silicon that increases with increasing WD mass, and becomes significant only when the WD mass becomes near-Ch; for our $\dot{M}_{\rm acc}=10^{-7}M_\odot\rm yr^{-1}$ model we see substantial amounts of silicon when the WD mass becomes $\gtrapprox1.1M_\odot$; and we see trace amounts again for $\dot{M}_{\rm acc}\lesssim10^{-7}M_\odot\rm yr^{-1}$. 
 (See also Figures \ref{fig:comp_nova}, \ref{fig:comp_novazoom} and
\ref{fig:comp_nova_065}$-$
\ref{fig:comp_novazoom_100}.)
This reversing trend is the result of competing processes: (1) the depth of the TNR, for which if it occurs at a deeper point there is a richer reservoir of CO in the envelope; (2) the increased sub-surface density which is higher for higher WD masses; and (3) the temperature during the eruption, $T_{\rm max}$, which is about $3-6\times10^8K$ --- higher for lower accretion rates (see Figure \ref{fig:Tmax080mdots}), thus enhancing the silicon production rate. The very high accretion rates are poor in CO due to a short mixing time, and are at the lower end of $T_{\rm max}$, so the silicon production is not efficient for these models, even though the WDs mass significantly increases. At intermediate accretion rates, the longer inter-eruptive period allows more efficient mixing of WD core material into the envelope, enhancing the available CO for silicon production. Very low accretion rates ($\lessapprox10^{-7}M_\odot\rm yr^{-1}$) result in $m_{\rm ej}\sim m_{\rm acc}$, so the WD does not grow significantly in mass over time. Although, for a fixed WD mass, lower accretion rates are expected to produce higher peak temperatures, in these models the mass remains low even after long evolutionary times. In contrast, systems with intermediate accretion rates secularly increase the WD mass, and because their accretion is not excessively high, they still achieve high eruption temperatures. Consequently, the low accretion rate models never attain comparable WD masses and therefore do not reach the higher densities and peak temperatures achieved by the grown intermediate-rate models, so they remain below the temperature-density conditions required for significant silicon production, resulting in substantial silicon production only for intermediate accretion rates within the HN type regime.

\begin{figure*}
	\begin{center}
{\includegraphics[trim={1.0cm 0.7cm 1.3cm 0.5cm}, clip, width=1.99\columnwidth]{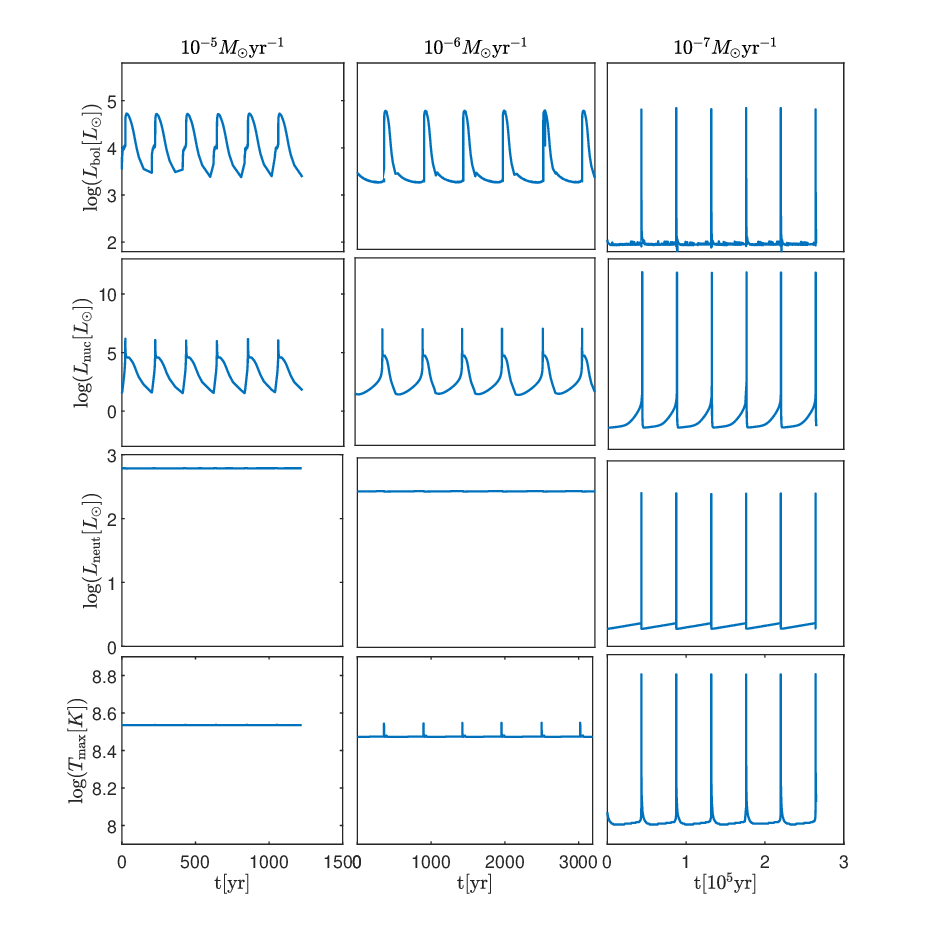}}
\caption{Bolometric, nuclear and neutrino luminosities ($L_{\rm bol}$, $L_{\rm nuc}$ and $L_{\rm neut}$ respectively) and maximum temperature ($T_{\rm max}$) over five cycles at the evolutionary point for which $M_{\rm WD}\approx1.1M_\odot$ for models of $M_{\rm WD,i}=0.8M_\odot$ and three accretion rates ($\dot{M}_{\rm acc}$): $10^{-5}$, $10^{-6}$ and $10^{-7}M_\odot \rm yr^{-1}$ (HN type models $\#$29, 32 and 33 respectively).}\label{fig:Ls080all}
	\end{center}
\end{figure*}

\begin{figure*}
	\begin{center}
{\includegraphics[trim={5.0cm 0.0cm 5.0cm 0.0cm}, clip, width=1.99\columnwidth]{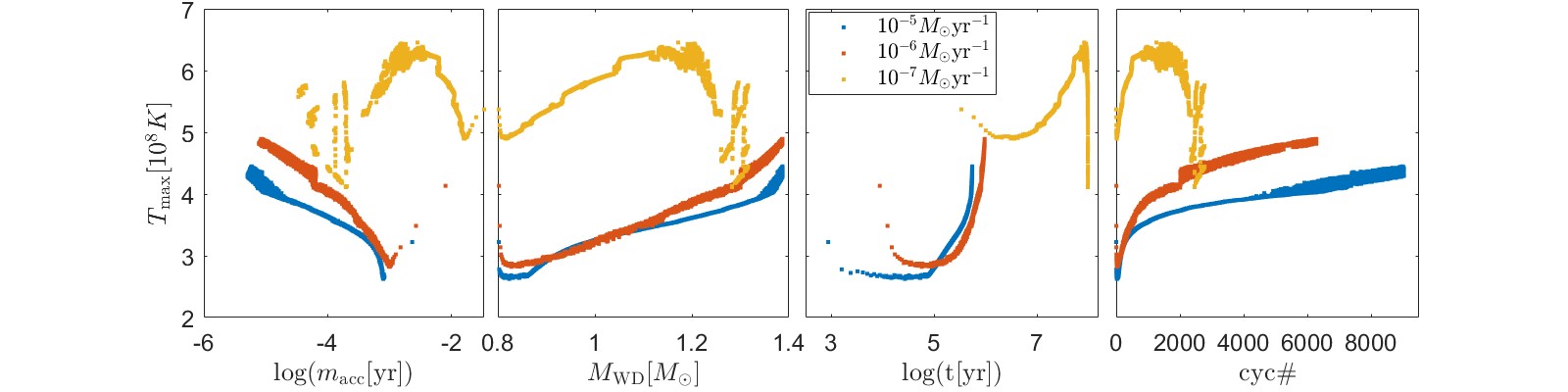}}
\caption{Evolution of maximum temperature ($T_{\rm max}$) per cycle vs. accreted mass per cycle ($m_{\rm acc}$) (top left); WD mass ($M_{\rm WD}$) (top right); evolutionary time (bottom left); and cycle number (bottom right); for the models shown in Figure \ref{fig:Ls080all}: $M_{\rm WD}=0.8M_\odot$ and three accretion rates ($\dot{M}_{\rm acc}$): $10^{-5}$, $10^{-6}$ and $10^{-7}M_\odot \rm yr^{-1}$ (models $\#$29, 32 and 33 respectively).}\label{fig:Tmax080mdots}
	\end{center}
\end{figure*}

To continue investigating the influence of the accretion rate on the evolution, we compare between the luminosities of 
three different accretion rates ($10^{-5}$, $10^{-6}$ and $10^{-7}M_\odot\rm yr^{-1}$) for a given initial WD mass of $0.8M_\odot$ (models $\#$ 29, 32 and 33). This is demonstrated in Figure \ref{fig:Ls080all} for a few consecutive cycles at the evolutionary point where the initial $0.8M_\odot$ WD has grown to $\sim1.1M_\odot$, 
exhibiting consistent trends: for a decreasing accretion rate, the bolometric luminosity ($L_{\rm bol}$) between eruptions decreases as does the nuclear luminosity ($L_{\rm nuc}$), while it increases during eruption due to the eruption being more energetic and ejecting more mass; the neutrino production (neutrino luminosity, $L_{\rm neut}$) increases very slowly over evolution and remains high for the $\dot{M}_{\rm acc}=10^{-5}M_\odot \rm yr^{-1}$ model, lower for the $\dot{M}_{\rm acc}=10^{-6}M_\odot \rm yr^{-1}$ model, while for the $\dot{M}_{\rm acc}=10^{-7}M_\odot \rm yr^{-1}$ model it is low during the long accretion phases and spikes during the powerful eruptions. Since the other two models have higher accretion rates, they experience much more moderate eruptions than the $\dot{M}_{\rm acc}=10^{-7}M_\odot\rm yr^{-1}$ model. In fact, at this point in evolution, the $\dot{M}_{\rm acc}=10^{-5}M_\odot \rm yr^{-1}$ model does not eject mass at all, the $\dot{M}_{\rm acc}=10^{-6}M_\odot \rm yr^{-1}$ model ejects mass of order $m_{\rm ej}\sim10^{-6}M_\odot$, and the $\dot{M}_{\rm acc}=10^{-7}M_\odot \rm yr^{-1}$ model ejects mass of order $m_{\rm ej}\sim10^{-3}M_\odot$ explaining the stark difference between their luminosities. 
Figure \ref{fig:Ls080all} also shows the maximum temperature ($T_{\rm max}$) below the accreted envelope vs. time for these three models, which is basically constant and high for the rapid accretion rate of the $\dot{M}_{\rm acc}=10^{-5}M_\odot \rm yr^{-1}$ model; exhibits a lower baseline temperature with small peaks indicating the moderate eruptions of the $\dot{M}_{\rm acc}=10^{-6}M_\odot \rm yr^{-1}$ model, and an even lower baseline with strong peaks for the $\dot{M}_{\rm acc}=10^{-7}M_\odot \rm yr^{-1}$ model. 
This temperature trend is consistent with the behavior that has been established for hydrogen nova via modeling \cite[e.g.,][]{Yaron2005,Epelstain2007} and explained by the short time between eruptions preventing the WD from cooling to its previous core temperature during the quiescent accretion phase, thus each eruption occurs at a higher temperature. Our work here demonstrates that this general explanation is valid for helium novae as well, while  
there is a difference, explained below. In hydrogen nova simulations, a temperature peak forms at the base of the accreted hydrogen shell as a TNR ignites to initiate a hydrogen nova eruption, however the peak temperature remains of order $1-2\times10^8$K, i.e., just below the helium burning threshold, and subsides with the relaxation of the nova eruption \cite[e.g.,][]{Yaron2005}. For helium novae, we see the same general behavior, but the temperature is expected to surpass the helium burning threshold as the TNR ignites \cite[]{Hillman2016}. We show in Figure \ref{fig:Tmax080mdots} the maximum temperature $T_{\rm max}$ attained per eruption for the same models shown in Figure \ref{fig:Ls080all}, 
exhibiting maximum temperatures roughly in the range $\sim2.6-6.5\times10^8$K --- suitable for helium fusion. Moreover, these plots show that the maximum temperature is generally higher for lower accretion rates, but interestingly, they are not linearly correlated, i.e., reducing the accretion rate by a factor of ten from $10^{-5}M_\odot \rm yr^{-1}$ to $10^{-6}M_\odot \rm yr^{-1}$ changes $T_{\rm max}$ only slightly, while, reducing it again by a factor of ten, from $10^{-6}M_\odot \rm yr^{-1}$ to $10^{-7}M_\odot \rm yr^{-1}$ leads to a substantial change in $T_{\rm max}$. This is related to the mixing of the accreted material with underlying WD core material being more extensive for lower rates. More time for mixing results in the TNR igniting deeper below the surface, meaning that more mass will be ejected. This type of behavior is characteristic of hydrogen novae as well, however for hydrogen novae, the amount of accreted mass required to trigger the TNR is only loosely dependent on the accretion rate, thus, the WD mass is the primary player that sets the triggering mass \cite[e.g.,][]{Prikov1995,Yaron2005,Hillman2022a}. This is because for hydrogen novae the temperature only has to rise until $1-2\times10^8$K to attain sufficient nuclear energy that converts to enough kinetic energy to unbind the accreted shell \cite[e.g.,][]{Yaron2005}. In contrast with this, our helium novae show that the triggering mass is \textit{highly dependent} not only on the WD mass but also \textit{on the accretion rate} as exhibited in Figure \ref{fig:macc_MWD_Tc_070}. 
This is because the required conditions for triggering a helium-fusion TNR are different than needed for triggering a hydrogen-fusion TNR. 
As the accretion rate becomes lower, its timescale becomes more competitive with the thermal timescale, so the heating due to accretion occurs at a less efficient rate. For this reason, for lower accretion rates, the WD manages to hold more helium before reaching the ignition temperature. This is a trend that is strong here, but weak in hydrogen novae.

The trend of lower accretion rates leading to a slower temperature rise is a continuous trend, shifting into the regime of prolonged-accretion models by allowing more and more mass to be accreted before heating to a temperature that is hot enough to start fusing helium. When a TNR finally ignites with a very massive helium shell, it must continue to heat and increase the nuclear fusion rate until attaining sufficient energy to gravitationally lift the helium shell, but by this time, heavy elements have begun to fuse indicating the initiation of an eruption much more powerful than a helium nova eruption. 

\subsection{Uninterrupted helium accumulation} \label{sec:nonerup}

\begin{figure}
	\begin{center}
{\includegraphics[trim={1.5cm 0.8cm 2.0cm 1.0cm}, clip, width=0.99\columnwidth]
{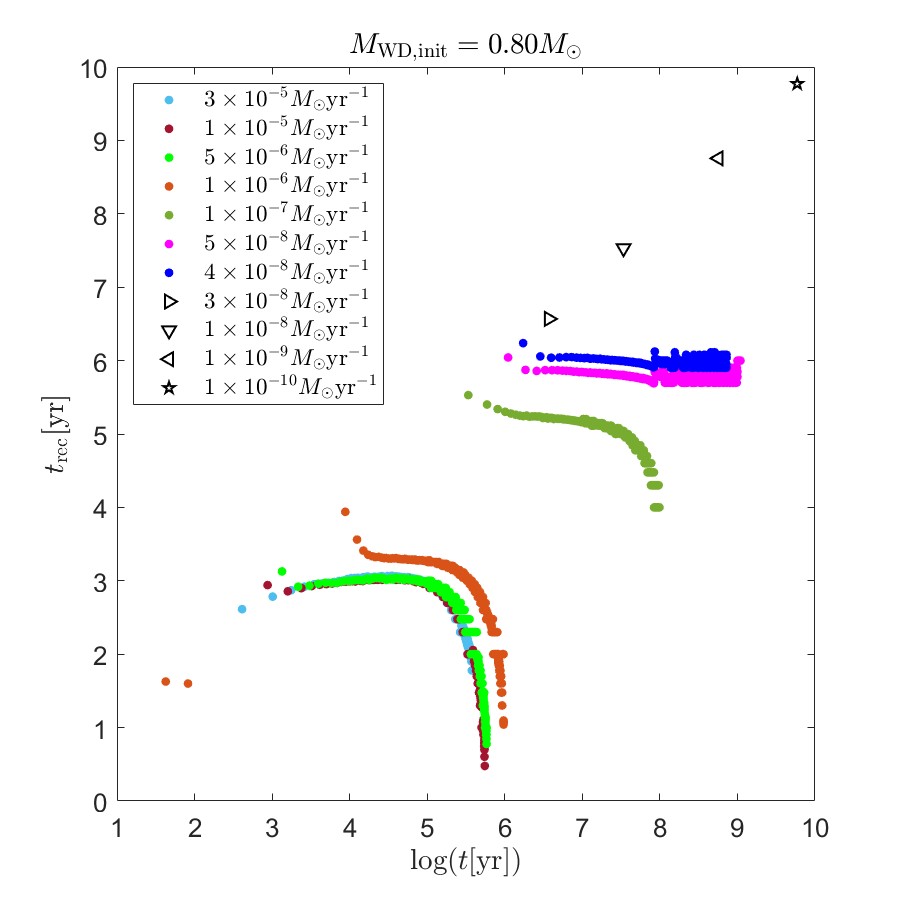}}
\caption{For $M_{\rm WD,init}=0.8M_\odot$: Recurrence period ($t_{\rm rec}$) vs. time for periodic helium-nova models (HN type models, color circles); uninterrupted accretion models w/o signs of SNIa ignition (UT type models, black full shapes (see Figure \ref{fig:trecs_A})); and uninterrupted accretion models that showed distinct signs of SNIa ignition (SN type models, black outlined shapes). Refer to Figure \ref{fig:trecs_A} for initial models of masses $0.65$, $0.7$ and $1.0M_\odot$.}\label{fig:trecs}
	\end{center}
\end{figure}

When using low accretion rates, less than a few times $10^{-8}M_\odot\rm{yr}^{-1}$, we find our models to endure prolonged quiescent accretion over \textit{hundreds of Myrs}, without producing nova eruptions\footnote{Specifically, models $\#8-11, 20-24, 37-40$ and $48-51$.}. These models experience a delayed TNR ignition due to the competing timescales of accretion and cooling. The cooling timescale of a WD can be expressed by the quotient of the thermal energy of the WD divided by its bolometric luminosity. The thermal energy can be taken as the thermal energy of a particle ($\frac{3}{2}kT_{\rm c}$, where $k$ is the Boltzmann constant) multiplied by the number of particles. The number of particles, assuming the WD to be of carbon and oxygen in equal parts, can be taken as roughly:
\begin{equation*}
    N_{\rm particles}\sim\frac{\frac{1}{2}M_{\rm WD}}{A({\rm C})m_H}+\frac{\frac{1}{2}M_{\rm WD}}{A({\rm O})m_{\rm H}}
\end{equation*}
where $A(\rm C)=12$, $A(\rm O)=16$ and $m_{\rm H}$ is the mass of a proton. The WD luminosity in quiescence can be taken as roughly of order $10^{-2}L_\odot$. 
Combining all this yields a cooling timescale of roughly $10^8$ years. Comparing this with the recurrence times in Figure \ref{fig:trecs} clearly shows that for the eruptive cases, the time between eruptions is much shorter than the cooling time --- thus the WD heats --- while the accretion timescale for the prolonged accretion cases competes with the cooling timescale, thus the WD heats at a much slower rate, and in some cases may cool. This means that since lower accretion rates mean more time for cooling, we would expect more mass to be accreted for these models, which is \textit{exactly} what we obtain. This may be seen as the empty markers in Figure \ref{fig:macc_MWD_Tc_070}, showing more accreted mass for lower accretion rates.

We note that this regime of uninterrupted accretion models comprises two sub-regimes. One regime, which we marked in Table \ref{tab:mdls} as SN type models, is of models that show \textit{distinct SNIa ignition signatures} --- rapidly rising temperatures to a few times $10^9\rm K$ resulting in a rapid increase of the nuclear fusion and neutrino emission; production of neutrons, increased amounts of heavy elements, and most importantly, a huge amount of silicon. These signatures are in excellent agreement with previous results of \textit{hydrogen} nova modeling that led a near-Chandrasekhar mass WD to the onset of a SNIa \cite[]{Hillman2015}. 

The second sub-regime, which we marked in Table \ref{tab:mdls} as UT type models, begins to show these signs, does not go into a runaway process, and does not produce the same amount of nuclear products, but the simulation runs into numerically small timesteps and terminates. While the small timesteps indicate that the code may have been initiating a SNIa producing TNR, since we do not see compelling evidence, we do not include these models (a total of four\footnote{Models \#8, 9, 20 and 21.}) in our SN type models, but instead we regard them as "undetermined transients", i.e., UT type models.

\begin{figure*}
	\begin{center}
{\includegraphics[trim={1.0cm 0.0cm 11.8cm 0.0cm}, clip, width=1.99\columnwidth]
{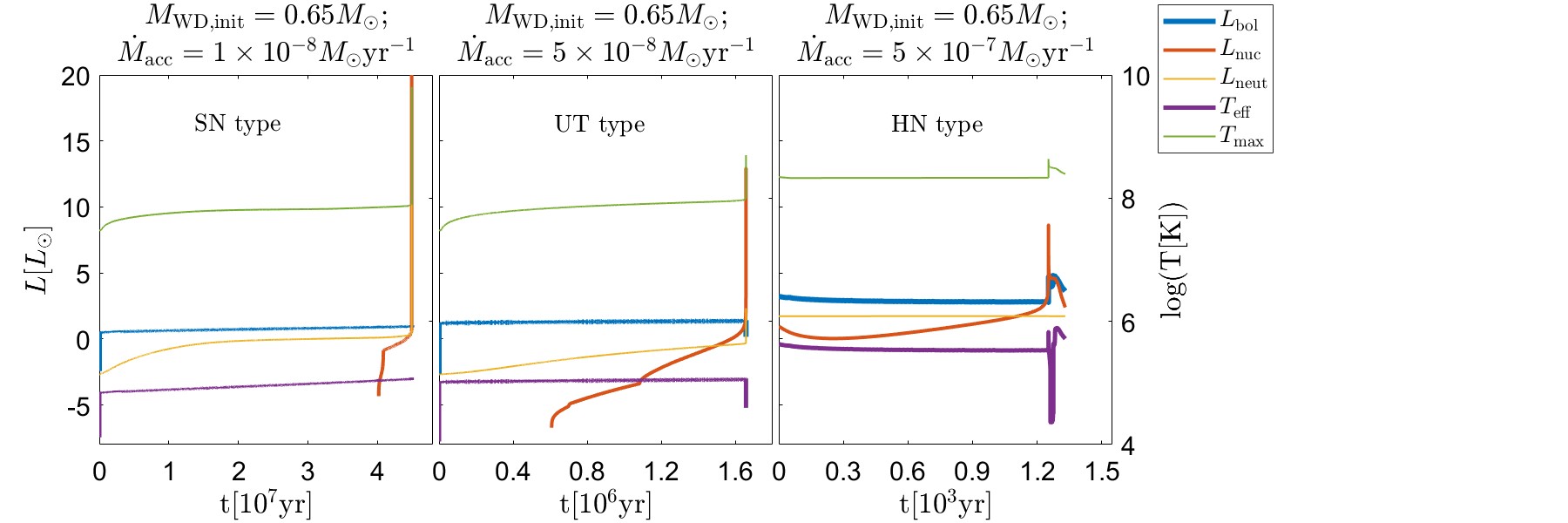}}\\
\caption{Luminosities (bolometric ($L_{\rm bol}$), nuclear ($L_{\rm nuc}$) and neutrino ($L_{\rm neut}$)) and effective temperature ($T_{\rm eff}$) and maximum temperature ($T_{\rm max}$) vs. time for a SN type model (model $\#11$, left); for a UT type model (model $\#8$, center); and for a HN type model (model $\#5$, right). 
}\label{fig:065Ls}
	\end{center}
\end{figure*}

We show in Figure \ref{fig:065Ls} the evolution of the bolometric, nuclear and neutrino luminosities ($L_{\rm bol}$, $L_{\rm nuc}$ and $L_{\rm neut}$ respectively) as well as the effective and maximum temperatures ($T_{\rm eff}$ and $T_{\rm max}$ respectively) for three sample models with an initial $0.65M_\odot$ WD --- one that showed distinct SNIa initiation (SN type model), one that underwent prolonged accretion but did not show distinct signs of SNIa initiation (UT type model) and one that produced periodic helium novae (HN type model). For the two prolonged accretion models, we show the entire evolution, and for the eruptive model we show one cycle of accretion and eruption\footnote{cycle \#100, randomly chosen.}.
The figure clearly shows that the HN type case behaves very differently from the other two cases, while the two other cases show distinct differences as well. While $L_{\rm bol}$ reaches a maximum of order $<10^5L_\odot$ for the HN type model, for the UT type model and the SN type model, it barely rises above quiescence. This is because for eruptive models, the envelope drastically expands as the WD ejects mass, as indicated by the sharp decrease in $T_{\rm eff}$, while the other two model types do not reach the point of mass ejection so they do not expand and therefore $T_{\rm eff}$ does not show a drastic decrease typical of nova eruptions. In contrast with $L_{\rm bol}$, $L_{\rm nuc}$ starts low for all three models, but reaches $\sim10^7L_\odot$, $\sim10^{12}L_\odot$ and $>10^{20}L_\odot$ for the HN, UT and SN type models respectively. The $L_{\rm neut}$ shows the same trend, showing no substantial increase in production for the HN model, a hint of the beginning of a runaway increase for the UT model, and a sharp runaway for the SN model. For $T_{\rm max}$ we see a sharp rise during the nova in the HN model, but then a recovery --- relaxation toward the next accretion phase, while the UT type begins to show signs of rising, and the SN type shows a runaway increase, indicating a possible SN ignition.  

\begin{table}[!h]
	\begin{center}
\begin{tabular}{|c|c|c|c|}
\hline
{A}&{Model \#5}&{Model \#8}&{Model \#11}\\
{}&{$[M_\odot]$}&{$[M_\odot]$}&{$[M_\odot]$}\\
			\hline\hline
{H}&{4.64e-23}&{2.74e-17}&{3.47e-06}\\
{n}&{4.89e-25}&{1.41e-18}&{1.01e-06}\\
{He4}&{1.30e-03}&{6.95e-02}&{4.27e-01}\\
{He3}&{0.00e-00}&{0.00e-00}&{7.97e-25}\\
{C12}&{4.56e-01}&{3.37e-01}&{3.28e-01}\\
{C13}&{1.53e-09}&{1.35e-09}&{1.58e-09}\\
{C14}&{1.18e-08}&{1.98e-07}&{1.94e-09}\\
{N13}&{0.00e-00}&{1.27e-13}&{8.81e-11}\\
{N14}&{3.98e-18}&{2.77e-05}&{6.02e-04}\\
{N15}&{5.13e-13}&{9.65e-10}&{1.71e-09}\\
{O16}&{3.87e-01}&{3.26e-01}&{3.31e-01}\\
{O17}&{2.80e-07}&{2.13e-08}&{1.13e-09}\\
{O18}&{4.64e-08}&{3.22e-08}&{1.07e-06}\\
{F17}&{0.00e-00}&{2.45e-17}&{0.00e-00}\\
{F18}&{0.00e-00}&{4.52e-14}&{4.92e-07}\\
{F19}&{1.87e-08}&{5.53e-07}&{0.00e-00}\\
{Ne20}&{1.85e-04}&{1.01e-04}&{6.59e-08}\\
{Ne21}&{1.34e-05}&{5.14e-05}&{5.36e-10}\\
{Ne22}&{4.91e-04}&{2.10e-05}&{1.97e-06}\\
{Na22}&{5.42e-18}&{5.02e-11}&{0.00e-00}\\
{Na23}&{4.19e-06}&{2.26e-07}&{9.67e-09}\\
{Mg24}&{3.26e-06}&{3.72e-05}&{1.22e-06}\\
{Mg25}&{8.71e-05}&{5.00e-06}&{1.21e-07}\\
{Mg26}&{5.76e-05}&{2.55e-07}&{8.78e-08}\\
{Al24}&{0.00e-00}&{0.00e-00}&{6.45e-08}\\
{Al25}&{0.00e-00}&{7.35e-17}&{3.27e-10}\\
{Al26}&{2.78e-15}&{1.13e-11}&{7.77e-09}\\
{Al27}&{4.64e-08}&{4.05e-09}&{1.55e-06}\\
{Si27}&{0.00e-00}&{0.00e-00}&{9.94e-09}\\
{Si28}&{3.39e-09}&{7.90e-09}&{1.22e-02}\\
{Si29}&{1.30e-09}&{2.43e-11}&{3.59e-05}\\
{Si30}&{1.40e-11}&{1.78e-13}&{2.89e-06}\\
{P29}&{0.00e-00}&{0.00e-00}&{5.96e-07}\\
{P30}&{0.00e-00}&{0.00e-00}&{7.53e-06}\\
{P31}&{0.00e-00}&{0.00e-00}&{3.54e-05}\\
\hline

  \end{tabular}
\caption{Composition for the three sample models shown in Figure \ref{fig:065Ls}: A HN type model ($\#5$); A UT type model ($\#8$); and a SN type model ($\#11$).}   
\label{tab:composition}
	\end{center}
\end{table}

To demonstrate the internal differences between the three types of results, we show in Table \ref{tab:composition} the composition of the WD for the three cases shown in Figure \ref{fig:065Ls}, demonstrating a general trend of the UT model producing more heavy elements than the HN model, and the SN model producing more heavy elements than the UT model. In particular, we note the non-negligible amount of hydrogen and neutrons that are produced only for the SN model and the non-negligible amount of nitrogen-14 that is produced for both the UT and the SN models but not for the HN model. But by far, the most prominent signature of an imminent SNIa is the silicon-28 production of order $10^{-2}M_\odot$ that is negligible for the HN and UT cases. We stress that these huge amounts of silicon-28 and neutrons are the result of a runaway fusion process that caused the temperature within the WD to rise to $>10^9\rm K$ as shown in Figure \ref{fig:T_t_SNIa_070}. At this temperature range, the elements should fuse to heavier nuclei, such as nickel and cobalt, however, this numerical code is designed to produce nova eruptions and thus does not include these reactions in its network. Nevertheless, the production of these nuclear products, along with the high temperature, are indicative of the onset of a SNIa.
We also note that Figure \ref{fig:T_t_SNIa_070} shows the temperature for the UT models to rise up to only $\sim5\times10^8\rm K$ which is of order the same maximum temperature reached in a helium nova eruption (Figure \ref{fig:macc_mej}), supporting our conclusion that the four UT type models are most probably intermediate, defining the limit between the regime of helium nova eruptions and SNIa.

\begin{table}[!h]
\begin{center}
\begin{tabular}{|c|c|c|c|c|}
\hline
{Model}&{$\dot{M}_{\rm acc}$}&{$M_{\rm WD,i}$}&{$M_{\rm WD,f}$}&{$\rm He$}\\

{\#}&{$[M_\odot\rm yr^{-1}]$}&{$[M_\odot]$}&{$[M_\odot]$}&{$[M_\odot]$}\\
\hline\hline

{8*}&{$5e{-}8$}&{0.65}&{0.733}&{0.081}\\

{9*}&{$3e{-}8$}&{0.65}&{0.777}&{0.124}\\

{10}&{$2e{-8}$}&{0.65}&{0.896}&{0.241}\\

{11}&{$1e{-8}$}&{0.65}&{1.099}&{0.440}\\

\hline
{20*}&{$5e{-8}$}&{0.7}&{0.771}&{0.069}\\

{21*}&{$3e{-8}$}&{0.7}&{0.820}&{0.118}\\

{22}&{$2e{-8}$}&{0.7}&{0.920}&{0.216}\\

{23}&{$1e{-8}$}&{0.7}&{1.102}&{0.394}\\

{24}&{$1e{-10}$}&{0.7}&{1.429}&{0.715}\\

\hline
{37}&{$3e{-8}$}&{0.8}&{0.910}&{0.108}\\

{38}&{$1e{-8}$}&{0.8}&{1.143}&{0.336}\\

{39}&{$1e{-9}$}&{0.8}&{1.373}&{0.561}\\

{40}&{$1e{-10}$}&{0.8}&{1.386}&{0.574}\\

\hline
{48}&{$4e{-8}$}&{1.0}&{1.050}&{0.049}\\

{49}&{$3e{-8}$}&{1.0}&{1.075}&{0.073}\\

{50}&{$1e{-8}$}&{1.0}&{1.210}&{0.206}\\

{51}&{$1e{-9}$}&{1.0}&{1.392}&{0.384}\\

\hline
  \end{tabular}
\caption{From left to right: Model number; given accretion rate; initial WD mass; final WD mass; accreted helium; for all our prolonged accretion models --- UT and SN types. The UT models are marked with an asterisk.}   
\label{tab:composition_SN}
	\end{center}
\end{table}

The models that show distinct SNIa initiation signatures (SN type) each reached this point at a different WD mass, the general trend being that a lower accretion rate allowed more mass to accumulate --- continuing the trend from the HN type models through the UT type models. The final WD masses for all the UT and SN models are marked in Figure \ref{fig:macc_MWD_Tc_070} (top, left) with gray markers, which clearly moves to the right with decreasing accretion rate, igniting a TNR (UT or SN) at WD masses ranging from $\sim0.8$ for an accretion rate of $5\times10^{-8}M_\odot\rm yr^{-1}$ to the Chandrasekhar mass ($\sim1.4M_\odot$) for an accretion rate of $10^{-10}M_\odot\rm yr^{-1}$. The final WD mass is roughly dependent on the initial WD mass because to reach a certain final WD mass, a lower initial WD mass will have to retain more helium than a more massive WD. We also sum these models in Table \ref{tab:composition_SN} specifying the mass of the accreted helium shell, which starts at a minimum of $10^{-2}-10^{-1}M_\odot$ for higher accretion rates in the UT and SN regime, and significantly increases with decreasing accretion rate. 

\begin{figure}
	\begin{center}
{\includegraphics[trim={0.5cm 0.5cm 1.0cm 0.8cm}, clip, width=0.99\columnwidth]
{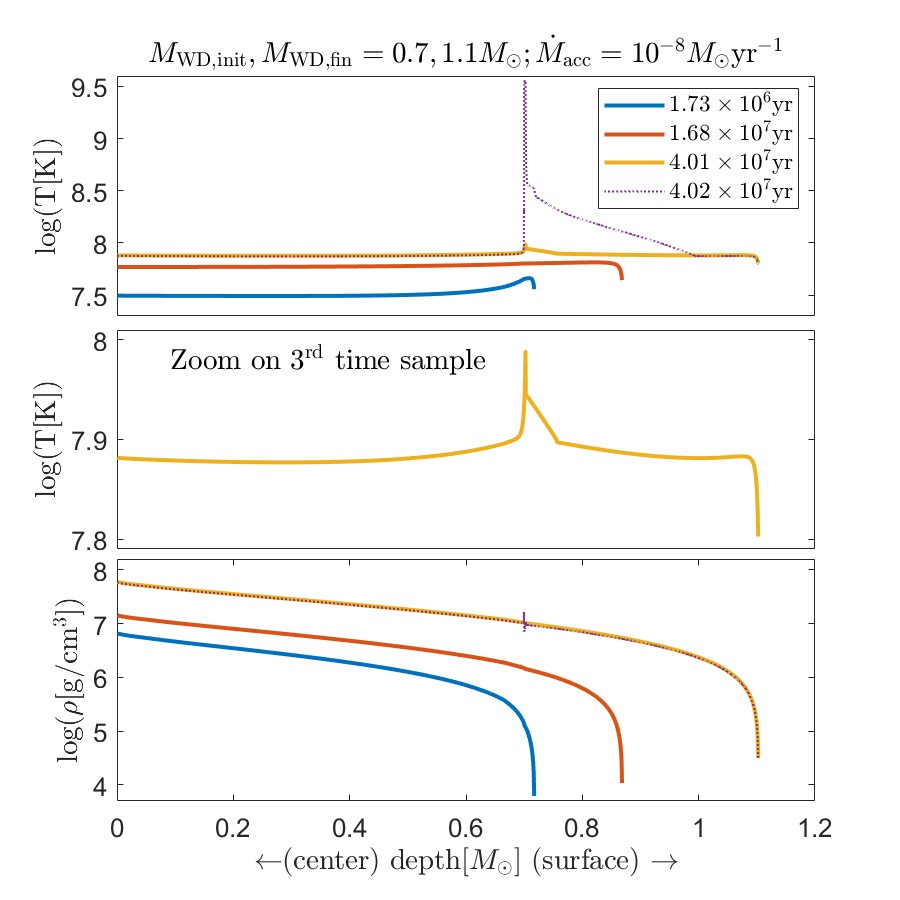}}
	\caption{Temperature and density profiles for the SN type model $\#23$ at four evolutionary points: Near the beginning of the simulation (blue); after significant evolutionary time (red); at the point where the TNR is beginning to ignite, thus the accreted mass has reached its limit (yellow); the next recorded data following the third point, indicating the ensuing of instability (dotted purple). The last point occurs $\sim10^5$ years after the third point. The initial WD mass is 0.7$M_\odot$, the final WD mass is $\sim1.1M_\odot$, and the accretion rate for this model is $10^{-8}M_\odot \rm yr^{-1}$.}\label{fig:Trho0701e8}
	\end{center}
\end{figure}

\begin{figure}
	\begin{center}
{\includegraphics[trim={0.5cm 0.5cm 1.0cm 0.0cm}, clip, width=0.99\columnwidth]{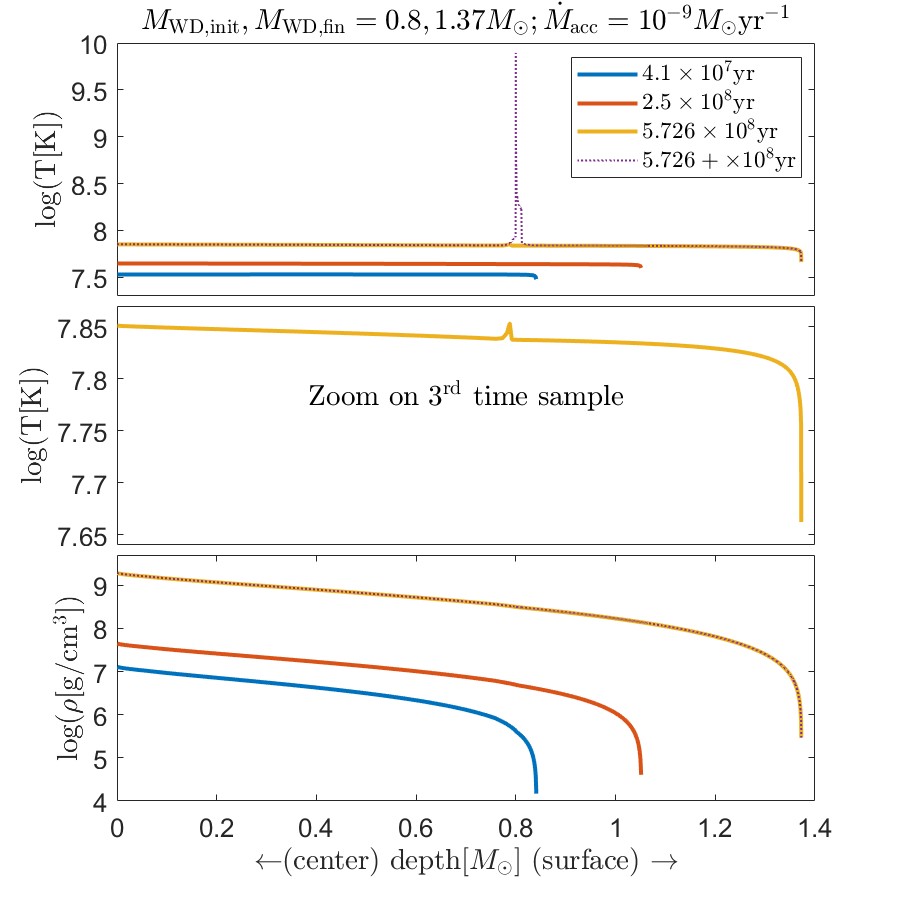}}
\caption{Description as in Figure \ref{fig:Trho0701e8} for the SN type model $\#39$ with an initial WD mass of 0.8$M_\odot$, final WD mass of $\sim1.37M_\odot$ and an accretion rate of $10^{-9}M_\odot \rm yr^{-1}$ and for which the instability (fourth time point) occurs of order one year after the third time point.}\label{fig:Trho0801e9}
	\end{center}

\end{figure}

To further understand the behavior of these SNIa signature systems, as the runaway of heavy elements production and rapidly rising temperature begins, we plot in Figures \ref{fig:Trho0701e8} and \ref{fig:Trho0801e9} the temperature profiles at a few time frames spanning the evolutionary time of two SN type models $\#$23 and $\#$39 respectively, the former having a lower initial WD mass ($0.7M_\odot$) and a higher accretion rate ($10^{-8}M_\odot\rm yr^{-1}$), reaches instability at the sub-Chandrasekhar mass of $M_{\rm WD}\approx1.1M_\odot$ and the latter, having a higher initial WD mass ($0.8M_\odot$) and a lower accretion rate ($10^{-9}M_\odot\rm yr^{-1}$), reaches instability at a near-Chandrasekhar mass WD, i.e., $M_{\rm WD}\approx1.37M_\odot$. Both models exhibit smooth temperature profiles until the WD ceases to accrete and the temperature begins to develop a sharp peak, of order $10^8$K, at the base of the accreted shell, indicating the conditions ripening for mass loss\footnote{We note that the difference in time delay between the model in Figure \ref{fig:Trho0701e8} to that in Figure \ref{fig:Trho0801e9} is due to numerical print out definitions.}. However, instead of transitioning to dynamical mass loss, and initiating a helium nova eruption, the temperature continues to rise because the helium shell is very massive, thus the conditions do not yet have the energy required to lift it, so the temperature rapidly rises, reaching $>10^9$K, heavy elements begin to form, and finally the simulation stops due to numerical reasons as explained earlier. 
 
The temperature spike occurs at the base of the accreted envelope for model $\#$23 after $\sim4\times10^7$ years, while for model $\#$39 --- which has a lower accretion rate --- after $\sim5.7\times10^8$ years. Model $\#$23 reaches instability at a lower  WD mass --- $\sim1.1M_\odot$, as mentioned earlier after accreting $\sim0.4M_\odot$, i.e., the ratio of the helium shell to the CO core is $4/7$ --- more than 50$\%$ of the WD mass! Model $\#$39 manages to accrete $\sim0.57M_\odot$ before becoming unstable at the total WD mass of $\sim1.37M_\odot$, the ratio of the helium shell to the CO core being roughly $3/4$ --- an even larger ratio than for model $\#$23. 
We conducted a detailed analysis of the dynamics and nuclear products of SN-type model $\#$23, and find that it leads to a sub-Chandrasekhar mass double-detonation SNIa. The full analysis is presented in Michaelis et al. (2025) (hereafter, Paper 2). 

We note that model $\#$24 reaches $\sim1.429M_\odot$, essentially at the Chandrasekhar limit, where central carbon ignition via pycnonuclear reactions could in principle occur \cite[]{Nomoto1982}. Since our code does not include pycnonuclear reaction rates, we cannot currently determine its final fate.

\section{Discussion} \label{sec:discussion}

We have spanned the entire range of accumulation rates of helium onto a CO WD and found it to comprise three basic regimes. The first regime is of models that were given a mass transfer rate too high to be accreted quiescently for a significant period of time, so instead, the transferred mass is simply pushed away by accretion radiation, leading to a red-giant-like formation, which halts our simulation.

The second regime produced periodic helium-nova eruptions, following the same basic principals known from hydrogen nova --- since lower accretion rates allow more time to accrete, the accreted mass has more time to mix into deeper layers, thus the TNR ignites at a deeper point, resulting in more ejected mass. However, this is where the similarity stops. While for hydrogen novae the amount of accreted mass required to trigger the TNR is predominantly determined by the WD mass, with the accretion rate having a minor influence, for our helium novae we find both the WD mass \textit{and the accretion rate} to be key parameters in determining the triggering accreted mass.
Moreover, we find the retention efficiency to not only generally be lower for lower accretion rates, but also to demonstrate significant changes as the system ages. These changes are correlated with the underlying core temperature, as explained in \cite{Hillman2016} to be due to higher core temperatures incurring a degeneracy decrease, thus enabling less violent burning. The core temperature rises because the eruption releases thermal energy that heats the underlying core. Over many eruptions that occur at a frequency that has a shorter timescale than the thermal timescale, the core temperature secularly increases, reaching a few times $10^8[K]$. We note that on a shorter timescale (of order a cycle) in addition to mass ejection, the eruption also leaves behind hot heavy elements (in particular carbon and oxygen) giving a head-start on the burning rate of the next cycle by allowing some quiescent burning during accretion, leading to less ejected mass. While the latter has a cyclic influence, explaining fluctuations, the former is long term, explaining the trend following that of the core temperature.

Additionally, we find that the range of accretion rates here is different than the typical range within which hydrogen novae reside, namely, for the same order of recurrence, the helium accretion rate for helium novae is of order $10-100$ times higher than the hydrogen accretion rate for hydrogen novae. This is because for a given accretion rate, an order of $10-100$ times more helium is required to trigger a helium TNR than the amount of hydrogen required to trigger a hydrogen TNR.

\begin{figure*}
	\begin{center}
{\includegraphics[trim={0.8cm 0.75cm 2.0cm 2.0cm}, clip, width=1.99\columnwidth]{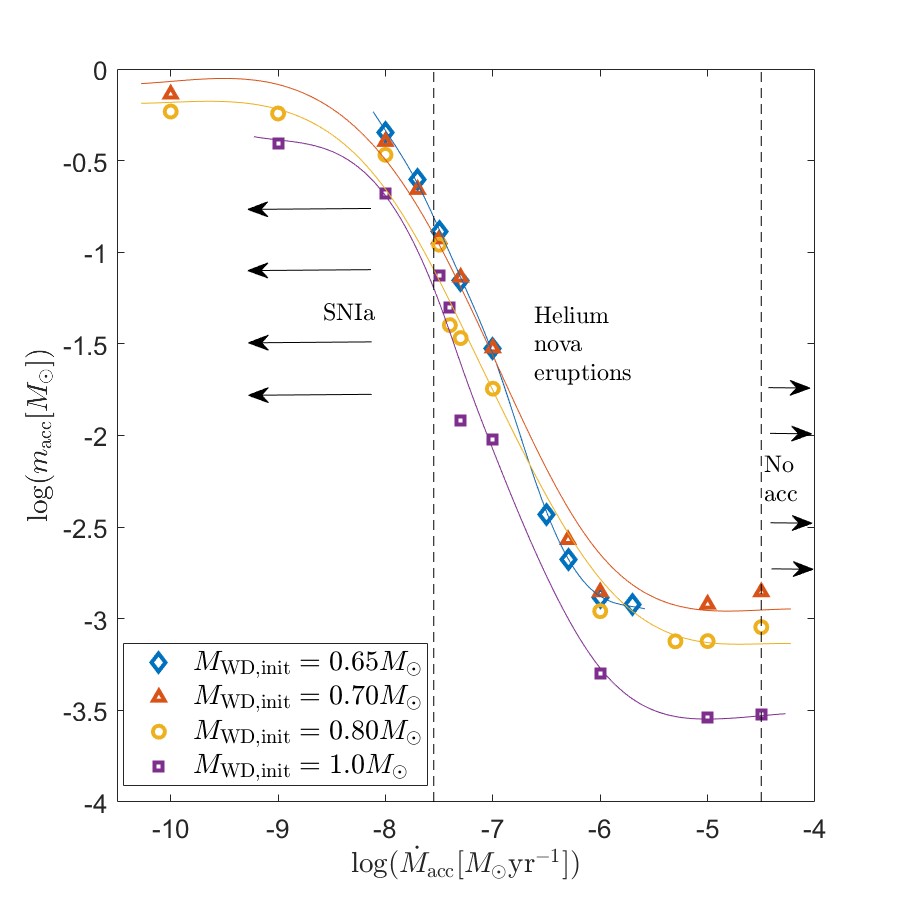}}
	\caption{Accretion regimes: SN (and UT) type (left); HN type (center); Eddington accretion (right).}\label{fig:acc_regimes}
	\end{center}
\end{figure*}

However, this trend does not follow through the entire feasible accretion rate range. Unlike hydrogen novae, low accretion rates do not simply lead to more energetic and less frequent novae that erode the WD secularly. We find that for helium accretion, low accretion rates do not lead to novae eruptions or nuclear burning \textit{at all} because lower rates shift the evolution into the third accretion rate regime --- uninterrupted prolonged accretion. In this regime, the accretion continues secularly until, eventually, a huge amount of helium is accumulated, at which point the ensuing TNR is expected to tear apart the WD altogether (Paper 2).
We illustrate the three basic regimes in Figure \ref{fig:acc_regimes}, showing the connection between the accretion rate, the initial WD mass, the amount of accreted mass resulting from this combination (per cycle for the HN regime and total for the UT and SN types), and the relevant regime.

Studies by \cite{Yoon2004} and \cite{Neunteufel2017} used accretion rates of $10^{-7}-10^{-8}M_\odot\rm yr^{-1}$ onto a range of WD masses that are spun up by the accreted material. They find that such models can retain up to 50$\%$ more helium before ignition. The implication for our results would be a slight shift in our ignition thresholds. For example, model $\#8$ would behave more like model $\#9$, i.e., a $0.65M_\odot$ WD that accreted helium at a rate of $5\times 10^{-8}M_\odot\rm yr^{-1}$, would resemble the same WD mass accreting  at the slightly lower rate of $3\times 10^{-8}M_\odot\rm yr^{-1}$. Similarly, model $\#21$ would behave like model $\#37$, i.e., an accretion rate of $3\times 10^{-8}M_\odot\rm yr^{-1}$, but with a slightly more massive WD (0.8$M_\odot$ instead of 0.7$M_\odot$). This could slightly alter the limits between our HN and UT/SN regimes or push a UT type system into the SN type territory. However, since our regime definitions are deliberately soft and depend on both accretion rate and WD mass, the overall structure of our results remains robust. Moreover, the accretion rates studied by these authors are only marginally within our helium nova regime, and thus does not account for possible angular momentum losses due to mass ejection, which could counteract the spin-up effect. Accounting for all this, we conclude that while rotation may introduce small quantitative shifts, it does not alter our qualitative findings, and the trends we identify in this work remain valid.

We turn to discuss the nature of stellar systems that may experience the sort of helium accumulation studied here. 

Examining CVs that endure hydrogen novae for which helium is the by-product, we refer to previous works that have deduced an effective rate of helium accumulation via hydrogen novae. These works have found the ratio of helium accumulation rate to hydrogen accretion rate to be roughly $<0.5$ \cite[]{Newsham2013,Hillman2016}. 
We deduce that our regime of prolonged accretion that does not produce helium nova eruptions ($\dot{M}_{\rm acc}\lessapprox10^{-8}M_\odot \rm yr^{-1}$) would be possible only in cataclysmic variables (CVs). Such slow helium accumulation for long periods would have to come from a binary in Roche-lobe overflow (RLOF), i.e., a low to moderate-mass red dwarf transferring primarily hydrogen. The hydrogen would lead to novae that leave a residue of helium. This helium would secularly accumulate. Since for low to moderate mass RDs the typical mass transfer rate is relatively low ($\sim10^{-9}-10^{-11} M_\odot \rm yr^{-1}$) \cite[]{Hillman2021b} and only a fraction of it may burn into helium and remain on the WD's surface at the end of a nova eruption \cite[]{Hillman2016}, the accumulation rate of helium will effectively be even lower. 
However, such hydrogen accretion rates lead to a net mass loss, so even if some helium is left behind, the long diffusion time prior to the hydrogen nova eruption will have resulted in $m_{\rm ej}>m_{\rm acc}$ thus the WD mass will be secularly decreasing. This means that while CVs with low to moderate mass RD donors may be a source of CO WDs with substantial, increasing helium envelopes, if they were to lead to a SNIa, we conclude the WD to have to be initially close to the Chandrasekhar mass. On the other hand, if the system were to produce recurrent hydrogen novae (RN), i.e., hydrogen-rich accretion rates of order $\sim10^{-7}-10^{-6}M_\odot\rm yr^{-1}$ the WD mass will increase, leave a substantial helium residue on the WD's surface, and require a lower helium triggering mass, thus even intermediate mass WDs may eventually lead to a sub-Chandrasekhar SNIa in this scenario. To date, there are ten known recurrent novae in our Galaxy: two in CVs, four with sub-giant donors, and four with red giant donors, so we do not limit this scenario to CVs. 
Nonetheless, the upper end of this regime --- the RN regime of hydrogen novae --- is a possible candidate path to SNIa on its own \cite[]{Hillman2015}, thus, this sub-regime of accretion rates has two methods of ending as a SNIa.

The high accretion rates required to produce the two known Galactic RNe in CVs is still difficult to explain via current nova theory while symbiotic systems, i.e., giant donors, have multiple mass transfer mechanisms, thus opening more possibilities for a high mass transfer rate \cite[]{Vathachira2025}.
The intermediate helium accretion rate regime, of order $\sim10^{-5}-10^{-8}M_\odot\rm yr^{-1}$, might be possible from a red giant (RGB) or asymptotic giant branch (AGB) donor, in RLOF or via Bondi-Hoyle-Littleton (BHL) accretion provided the wind rate from the AGB is high enough, and a large enough fraction of it is gravitationally captured by the WD \cite[]{Mikolajewska2008,Mikolajewska2010,Hillman2021a,Vathachira2024}. There is also the possibility of the giant's wind being gravitationally focused toward the WD; thus, the mass transfer would be more efficient \cite[]{Abate2013,Mohamed2015,Vathachira2025}. 

A giant donor, being more evolved than a RD and experiencing extensive mixing (in AGBs), will possibly increase the helium mass fraction in the transferred mass and increase the helium accumulation rate directly.
We note that such RGB or AGB donors in symbiotic systems may also transfer mass at a low rate, depending on many parameters (e.g., donor mass, WD mass, thermal pulses, wind rate, separation etc.). In such cases, the symbiotic system may belong to the low helium-accumulation regime described earlier.

\begin{figure}
	\begin{center}
{\includegraphics[trim={0.0cm 0.0cm 1.8cm 0.0cm}, clip, width=0.99\columnwidth]{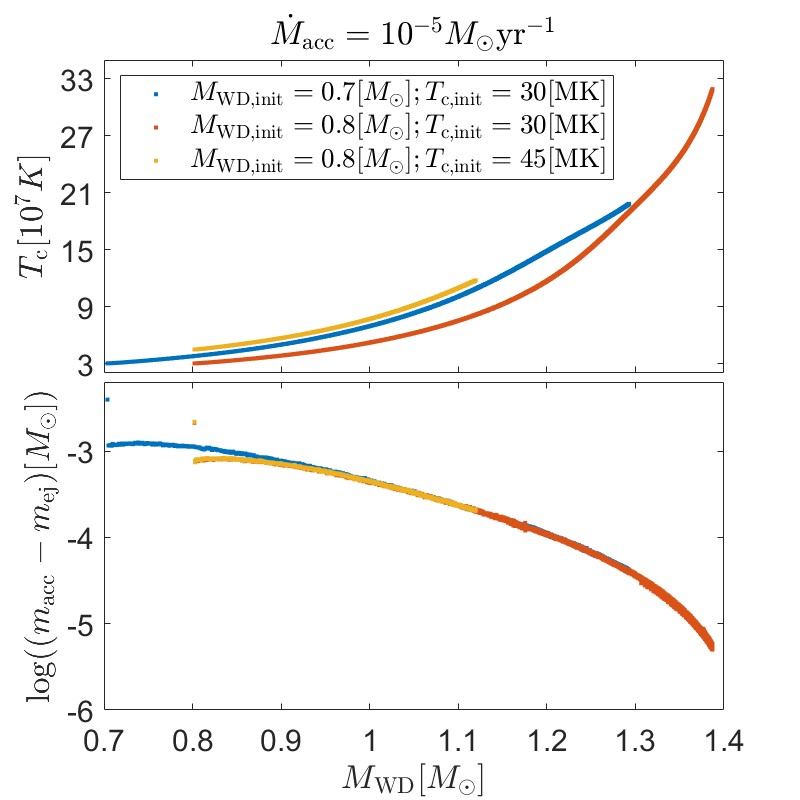}}
	\caption{$T_{\rm c}$ (top) and net accreted mass per cycle (bottom) vs., WD mass for models $\#$15 (blue), $\#29$ (red) and $\#$30 (yellow).}\label{fig:Tc_different}
	\end{center}
\end{figure}

Although this work addresses the evolution ruled by helium accumulation, the systems described above (CVs or symbiotic systems) primarily accrete hydrogen, and the helium is a by-product of hydrogen fusion. We note that it is likely that observations of such systems would not be absent of hydrogen signatures. 
Additionally, we point out a caveat regarding the simulations that stems from bypassing the hydrogen fusion with direct helium accretion. Hydrogen nova have a heating effect on the outer layers of the WD that may have a secular influence on the evolution of the system in such a way that the WD's cooling timescale would be extended. We do not expect this to have a significant effect since nova models have shown that the core temperature has only a secondary effect on the outcome. We support this with the two models that we produced with a higher core temperature (45MK, models $\#30$ and 34) and obtained insignificant differences in the results. Furthermore, following the evolution of our HN type models shows a consistent increase in the core temperature, which allows us to compare between models at higher temperatures as well. To demonstrate this we have plotted in Figure \ref{fig:Tc_different} the net accreted mass per cycle over time for three models: An initial 0.70$M_\odot$ WD with the initial core temperature of 30MK (model $\#$15), an initial 0.80$M_\odot$ WD with the initial core temperature of 30MK (model $\#$29), and an initial 0.80$M_\odot$ WD with the initial core temperature of 45MK (model $\#$30), all three models with the accretion rate of $10^{-5}M_\odot\rm yr^{-1}$. The the net accreted mass curves for these three models overlap, demonstrating the robustness of the results. Moreover, this figure allows to compare between models $\#$15 (blue curve) and $\#$30 (yellow curve), for which the former has a lower mass and temperature than the latter, but by the time the lower mass WD (0.7$M_\odot$) reaches the initial mass of the latter (0.8$M_\odot$), its temperature reaches 45MK, leading to the convergence of the blue and yellow temperature curves. The red curve, representing an initial WD mass of 0.8$M_\odot$, since began with the core temperature of 30MK, is slightly shifted from the other two curve. This implies that, say, an initial 0.8$M_\odot$ WD with a very hot initial core temperature of 60$-$70MK would be slightly more shifted compared to the 30MK model. We conclude that the initial core temperature has a minor effect on the evolution and thus, the robustness of our results holds.

On the other hand, the donor may be helium-rich, i.e., a helium star, in which case the \textit{helium is transferred directly} and then the accretion rate would depend on the system parameters, such as the separation and the donor radius, in principle leading to any of the accretion regimes. 

For the extremely high accretion rate regime ($\gtrapprox10^{-5}M_\odot\rm yr^{-1}$), since the Eddington limit prevents the mass from being accreted quiescently, we suspect the only feasible mass transfer mechanism to be a merger of a double WD (DWD) where the less massive WD is helium-rich, and is deformed and stretched around the accreting WD to be merged. \cite[e.g.,][]{Zenati2023}.

In this work we adopted a simplified helium-dominated composition ($98\%$ helium, $2\%$ Solar Z) accreted at constant rates for all models. We note that, depending on the donor type, the mass being accreted could be N-enhanced from prior nuclear processing and this could slightly alter the ignition mass. Nonetheless, previous studies \cite[]{Piersanti2001,Shen2009} show this effect to be modest, so the general trends and accretion regimes identified here remain robust.



\section{V445 Puppis}\label{sec:V445pup}
The only confirmed helium nova to date is V445 Puppis, which is known to have erupted once in late 2000. The spectra were reported to be very different than that of classical novae by being void of hydrogen lines while showing many helium and carbon lines, indicating a helium star donor \cite[e.g.,][and references within]{Ashok2003,Lynch2004,Banerjee2023}. This means that the helium in this system is accreted directly (and not as a by-product of hydrogen fusion).
By analyzing the SED of dust production in the 2000 eruption, \cite{Banerjee2023} estimated the amount of mass ejected to be of order $10^{-2}-10^{-1}M_\odot$. This is in coincidence with our models with accretion rates of order $0.5-1\times10^{-7}M_\odot\rm yr^{-1}
$, and WD masses of up to $\sim1.0M_\odot$. This range of accretion rates is very constrained because a higher rate leads to less ejected mass --- below the limit determined by \cite{Banerjee2023} --- while a lower rate leads to a SN. This narrow range of accretion rates, is the same regime of models for which we obtain $m_{\rm ej}\gtrapprox m_{\rm acc}$ (see Figure \ref{fig:macc_mej}), i.e., these models retain only a small amount of mass after each eruption, so their net mass growth is very slow, if any.

\begin{figure}
	\begin{center}
{\includegraphics[trim={1.0cm 2.0cm 1.8cm 0.8cm}, clip, width=0.99\columnwidth]{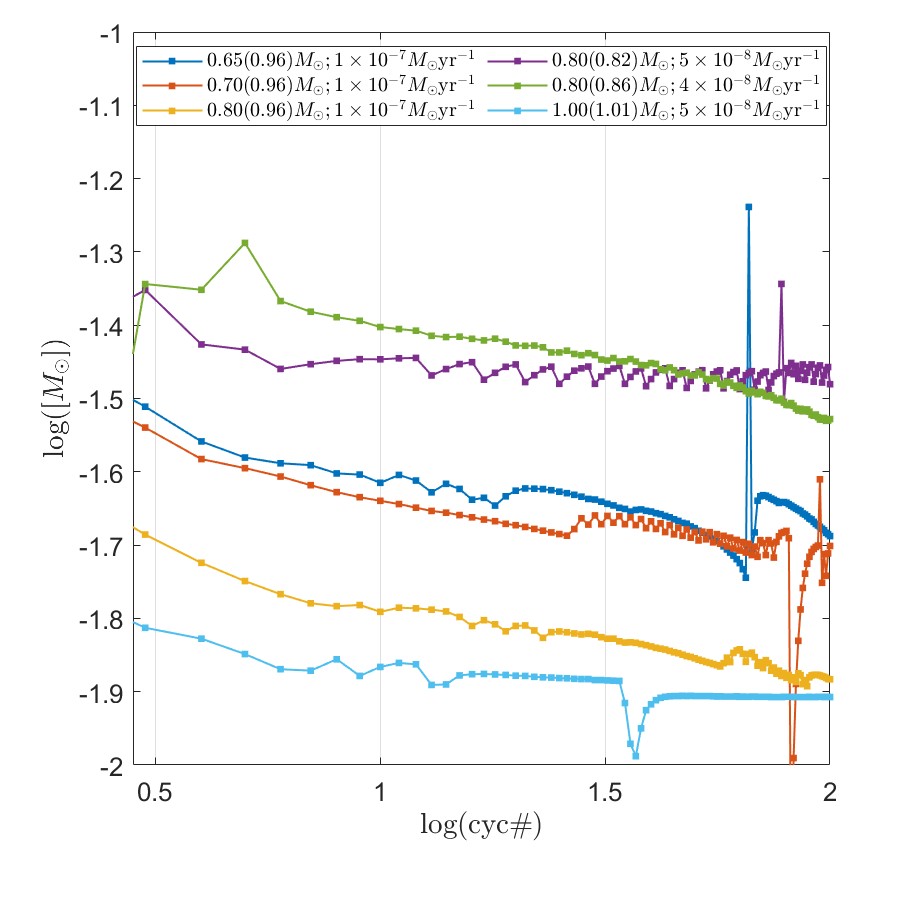}}
	\caption{Models with ejected mass of order $10^{-2}-10^{-1}M_\odot \rm yr^{-1}$. The legend details the initial WD mass along with (in parentheses) the WD masses at the evolutionary point shown in the figure.}\label{fig:mejV445pup}
	\end{center}
\end{figure}

In contrast with the strong constraint that our models place on the accretion rate, the WD mass could be anything below $\sim1.0M_\odot$. We show in Figure \ref{fig:mejV445pup} the possible HN models with $m_{\rm ej}$ of order the estimate for V445 Puppis determined by \cite{Banerjee2023}.
\cite{Piersanti2014} explored models of WD masses in the range $0.6-1.0M_\odot$ with accretion of helium at rates in the range $10^{-9}-10^{-5}M_\odot\rm yr^{-1}$. They obtain what they refer to as mild flashes and dynamical helium flashes for higher and lower accretion rates (within their explored range), respectively. For the lowest end of their accretion rate range, they report the occurrence of helium detonation. This roughly coincides with our three-regime results --- non-accretive models, helium nova models, and uninterrupted accretion that led to SNIa signature models (and UT type) --- while there are some differences, the main one being that they obtained helium novae for accretion rates that reside in our regime of uninterrupted accretion that led to SNIa-ignition signatures.
However, considering the many differences between our codes --- 
such as their treatment of the envelope as quasi-static layers with simplified mass loss, while we follow dynamically expanding shells and hydrodynamic ejection;
their simulations spanning a few consecutive cycles while ours carries on for thousands of cycles while discarding the initial few as they reflect the effects of initial conditions; and our inclusion of a more comprehensive nuclear network \cite[]{Prikov1995} along with updated OPAL opacities \cite[]{Iglesias1996} --- the agreement is remarkably good.

Following this, we can calculate a rough estimated prediction of the next eruption simply by multiplying the average cyclic accretion rate by the amount of accreted mass, and since this regime of accretion rates dictates $m_{\rm acc}\approx m_{\rm ej}$ we limit the accreted mass to $10^{-2}-10^{-1}M_\odot$ as well. This yields a recurrence period of order $10^5-10^6$ years. 





\cite{Banerjee2023} expresses that the huge amount of dust production (from which they deduced the ejected mass of order $10^{-2}-10^{-1}M_\odot$) should have triggered a SNIa. Our results here show that just a slightly higher accreted mass would have! The models show an anti-correlation between the accretion rate and the amount of ejected mass. The amount of mass ejected by V445 Pup is right on the edge of the nova-producing regime. A lower accretion rate would lead to a more massive accreted envelope and a higher ejected mass that would trigger a SNIa. A few of our models that are just past this limit terminate without producing a nova eruption. Decreasing the accretion rate just a little more initiates a SNIa.

\section{Conclusions}\label{sec:conclusions}
We have carried out extensive simulations of the accumulation of helium onto the surface of WDs with initial masses in the range $0.65-1.0M_\odot$ covering the entire range of feasible accumulation rates ($10^{-4}-10^{-10}M_\odot\rm yr^{-1}$). We find that the amount of helium that a WD can hold depends not only on the WD mass but also strongly on the accretion rate, decreasing with the increase of either one of these two key parameters.
While we find this to be a \textit{continuous trend}, we also find there to be three regimes of helium accumulation rates, each leading to \textit{a fundamentally different evolutionary outcome}.
We define the three regimes as:

\begin{itemize}
\item[i)] 
$\gtrapprox10^{-5}M_\odot\rm yr^{-1}$: super Eddington regime with no accretion
\item[ii)] $\approx10^{-5}-10^{-8}M_\odot\rm yr^{-1}$: periodic helium-nova-producing regime --- HN type.
\item[iii)] $\lessapprox10^{-8}M_\odot\rm yr^{-1}$: prolonged quiescent accumulation leading to a single mighty explosion --- UT and SN types.%
\end{itemize}
While there is some variance in these regimes depending on the WD mass, we refer to Table \ref{tab:mdls} for a more resolved range.

The first regime is of high accumulation rates for which the helium cannot be sustained due to the high rate producing Eddington accretion. 

The second regime is an intermediate range of accumulation rates for which we obtained periodic helium novae --- HN type models. An important result here is that all the models (but the lowest rates) led the WD to a \textit{net mass growth} while within this regime the lower rates resulted in an ejected-to-accreted mass ratio of close to unity, while the higher accretion rates led in some cases to non-ejective, mild novae. This is in contrast with hydrogen novae for which there \textit{are} hydrogen accumulation rates (low rates) that lead the WD to a net mass loss.

The third regime is of low accumulation rates for which we obtained \textit{quiescent secular uninterrupted helium accretion} that, for most cases, culminated with \textit{distinct signs} of the ignition of a SNIa --- SN type model. A dynamical simulation of the explosion and nuclear yields of a chosen SN-type model showed the end result to be a double-detonation SNIa (Paper 2). These signatures were obtained for all the models in this regime except for the highest rates in this regime, for which we did not see clear signs of an SNIa ignition. We defined these models as undetermined transients (UT type models). 

A key conclusion stemming from this work is that SNIae might be able to occur at WD masses that are significantly less than the Chandrasekhar mass, of order $\lessapprox1.0M_\odot$, as the result of slow, secular helium accumulation over \textit{hundreds of millions of years}, leading to signs of ignition, provided the input conditions we defined here can occur in reality. The basic condition is an accumulation rate towards the higher edge of the lowest rate regime, i.e., of order $\sim10^{-8}M_\odot\rm yr^{-1}$. We have deduced the possible types of systems that may produce such conditions to be either a CV or symbiotic system in which the donor (RD, RGB or AGB) transfers hydrogen-rich mass to the WD at a rate of order $\sim10^{-7}M_\odot\rm yr^{-1}$, i.e., the type that may produce RNe, or a helium donor transferring mass at the required rate, similar to the conditions found in V445 Puppis. We find this system to experience conditions very close to those required to ignite a SNIa  --- an average cyclic rate of order $0.5-1\times10^{-7}M_\odot\rm yr^{-1}$. The results here indicate that if it were a little bit lower, it would be in our regime of uninterrupted accretion, meaning that it may have been a UT or SN type and could have ignited a sub-Chandrasekhar SNIa, possibly a double detonation. We speculate that the few models that we have defined as UT type might be possible double detonation candidates \cite[as proposed by, e.g.,][]{Shen2009,Maoz2014,Starrfield2021}. We defer further analysis of the UT type models to future work.

An important finding here is that while it has been shown for hydrogen novae that although the accretion rate changes between eruptions, the average-cyclic accretion rate determines the outcome via the time to accrete the required triggering mass for the given $M_{\rm WD}$ \cite[]{Hillman2020,Hillman2021b,Hillman2021a}, here, for helium novae, we have shown that the \textit{actual accretion rate} is important, in agreement with the general trend found by \cite{Piersanti2014}. This means that a temporary high rate may trigger a helium nova at less mass than expected from the average-cyclic rate.

We stress that if CVs in the regime of RNe may be able to culminate as a SNIa as a result of helium accumulation via hydrogen accretion, this constitutes an \textit{additional method} for these systems to do this, the other being the net mass growth via RN eruptions. 

Since the WD masses that reside in the regime of RNe is $\gtrsim1.0M_\odot$ \cite[e.g.,][]{Yaron2005}, while we find in this work that $M_{\rm WD,i}\lesssim1.0M_\odot$, may end as SNIa via direct helium accretion, it turns out that these two mechanisms compliment each other, and together they essentially cover all possibilities of WD masses. A caveat to this conclusion is the absence of $M_{\rm WD,i}>1.0M_\odot$ in this work. If they were to show the possibility of igniting SNIae via a helium donor, that would mean that any WD mass could lead to a SNIa via helium accumulation. We reserve this aspect of the research for future exploration.

\section*{Acknowledgements}
We acknowledge support for this project from the European Union's Horizon 2020 research and innovation program under grant agreement No 865932-ERC-SNeX. This work was also supported by the Azrieli College of Engineering – Jerusalem Research Fund.
We acknowledge the Ariel HPC Center at Ariel University for providing computing resources that have contributed to the research results reported in this paper.

\appendix

\section{Additional figures}
\label{sec:A_adtnl_figs}
\begin{figure}
	\begin{center}
{\includegraphics[trim={4.0cm 1.5cm 6.5cm 0.0cm}, clip, width=0.99\columnwidth]{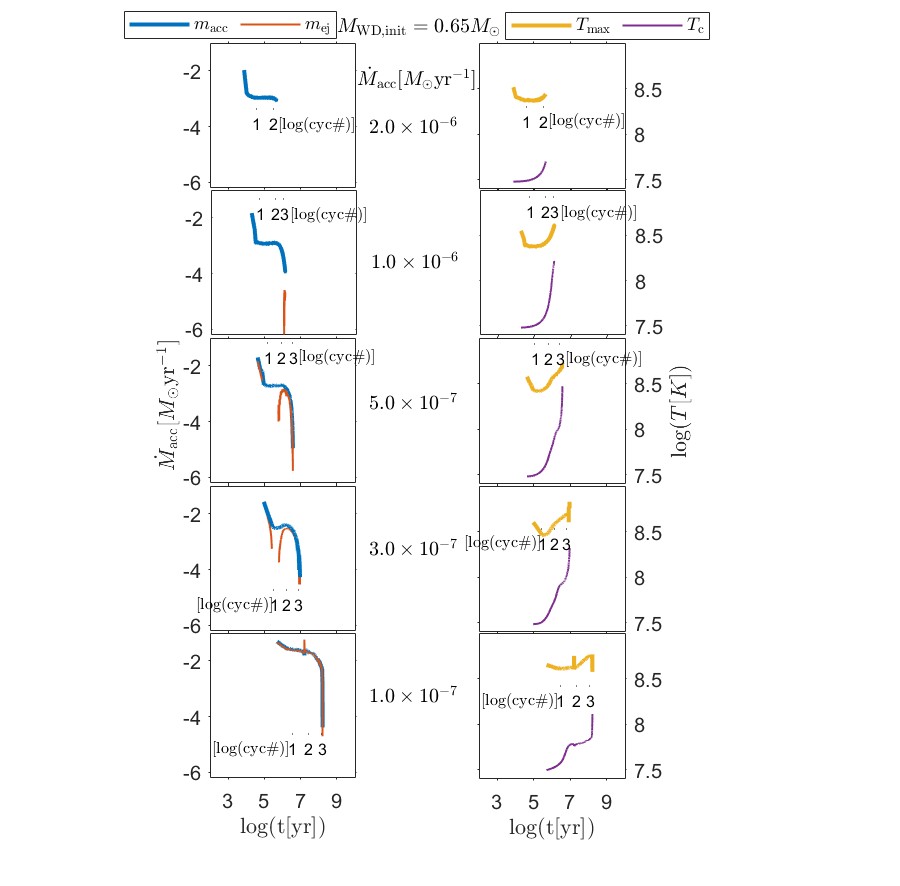}}
\caption{Accreted and ejected mass ($m_{\rm acc}$ and $m_{\rm ej}$ respectively) and core and maximum temperatures ($T_{\rm c}$ and $T_{\rm max}$ respectively) per eruption vs. evolutionary time and vs. cycle number on logarithmic scales for our periodic helium-nova-producing models with initial WD masses of $0.65M_\odot$.}\label{fig:macc_mej_065_A}
\end{center}
\end{figure}

\begin{figure}
	\begin{center}
{\includegraphics[trim={4.0cm 14.2cm 6.5cm 0.0cm}, clip, width=0.99\columnwidth]{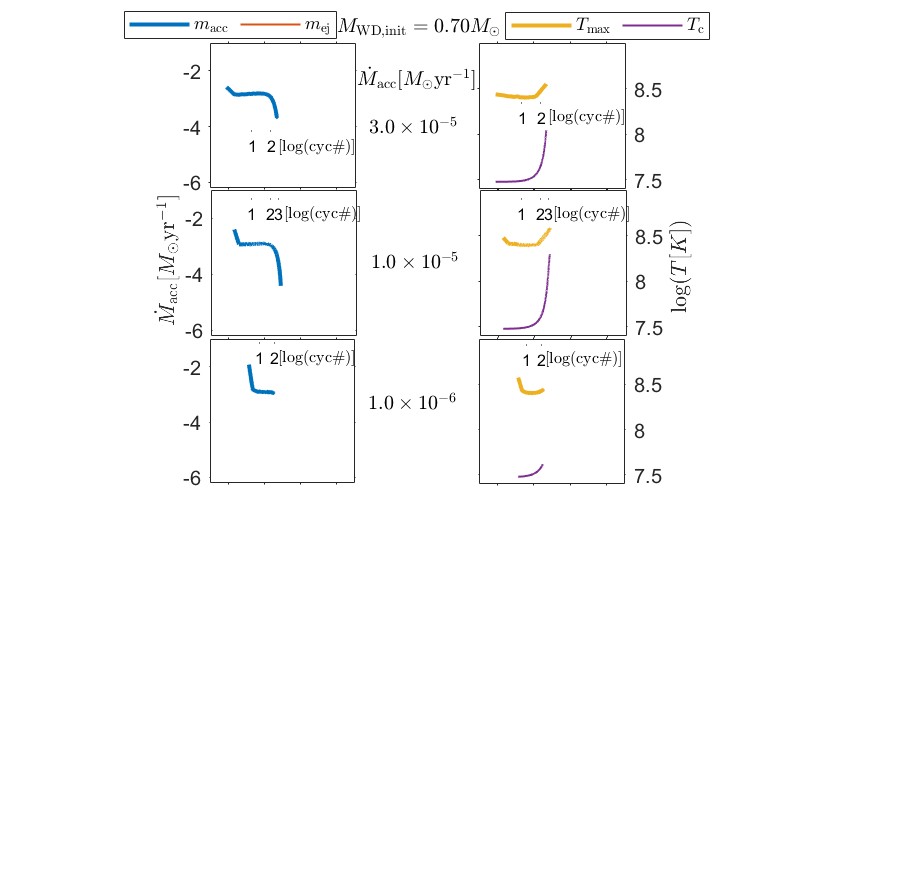}}
{\includegraphics[trim={4.0cm 12.0cm 6.5cm 1.56cm}, clip, width=0.99\columnwidth]{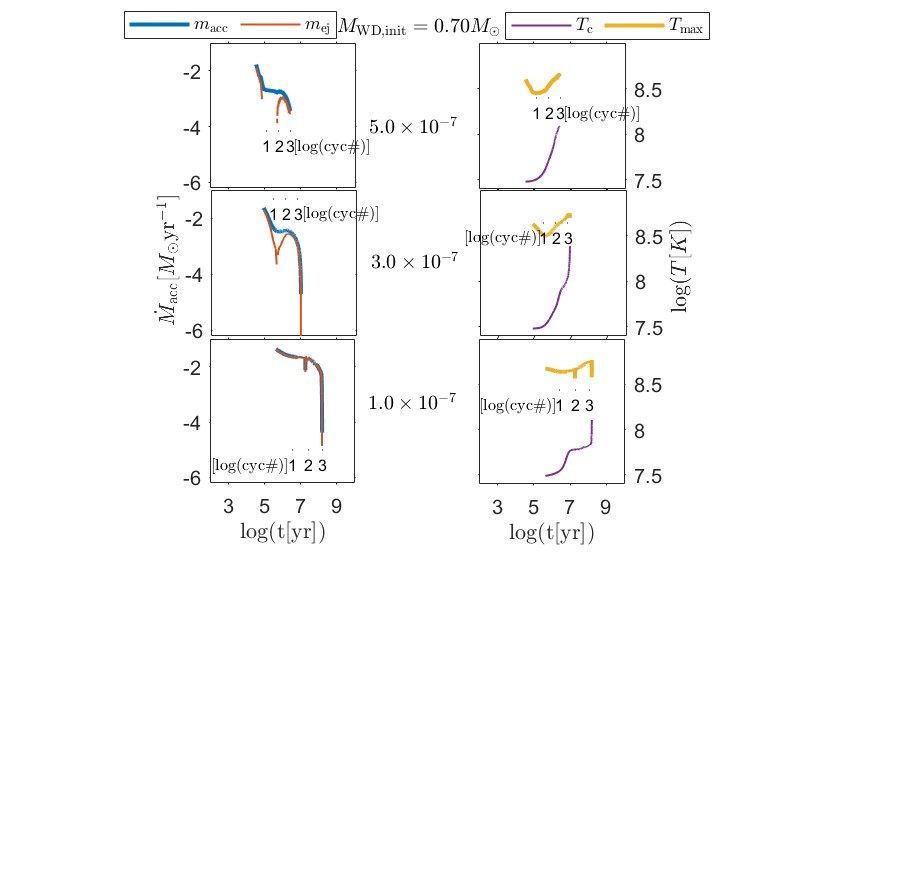}}
\caption{Description as in Figure \ref{fig:macc_mej_065_A} for initial WD masses of $0.70M_\odot$. }\label{fig:macc_mej_070_A}
	\end{center}
\end{figure}

\begin{figure}
	\begin{center}
{\includegraphics[trim={4.0cm 8.9cm 6.5cm 0.0cm}, clip, width=0.99\columnwidth]{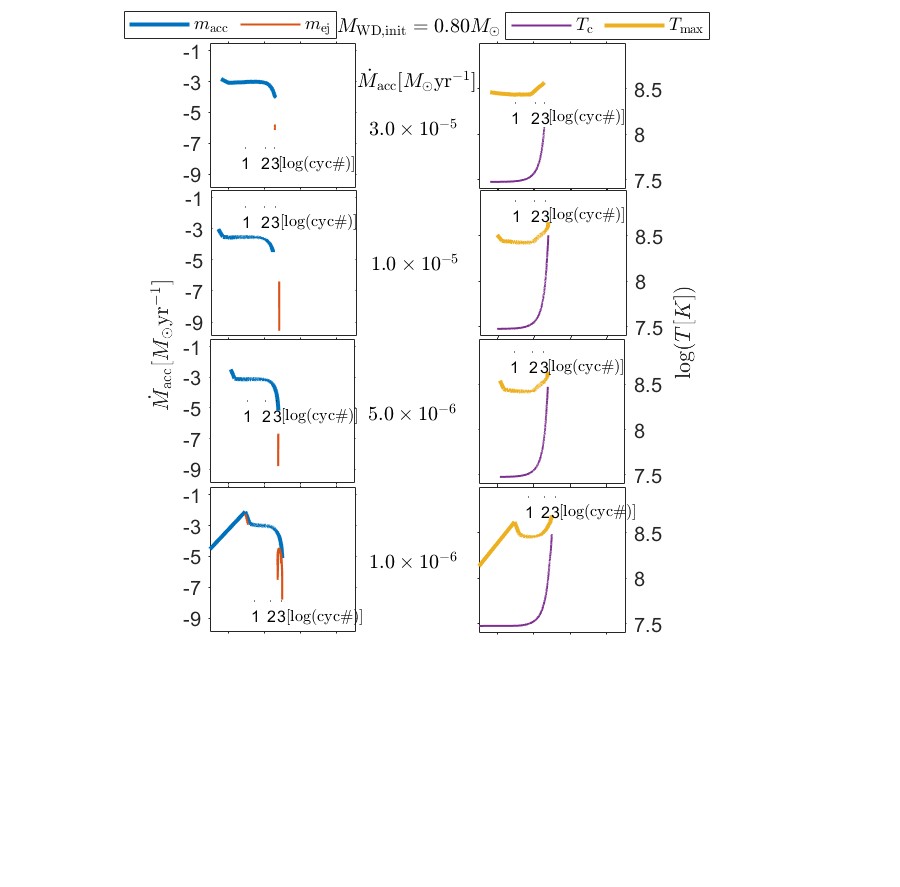}}
{\includegraphics[trim={4.0cm 12.0cm 6.5cm 1.55cm}, clip, width=0.99\columnwidth]{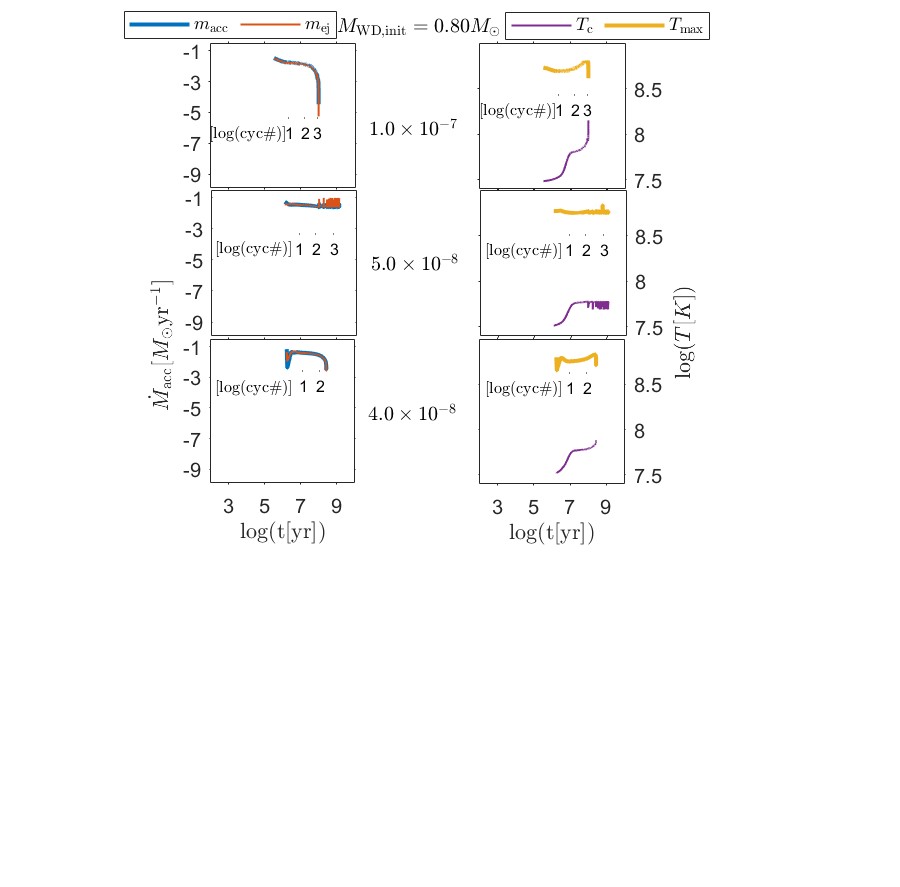}}

\caption{Description as in Figure \ref{fig:macc_mej_065_A} for initial WD masses of $0.80M_\odot$.}\label{fig:macc_mej_080_A}
	\end{center}
\end{figure}

\begin{figure}
	\begin{center}
{\includegraphics[trim={4.0cm 1.5cm 6.5cm 0.0cm}, clip, width=0.99\columnwidth]{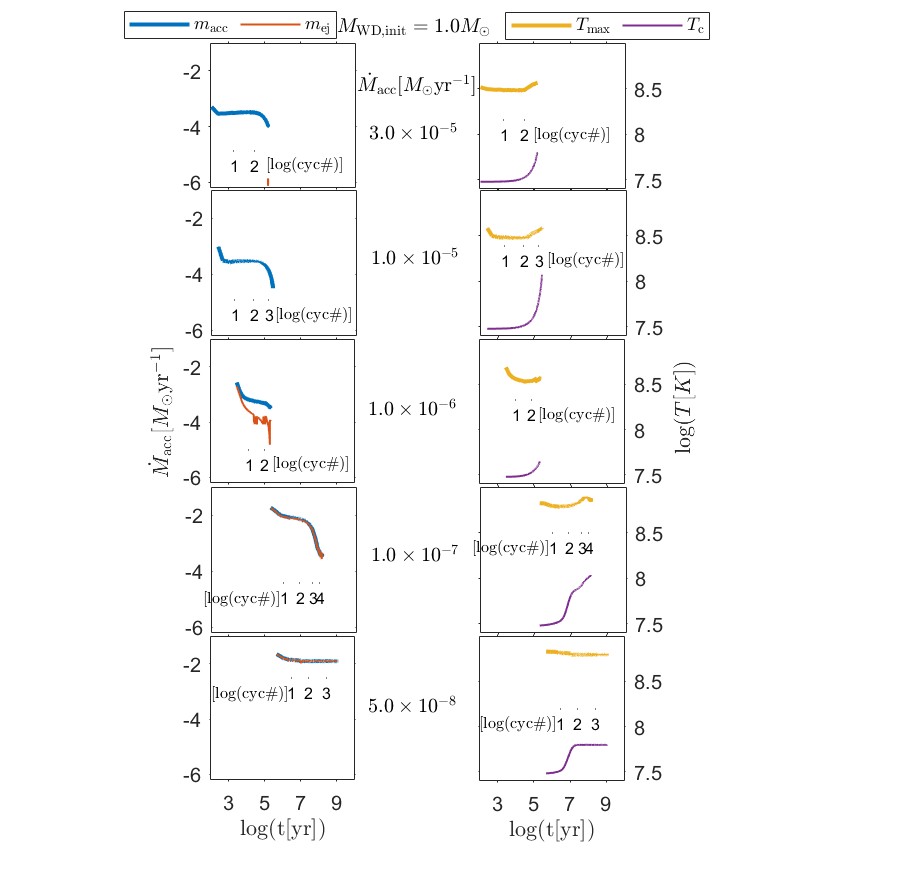}}
\caption{Description as in Figure \ref{fig:macc_mej_065_A} for initial WD masses of $1.0M_\odot$.}\label{fig:macc_mej_100_A}
	\end{center}
\end{figure}

\begin{figure*}
	\begin{center}
{\includegraphics[trim={0.8cm 0.0cm 1.8cm 0.0cm}, clip, width=0.65\columnwidth]{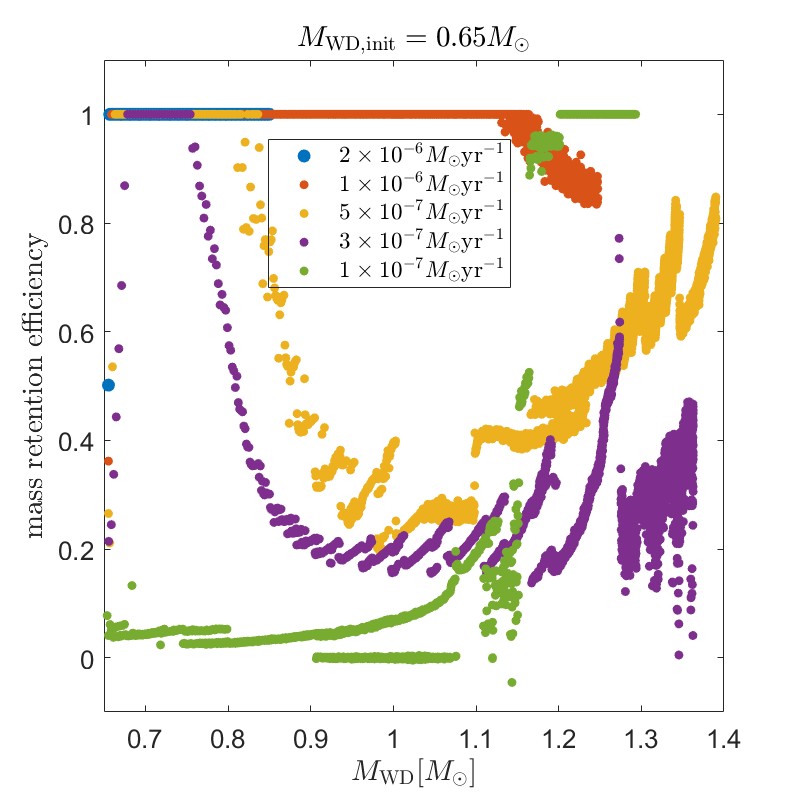}}
{\includegraphics[trim={0.8cm 0.0cm 1.8cm 0.0cm}, clip, width=0.65\columnwidth]{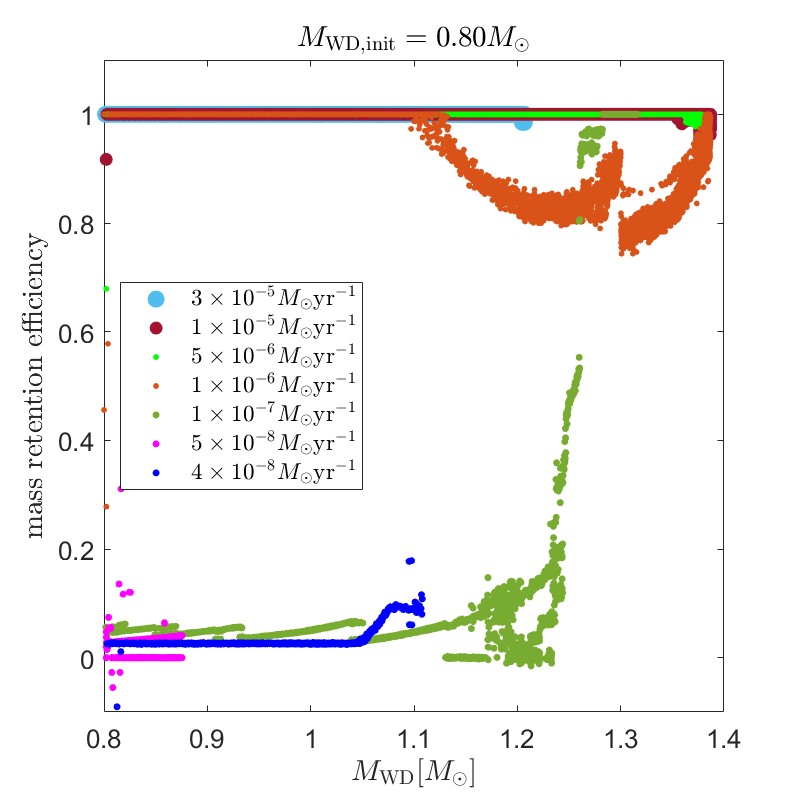}}
{\includegraphics[trim={0.8cm 0.0cm 1.8cm 0.0cm}, clip, width=0.65\columnwidth]{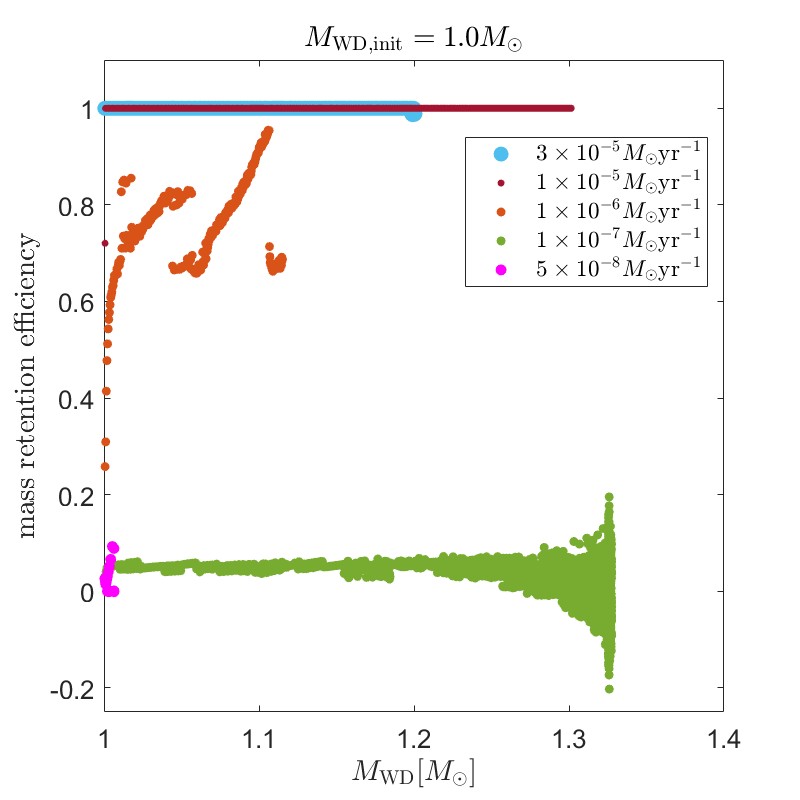}}
\caption{Description as in Figure \ref{fig:retention} for initial WD masses of $0.65$, $0.80$ and $1.0M_\odot$.}\label{fig:retention_A}
	\end{center}
\end{figure*}

Figures \ref{fig:macc_mej_065_A}, \ref{fig:macc_mej_070_A}, \ref{fig:macc_mej_080_A} and \ref{fig:macc_mej_100_A} show the accreted and ejected masses, and the core and maximum temperatures ($m_{\rm acc}$, $m_{\rm ej}$, $T_{\rm c}$ and $T_{\rm max}$ respectively) for our 0.65, 0.70, 0.80 and 1.0$M_\odot$ models respectively as described in Figure \ref{fig:macc_mej},and their respective mass retention efficiencies are shown in Figure \ref{fig:retention_A}.

\begin{figure*}
	\begin{center}
{\includegraphics[trim={1.7cm 1.0cm 1.2cm 0.cm}, clip, width=0.68\columnwidth]{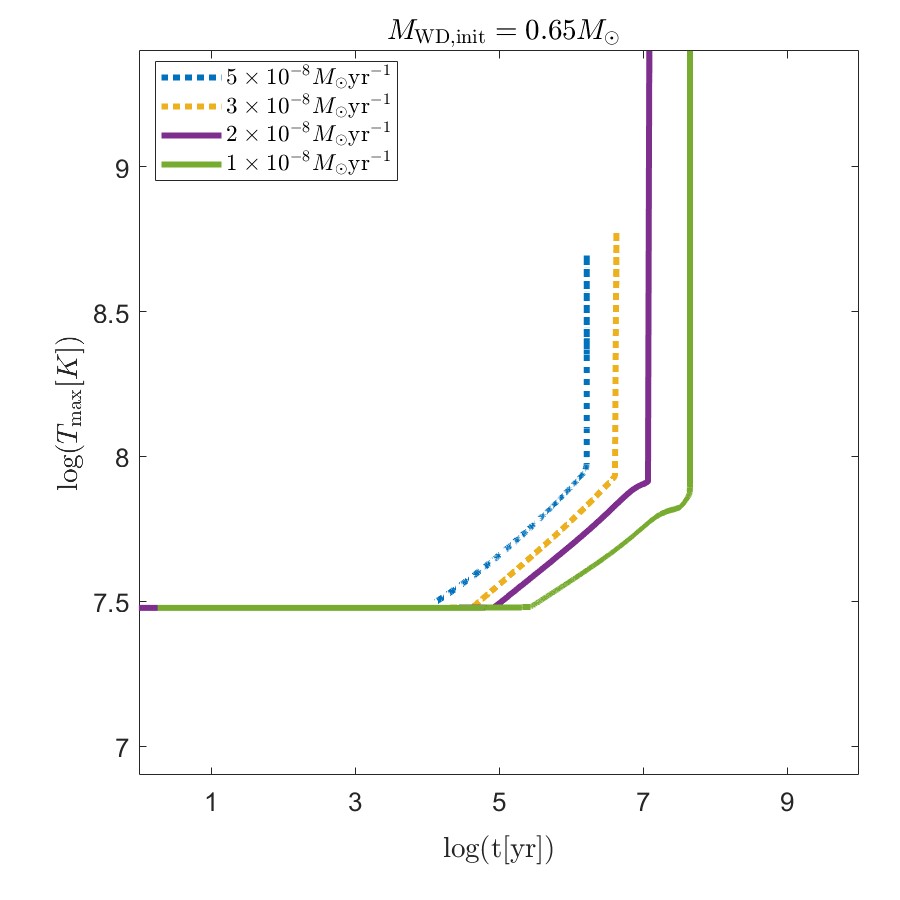}}
{\includegraphics[trim={1.7cm 1.0cm 1.2cm 0.cm}, clip, width=0.68\columnwidth]{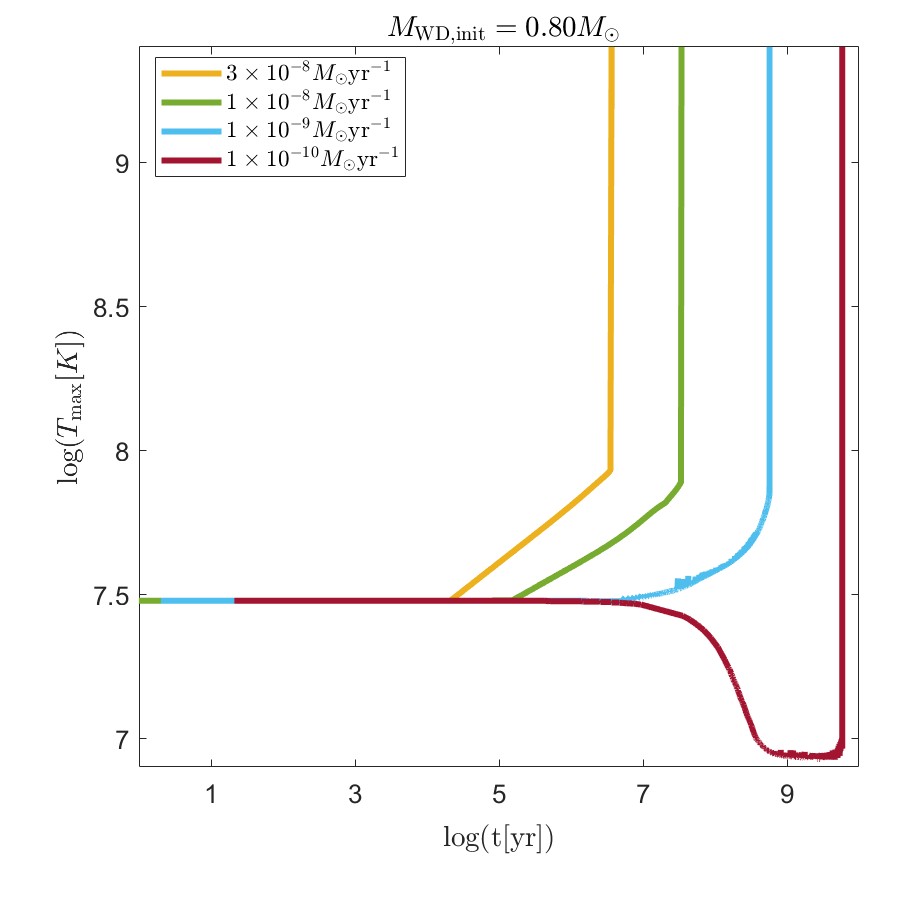}}
{\includegraphics[trim={1.7cm 1.0cm 1.2cm 0.cm}, clip, width=0.68\columnwidth]{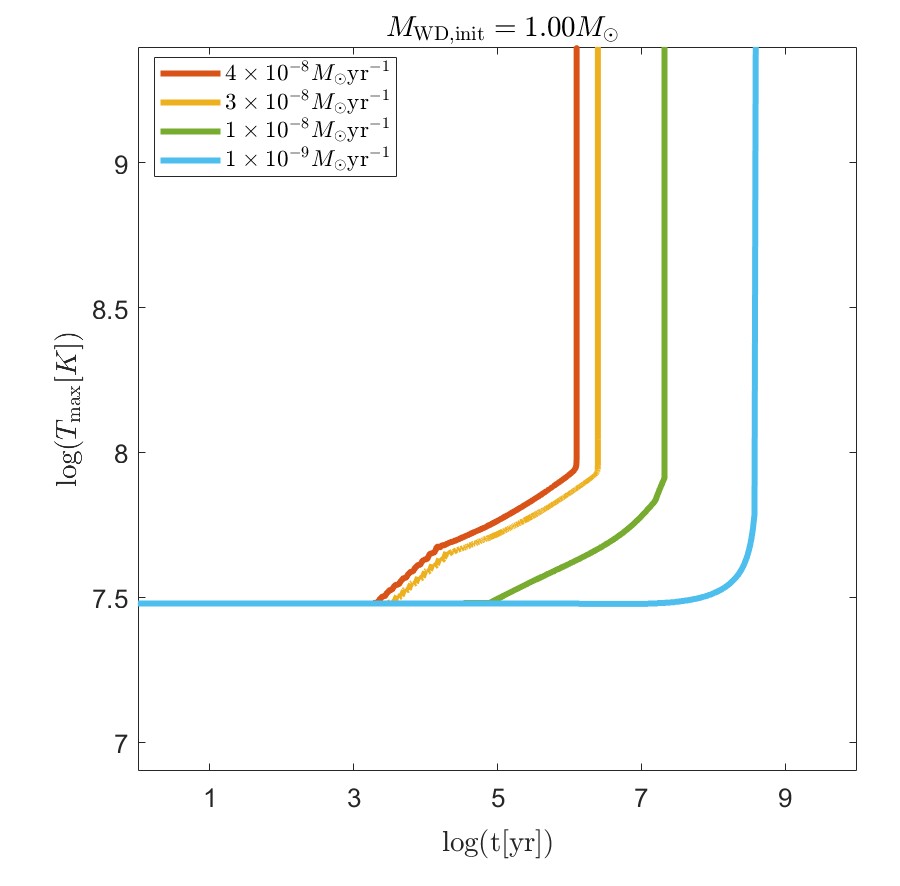}}
\caption{Description as in Figure \ref{fig:T_t_SNIa_070}
 for our $0.65M_\odot$ models (left); $0.80M_\odot$ models (center); and $1.0M_\odot$ models (right).}\label{fig:App_T_t_SNIa}
	\end{center}
\end{figure*}

We show in Figure \ref{fig:App_T_t_SNIa} The evolution of the maximum temperature ($T_{\rm max}$) over time for our UT and SN type models with initial WD masses of $0.65$, $0.80$ and $1.0M_\odot$ as described for our initial $0.70M_\odot$ models in Figure \ref{fig:T_t_SNIa_070}.

\begin{figure*}
	\begin{center}
{\includegraphics[trim={0.5cm 0.cm 0.5cm 0.cm}, clip, width=0.99\columnwidth]{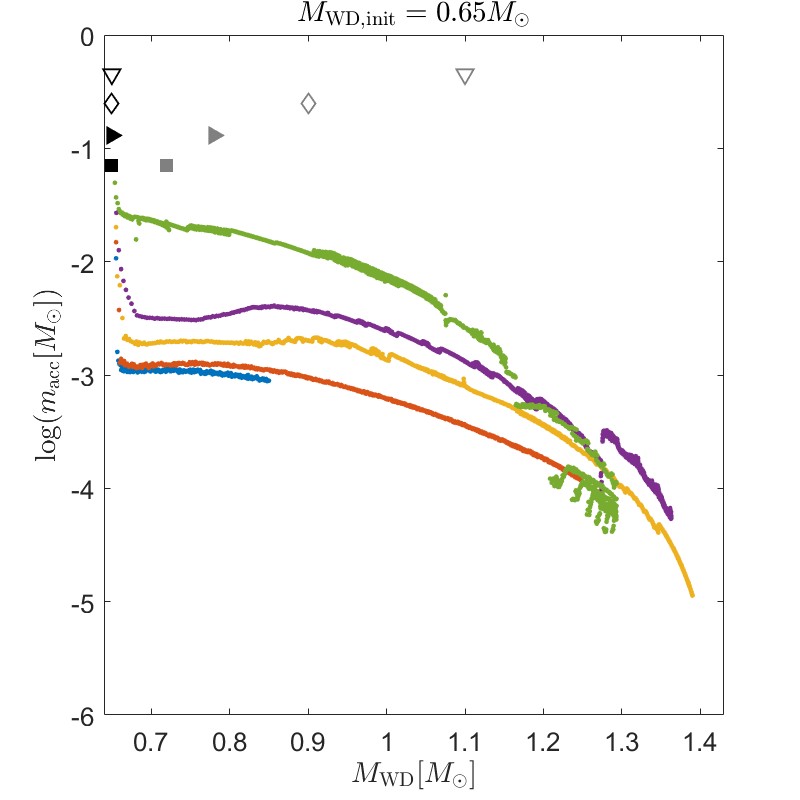}}
{\includegraphics[trim={0.5cm 0.0cm 0.5cm 0.0cm}, clip, width=0.99\columnwidth]{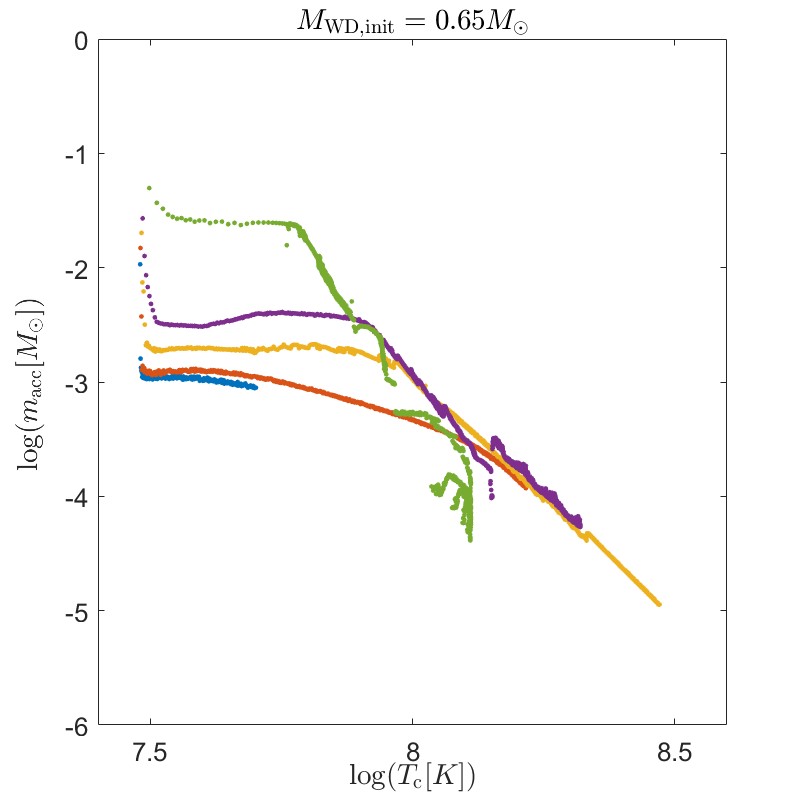}}\\
{\includegraphics[trim={0.5cm 0.cm 0.5cm 0.cm}, clip, width=0.99\columnwidth]{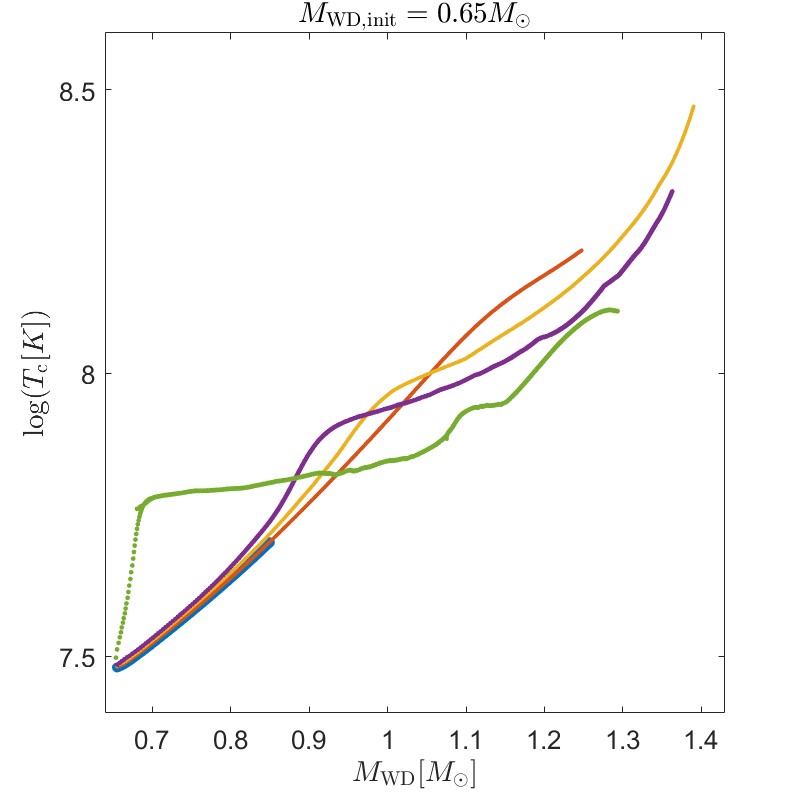}}
{\includegraphics[trim={-0.8cm -4.5cm -4.0cm 0.0cm}, clip, width=0.99\columnwidth]{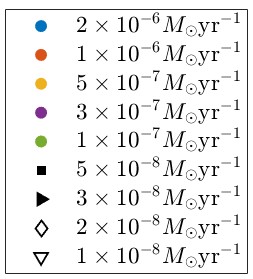}}\\
\caption{Description as in Figure \ref{fig:macc_MWD_Tc_070} for initial WD masses of $0.65M_\odot$.
}\label{fig:macc_MWD_TC_065_A}
\end{center}
\end{figure*}

\begin{figure*}
	\begin{center}
{\includegraphics[trim={0.5cm 0.5cm 0.5cm 0.5cm}, clip, width=0.99\columnwidth]{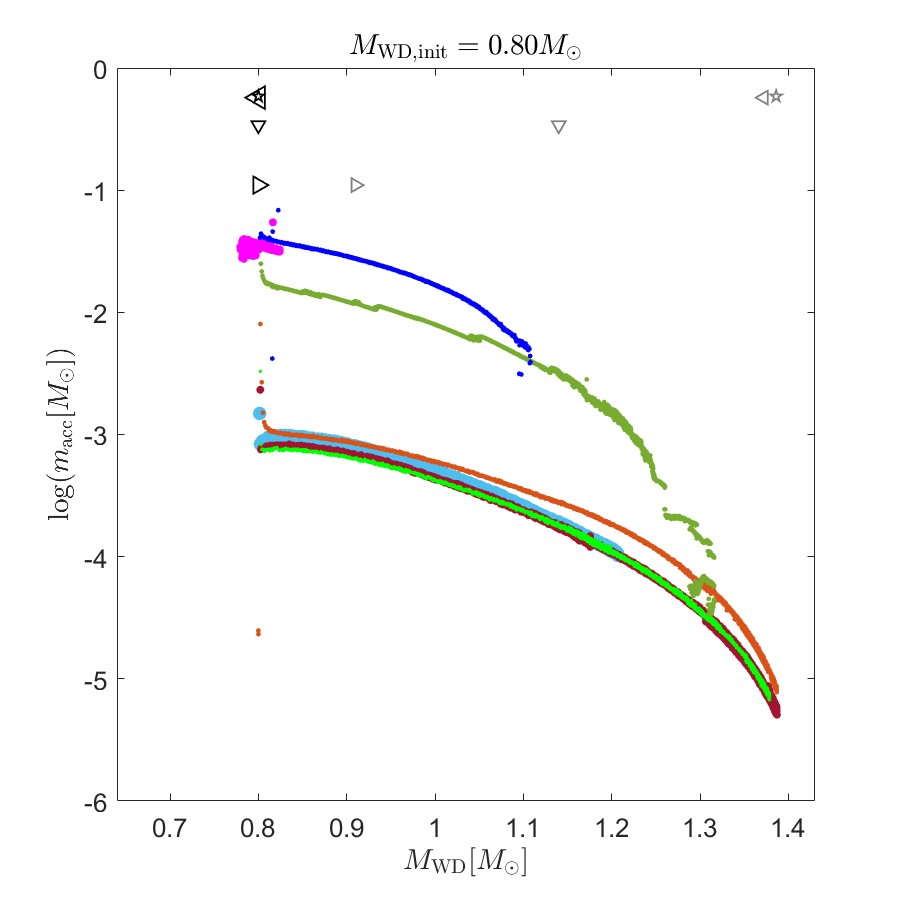}}
{\includegraphics[trim={0.5cm 0.0cm 0.5cm 0.5cm}, clip, width=0.99\columnwidth]{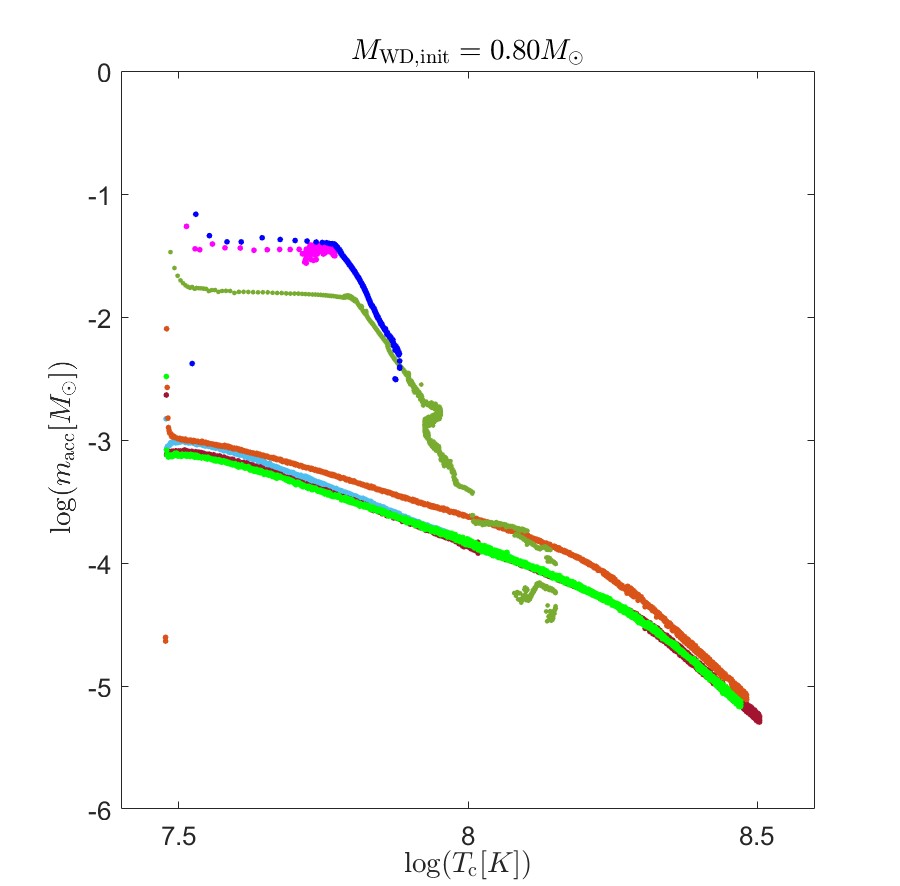}}\\
{\includegraphics[trim={0.5cm 0.5cm 0.5cm 0.5cm}, clip, width=0.99\columnwidth]{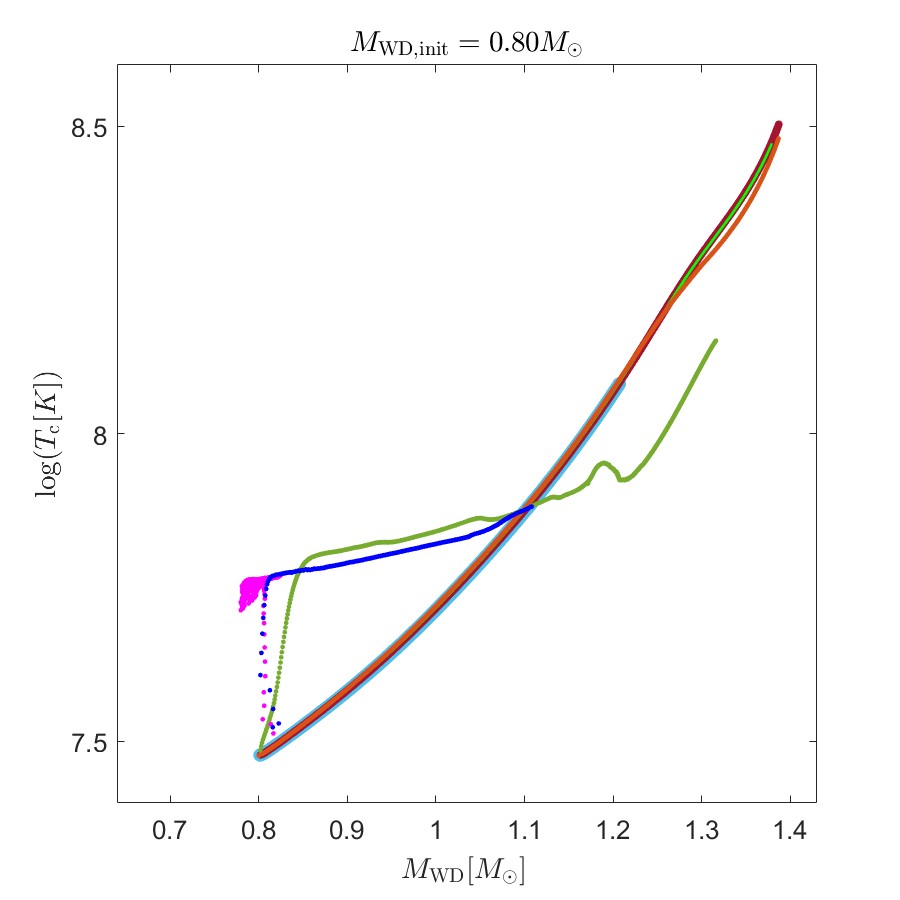}}
{\includegraphics[trim={-1.0cm -3.3cm -4.0cm 0.0cm}, clip, width=0.99\columnwidth]{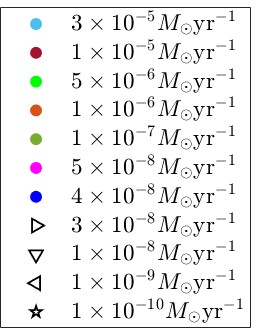}}\\
\caption{Description as in Figure \ref{fig:macc_MWD_Tc_070} for initial WD masses of $0.8M_\odot$.
}\label{fig:macc_MWD_TC_080_A}
\end{center}
\end{figure*}

\begin{figure*}
	\begin{center}
{\includegraphics[trim={0.5cm 0.5cm 0.5cm 0.5cm}, clip, width=0.99\columnwidth]{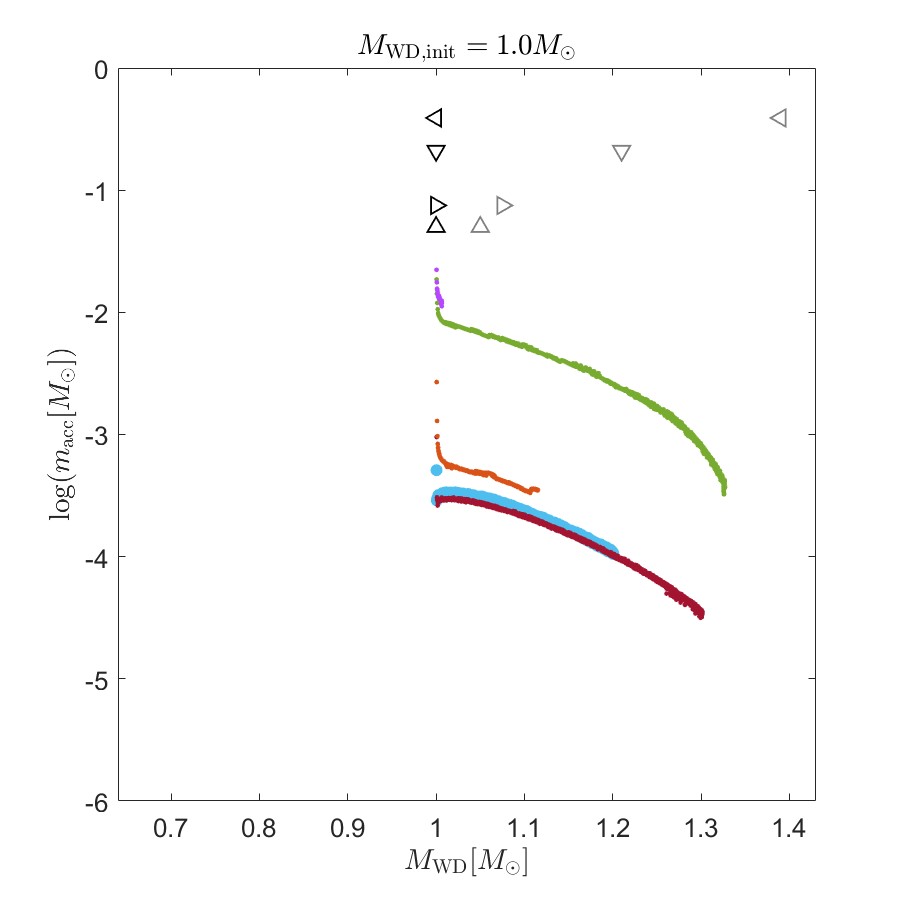}}
{\includegraphics[trim={0.0cm 0.2cm 0.6cm 0.5cm}, clip, width=0.99\columnwidth]{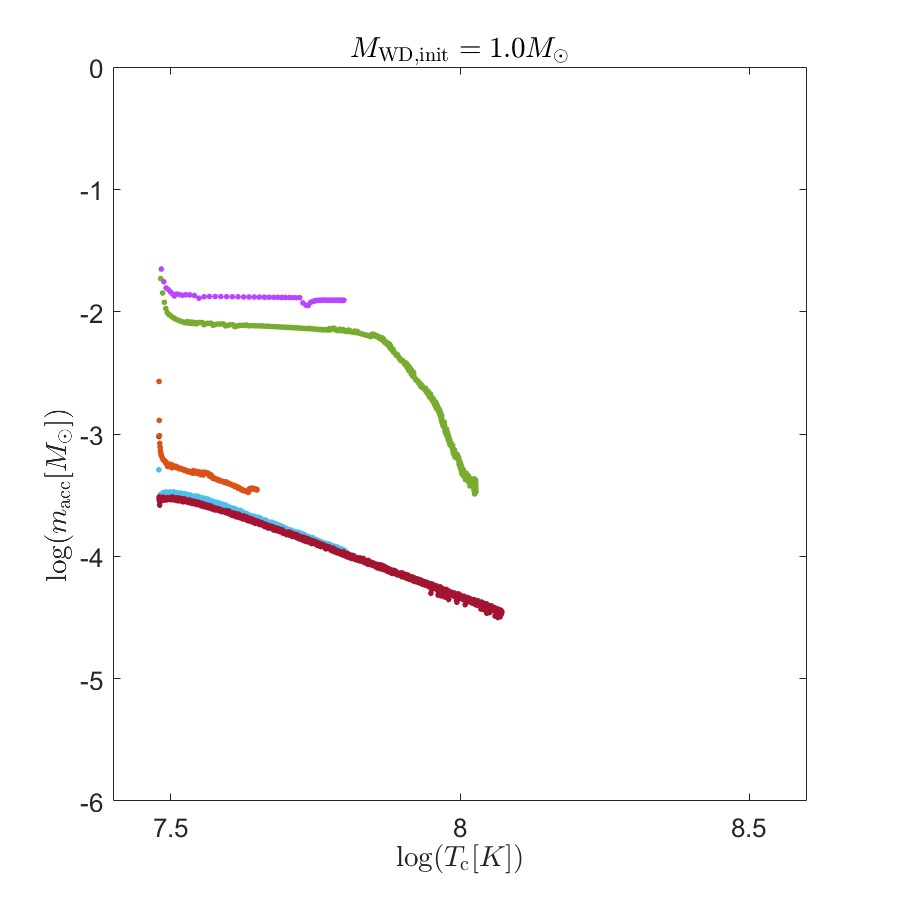}}\\
{\includegraphics[trim={0.5cm 0.5cm 0.5cm 0.5cm}, clip, width=0.99\columnwidth]{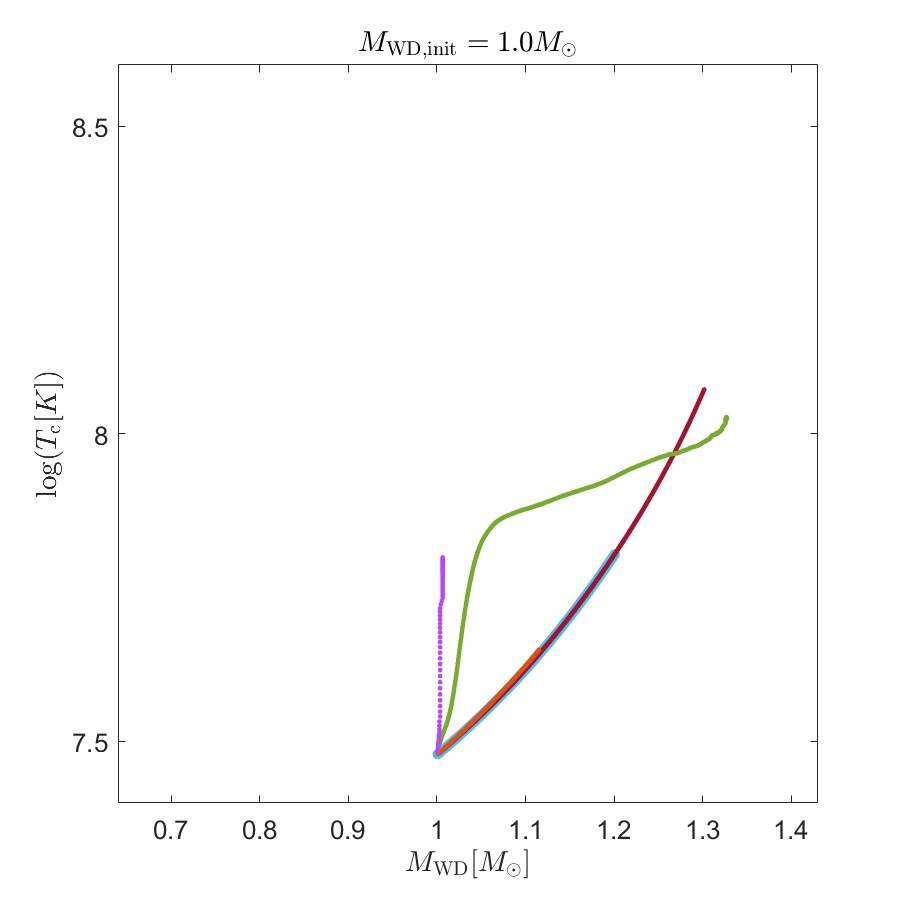}}
{\includegraphics[trim={-1.0cm -3.8cm -3.7cm 0.0cm}, clip, width=0.99\columnwidth]{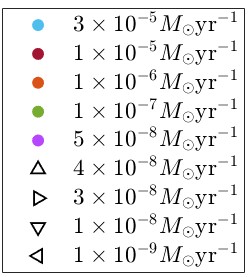}}\\

\caption{Description as in Figure \ref{fig:macc_MWD_Tc_070} for initial WD masses of $1.0M_\odot$.
}\label{fig:macc_MWD_TC_100_A}
	\end{center}
\end{figure*}

In Figures \ref{fig:macc_MWD_TC_065_A}, \ref{fig:macc_MWD_TC_080_A} and \ref{fig:macc_MWD_TC_100_A}we present the accreted mass ($m_{\rm acc}$) vs. WD mass ($M_{\rm WD}$); accreted mass ($m_{\rm acc}$) vs. core temperature ($T_{\rm c}$); and core temperature ($T_{\rm c}$) vs. WD mass
($M_{\rm WD}$) for our 0.65, 0.80 and 1.0$M_\odot$ models respectively as described for our initial $0.70M_\odot$ models in Figure \ref{fig:macc_MWD_Tc_070}.

\begin{figure}
	\begin{center}
{\includegraphics[trim={1.0cm 0.0cm 1.5cm 0.0cm}, clip, width=0.62\columnwidth]{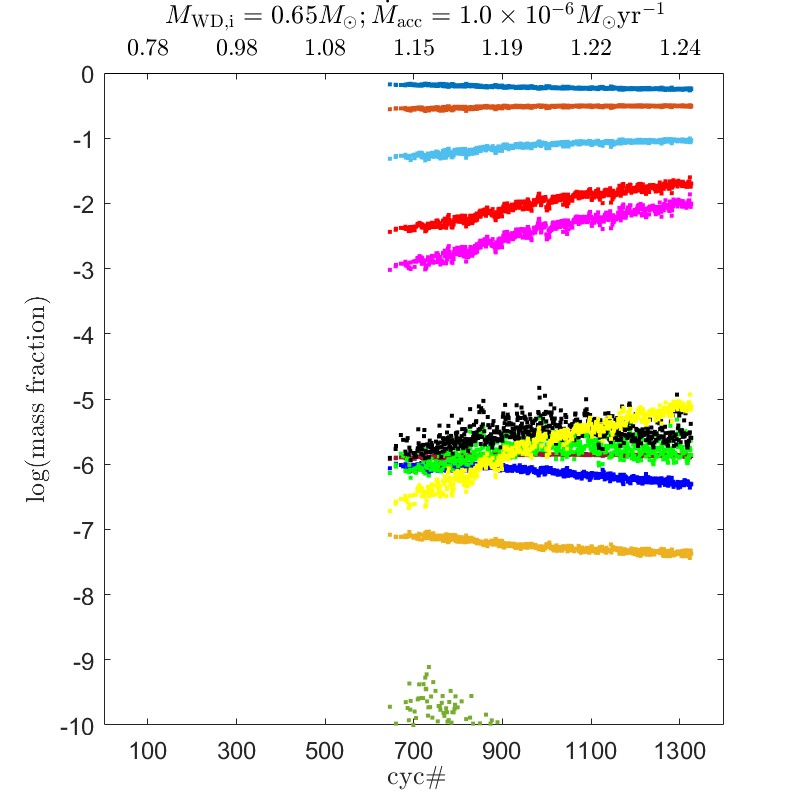}}
{\includegraphics[trim={22.5cm 5.6cm 0.7cm 4.0cm}, clip, width=0.31\columnwidth]{comp_for_legend.jpg}}\\

{\includegraphics[trim={1.0cm 0.0cm 1.5cm 0.0cm}, clip, width=0.62\columnwidth]{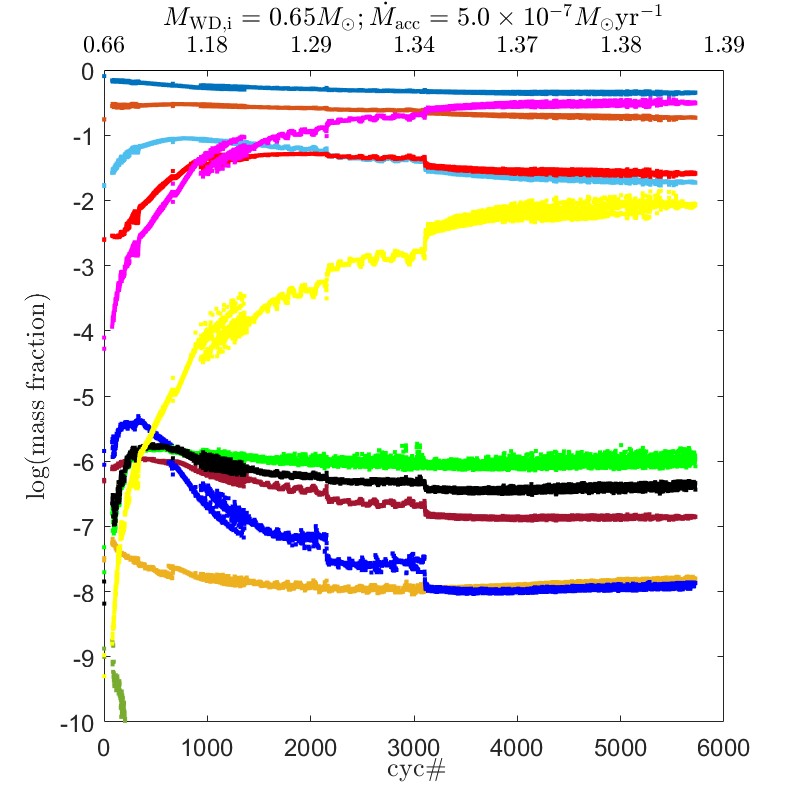}}
{\includegraphics[trim={22.5cm 5.6cm 0.7cm 4.0cm}, clip, width=0.31\columnwidth]{comp_for_legend.jpg}}\\
{\includegraphics[trim={1.0cm 0.0cm 1.5cm 0.0cm}, clip, width=0.62\columnwidth]{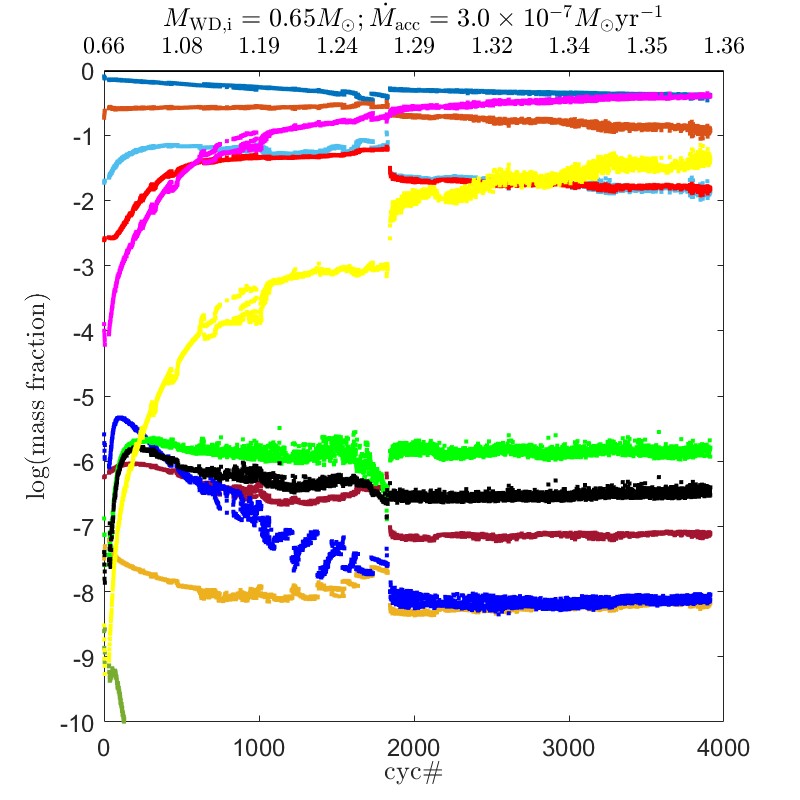}}
{\includegraphics[trim={22.5cm 5.6cm 0.7cm 4.0cm}, clip, width=0.31\columnwidth]{comp_for_legend.jpg}}\\
{\includegraphics[trim={1.0cm 0.0cm 1.5cm 0.0cm}, clip, width=0.6\columnwidth]{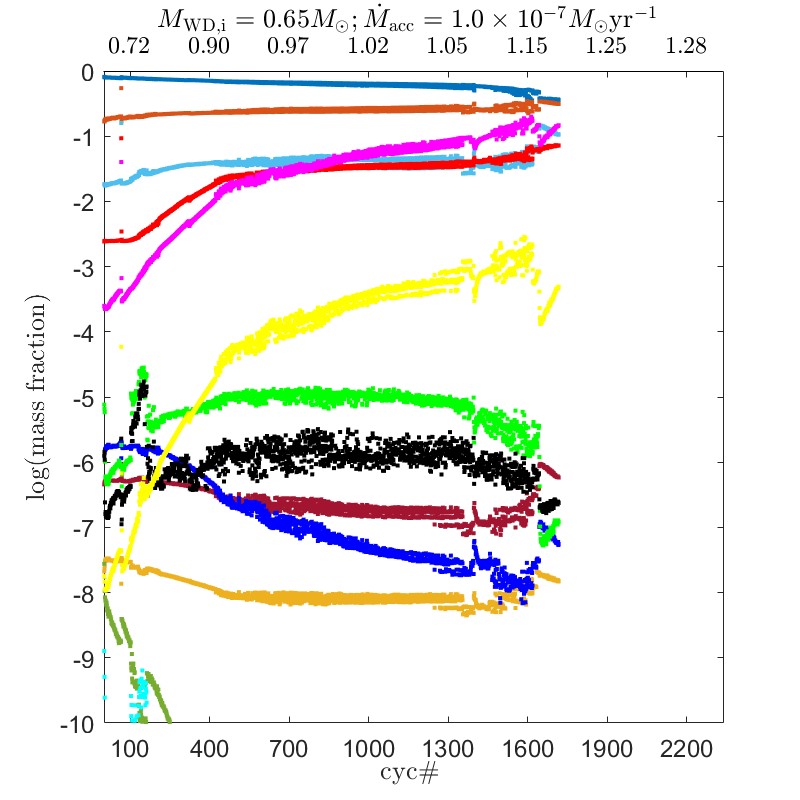}}
{\includegraphics[trim={22.5cm 5.6cm 0.4cm 4.0cm}, clip, width=0.31\columnwidth]{comp_for_legend.jpg}}\\
\caption{ Composition of ejected material per cycle for the ejective periodic helium nova models (HN type) with initial WD masses of $0.65$: Models $\#4-7$. (Model $\#3$, with a higher $\dot{M}_{\rm acc}$ experienced only non-ejective novae.)} \label{fig:comp_nova_065}
 \end{center}
\end{figure}

\begin{figure}
	\begin{center}
{\includegraphics[trim={1.0cm 0.0cm 1.5cm 0.0cm}, clip, width=0.62\columnwidth]{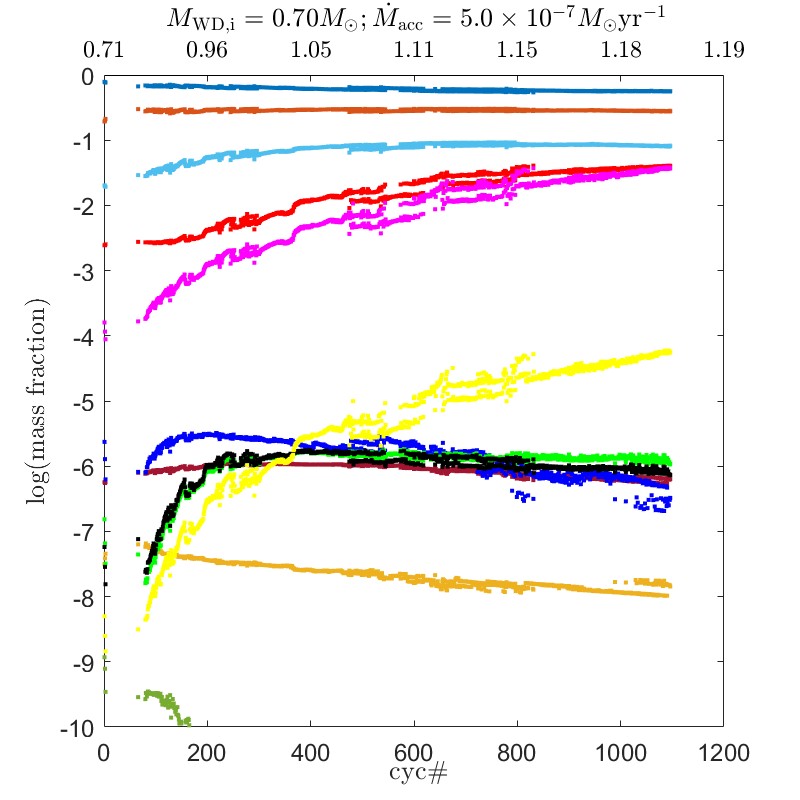}}
{\includegraphics[trim={22.5cm 5.6cm 0.7cm 4.0cm}, clip, width=0.31\columnwidth]{comp_for_legend.jpg}}\\
{\includegraphics[trim={1.0cm 0.0cm 1.5cm 0.0cm}, clip, width=0.62\columnwidth]{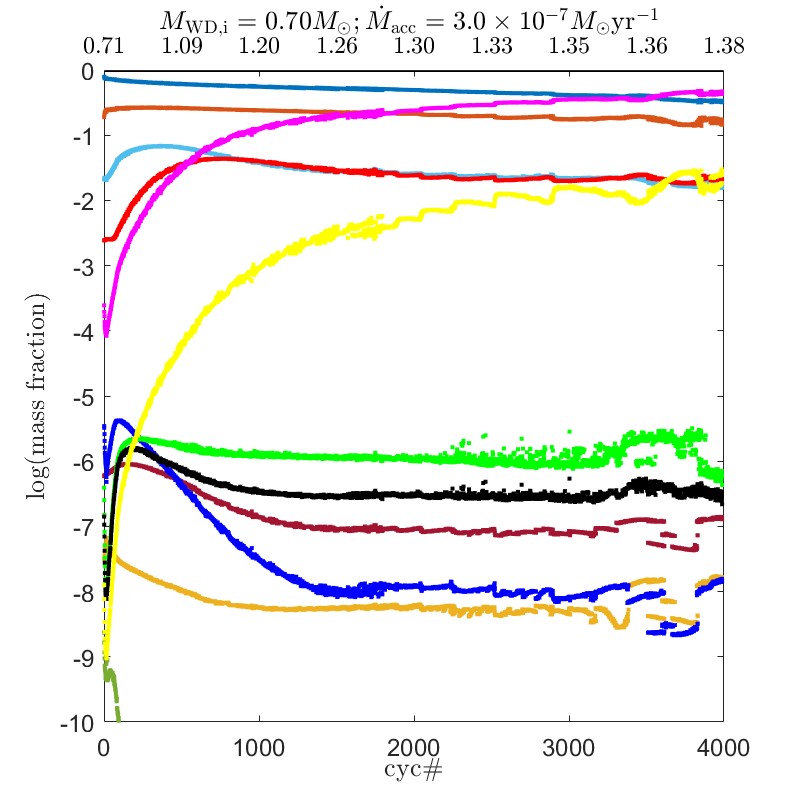}}
{\includegraphics[trim={22.5cm 5.6cm 0.7cm 4.0cm}, clip, width=0.31\columnwidth]{comp_for_legend.jpg}}\\
{\includegraphics[trim={1.0cm 0.0cm 1.5cm 0.0cm}, clip, width=0.62\columnwidth]{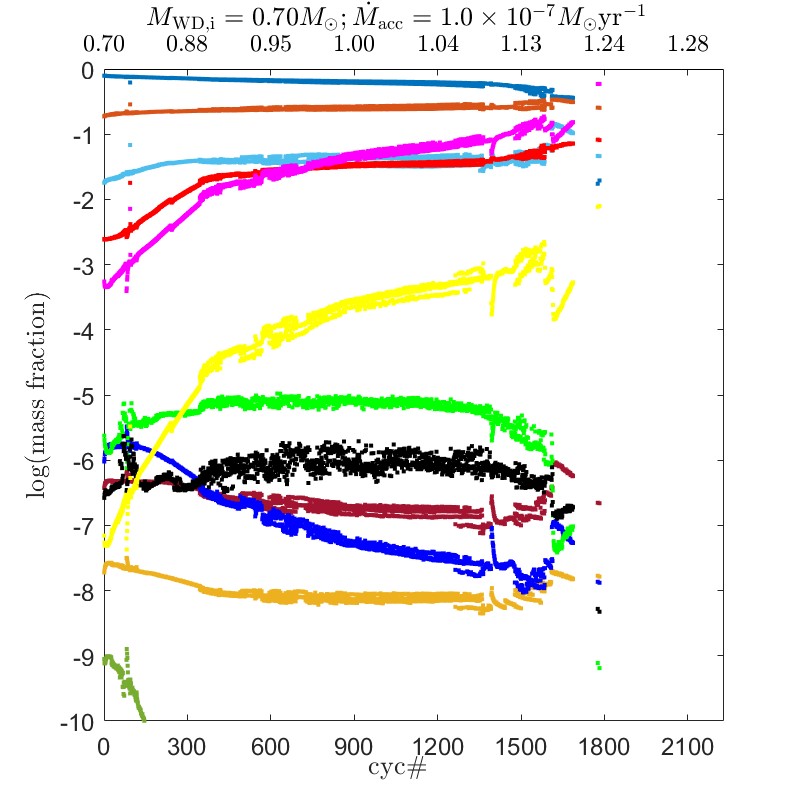}}
{\includegraphics[trim={22.5cm 5.6cm 0.7cm 4.0cm}, clip, width=0.31\columnwidth]{comp_for_legend.jpg}}\\
\caption{Description as in Figure \ref{fig:comp_nova_065} for initial WD masses of $0.7$: Models $\#17-19$. (Models $\#14-16$, with higher accretion rates, experienced only non-ejective novae.)}\label{fig:comp_nova_070}
	\end{center}
\end{figure}

\begin{figure}
	\begin{center}
{\includegraphics[trim={1.0cm 0.0cm 1.5cm 0.0cm}, clip, width=0.62\columnwidth]{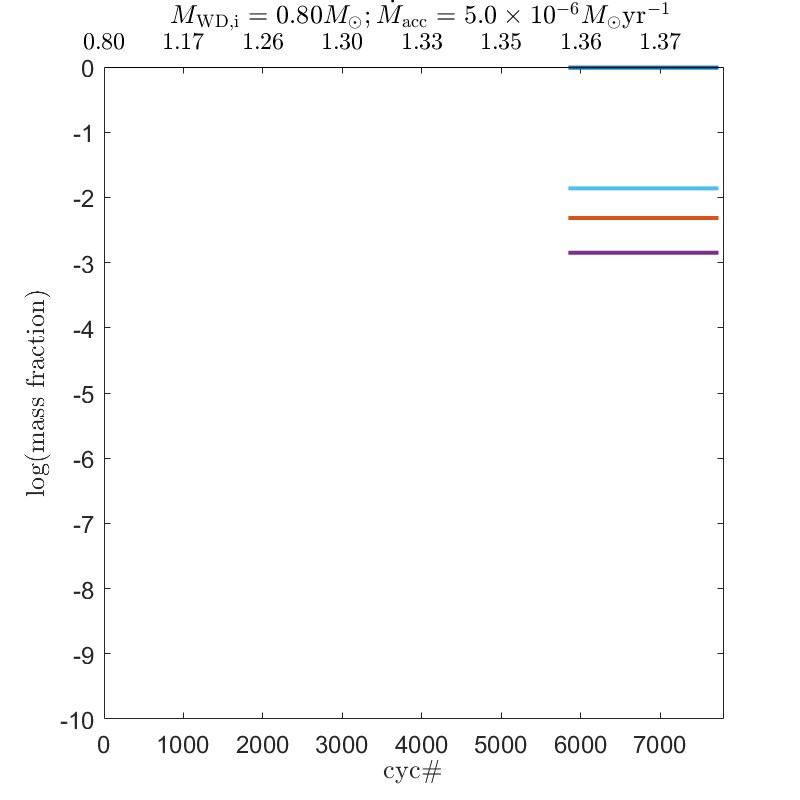}}
{\includegraphics[trim={22.5cm 5.6cm 0.7cm 4.0cm}, clip, width=0.31\columnwidth]{comp_for_legend.jpg}}\\
{\includegraphics[trim={1.0cm 0.0cm 1.5cm 0.0cm}, clip, width=0.62\columnwidth]{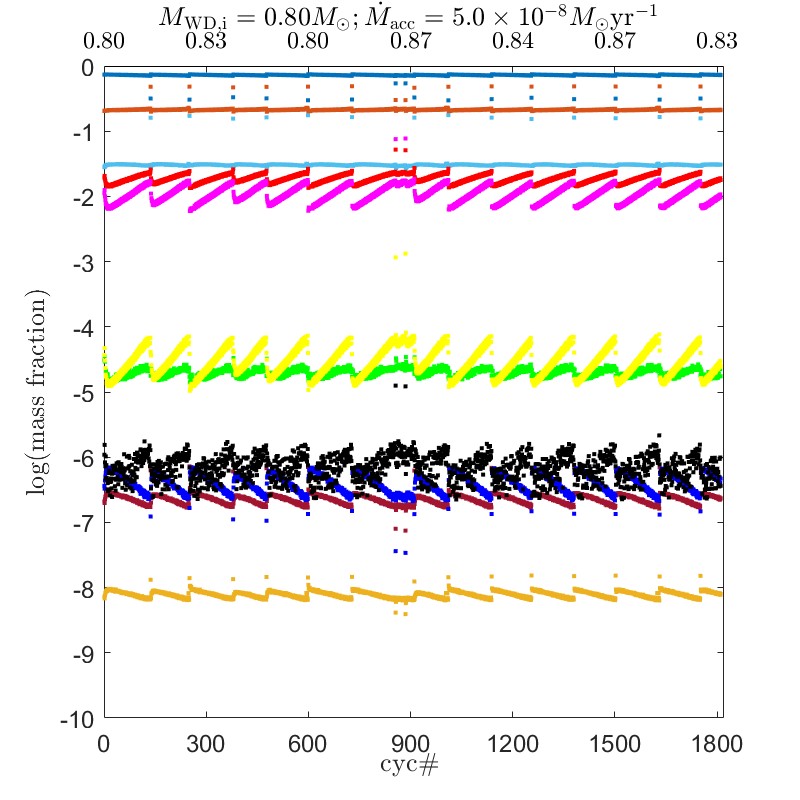}}
{\includegraphics[trim={22.5cm 5.6cm 0.7cm 4.0cm}, clip, width=0.31\columnwidth]{comp_for_legend.jpg}}\\
{\includegraphics[trim={1.0cm 0.0cm 1.5cm 0.0cm}, clip, width=0.62\columnwidth]{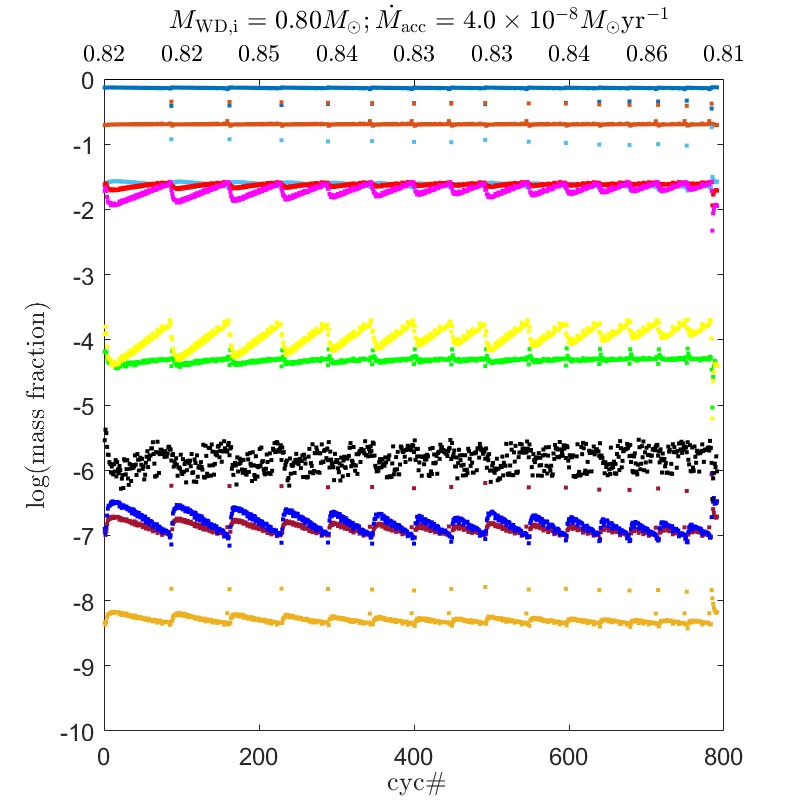}}
{\includegraphics[trim={22.5cm 5.6cm 0.7cm 4.0cm}, clip, width=0.31\columnwidth]{comp_for_legend.jpg}}
\caption{Description as in Figure \ref{fig:comp_nova_065} for initial WD masses of $0.8$ that are not shown in Figure \ref{fig:comp_nova}.: Models $\#31, 35-36$. (Model $\#28$, with a higher $\dot{M}_{\rm acc}$, experienced only non-ejective novae.)}\label{fig:comp_nova_080}
	\end{center}
\end{figure}

\begin{figure}
	\begin{center}
{\includegraphics[trim={1.0cm 0.0cm 1.5cm 0.0cm}, clip, width=0.62\columnwidth]{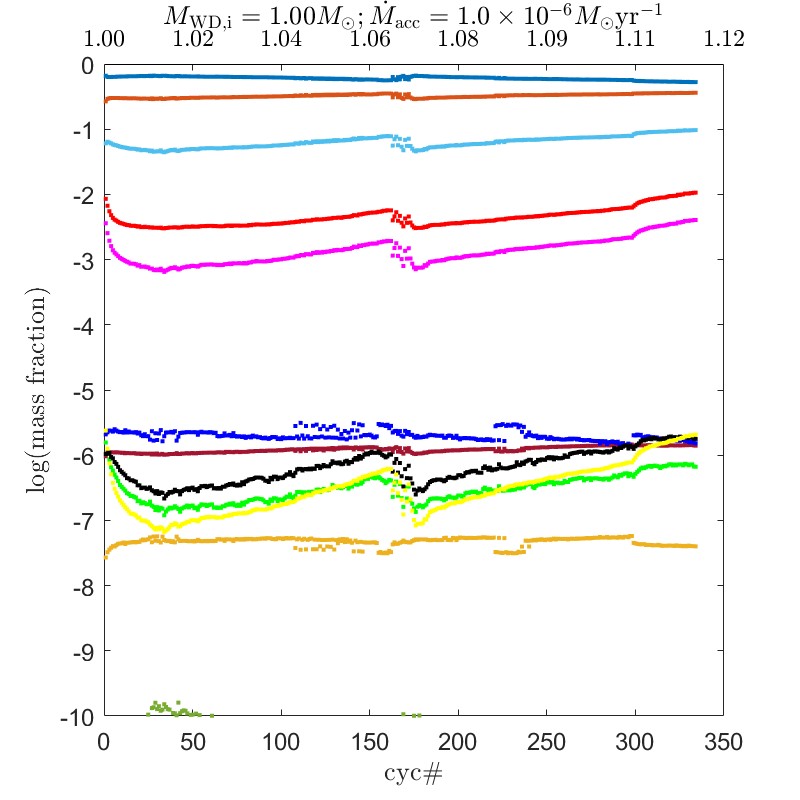}}
{\includegraphics[trim={22.5cm 5.6cm 0.7cm 4.0cm}, clip, width=0.31\columnwidth]{comp_for_legend.jpg}}\\
{\includegraphics[trim={1.0cm 0.0cm 1.5cm 0.0cm}, clip, width=0.62\columnwidth]{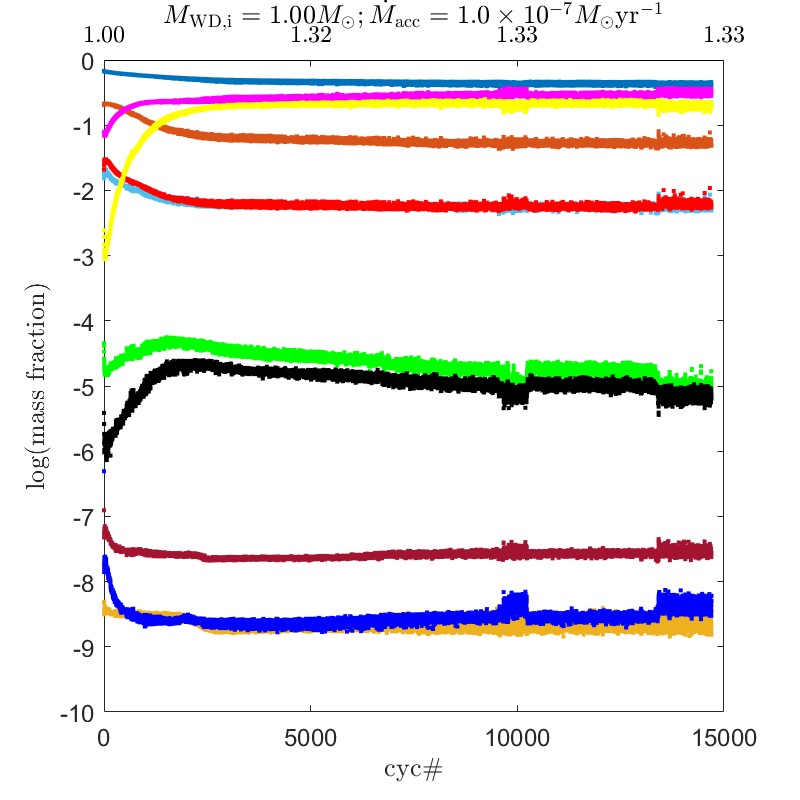}}
{\includegraphics[trim={22.5cm 5.6cm 0.7cm 4.0cm}, clip, width=0.31\columnwidth]{comp_for_legend.jpg}}\\
{\includegraphics[trim={1.0cm 0.0cm 1.5cm 0.0cm}, clip, width=0.62\columnwidth]{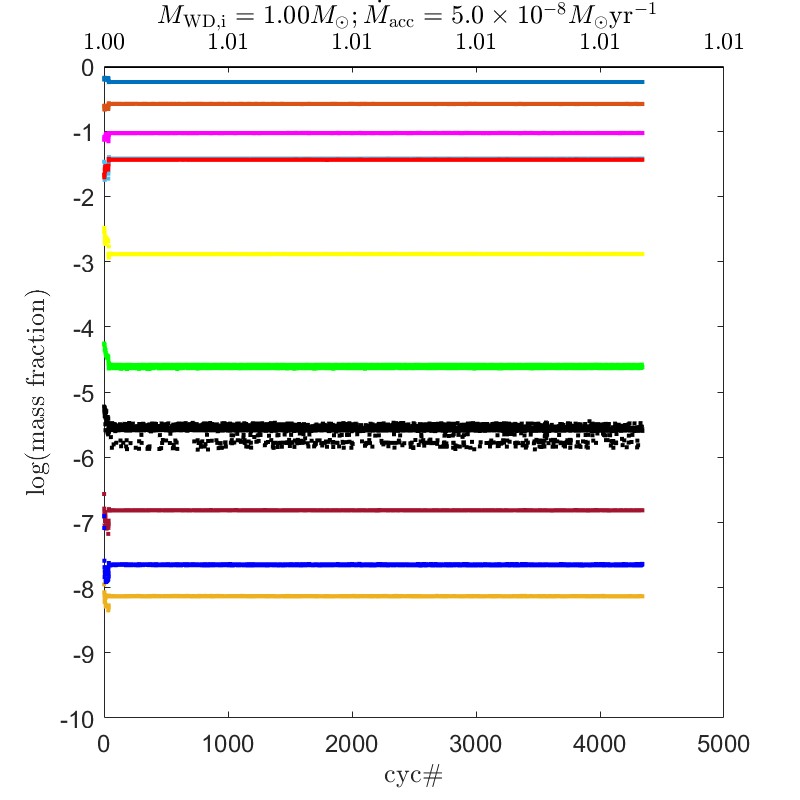}}
{\includegraphics[trim={22.5cm 5.6cm 0.7cm 4.0cm}, clip, width=0.31\columnwidth]{comp_for_legend.jpg}}
\caption{Description as in Figure \ref{fig:comp_nova_065} for initial WD masses of $1.0$: Models $\#45-47$. (Models $\#43-44$, with a higher $\dot{M}_{\rm acc}$, experienced only non-ejective novae.}\label{fig:comp_nova_100}
	\end{center}
\end{figure}

\begin{figure}
	\begin{center}
{\includegraphics[trim={1.0cm 0.0cm 1.5cm 0.0cm}, clip, width=0.62\columnwidth]{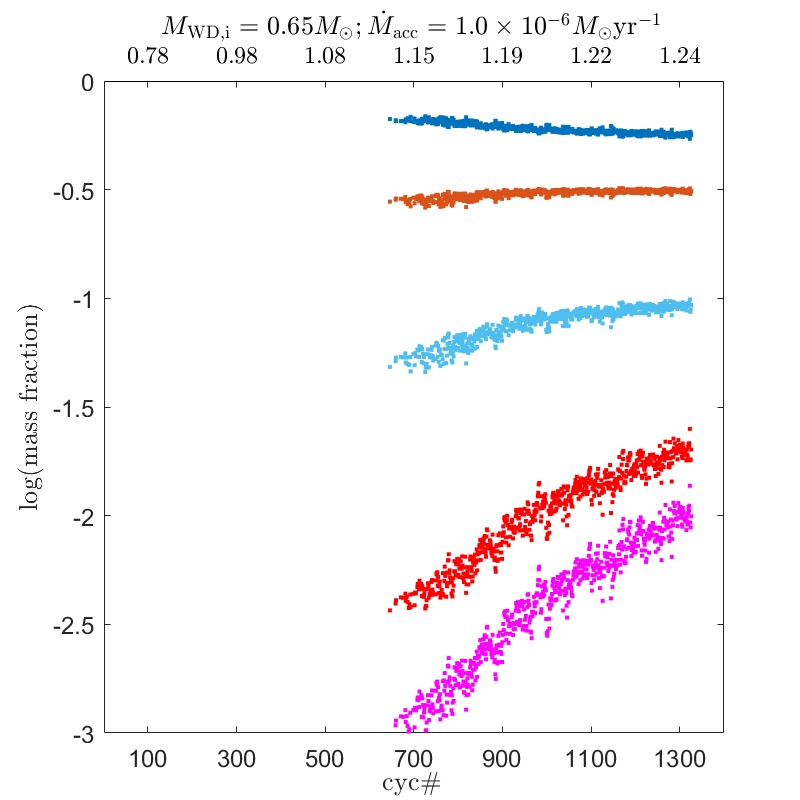}}
{\includegraphics[trim={22.5cm 5.6cm 0.7cm 4.0cm}, clip, width=0.31\columnwidth]{comp_for_legend.jpg}}\\
{\includegraphics[trim={1.0cm 0.0cm 1.5cm 0.0cm}, clip, width=0.62\columnwidth]{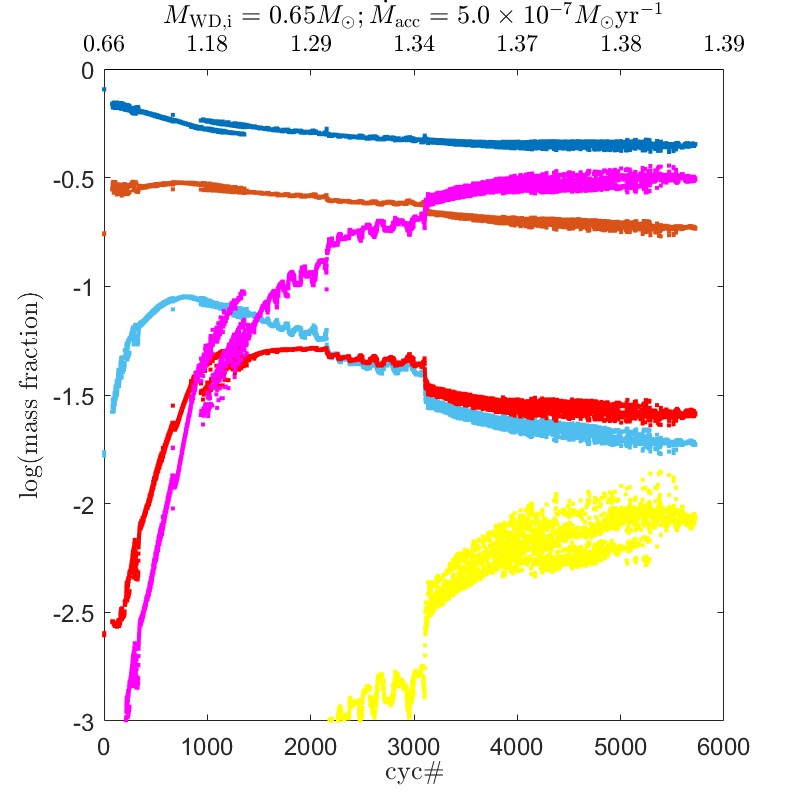}}
{\includegraphics[trim={22.5cm 5.6cm 0.7cm 4.0cm}, clip, width=0.31\columnwidth]{comp_for_legend.jpg}}\\
{\includegraphics[trim={1.0cm 0.0cm 1.5cm 0.0cm}, clip, width=0.62\columnwidth]{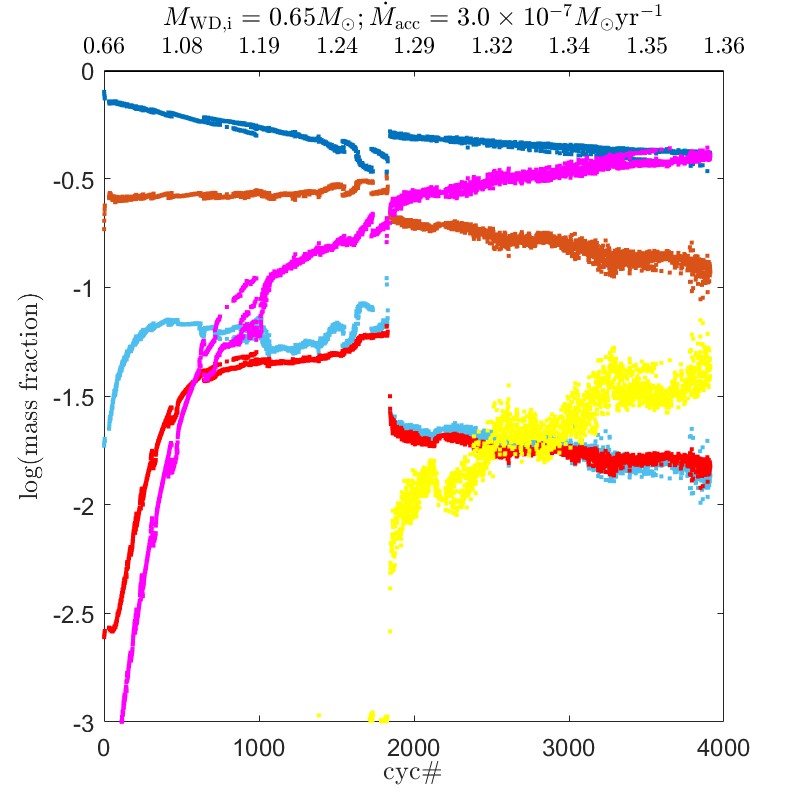}}
{\includegraphics[trim={22.5cm 5.6cm 0.7cm 4.0cm}, clip, width=0.31\columnwidth]{comp_for_legend.jpg}}\\
{\includegraphics[trim={1.0cm 0.0cm 1.5cm 0.0cm}, clip, width=0.62\columnwidth]{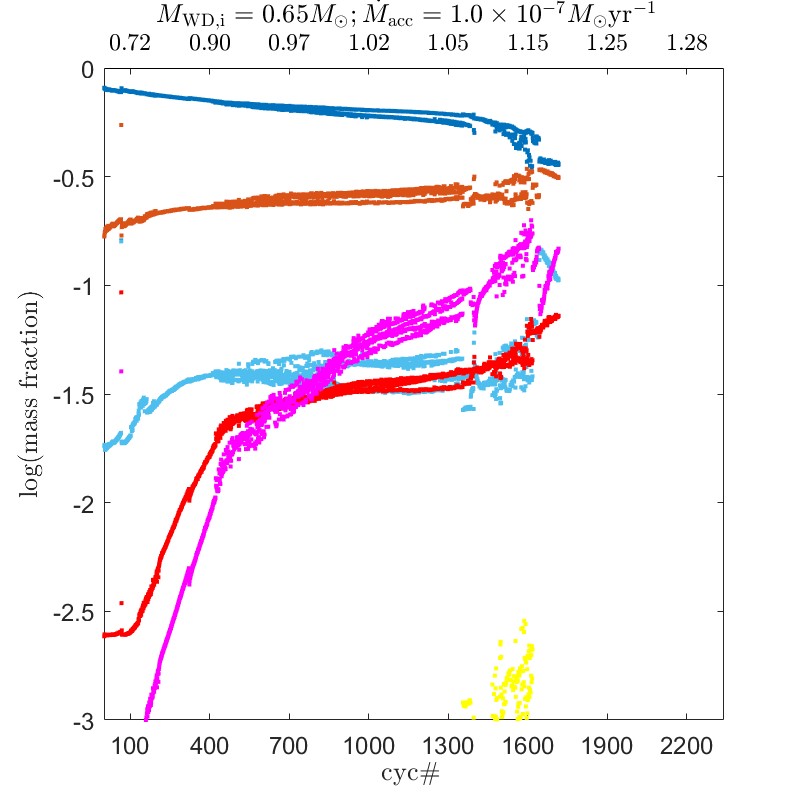}}
{\includegraphics[trim={22.5cm 5.6cm 0.7cm 4.0cm}, clip, width=0.31\columnwidth]{comp_for_legend.jpg}}
\caption{A close-up of Figure \ref{fig:comp_nova_065}, for mass fractions $\ge0.1\%$.}\label{fig:comp_novazoom_065}
	\end{center}
\end{figure}

\begin{figure}
	\begin{center}
{\includegraphics[trim={1.0cm 0.0cm 1.5cm 0.0cm}, clip, width=0.62\columnwidth]{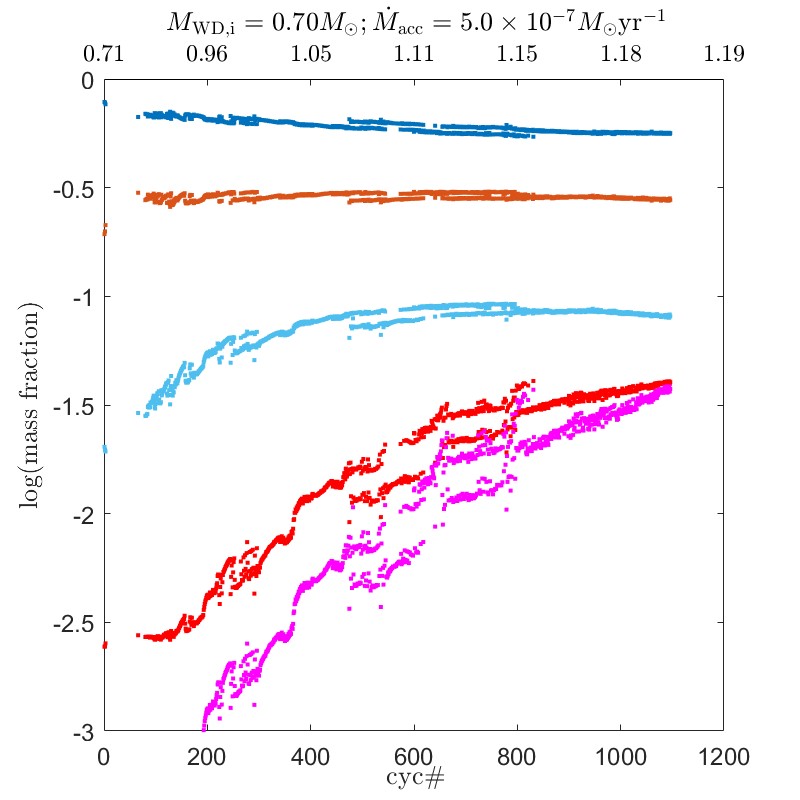}}
{\includegraphics[trim={22.5cm 5.6cm 0.7cm 4.0cm}, clip, width=0.31\columnwidth]{comp_for_legend.jpg}}\\
{\includegraphics[trim={1.0cm 0.0cm 1.5cm 0.0cm}, clip, width=0.62\columnwidth]{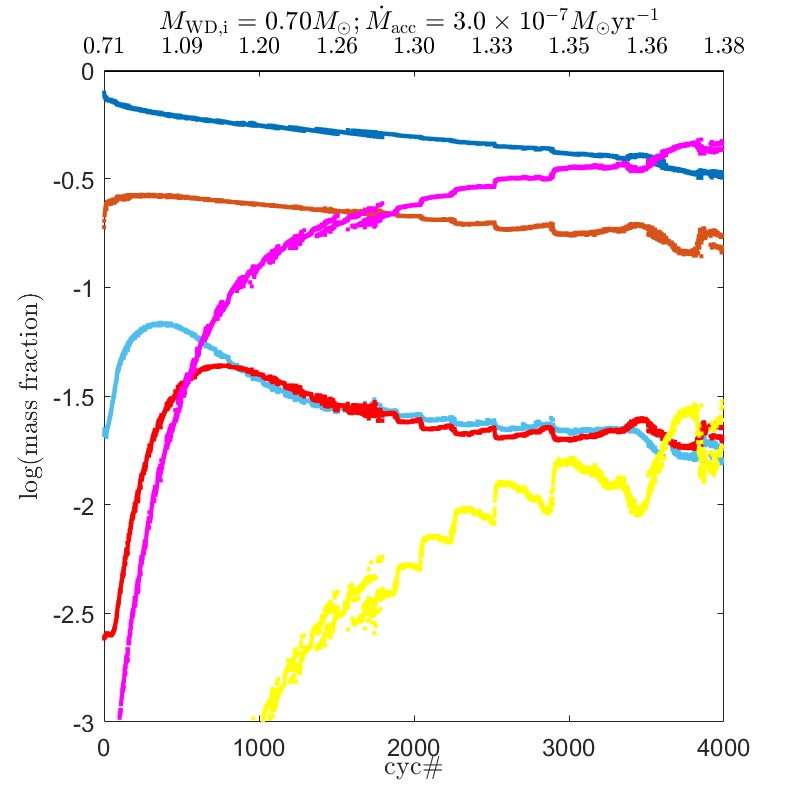}}
{\includegraphics[trim={22.5cm 5.6cm 0.7cm 4.0cm}, clip, width=0.31\columnwidth]{comp_for_legend.jpg}}\\
{\includegraphics[trim={1.0cm 0.0cm 1.5cm 0.0cm}, clip, width=0.62\columnwidth]{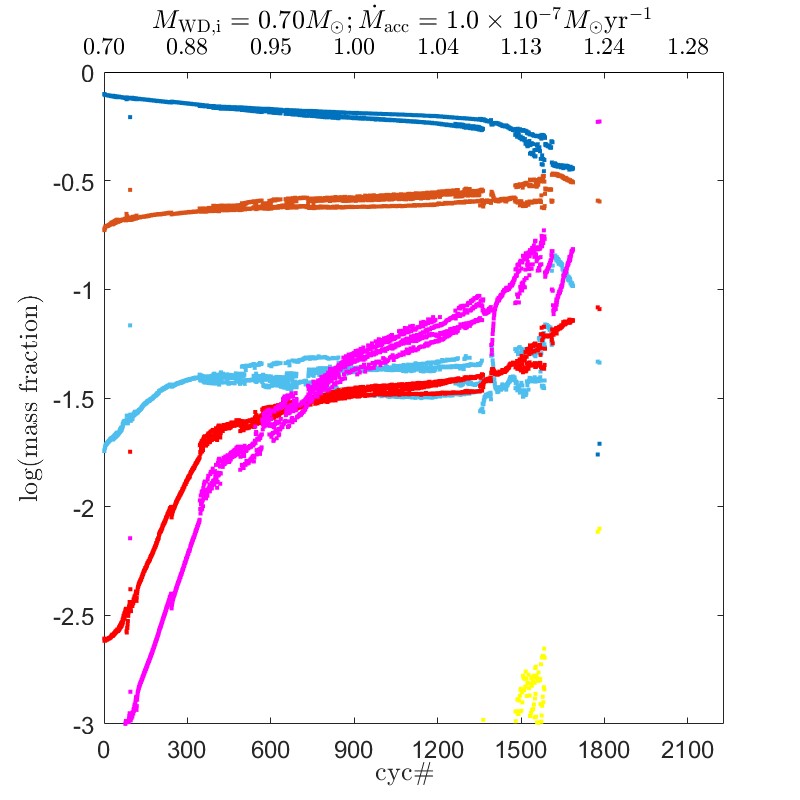}}
{\includegraphics[trim={22.5cm 5.6cm 0.7cm 4.0cm}, clip, width=0.31\columnwidth]{comp_for_legend.jpg}}
\caption{A close-up of Figure \ref{fig:comp_nova_070}, for mass fractions $\ge0.1\%$.}\label{fig:comp_novazoom_070}
	\end{center}
\end{figure}

\begin{figure}
	\begin{center}
{\includegraphics[trim={1.0cm 0.0cm 1.5cm 0.0cm}, clip, width=0.62\columnwidth]{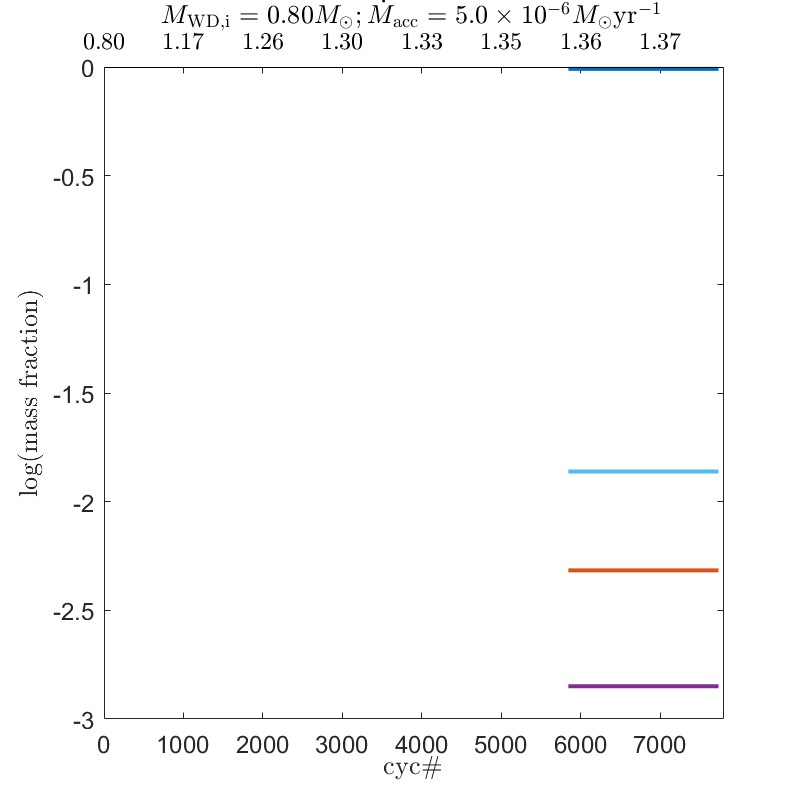}}
{\includegraphics[trim={22.5cm 5.6cm 0.7cm 4.0cm}, clip, width=0.31\columnwidth]{comp_for_legend.jpg}}\\
{\includegraphics[trim={1.0cm 0.0cm 1.5cm 0.0cm}, clip, width=0.62\columnwidth]{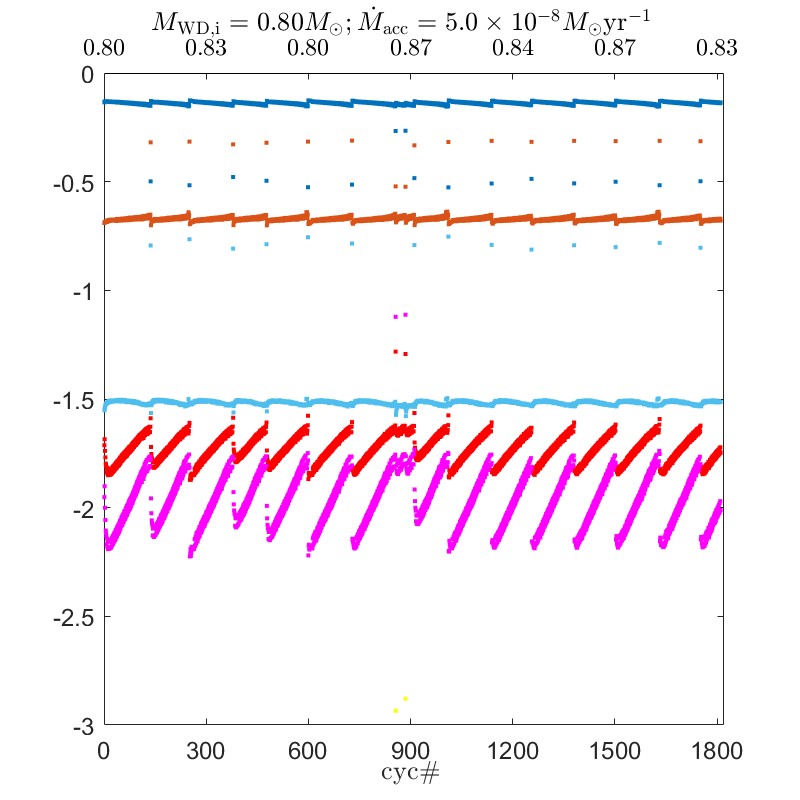}}
{\includegraphics[trim={22.5cm 5.6cm 0.7cm 4.0cm}, clip, width=0.31\columnwidth]{comp_for_legend.jpg}}\\
{\includegraphics[trim={1.0cm 0.0cm 1.5cm 0.0cm}, clip, width=0.62\columnwidth]{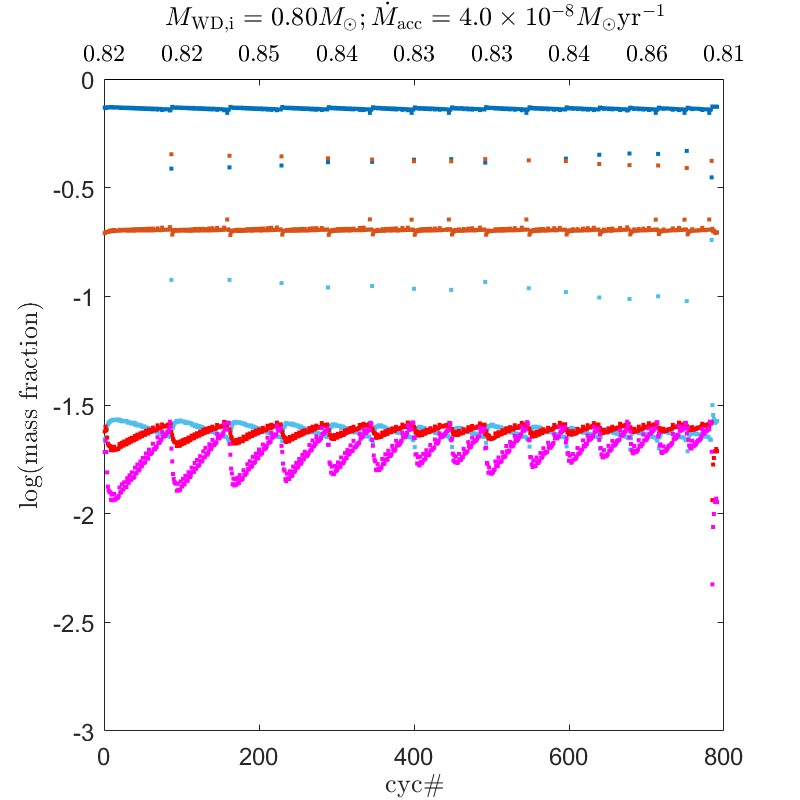}}
{\includegraphics[trim={22.5cm 5.6cm 0.7cm 4.0cm}, clip, width=0.31\columnwidth]{comp_for_legend.jpg}}
\caption{A close-up of Figure \ref{fig:comp_nova_080}, for mass fractions $\ge0.1\%$.}\label{fig:comp_novazoom_080}
	\end{center}
\end{figure}

\begin{figure}
	\begin{center}
{\includegraphics[trim={1.0cm 0.0cm 1.5cm 0.0cm}, clip, width=0.62\columnwidth]{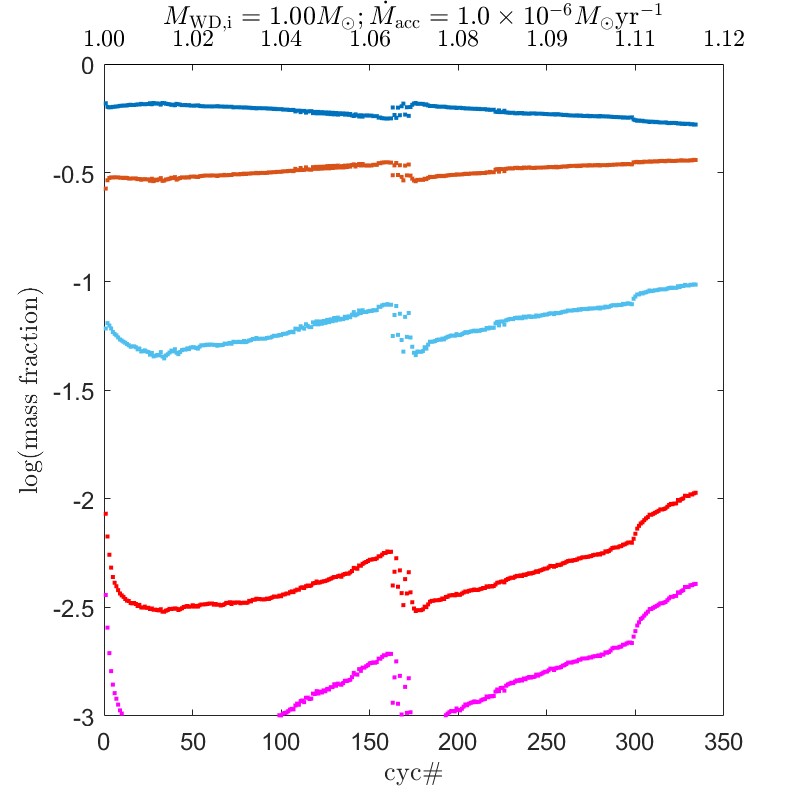}}
{\includegraphics[trim={22.5cm 5.6cm 0.7cm 4.0cm}, clip, width=0.31\columnwidth]{comp_for_legend.jpg}}\\
{\includegraphics[trim={1.0cm 0.0cm 1.5cm 0.0cm}, clip, width=0.62\columnwidth]{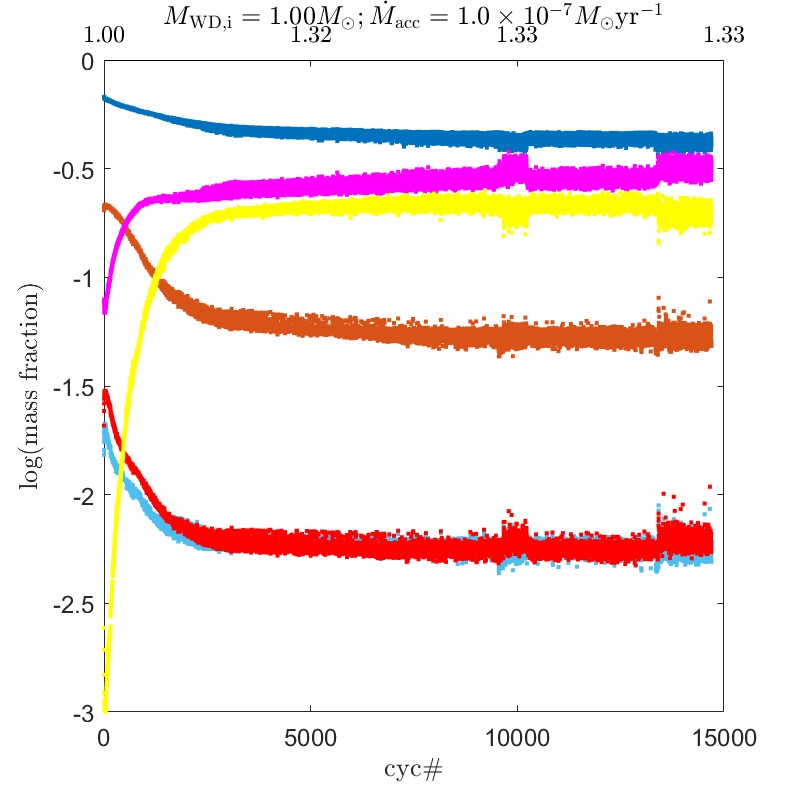}}
{\includegraphics[trim={22.5cm 5.6cm 0.7cm 4.0cm}, clip, width=0.31\columnwidth]{comp_for_legend.jpg}}\\
{\includegraphics[trim={1.0cm 0.0cm 1.5cm 0.0cm}, clip, width=0.62\columnwidth]{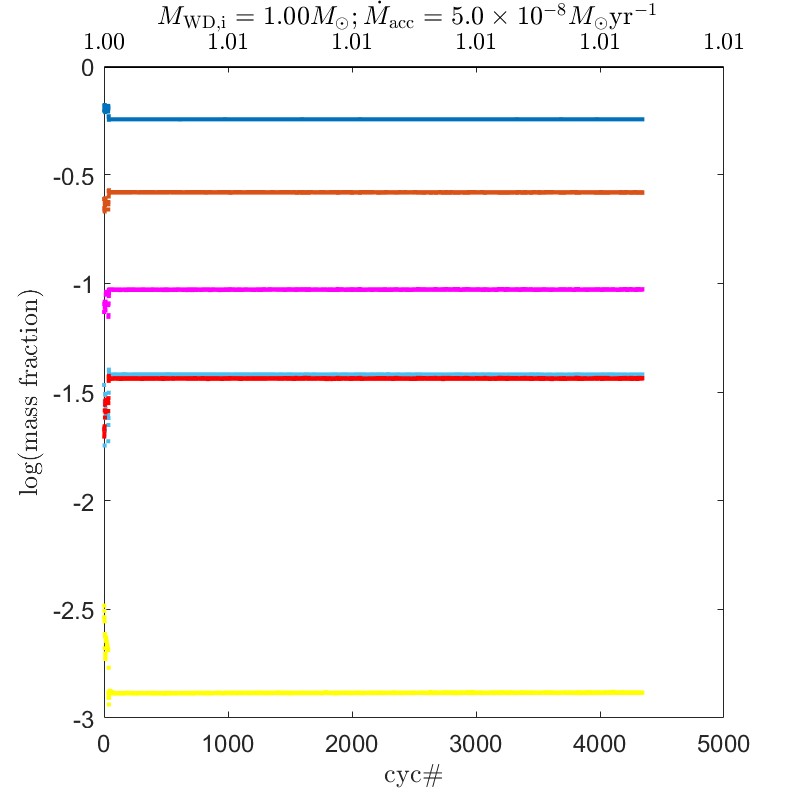}}
{\includegraphics[trim={22.5cm 5.6cm 0.7cm 4.0cm}, clip, width=0.31\columnwidth]{comp_for_legend.jpg}}
\caption{A close-up of Figure \ref{fig:comp_nova_100}, for mass fractions $\ge0.1\%$.}\label{fig:comp_novazoom_100}
	\end{center}
\end{figure}

The composition of the ejected mass for our HN type models as described in Figures \ref{fig:comp_nova} and \ref{fig:comp_novazoom} (shown there for a sample of models) is shown for the four initial WD mass models in Figures \ref{fig:comp_nova_065}$-$\ref{fig:comp_nova_100} and their closeups for mass fractions $>0.1\%$ are shown in Figures \ref{fig:comp_novazoom_065}$-$\ref{fig:comp_novazoom_100}.

\begin{figure*}
	\begin{center}
{\includegraphics[trim={1.5cm 0.8cm 2.5cm 1.0cm}, clip, width=0.65\columnwidth]{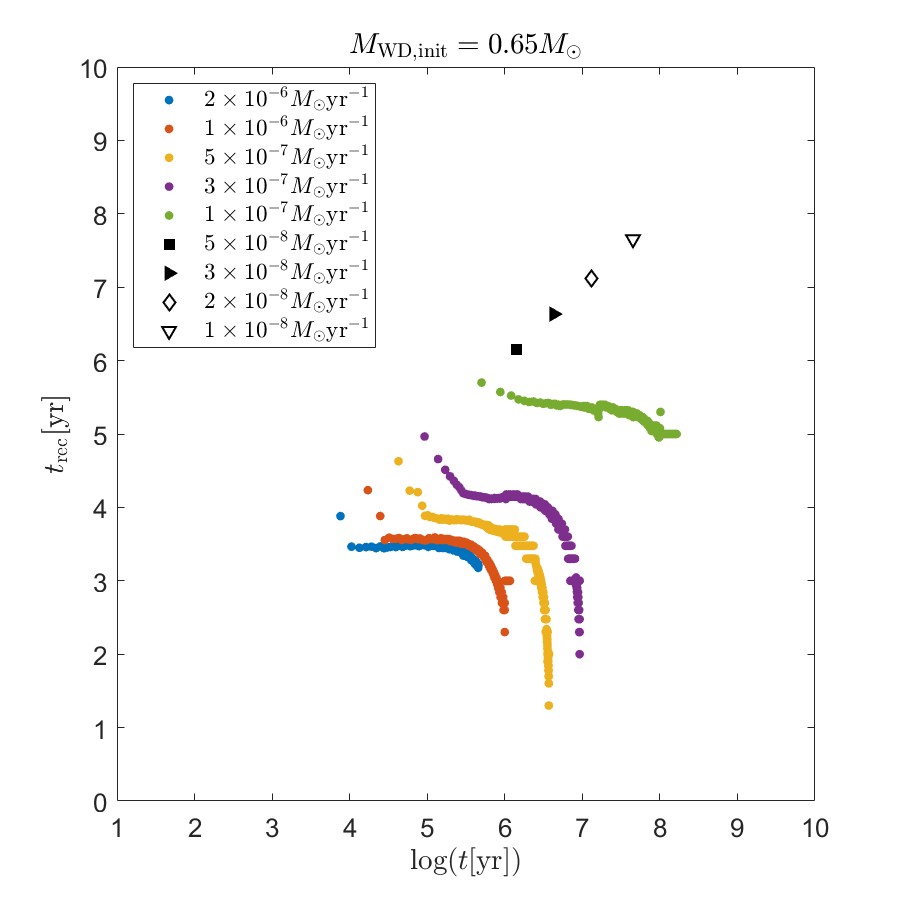}}
{\includegraphics[trim={1.5cm 0.8cm 2.5cm 1.0cm}, clip, width=0.65\columnwidth]{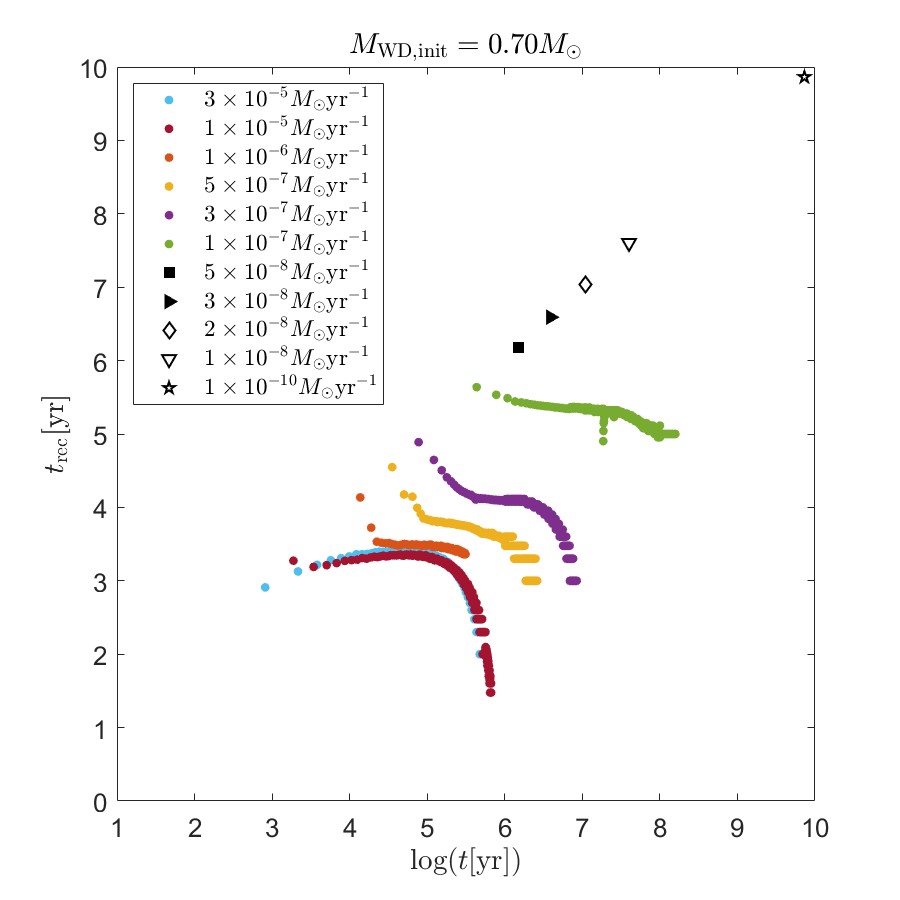}}
{\includegraphics[trim={1.5cm 0.8cm 2.5cm 1.0cm}, clip, width=0.65\columnwidth]
{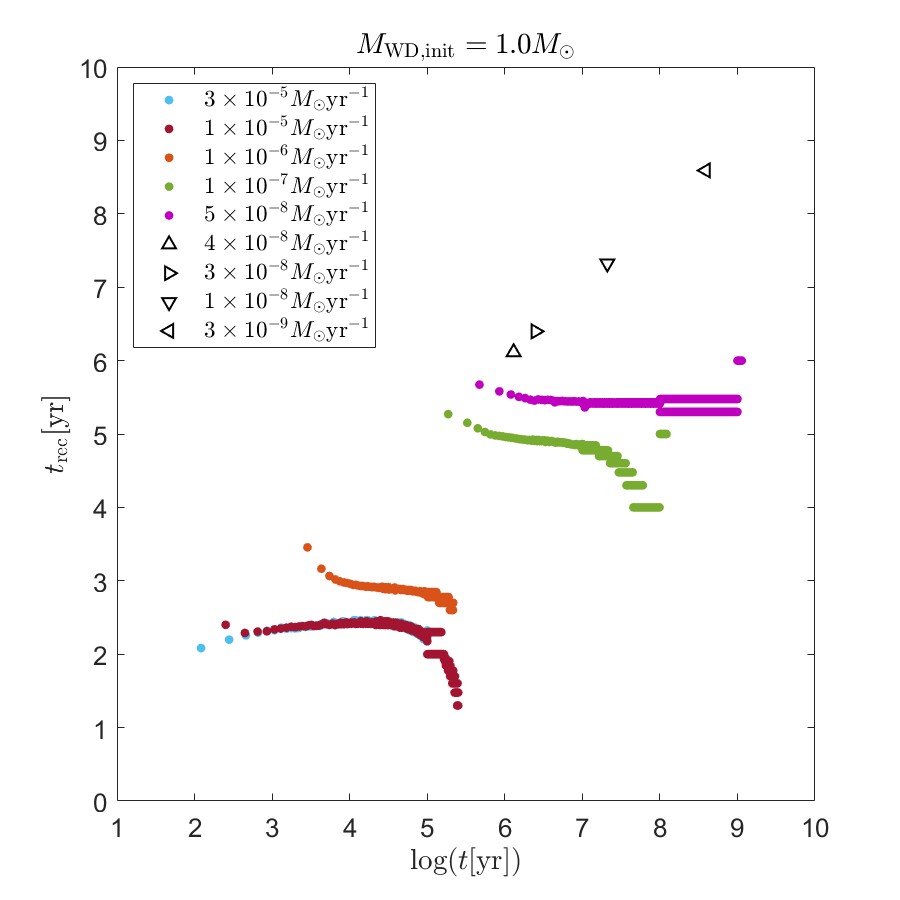}}
\caption{Recurrence period ($t_{\rm rec}$) vs. time for HN, UT and SN type models as described for our initial $0.80Modot$ models in Figure \ref{fig:trecs}. Left: $M_{\rm WD,init}=0.65M_\odot$; Center: $M_{\rm WD,init}=0.70M_\odot$; Right: $M_{\rm WD,init}=1.0M_\odot$.}\label{fig:trecs_A}
	\end{center}
\end{figure*}

The evolution of the recurrence period ($t_{\rm rec}$) over time is shown in Figure \ref{fig:trecs_A} for our initial $0.65$, $0.70$ and $1.0M_\odot$ models as described in Figure \ref{fig:trecs} for our $0.70M_\odot$ models. 

\section{Additional tables}\label{sec:B_adtnl_tbl}

In Table \ref{tab:for_fig_1_A} we show selected data from which Figures \ref{fig:macc_mej} and \ref{fig:macc_mej_065_A}$-$\ref{fig:macc_mej_100_A} were created, as well as the ejecta enrichment.

\onecolumn
\setlength{\LTcapwidth}{\textwidth}
\begin{longtable}{|c|c|c|c|c|c|c|c|c|c|}
\caption{Data for selected HN type cycles. From left to right: Model number and data as given in Table \ref{tab:mdls};WD mass; accreted mass; ejected mass; mass fraction of helium in ejected mass; mass fraction of heavy elements in ejected mass; maximum temperature per eruption; core temperature per eruption; evolutionary time; and, cycle number.}   
\label{tab:for_fig_1_A}
\\
\hline
{Model}&{$M_{\rm WD}$}&{$m_{\rm acc}$}&{$m_{\rm ej}$}&{$Y_{\rm ej}$}&{$Z_{\rm ej}$}&{$T_{\rm max}$}&{$T_{\rm c}$}&{$t$}&{cycle}\\
{\#;$M_{\rm WD,init}$;$\dot{M}_{\rm acc}$}&{$[M_\odot]$}&{$[M_\odot]$}&{$[M_\odot]$}&{}&{}&{$[10^8\rm K]$}&{$[10^7\rm K]$}&{[yr]}&{[\#]}\\
\hline\hline
\endfirsthead
\hline
{Model}&{$M_{\rm WD}$}&{$m_{\rm acc}$}&{$m_{\rm ej}$}&{$Y_{\rm ej}$}&{$Z_{\rm ej}$}&{$T_{\rm max}$}&{$T_{\rm c}$}&{$t$}&{cycle}\\
{\#;$M_{\rm WD,init}$;$\dot{M}_{\rm acc}$}&{$[M_\odot]$}&{$[M_\odot]$}&{$[M_\odot]$}&{}&{}&{$[10^8\rm K]$}&{$[10^7\rm K]$}&{[yr]}&{[\#]}\\
\hline\hline
\endhead

\hline
\multicolumn{10}{|c|}{\emph{Continued on next page}} \\
\hline
\endfoot

\hline
\endlastfoot

\multirow{18}{*}{\begin{tabular}{@{}c@{}} 3\\\\$0.65[M_\odot]$\\\\$2\times10^{-6}[M_\odot\rm yr^{-1}]$\end{tabular}}

& $0.6663$ & $1.09e-03$ & $0.00e+00$ & $0.00e+00$ & $0.00e+00$ & $2.3490$ & $3.0770$ & $3.34e+04$ & $10$ \\
& $0.6772$ & $1.06e-03$ & $0.00e+00$ & $0.00e+00$ & $0.00e+00$ & $2.3270$ & $3.1510$ & $6.32e+04$ & $20$ \\
& $0.6881$ & $1.07e-03$ & $0.00e+00$ & $0.00e+00$ & $0.00e+00$ & $2.3330$ & $3.2340$ & $9.39e+04$ & $30$ \\
& $0.6992$ & $1.08e-03$ & $0.00e+00$ & $0.00e+00$ & $0.00e+00$ & $2.3740$ & $3.3230$ & $1.24e+05$ & $40$ \\
& $0.7102$ & $1.08e-03$ & $0.00e+00$ & $0.00e+00$ & $0.00e+00$ & $2.4070$ & $3.4170$ & $1.54e+05$ & $50$ \\
& $0.7211$ & $1.11e-03$ & $0.00e+00$ & $0.00e+00$ & $0.00e+00$ & $2.3810$ & $3.5130$ & $1.83e+05$ & $60$ \\
& $0.7322$ & $1.12e-03$ & $0.00e+00$ & $0.00e+00$ & $0.00e+00$ & $2.4620$ & $3.6140$ & $2.11e+05$ & $70$ \\
& $0.7430$ & $1.08e-03$ & $0.00e+00$ & $0.00e+00$ & $0.00e+00$ & $2.4110$ & $3.7170$ & $2.37e+05$ & $80$ \\
& $0.7538$ & $1.11e-03$ & $0.00e+00$ & $0.00e+00$ & $0.00e+00$ & $2.4720$ & $3.8250$ & $2.63e+05$ & $90$ \\
& $0.7646$ & $1.06e-03$ & $0.00e+00$ & $0.00e+00$ & $0.00e+00$ & $2.4740$ & $3.9370$ & $2.88e+05$ & $100$ \\
& $0.7751$ & $1.04e-03$ & $0.00e+00$ & $0.00e+00$ & $0.00e+00$ & $2.5690$ & $4.0510$ & $3.11e+05$ & $110$ \\
& $0.7855$ & $1.02e-03$ & $0.00e+00$ & $0.00e+00$ & $0.00e+00$ & $2.5410$ & $4.1670$ & $3.33e+05$ & $120$ \\
& $0.7957$ & $1.02e-03$ & $0.00e+00$ & $0.00e+00$ & $0.00e+00$ & $2.5800$ & $4.2880$ & $3.55e+05$ & $130$ \\
& $0.8057$ & $9.80e-04$ & $0.00e+00$ & $0.00e+00$ & $0.00e+00$ & $2.6170$ & $4.4110$ & $3.75e+05$ & $140$ \\
& $0.8156$ & $9.45e-04$ & $0.00e+00$ & $0.00e+00$ & $0.00e+00$ & $2.6510$ & $4.5380$ & $3.94e+05$ & $150$ \\
& $0.8252$ & $9.55e-04$ & $0.00e+00$ & $0.00e+00$ & $0.00e+00$ & $2.6830$ & $4.6660$ & $4.13e+05$ & $160$ \\
& $0.8346$ & $9.72e-04$ & $0.00e+00$ & $0.00e+00$ & $0.00e+00$ & $2.7130$ & $4.7980$ & $4.30e+05$ & $170$ \\
& $0.8438$ & $9.26e-04$ & $0.00e+00$ & $0.00e+00$ & $0.00e+00$ & $2.7400$ & $4.9320$ & $4.47e+05$ & $180$ \\

\hline\hline

\multirow{44}{*}{\begin{tabular}{@{}c@{}} \vspace{-14.0cm}\\4\\\\$0.65[M_\odot]$\\\\$1\times10^{-6}[M_\odot\rm yr^{-1}]$\end{tabular}}
& $0.6935$ & $1.26e-03$ & $0.00e+00$ & $0.00e+00$ & $0.00e+00$ & $2.3950$ & $3.2810$ & $1.29e+05$ & $30$ \\
& $0.7304$ & $1.27e-03$ & $0.00e+00$ & $0.00e+00$ & $0.00e+00$ & $2.4600$ & $3.6000$ & $2.36e+05$ & $60$ \\
& $0.7682$ & $1.24e-03$ & $0.00e+00$ & $0.00e+00$ & $0.00e+00$ & $2.5930$ & $3.9770$ & $3.36e+05$ & $90$ \\
& $0.8051$ & $1.23e-03$ & $0.00e+00$ & $0.00e+00$ & $0.00e+00$ & $2.6630$ & $4.4070$ & $4.23e+05$ & $120$ \\
& $0.8401$ & $1.12e-03$ & $0.00e+00$ & $0.00e+00$ & $0.00e+00$ & $2.7710$ & $4.8840$ & $4.99e+05$ & $150$ \\
& $0.8727$ & $1.04e-03$ & $0.00e+00$ & $0.00e+00$ & $0.00e+00$ & $2.8540$ & $5.4020$ & $5.64e+05$ & $180$ \\
& $0.9023$ & $9.47e-04$ & $0.00e+00$ & $0.00e+00$ & $0.00e+00$ & $2.9520$ & $5.9370$ & $6.20e+05$ & $210$ \\
& $0.9289$ & $8.48e-04$ & $0.00e+00$ & $0.00e+00$ & $0.00e+00$ & $2.9880$ & $6.4780$ & $6.68e+05$ & $240$ \\
& $0.9531$ & $7.63e-04$ & $0.00e+00$ & $0.00e+00$ & $0.00e+00$ & $3.0800$ & $7.0230$ & $7.10e+05$ & $270$ \\
& $0.9749$ & $6.99e-04$ & $0.00e+00$ & $0.00e+00$ & $0.00e+00$ & $3.1570$ & $7.5630$ & $7.46e+05$ & $300$ \\
& $0.9948$ & $6.38e-04$ & $0.00e+00$ & $0.00e+00$ & $0.00e+00$ & $3.2060$ & $8.1010$ & $7.79e+05$ & $330$ \\
& $1.0130$ & $5.87e-04$ & $0.00e+00$ & $0.00e+00$ & $0.00e+00$ & $3.2560$ & $8.6310$ & $8.08e+05$ & $360$ \\
& $1.0297$ & $5.43e-04$ & $0.00e+00$ & $0.00e+00$ & $0.00e+00$ & $3.2560$ & $9.1510$ & $8.35e+05$ & $390$ \\
& $1.0451$ & $4.97e-04$ & $0.00e+00$ & $0.00e+00$ & $0.00e+00$ & $3.3870$ & $9.6590$ & $8.59e+05$ & $420$ \\
& $1.0593$ & $4.49e-04$ & $0.00e+00$ & $0.00e+00$ & $0.00e+00$ & $3.3250$ & $10.1500$ & $8.81e+05$ & $450$ \\
& $1.0724$ & $4.22e-04$ & $0.00e+00$ & $0.00e+00$ & $0.00e+00$ & $3.4280$ & $10.6200$ & $9.01e+05$ & $480$ \\
& $1.0846$ & $3.86e-04$ & $0.00e+00$ & $0.00e+00$ & $0.00e+00$ & $3.4770$ & $11.0700$ & $9.19e+05$ & $510$ \\
& $1.0960$ & $3.69e-04$ & $0.00e+00$ & $0.00e+00$ & $0.00e+00$ & $3.4720$ & $11.4900$ & $9.36e+05$ & $540$ \\
& $1.1066$ & $3.43e-04$ & $0.00e+00$ & $0.00e+00$ & $0.00e+00$ & $3.5340$ & $11.8800$ & $9.52e+05$ & $570$ \\
& $1.1166$ & $3.27e-04$ & $0.00e+00$ & $0.00e+00$ & $0.00e+00$ & $3.5290$ & $12.2400$ & $9.67e+05$ & $600$ \\
& $1.1260$ & $3.08e-04$ & $0.00e+00$ & $0.00e+00$ & $0.00e+00$ & $3.6200$ & $12.5600$ & $9.81e+05$ & $630$ \\
& $1.1348$ & $2.87e-04$ & $5.13e-06$ & $6.59e-01$ & $3.41e-01$ & $3.6180$ & $12.8700$ & $9.94e+05$ & $660$ \\
& $1.1432$ & $2.67e-04$ & $4.76e-06$ & $6.68e-01$ & $3.32e-01$ & $3.6210$ & $13.1400$ & $1.01e+06$ & $690$ \\
& $1.1510$ & $2.60e-04$ & $1.35e-05$ & $6.42e-01$ & $3.57e-01$ & $3.6860$ & $13.4000$ & $1.02e+06$ & $720$ \\
& $1.1584$ & $2.43e-04$ & $1.05e-05$ & $6.65e-01$ & $3.35e-01$ & $3.6680$ & $13.6300$ & $1.03e+06$ & $750$ \\
& $1.1653$ & $2.38e-04$ & $2.07e-05$ & $6.14e-01$ & $3.86e-01$ & $3.7690$ & $13.8500$ & $1.04e+06$ & $780$ \\
& $1.1719$ & $2.23e-04$ & $1.48e-05$ & $6.20e-01$ & $3.80e-01$ & $3.7610$ & $14.0500$ & $1.05e+06$ & $810$ \\
& $1.1781$ & $2.14e-04$ & $1.15e-05$ & $6.53e-01$ & $3.47e-01$ & $3.7190$ & $14.2300$ & $1.06e+06$ & $840$ \\
& $1.1839$ & $2.07e-04$ & $1.36e-05$ & $6.24e-01$ & $3.76e-01$ & $3.7860$ & $14.4100$ & $1.07e+06$ & $870$ \\
& $1.1895$ & $1.95e-04$ & $1.16e-05$ & $6.41e-01$ & $3.59e-01$ & $3.7730$ & $14.5700$ & $1.08e+06$ & $900$ \\
& $1.1948$ & $1.91e-04$ & $2.22e-05$ & $6.06e-01$ & $3.94e-01$ & $3.8630$ & $14.7300$ & $1.08e+06$ & $930$ \\
& $1.1998$ & $1.85e-04$ & $1.68e-05$ & $5.97e-01$ & $4.03e-01$ & $3.9010$ & $14.8800$ & $1.09e+06$ & $960$ \\
& $1.2047$ & $1.78e-04$ & $2.44e-05$ & $5.80e-01$ & $4.20e-01$ & $3.9160$ & $15.0200$ & $1.10e+06$ & $990$ \\
& $1.2093$ & $1.69e-04$ & $2.06e-05$ & $5.93e-01$ & $4.07e-01$ & $3.9100$ & $15.1600$ & $1.11e+06$ & $1020$ \\
& $1.2137$ & $1.64e-04$ & $1.60e-05$ & $6.15e-01$ & $3.85e-01$ & $3.8860$ & $15.3000$ & $1.11e+06$ & $1050$ \\
& $1.2180$ & $1.59e-04$ & $2.04e-05$ & $5.81e-01$ & $4.19e-01$ & $3.9590$ & $15.4300$ & $1.12e+06$ & $1080$ \\
& $1.2220$ & $1.51e-04$ & $1.86e-05$ & $5.81e-01$ & $4.19e-01$ & $3.9710$ & $15.5700$ & $1.13e+06$ & $1110$ \\
& $1.2259$ & $1.46e-04$ & $1.82e-05$ & $5.91e-01$ & $4.09e-01$ & $3.9560$ & $15.7000$ & $1.13e+06$ & $1140$ \\
& $1.2297$ & $1.43e-04$ & $2.02e-05$ & $5.65e-01$ & $4.35e-01$ & $4.0190$ & $15.8200$ & $1.14e+06$ & $1170$ \\
& $1.2333$ & $1.38e-04$ & $2.08e-05$ & $5.80e-01$ & $4.20e-01$ & $4.0050$ & $15.9500$ & $1.14e+06$ & $1200$ \\
& $1.2368$ & $1.32e-04$ & $1.52e-05$ & $5.69e-01$ & $4.31e-01$ & $4.0100$ & $16.0700$ & $1.15e+06$ & $1230$ \\
& $1.2402$ & $1.27e-04$ & $1.53e-05$ & $5.97e-01$ & $4.04e-01$ & $3.9820$ & $16.2000$ & $1.15e+06$ & $1260$ \\
& $1.2435$ & $1.25e-04$ & $1.85e-05$ & $5.55e-01$ & $4.45e-01$ & $4.0730$ & $16.3200$ & $1.16e+06$ & $1290$ \\
& $1.2466$ & $1.21e-04$ & $1.97e-05$ & $5.59e-01$ & $4.41e-01$ & $4.0650$ & $16.4300$ & $1.16e+06$ & $1320$ \\

\hline\hline

\multirow{57}{*}{\begin{tabular}{@{}c@{}} \vspace{-14.0cm}\\5\\\\$0.65[M_\odot]$\\\\$5\times10^{-7}[M_\odot\rm yr^{-1}]$\end{tabular}}
& $0.8451$ & $1.94e-03$ & $4.46e-04$ & $6.78e-01$ & $3.22e-01$ & $3.1520$ & $5.1150$ & $6.84e+05$ & $100$ \\
& $0.9352$ & $1.86e-03$ & $1.17e-03$ & $6.76e-01$ & $3.24e-01$ & $3.6760$ & $7.1490$ & $1.15e+06$ & $200$ \\
& $0.9867$ & $1.54e-03$ & $1.02e-03$ & $6.57e-01$ & $3.43e-01$ & $3.8820$ & $8.8120$ & $1.55e+06$ & $300$ \\
& $1.0255$ & $1.31e-03$ & $9.83e-04$ & $6.30e-01$ & $3.70e-01$ & $4.0880$ & $9.5990$ & $1.87e+06$ & $400$ \\
& $1.0568$ & $1.06e-03$ & $7.82e-04$ & $6.13e-01$ & $3.87e-01$ & $4.1610$ & $10.0300$ & $2.14e+06$ & $500$ \\
& $1.0829$ & $8.86e-04$ & $6.52e-04$ & $5.92e-01$ & $4.08e-01$ & $4.2390$ & $10.3700$ & $2.36e+06$ & $600$ \\
& $1.1085$ & $7.36e-04$ & $4.36e-04$ & $5.79e-01$ & $4.21e-01$ & $4.2780$ & $10.8000$ & $2.54e+06$ & $700$ \\
& $1.1360$ & $6.13e-04$ & $3.62e-04$ & $5.58e-01$ & $4.42e-01$ & $4.3740$ & $11.4300$ & $2.69e+06$ & $800$ \\
& $1.1588$ & $5.24e-04$ & $3.13e-04$ & $5.42e-01$ & $4.58e-01$ & $4.4710$ & $11.9600$ & $2.81e+06$ & $900$ \\
& $1.1785$ & $4.46e-04$ & $2.41e-04$ & $5.72e-01$ & $4.28e-01$ & $4.5130$ & $12.4500$ & $2.91e+06$ & $1000$ \\
& $1.1961$ & $3.78e-04$ & $2.02e-04$ & $5.69e-01$ & $4.31e-01$ & $4.5390$ & $12.9400$ & $3.00e+06$ & $1100$ \\
& $1.2118$ & $3.17e-04$ & $1.75e-04$ & $5.12e-01$ & $4.88e-01$ & $4.6060$ & $13.4200$ & $3.08e+06$ & $1200$ \\
& $1.2259$ & $2.77e-04$ & $1.41e-04$ & $5.59e-01$ & $4.41e-01$ & $4.5940$ & $13.8900$ & $3.14e+06$ & $1300$ \\
& $1.2386$ & $2.41e-04$ & $1.20e-04$ & $5.47e-01$ & $4.53e-01$ & $4.6290$ & $14.3600$ & $3.20e+06$ & $1400$ \\
& $1.2503$ & $2.07e-04$ & $9.44e-05$ & $5.36e-01$ & $4.64e-01$ & $4.6440$ & $14.8300$ & $3.25e+06$ & $1500$ \\
& $1.2607$ & $1.85e-04$ & $9.00e-05$ & $5.29e-01$ & $4.71e-01$ & $4.6870$ & $15.2900$ & $3.29e+06$ & $1600$ \\
& $1.2701$ & $1.64e-04$ & $7.72e-05$ & $5.20e-01$ & $4.80e-01$ & $4.7100$ & $15.7400$ & $3.33e+06$ & $1700$ \\
& $1.2788$ & $1.46e-04$ & $6.69e-05$ & $5.13e-01$ & $4.87e-01$ & $4.7280$ & $16.1800$ & $3.36e+06$ & $1800$ \\
& $1.2869$ & $1.25e-04$ & $4.27e-05$ & $5.20e-01$ & $4.80e-01$ & $4.6800$ & $16.6400$ & $3.39e+06$ & $1900$ \\
& $1.2941$ & $1.11e-04$ & $3.82e-05$ & $5.08e-01$ & $4.92e-01$ & $4.6880$ & $17.0700$ & $3.42e+06$ & $2000$ \\
& $1.3009$ & $1.02e-04$ & $3.80e-05$ & $5.01e-01$ & $4.98e-01$ & $4.7260$ & $17.5200$ & $3.44e+06$ & $2100$ \\
& $1.3071$ & $9.76e-05$ & $4.11e-05$ & $4.94e-01$ & $5.06e-01$ & $4.8620$ & $17.9300$ & $3.46e+06$ & $2200$ \\
& $1.3127$ & $8.85e-05$ & $3.50e-05$ & $4.92e-01$ & $5.08e-01$ & $4.8630$ & $18.3100$ & $3.48e+06$ & $2300$ \\
& $1.3179$ & $8.05e-05$ & $3.14e-05$ & $4.90e-01$ & $5.10e-01$ & $4.8720$ & $18.7000$ & $3.50e+06$ & $2400$ \\
& $1.3228$ & $7.19e-05$ & $2.19e-05$ & $4.89e-01$ & $5.11e-01$ & $4.8550$ & $19.1000$ & $3.52e+06$ & $2500$ \\
& $1.3273$ & $6.46e-05$ & $1.89e-05$ & $4.85e-01$ & $5.15e-01$ & $4.8330$ & $19.4800$ & $3.53e+06$ & $2600$ \\
& $1.3315$ & $6.03e-05$ & $1.91e-05$ & $4.77e-01$ & $5.22e-01$ & $4.8620$ & $19.8700$ & $3.54e+06$ & $2700$ \\
& $1.3356$ & $5.43e-05$ & $1.23e-05$ & $4.83e-01$ & $5.17e-01$ & $4.8400$ & $20.2900$ & $3.56e+06$ & $2800$ \\
& $1.3393$ & $4.94e-05$ & $1.22e-05$ & $4.78e-01$ & $5.22e-01$ & $4.8360$ & $20.6700$ & $3.57e+06$ & $2900$ \\
& $1.3429$ & $4.57e-05$ & $1.30e-05$ & $4.67e-01$ & $5.33e-01$ & $4.8390$ & $21.1000$ & $3.58e+06$ & $3000$ \\
& $1.3462$ & $4.10e-05$ & $7.82e-06$ & $4.69e-01$ & $5.31e-01$ & $4.8000$ & $21.4900$ & $3.59e+06$ & $3100$ \\
& $1.3491$ & $4.60e-05$ & $1.79e-05$ & $4.70e-01$ & $5.30e-01$ & $5.1410$ & $21.7900$ & $3.60e+06$ & $3200$ \\
& $1.3518$ & $4.31e-05$ & $1.63e-05$ & $4.68e-01$ & $5.32e-01$ & $5.1670$ & $22.0900$ & $3.61e+06$ & $3300$ \\
& $1.3544$ & $4.06e-05$ & $1.50e-05$ & $4.63e-01$ & $5.37e-01$ & $5.2080$ & $22.4000$ & $3.62e+06$ & $3400$ \\
& $1.3569$ & $3.76e-05$ & $1.34e-05$ & $4.64e-01$ & $5.36e-01$ & $5.1860$ & $22.7100$ & $3.63e+06$ & $3500$ \\
& $1.3592$ & $3.52e-05$ & $1.27e-05$ & $4.55e-01$ & $5.45e-01$ & $5.1620$ & $23.0200$ & $3.64e+06$ & $3600$ \\
& $1.3614$ & $3.33e-05$ & $1.16e-05$ & $4.52e-01$ & $5.48e-01$ & $5.1800$ & $23.3300$ & $3.64e+06$ & $3700$ \\
& $1.3635$ & $3.15e-05$ & $1.06e-05$ & $4.50e-01$ & $5.50e-01$ & $5.1940$ & $23.6500$ & $3.65e+06$ & $3800$ \\
& $1.3655$ & $2.95e-05$ & $9.99e-06$ & $4.46e-01$ & $5.54e-01$ & $5.1960$ & $23.9600$ & $3.66e+06$ & $3900$ \\
& $1.3674$ & $2.77e-05$ & $8.74e-06$ & $4.63e-01$ & $5.37e-01$ & $5.2130$ & $24.2800$ & $3.66e+06$ & $4000$ \\
& $1.3693$ & $2.62e-05$ & $8.34e-06$ & $4.44e-01$ & $5.56e-01$ & $5.2060$ & $24.6000$ & $3.67e+06$ & $4100$ \\
& $1.3710$ & $2.47e-05$ & $7.44e-06$ & $4.49e-01$ & $5.51e-01$ & $5.1910$ & $24.9100$ & $3.67e+06$ & $4200$ \\
& $1.3727$ & $2.32e-05$ & $6.81e-06$ & $4.61e-01$ & $5.39e-01$ & $5.2370$ & $25.2300$ & $3.68e+06$ & $4300$ \\
& $1.3743$ & $2.20e-05$ & $6.04e-06$ & $4.65e-01$ & $5.35e-01$ & $5.2460$ & $25.5500$ & $3.69e+06$ & $4400$ \\
& $1.3758$ & $2.08e-05$ & $5.72e-06$ & $4.61e-01$ & $5.39e-01$ & $5.2460$ & $25.8700$ & $3.69e+06$ & $4500$ \\
& $1.3773$ & $1.97e-05$ & $5.04e-06$ & $4.64e-01$ & $5.36e-01$ & $5.2850$ & $26.1800$ & $3.69e+06$ & $4600$ \\
& $1.3787$ & $1.86e-05$ & $4.86e-06$ & $4.58e-01$ & $5.42e-01$ & $5.2680$ & $26.5000$ & $3.70e+06$ & $4700$ \\
& $1.3801$ & $1.77e-05$ & $4.37e-06$ & $4.56e-01$ & $5.44e-01$ & $5.2610$ & $26.8100$ & $3.70e+06$ & $4800$ \\
& $1.3813$ & $1.67e-05$ & $4.26e-06$ & $4.28e-01$ & $5.72e-01$ & $5.2520$ & $27.1300$ & $3.71e+06$ & $4900$ \\
& $1.3826$ & $1.61e-05$ & $4.03e-06$ & $4.44e-01$ & $5.56e-01$ & $5.2230$ & $27.4300$ & $3.71e+06$ & $5000$ \\
& $1.3838$ & $1.52e-05$ & $3.67e-06$ & $4.44e-01$ & $5.56e-01$ & $5.2230$ & $27.7400$ & $3.71e+06$ & $5100$ \\
& $1.3849$ & $1.44e-05$ & $3.11e-06$ & $4.54e-01$ & $5.46e-01$ & $5.2320$ & $28.0400$ & $3.72e+06$ & $5200$ \\
& $1.3860$ & $1.38e-05$ & $2.96e-06$ & $4.40e-01$ & $5.60e-01$ & $5.2400$ & $28.3500$ & $3.72e+06$ & $5300$ \\
& $1.3871$ & $1.32e-05$ & $2.75e-06$ & $4.44e-01$ & $5.56e-01$ & $5.2350$ & $28.6500$ & $3.72e+06$ & $5400$ \\
& $1.3881$ & $1.25e-05$ & $2.61e-06$ & $4.45e-01$ & $5.55e-01$ & $5.2360$ & $28.9500$ & $3.73e+06$ & $5500$ \\
& $1.3891$ & $1.19e-05$ & $2.37e-06$ & $4.39e-01$ & $5.61e-01$ & $5.2300$ & $29.2500$ & $3.73e+06$ & $5600$ \\
& $1.3900$ & $1.12e-05$ & $2.18e-06$ & $4.51e-01$ & $5.49e-01$ & $5.2400$ & $29.5500$ & $3.73e+06$ & $5700$ \\

\hline\hline

\multirow{39}{*}{\begin{tabular}{@{}c@{}} \vspace{-10.0cm}\\6\\\\$0.65[M_\odot]$\\\\$3\times10^{-7}[M_\odot\rm yr^{-1}]$\end{tabular}}
& $0.8672$ & $3.99e-03$ & $2.99e-03$ & $7.02e-01$ & $2.98e-01$ & $3.9290$ & $5.9480$ & $1.56e+06$ & $100$ \\
& $0.9431$ & $3.24e-03$ & $2.64e-03$ & $6.69e-01$ & $3.31e-01$ & $4.3420$ & $8.1700$ & $2.82e+06$ & $200$ \\
& $0.9969$ & $2.47e-03$ & $1.97e-03$ & $6.45e-01$ & $3.55e-01$ & $4.5190$ & $8.6390$ & $3.80e+06$ & $300$ \\
& $1.0392$ & $1.98e-03$ & $1.57e-03$ & $6.23e-01$ & $3.77e-01$ & $4.6660$ & $9.0050$ & $4.58e+06$ & $400$ \\
& $1.0753$ & $1.63e-03$ & $1.34e-03$ & $6.13e-01$ & $3.87e-01$ & $4.7800$ & $9.4190$ & $5.20e+06$ & $500$ \\
& $1.1053$ & $1.25e-03$ & $9.49e-04$ & $5.85e-01$ & $4.15e-01$ & $4.8150$ & $9.8130$ & $5.71e+06$ & $600$ \\
& $1.1301$ & $1.01e-03$ & $8.07e-04$ & $5.66e-01$ & $4.34e-01$ & $4.8690$ & $10.1700$ & $6.12e+06$ & $700$ \\
& $1.1510$ & $8.81e-04$ & $6.78e-04$ & $5.82e-01$ & $4.18e-01$ & $4.8430$ & $10.5100$ & $6.47e+06$ & $800$ \\
& $1.1705$ & $6.94e-04$ & $4.93e-04$ & $5.70e-01$ & $4.30e-01$ & $4.7880$ & $10.9200$ & $6.75e+06$ & $900$ \\
& $1.1885$ & $5.12e-04$ & $3.14e-04$ & $5.44e-01$ & $4.56e-01$ & $4.6630$ & $11.3800$ & $6.99e+06$ & $1000$ \\
& $1.2007$ & $5.84e-04$ & $4.83e-04$ & $5.29e-01$ & $4.71e-01$ & $5.0170$ & $11.5900$ & $7.20e+06$ & $1100$ \\
& $1.2107$ & $5.08e-04$ & $4.09e-04$ & $5.10e-01$ & $4.90e-01$ & $4.9870$ & $11.7400$ & $7.40e+06$ & $1200$ \\
& $1.2207$ & $4.44e-04$ & $3.40e-04$ & $4.95e-01$ & $5.05e-01$ & $4.9570$ & $11.9500$ & $7.58e+06$ & $1300$ \\
& $1.2307$ & $3.85e-04$ & $2.84e-04$ & $4.76e-01$ & $5.24e-01$ & $4.9400$ & $12.2200$ & $7.74e+06$ & $1400$ \\
& $1.2407$ & $3.18e-04$ & $2.17e-04$ & $4.43e-01$ & $5.57e-01$ & $4.8410$ & $12.5600$ & $7.88e+06$ & $1500$ \\
& $1.2507$ & $2.72e-04$ & $1.71e-04$ & $4.50e-01$ & $5.50e-01$ & $4.8060$ & $12.9500$ & $8.00e+06$ & $1600$ \\
& $1.2607$ & $2.08e-04$ & $1.06e-04$ & $3.56e-01$ & $6.44e-01$ & $4.6290$ & $13.4100$ & $8.10e+06$ & $1700$ \\
& $1.2707$ & $1.79e-04$ & $7.88e-05$ & $4.02e-01$ & $5.98e-01$ & $4.6180$ & $13.9400$ & $8.20e+06$ & $1800$ \\
& $1.2792$ & $3.11e-04$ & $2.51e-04$ & $5.09e-01$ & $4.91e-01$ & $5.6720$ & $14.3500$ & $8.29e+06$ & $1900$ \\
& $1.2860$ & $2.71e-04$ & $2.13e-04$ & $4.93e-01$ & $5.07e-01$ & $5.6850$ & $14.5800$ & $8.39e+06$ & $2000$ \\
& $1.2921$ & $2.73e-04$ & $2.21e-04$ & $4.90e-01$ & $5.10e-01$ & $5.8080$ & $14.8000$ & $8.48e+06$ & $2100$ \\
& $1.2985$ & $2.10e-04$ & $1.41e-04$ & $4.78e-01$ & $5.22e-01$ & $5.6500$ & $15.1100$ & $8.56e+06$ & $2200$ \\
& $1.3049$ & $1.92e-04$ & $1.25e-04$ & $4.83e-01$ & $5.17e-01$ & $5.6500$ & $15.5000$ & $8.63e+06$ & $2300$ \\
& $1.3106$ & $1.87e-04$ & $1.26e-04$ & $4.73e-01$ & $5.27e-01$ & $5.7460$ & $15.8600$ & $8.70e+06$ & $2400$ \\
& $1.3157$ & $1.73e-04$ & $1.16e-04$ & $4.65e-01$ & $5.35e-01$ & $5.8310$ & $16.1700$ & $8.76e+06$ & $2500$ \\
& $1.3200$ & $1.63e-04$ & $1.24e-04$ & $4.62e-01$ & $5.38e-01$ & $5.8570$ & $16.4300$ & $8.82e+06$ & $2600$ \\
& $1.3244$ & $1.45e-04$ & $9.67e-05$ & $4.60e-01$ & $5.40e-01$ & $5.8170$ & $16.7200$ & $8.87e+06$ & $2700$ \\
& $1.3286$ & $1.27e-04$ & $8.13e-05$ & $4.47e-01$ & $5.53e-01$ & $5.7930$ & $17.0500$ & $8.92e+06$ & $2800$ \\
& $1.3329$ & $1.14e-04$ & $6.94e-05$ & $4.52e-01$ & $5.48e-01$ & $5.7480$ & $17.4500$ & $8.96e+06$ & $2900$ \\
& $1.3368$ & $1.08e-04$ & $6.74e-05$ & $4.11e-01$ & $5.89e-01$ & $5.7980$ & $17.8200$ & $9.00e+06$ & $3000$ \\
& $1.3404$ & $1.05e-04$ & $7.02e-05$ & $4.35e-01$ & $5.65e-01$ & $5.8900$ & $18.1700$ & $9.03e+06$ & $3100$ \\
& $1.3436$ & $1.06e-04$ & $6.49e-05$ & $4.34e-01$ & $5.66e-01$ & $5.9970$ & $18.4800$ & $9.07e+06$ & $3200$ \\
& $1.3464$ & $9.06e-05$ & $5.96e-05$ & $4.34e-01$ & $5.66e-01$ & $5.9120$ & $18.7400$ & $9.10e+06$ & $3300$ \\
& $1.3494$ & $8.18e-05$ & $5.53e-05$ & $4.25e-01$ & $5.75e-01$ & $5.8490$ & $19.0600$ & $9.13e+06$ & $3400$ \\
& $1.3522$ & $7.41e-05$ & $4.40e-05$ & $4.24e-01$ & $5.76e-01$ & $5.8710$ & $19.3900$ & $9.15e+06$ & $3500$ \\
& $1.3550$ & $6.65e-05$ & $3.73e-05$ & $4.26e-01$ & $5.74e-01$ & $5.8180$ & $19.7700$ & $9.18e+06$ & $3600$ \\
& $1.3577$ & $6.14e-05$ & $3.88e-05$ & $4.17e-01$ & $5.83e-01$ & $5.8270$ & $20.1600$ & $9.20e+06$ & $3700$ \\
& $1.3603$ & $5.94e-05$ & $3.27e-05$ & $4.13e-01$ & $5.87e-01$ & $5.8890$ & $20.5300$ & $9.22e+06$ & $3800$ \\
& $1.3625$ & $5.67e-05$ & $3.11e-05$ & $4.15e-01$ & $5.85e-01$ & $5.9260$ & $20.8700$ & $9.24e+06$ & $3900$ \\

\hline\hline

\multirow{46}{*}{\begin{tabular}{@{}c@{}} \vspace{-5.0cm}\\7\\\\$0.65[M_\odot]$\\\\$1\times10^{-7}[M_\odot\rm yr^{-1}]$\end{tabular}}
& $0.7078$ & $2.21e-02$ & $2.10e-02$ & $7.84e-01$ & $2.16e-01$ & $4.2280$ & $6.0570$ & $1.28e+07$ & $50$ \\
& $0.7168$ & $2.15e-02$ & $2.05e-02$ & $7.77e-01$ & $2.23e-01$ & $4.2900$ & $6.0860$ & $2.42e+07$ & $100$ \\
& $0.7618$ & $1.91e-02$ & $1.86e-02$ & $7.66e-01$ & $2.34e-01$ & $4.5370$ & $6.1970$ & $3.45e+07$ & $150$ \\
& $0.7958$ & $1.91e-02$ & $1.81e-02$ & $7.50e-01$ & $2.49e-01$ & $4.8260$ & $6.2410$ & $4.42e+07$ & $200$ \\
& $0.8213$ & $1.73e-02$ & $1.68e-02$ & $7.38e-01$ & $2.62e-01$ & $4.9670$ & $6.2830$ & $5.34e+07$ & $250$ \\
& $0.8463$ & $1.54e-02$ & $1.49e-02$ & $7.26e-01$ & $2.75e-01$ & $5.0830$ & $6.3890$ & $6.18e+07$ & $300$ \\
& $0.8713$ & $1.39e-02$ & $1.34e-02$ & $7.20e-01$ & $2.80e-01$ & $5.1660$ & $6.4760$ & $6.92e+07$ & $350$ \\
& $0.8963$ & $1.21e-02$ & $1.16e-02$ & $7.08e-01$ & $2.92e-01$ & $5.2440$ & $6.5880$ & $7.59e+07$ & $400$ \\
& $0.9168$ & $1.21e-02$ & $1.16e-02$ & $6.94e-01$ & $3.06e-01$ & $5.4060$ & $6.6540$ & $8.19e+07$ & $450$ \\
& $0.9288$ & $1.12e-02$ & $1.12e-02$ & $6.80e-01$ & $3.20e-01$ & $5.4920$ & $6.6350$ & $8.78e+07$ & $500$ \\
& $0.9378$ & $9.95e-03$ & $9.95e-03$ & $6.78e-01$ & $3.22e-01$ & $5.4320$ & $6.6420$ & $9.36e+07$ & $550$ \\
& $0.9498$ & $1.00e-02$ & $1.00e-02$ & $6.66e-01$ & $3.34e-01$ & $5.5700$ & $6.7290$ & $9.88e+07$ & $600$ \\
& $0.9568$ & $9.84e-03$ & $9.34e-03$ & $6.67e-01$ & $3.33e-01$ & $5.5570$ & $6.7170$ & $1.04e+08$ & $650$ \\
& $0.9648$ & $9.32e-03$ & $9.32e-03$ & $6.52e-01$ & $3.48e-01$ & $5.5700$ & $6.7600$ & $1.09e+08$ & $700$ \\
& $0.9733$ & $8.90e-03$ & $8.90e-03$ & $6.40e-01$ & $3.60e-01$ & $5.6010$ & $6.8120$ & $1.14e+08$ & $750$ \\
& $0.9833$ & $7.80e-03$ & $7.30e-03$ & $6.58e-01$ & $3.42e-01$ & $5.5260$ & $6.8760$ & $1.18e+08$ & $800$ \\
& $0.9943$ & $7.54e-03$ & $7.04e-03$ & $6.52e-01$ & $3.48e-01$ & $5.5960$ & $6.9670$ & $1.22e+08$ & $850$ \\
& $1.0028$ & $6.77e-03$ & $6.77e-03$ & $6.08e-01$ & $3.93e-01$ & $5.5450$ & $7.0030$ & $1.26e+08$ & $900$ \\
& $1.0098$ & $6.91e-03$ & $6.40e-03$ & $6.43e-01$ & $3.57e-01$ & $5.6500$ & $7.0150$ & $1.29e+08$ & $950$ \\
& $1.0163$ & $6.48e-03$ & $6.48e-03$ & $6.05e-01$ & $3.95e-01$ & $5.6380$ & $7.0480$ & $1.33e+08$ & $1000$ \\
& $1.0213$ & $6.56e-03$ & $6.56e-03$ & $5.96e-01$ & $4.04e-01$ & $5.7110$ & $7.0570$ & $1.36e+08$ & $1050$ \\
& $1.0263$ & $6.32e-03$ & $6.31e-03$ & $5.91e-01$ & $4.09e-01$ & $5.7160$ & $7.0790$ & $1.40e+08$ & $1100$ \\
& $1.0318$ & $6.11e-03$ & $5.61e-03$ & $6.30e-01$ & $3.70e-01$ & $5.7300$ & $7.1230$ & $1.43e+08$ & $1150$ \\
& $1.0373$ & $5.75e-03$ & $5.74e-03$ & $5.77e-01$ & $4.23e-01$ & $5.7080$ & $7.1690$ & $1.46e+08$ & $1200$ \\
& $1.0443$ & $5.33e-03$ & $5.34e-03$ & $5.70e-01$ & $4.30e-01$ & $5.6800$ & $7.2500$ & $1.49e+08$ & $1250$ \\
& $1.0523$ & $5.08e-03$ & $4.58e-03$ & $6.14e-01$ & $3.86e-01$ & $5.6910$ & $7.3480$ & $1.51e+08$ & $1300$ \\
& $1.0608$ & $4.55e-03$ & $4.56e-03$ & $5.45e-01$ & $4.55e-01$ & $5.6280$ & $7.4600$ & $1.54e+08$ & $1350$ \\
& $1.0763$ & $2.96e-03$ & $2.46e-03$ & $5.86e-01$ & $4.14e-01$ & $5.3260$ & $7.8070$ & $1.56e+08$ & $1400$ \\
& $1.1013$ & $2.59e-03$ & $2.10e-03$ & $5.65e-01$ & $4.35e-01$ & $5.4960$ & $8.5790$ & $1.57e+08$ & $1450$ \\
& $1.1208$ & $2.35e-03$ & $1.98e-03$ & $5.50e-01$ & $4.50e-01$ & $5.6850$ & $8.7170$ & $1.59e+08$ & $1500$ \\
& $1.1353$ & $1.77e-03$ & $1.57e-03$ & $4.82e-01$ & $5.18e-01$ & $5.2280$ & $8.7680$ & $1.60e+08$ & $1550$ \\
& $1.1472$ & $1.48e-03$ & $1.13e-03$ & $5.12e-01$ & $4.88e-01$ & $5.3330$ & $8.8380$ & $1.61e+08$ & $1600$ \\
& $1.1678$ & $5.30e-04$ & $4.29e-05$ & $3.71e-01$ & $6.29e-01$ & $4.1540$ & $9.3630$ & $1.61e+08$ & $1650$ \\
& $1.1928$ & $5.29e-04$ & $3.04e-05$ & $3.66e-01$ & $6.34e-01$ & $4.4060$ & $10.2600$ & $1.62e+08$ & $1700$ \\
& $1.2149$ & $1.24e-04$ & $0.00e+00$ & $0.00e+00$ & $0.00e+00$ & $4.8340$ & $11.1300$ & $1.62e+08$ & $1750$ \\
& $1.2279$ & $1.43e-04$ & $0.00e+00$ & $0.00e+00$ & $0.00e+00$ & $5.0300$ & $11.6100$ & $1.62e+08$ & $1800$ \\
& $1.2396$ & $3.36e-04$ & $0.00e+00$ & $0.00e+00$ & $0.00e+00$ & $5.1240$ & $12.0100$ & $1.63e+08$ & $1850$ \\
& $1.2482$ & $1.32e-04$ & $0.00e+00$ & $0.00e+00$ & $0.00e+00$ & $4.4110$ & $12.2700$ & $1.63e+08$ & $1900$ \\
& $1.2563$ & $1.04e-04$ & $0.00e+00$ & $0.00e+00$ & $0.00e+00$ & $4.6190$ & $12.4900$ & $1.63e+08$ & $1950$ \\
& $1.2627$ & $1.13e-04$ & $0.00e+00$ & $0.00e+00$ & $0.00e+00$ & $4.5160$ & $12.6300$ & $1.63e+08$ & $2000$ \\
& $1.2686$ & $1.07e-04$ & $0.00e+00$ & $0.00e+00$ & $0.00e+00$ & $4.5480$ & $12.7600$ & $1.63e+08$ & $2050$ \\
& $1.2736$ & $1.07e-04$ & $0.00e+00$ & $0.00e+00$ & $0.00e+00$ & $4.6370$ & $12.8400$ & $1.63e+08$ & $2100$ \\
& $1.2783$ & $6.50e-05$ & $0.00e+00$ & $0.00e+00$ & $0.00e+00$ & $4.9600$ & $12.8900$ & $1.64e+08$ & $2150$ \\
& $1.2825$ & $1.27e-04$ & $0.00e+00$ & $0.00e+00$ & $0.00e+00$ & $4.4440$ & $12.9200$ & $1.64e+08$ & $2200$ \\
& $1.2866$ & $7.58e-05$ & $0.00e+00$ & $0.00e+00$ & $0.00e+00$ & $4.9220$ & $12.9000$ & $1.64e+08$ & $2250$ \\
& $1.2903$ & $5.64e-05$ & $0.00e+00$ & $0.00e+00$ & $0.00e+00$ & $5.0660$ & $12.8800$ & $1.64e+08$ & $2300$ \\

\hline\pagebreak \hline

\multirow{34}{*}{\begin{tabular}{@{}c@{}} 14\\\\$0.70[M_\odot]$\\\\$3\times10^{-5}[M_\odot\rm yr^{-1}]$\end{tabular}}
& $0.7311$ & $1.56e-03$ & $0.00e+00$ & $0.00e+00$ & $0.00e+00$ & $2.5460$ & $3.1790$ & $4.43e+04$ & $20$ \\
& $0.7619$ & $1.50e-03$ & $0.00e+00$ & $0.00e+00$ & $0.00e+00$ & $2.5770$ & $3.4170$ & $9.42e+04$ & $40$ \\
& $0.7914$ & $1.42e-03$ & $0.00e+00$ & $0.00e+00$ & $0.00e+00$ & $2.6340$ & $3.6750$ & $1.39e+05$ & $60$ \\
& $0.8185$ & $1.30e-03$ & $0.00e+00$ & $0.00e+00$ & $0.00e+00$ & $2.7800$ & $3.9410$ & $1.78e+05$ & $80$ \\
& $0.8434$ & $1.17e-03$ & $0.00e+00$ & $0.00e+00$ & $0.00e+00$ & $2.8970$ & $4.2130$ & $2.10e+05$ & $100$ \\
& $0.8660$ & $1.09e-03$ & $0.00e+00$ & $0.00e+00$ & $0.00e+00$ & $2.9880$ & $4.4900$ & $2.39e+05$ & $120$ \\
& $0.8865$ & $9.85e-04$ & $0.00e+00$ & $0.00e+00$ & $0.00e+00$ & $3.0600$ & $4.7690$ & $2.63e+05$ & $140$ \\
& $0.9052$ & $9.01e-04$ & $0.00e+00$ & $0.00e+00$ & $0.00e+00$ & $3.1170$ & $5.0480$ & $2.84e+05$ & $160$ \\
& $0.9221$ & $7.95e-04$ & $0.00e+00$ & $0.00e+00$ & $0.00e+00$ & $3.1630$ & $5.3220$ & $3.02e+05$ & $180$ \\
& $0.9375$ & $7.45e-04$ & $0.00e+00$ & $0.00e+00$ & $0.00e+00$ & $3.2030$ & $5.5930$ & $3.19e+05$ & $200$ \\
& $0.9518$ & $6.85e-04$ & $0.00e+00$ & $0.00e+00$ & $0.00e+00$ & $3.2380$ & $5.8600$ & $3.33e+05$ & $220$ \\
& $0.9649$ & $6.33e-04$ & $0.00e+00$ & $0.00e+00$ & $0.00e+00$ & $3.2680$ & $6.1230$ & $3.46e+05$ & $240$ \\
& $0.9771$ & $5.92e-04$ & $0.00e+00$ & $0.00e+00$ & $0.00e+00$ & $3.2950$ & $6.3820$ & $3.58e+05$ & $260$ \\
& $0.9884$ & $5.47e-04$ & $0.00e+00$ & $0.00e+00$ & $0.00e+00$ & $3.3190$ & $6.6340$ & $3.69e+05$ & $280$ \\
& $0.9990$ & $5.24e-04$ & $0.00e+00$ & $0.00e+00$ & $0.00e+00$ & $3.3390$ & $6.8840$ & $3.79e+05$ & $300$ \\
& $1.0089$ & $4.85e-04$ & $0.00e+00$ & $0.00e+00$ & $0.00e+00$ & $3.3580$ & $7.1270$ & $3.88e+05$ & $320$ \\
& $1.0182$ & $4.56e-04$ & $0.00e+00$ & $0.00e+00$ & $0.00e+00$ & $3.3750$ & $7.3690$ & $3.96e+05$ & $340$ \\
& $1.0271$ & $4.32e-04$ & $0.00e+00$ & $0.00e+00$ & $0.00e+00$ & $3.3870$ & $7.6090$ & $4.04e+05$ & $360$ \\
& $1.0355$ & $4.04e-04$ & $0.00e+00$ & $0.00e+00$ & $0.00e+00$ & $3.4000$ & $7.8450$ & $4.12e+05$ & $380$ \\
& $1.0434$ & $3.82e-04$ & $0.00e+00$ & $0.00e+00$ & $0.00e+00$ & $3.4150$ & $8.0750$ & $4.18e+05$ & $400$ \\
& $1.0509$ & $3.74e-04$ & $0.00e+00$ & $0.00e+00$ & $0.00e+00$ & $3.4310$ & $8.3010$ & $4.25e+05$ & $420$ \\
& $1.0580$ & $3.50e-04$ & $0.00e+00$ & $0.00e+00$ & $0.00e+00$ & $3.4430$ & $8.5250$ & $4.31e+05$ & $440$ \\
& $1.0648$ & $3.32e-04$ & $0.00e+00$ & $0.00e+00$ & $0.00e+00$ & $3.4560$ & $8.7460$ & $4.37e+05$ & $460$ \\
& $1.0713$ & $3.20e-04$ & $0.00e+00$ & $0.00e+00$ & $0.00e+00$ & $3.4680$ & $8.9640$ & $4.42e+05$ & $480$ \\
& $1.0776$ & $3.04e-04$ & $0.00e+00$ & $0.00e+00$ & $0.00e+00$ & $3.4780$ & $9.1820$ & $4.47e+05$ & $500$ \\
& $1.0836$ & $2.92e-04$ & $0.00e+00$ & $0.00e+00$ & $0.00e+00$ & $3.4850$ & $9.3970$ & $4.52e+05$ & $520$ \\
& $1.0893$ & $2.78e-04$ & $0.00e+00$ & $0.00e+00$ & $0.00e+00$ & $3.4940$ & $9.6080$ & $4.57e+05$ & $540$ \\
& $1.0948$ & $2.65e-04$ & $0.00e+00$ & $0.00e+00$ & $0.00e+00$ & $3.5000$ & $9.8160$ & $4.62e+05$ & $560$ \\
& $1.1001$ & $2.60e-04$ & $0.00e+00$ & $0.00e+00$ & $0.00e+00$ & $3.5080$ & $10.0200$ & $4.66e+05$ & $580$ \\
& $1.1052$ & $2.51e-04$ & $0.00e+00$ & $0.00e+00$ & $0.00e+00$ & $3.5150$ & $10.2300$ & $4.70e+05$ & $600$ \\
& $1.1101$ & $2.42e-04$ & $0.00e+00$ & $0.00e+00$ & $0.00e+00$ & $3.5220$ & $10.4300$ & $4.74e+05$ & $620$ \\
& $1.1148$ & $2.29e-04$ & $0.00e+00$ & $0.00e+00$ & $0.00e+00$ & $3.5280$ & $10.6300$ & $4.78e+05$ & $640$ \\
& $1.1194$ & $2.20e-04$ & $0.00e+00$ & $0.00e+00$ & $0.00e+00$ & $3.5360$ & $10.8200$ & $4.82e+05$ & $660$ \\
& $1.1238$ & $2.16e-04$ & $0.00e+00$ & $0.00e+00$ & $0.00e+00$ & $3.5450$ & $11.0100$ & $4.85e+05$ & $680$ \\

\hline\hline

\multirow{53}{*}{\begin{tabular}{@{}c@{}} \vspace{-15.0cm}\\15\\\\$0.70[M_\odot]$\\\\$1\times10^{-5}[M_\odot\rm yr^{-1}]$\end{tabular}}
& $0.7632$ & $1.23e-03$ & $0.00e+00$ & $0.00e+00$ & $0.00e+00$ & $2.4940$ & $3.4340$ & $1.06e+05$ & $50$ \\
& $0.8197$ & $1.06e-03$ & $0.00e+00$ & $0.00e+00$ & $0.00e+00$ & $2.7290$ & $3.9650$ & $1.97e+05$ & $100$ \\
& $0.8680$ & $8.92e-04$ & $0.00e+00$ & $0.00e+00$ & $0.00e+00$ & $2.9300$ & $4.5370$ & $2.65e+05$ & $150$ \\
& $0.9083$ & $7.38e-04$ & $0.00e+00$ & $0.00e+00$ & $0.00e+00$ & $3.0560$ & $5.1220$ & $3.16e+05$ & $200$ \\
& $0.9422$ & $6.19e-04$ & $0.00e+00$ & $0.00e+00$ & $0.00e+00$ & $3.1410$ & $5.7100$ & $3.55e+05$ & $250$ \\
& $0.9713$ & $5.37e-04$ & $0.00e+00$ & $0.00e+00$ & $0.00e+00$ & $3.2050$ & $6.2900$ & $3.87e+05$ & $300$ \\
& $0.9961$ & $4.59e-04$ & $0.00e+00$ & $0.00e+00$ & $0.00e+00$ & $3.2570$ & $6.8500$ & $4.12e+05$ & $350$ \\
& $1.0177$ & $4.01e-04$ & $0.00e+00$ & $0.00e+00$ & $0.00e+00$ & $3.2980$ & $7.3920$ & $4.34e+05$ & $400$ \\
& $1.0368$ & $3.58e-04$ & $0.00e+00$ & $0.00e+00$ & $0.00e+00$ & $3.3300$ & $7.9190$ & $4.52e+05$ & $450$ \\
& $1.0540$ & $3.26e-04$ & $0.00e+00$ & $0.00e+00$ & $0.00e+00$ & $3.3550$ & $8.4380$ & $4.68e+05$ & $500$ \\
& $1.0692$ & $2.98e-04$ & $0.00e+00$ & $0.00e+00$ & $0.00e+00$ & $3.3790$ & $8.9360$ & $4.82e+05$ & $550$ \\
& $1.0831$ & $2.65e-04$ & $0.00e+00$ & $0.00e+00$ & $0.00e+00$ & $3.4020$ & $9.4230$ & $4.95e+05$ & $600$ \\
& $1.0959$ & $2.43e-04$ & $0.00e+00$ & $0.00e+00$ & $0.00e+00$ & $3.4200$ & $9.9030$ & $5.06e+05$ & $650$ \\
& $1.1075$ & $2.22e-04$ & $0.00e+00$ & $0.00e+00$ & $0.00e+00$ & $3.4370$ & $10.3600$ & $5.17e+05$ & $700$ \\
& $1.1182$ & $2.08e-04$ & $0.00e+00$ & $0.00e+00$ & $0.00e+00$ & $3.4510$ & $10.8100$ & $5.26e+05$ & $750$ \\
& $1.1281$ & $1.89e-04$ & $0.00e+00$ & $0.00e+00$ & $0.00e+00$ & $3.4700$ & $11.2500$ & $5.35e+05$ & $800$ \\
& $1.1373$ & $1.81e-04$ & $0.00e+00$ & $0.00e+00$ & $0.00e+00$ & $3.4920$ & $11.6700$ & $5.43e+05$ & $850$ \\
& $1.1461$ & $1.69e-04$ & $0.00e+00$ & $0.00e+00$ & $0.00e+00$ & $3.5110$ & $12.0800$ & $5.50e+05$ & $900$ \\
& $1.1542$ & $1.57e-04$ & $0.00e+00$ & $0.00e+00$ & $0.00e+00$ & $3.5220$ & $12.4800$ & $5.57e+05$ & $950$ \\
& $1.1619$ & $1.48e-04$ & $0.00e+00$ & $0.00e+00$ & $0.00e+00$ & $3.5280$ & $12.8600$ & $5.63e+05$ & $1000$ \\
& $1.1691$ & $1.39e-04$ & $0.00e+00$ & $0.00e+00$ & $0.00e+00$ & $3.5490$ & $13.2300$ & $5.69e+05$ & $1050$ \\
& $1.1760$ & $1.34e-04$ & $0.00e+00$ & $0.00e+00$ & $0.00e+00$ & $3.5660$ & $13.5900$ & $5.75e+05$ & $1100$ \\
& $1.1826$ & $1.26e-04$ & $0.00e+00$ & $0.00e+00$ & $0.00e+00$ & $3.5820$ & $13.9300$ & $5.80e+05$ & $1150$ \\
& $1.1889$ & $1.21e-04$ & $0.00e+00$ & $0.00e+00$ & $0.00e+00$ & $3.5940$ & $14.2600$ & $5.85e+05$ & $1200$ \\
& $1.1948$ & $1.15e-04$ & $0.00e+00$ & $0.00e+00$ & $0.00e+00$ & $3.6050$ & $14.5700$ & $5.90e+05$ & $1250$ \\
& $1.2004$ & $1.13e-04$ & $0.00e+00$ & $0.00e+00$ & $0.00e+00$ & $3.6150$ & $14.8600$ & $5.94e+05$ & $1300$ \\
& $1.2057$ & $1.00e-04$ & $0.00e+00$ & $0.00e+00$ & $0.00e+00$ & $3.6240$ & $15.1400$ & $5.99e+05$ & $1350$ \\
& $1.2107$ & $9.83e-05$ & $0.00e+00$ & $0.00e+00$ & $0.00e+00$ & $3.6350$ & $15.4100$ & $6.03e+05$ & $1400$ \\
& $1.2156$ & $9.70e-05$ & $0.00e+00$ & $0.00e+00$ & $0.00e+00$ & $3.6450$ & $15.6600$ & $6.07e+05$ & $1450$ \\
& $1.2203$ & $9.26e-05$ & $0.00e+00$ & $0.00e+00$ & $0.00e+00$ & $3.6530$ & $15.9000$ & $6.10e+05$ & $1500$ \\
& $1.2247$ & $8.56e-05$ & $0.00e+00$ & $0.00e+00$ & $0.00e+00$ & $3.6630$ & $16.1300$ & $6.14e+05$ & $1550$ \\
& $1.2290$ & $8.37e-05$ & $0.00e+00$ & $0.00e+00$ & $0.00e+00$ & $3.6750$ & $16.3500$ & $6.17e+05$ & $1600$ \\
& $1.2331$ & $7.94e-05$ & $0.00e+00$ & $0.00e+00$ & $0.00e+00$ & $3.6830$ & $16.5600$ & $6.20e+05$ & $1650$ \\
& $1.2371$ & $7.75e-05$ & $0.00e+00$ & $0.00e+00$ & $0.00e+00$ & $3.6890$ & $16.7700$ & $6.23e+05$ & $1700$ \\
& $1.2408$ & $7.10e-05$ & $0.00e+00$ & $0.00e+00$ & $0.00e+00$ & $3.6960$ & $16.9600$ & $6.26e+05$ & $1750$ \\
& $1.2445$ & $7.07e-05$ & $0.00e+00$ & $0.00e+00$ & $0.00e+00$ & $3.7010$ & $17.1400$ & $6.29e+05$ & $1800$ \\
& $1.2479$ & $6.96e-05$ & $0.00e+00$ & $0.00e+00$ & $0.00e+00$ & $3.7080$ & $17.3200$ & $6.32e+05$ & $1850$ \\
& $1.2513$ & $6.76e-05$ & $0.00e+00$ & $0.00e+00$ & $0.00e+00$ & $3.7160$ & $17.4900$ & $6.34e+05$ & $1900$ \\
& $1.2545$ & $6.17e-05$ & $0.00e+00$ & $0.00e+00$ & $0.00e+00$ & $3.7230$ & $17.6600$ & $6.37e+05$ & $1950$ \\
& $1.2577$ & $6.20e-05$ & $0.00e+00$ & $0.00e+00$ & $0.00e+00$ & $3.7320$ & $17.8300$ & $6.39e+05$ & $2000$ \\
& $1.2607$ & $6.14e-05$ & $0.00e+00$ & $0.00e+00$ & $0.00e+00$ & $3.7370$ & $17.9900$ & $6.42e+05$ & $2050$ \\
& $1.2637$ & $5.70e-05$ & $0.00e+00$ & $0.00e+00$ & $0.00e+00$ & $3.7460$ & $18.1400$ & $6.44e+05$ & $2100$ \\
& $1.2665$ & $5.49e-05$ & $0.00e+00$ & $0.00e+00$ & $0.00e+00$ & $3.7540$ & $18.3000$ & $6.46e+05$ & $2150$ \\
& $1.2693$ & $5.46e-05$ & $0.00e+00$ & $0.00e+00$ & $0.00e+00$ & $3.7610$ & $18.4500$ & $6.48e+05$ & $2200$ \\
& $1.2720$ & $5.26e-05$ & $0.00e+00$ & $0.00e+00$ & $0.00e+00$ & $3.7680$ & $18.5900$ & $6.50e+05$ & $2250$ \\
& $1.2746$ & $5.06e-05$ & $0.00e+00$ & $0.00e+00$ & $0.00e+00$ & $3.7770$ & $18.7400$ & $6.52e+05$ & $2300$ \\
& $1.2771$ & $5.03e-05$ & $0.00e+00$ & $0.00e+00$ & $0.00e+00$ & $3.7840$ & $18.8800$ & $6.54e+05$ & $2350$ \\
& $1.2795$ & $4.78e-05$ & $0.00e+00$ & $0.00e+00$ & $0.00e+00$ & $3.7910$ & $19.0200$ & $6.56e+05$ & $2400$ \\
& $1.2819$ & $4.68e-05$ & $0.00e+00$ & $0.00e+00$ & $0.00e+00$ & $3.7980$ & $19.1600$ & $6.58e+05$ & $2450$ \\
& $1.2842$ & $4.53e-05$ & $0.00e+00$ & $0.00e+00$ & $0.00e+00$ & $3.8020$ & $19.3000$ & $6.60e+05$ & $2500$ \\
& $1.2865$ & $4.46e-05$ & $0.00e+00$ & $0.00e+00$ & $0.00e+00$ & $3.8070$ & $19.4400$ & $6.61e+05$ & $2550$ \\
& $1.2887$ & $4.36e-05$ & $0.00e+00$ & $0.00e+00$ & $0.00e+00$ & $3.8100$ & $19.5700$ & $6.63e+05$ & $2600$ \\
& $1.2909$ & $4.28e-05$ & $0.00e+00$ & $0.00e+00$ & $0.00e+00$ & $3.8110$ & $19.7000$ & $6.65e+05$ & $2650$ \\

\hline\hline

\multirow{10}{*}{\begin{tabular}{@{}c@{}} 16\\\\$0.70[M_\odot]$\\\\$1\times10^{-6}[M_\odot\rm yr^{-1}]$\end{tabular}}
& $0.7180$ & $1.24e-03$ & $0.00e+00$ & $0.00e+00$ & $0.00e+00$ & $2.5300$ & $3.0980$ & $4.45e+04$ & $10$ \\
& $0.7304$ & $1.23e-03$ & $0.00e+00$ & $0.00e+00$ & $0.00e+00$ & $2.5290$ & $3.1860$ & $7.56e+04$ & $20$ \\
& $0.7426$ & $1.25e-03$ & $0.00e+00$ & $0.00e+00$ & $0.00e+00$ & $2.5640$ & $3.2800$ & $1.06e+05$ & $30$ \\
& $0.7548$ & $1.20e-03$ & $0.00e+00$ & $0.00e+00$ & $0.00e+00$ & $2.5670$ & $3.3790$ & $1.36e+05$ & $40$ \\
& $0.7668$ & $1.18e-03$ & $0.00e+00$ & $0.00e+00$ & $0.00e+00$ & $2.5750$ & $3.4820$ & $1.65e+05$ & $50$ \\
& $0.7788$ & $1.17e-03$ & $0.00e+00$ & $0.00e+00$ & $0.00e+00$ & $2.6050$ & $3.5890$ & $1.93e+05$ & $60$ \\
& $0.7907$ & $1.17e-03$ & $0.00e+00$ & $0.00e+00$ & $0.00e+00$ & $2.6500$ & $3.7000$ & $2.20e+05$ & $70$ \\
& $0.8026$ & $1.19e-03$ & $0.00e+00$ & $0.00e+00$ & $0.00e+00$ & $2.6560$ & $3.8170$ & $2.46e+05$ & $80$ \\
& $0.8142$ & $1.16e-03$ & $0.00e+00$ & $0.00e+00$ & $0.00e+00$ & $2.6880$ & $3.9380$ & $2.72e+05$ & $90$ \\
& $0.8257$ & $1.12e-03$ & $0.00e+00$ & $0.00e+00$ & $0.00e+00$ & $2.7050$ & $4.0630$ & $2.96e+05$ & $100$ \\

\hline\hline

\multirow{43}{*}{\begin{tabular}{@{}c@{}} \vspace{-15.0cm}\\17\\\\$0.70[M_\odot]$\\\\$5\times10^{-7}[M_\odot\rm yr^{-1}]$\end{tabular}}
& $0.7540$ & $1.97e-03$ & $0.00e+00$ & $0.00e+00$ & $0.00e+00$ & $2.8690$ & $3.3940$ & $2.10e+05$ & $25$ \\
& $0.8016$ & $1.87e-03$ & $0.00e+00$ & $0.00e+00$ & $0.00e+00$ & $2.9960$ & $3.8440$ & $3.52e+05$ & $50$ \\
& $0.8460$ & $1.70e-03$ & $0.00e+00$ & $0.00e+00$ & $0.00e+00$ & $3.0890$ & $4.3600$ & $4.78e+05$ & $75$ \\
& $0.8812$ & $1.76e-03$ & $5.91e-04$ & $6.96e-01$ & $3.04e-01$ & $3.2490$ & $4.8760$ & $5.90e+05$ & $100$ \\
& $0.9068$ & $1.80e-03$ & $9.50e-04$ & $6.62e-01$ & $3.38e-01$ & $3.4940$ & $5.3250$ & $6.99e+05$ & $125$ \\
& $0.9279$ & $1.80e-03$ & $1.04e-03$ & $6.60e-01$ & $3.40e-01$ & $3.6160$ & $5.7350$ & $8.02e+05$ & $150$ \\
& $0.9459$ & $1.72e-03$ & $1.05e-03$ & $6.85e-01$ & $3.15e-01$ & $3.6420$ & $6.1180$ & $9.01e+05$ & $175$ \\
& $0.9620$ & $1.69e-03$ & $1.13e-03$ & $6.47e-01$ & $3.53e-01$ & $3.8050$ & $6.4930$ & $9.94e+05$ & $200$ \\
& $0.9757$ & $1.59e-03$ & $1.03e-03$ & $6.73e-01$ & $3.27e-01$ & $3.7840$ & $6.8650$ & $1.08e+06$ & $225$ \\
& $0.9881$ & $1.52e-03$ & $9.33e-04$ & $6.64e-01$ & $3.36e-01$ & $3.8390$ & $7.2500$ & $1.17e+06$ & $250$ \\
& $1.0000$ & $1.47e-03$ & $9.77e-04$ & $6.51e-01$ & $3.49e-01$ & $3.9250$ & $7.6530$ & $1.25e+06$ & $275$ \\
& $1.0118$ & $1.34e-03$ & $8.35e-04$ & $6.50e-01$ & $3.50e-01$ & $3.9400$ & $8.0680$ & $1.33e+06$ & $300$ \\
& $1.0221$ & $1.29e-03$ & $8.80e-04$ & $6.38e-01$ & $3.62e-01$ & $4.0010$ & $8.4530$ & $1.40e+06$ & $325$ \\
& $1.0325$ & $1.18e-03$ & $7.57e-04$ & $6.34e-01$ & $3.66e-01$ & $4.0070$ & $8.8200$ & $1.47e+06$ & $350$ \\
& $1.0421$ & $1.20e-03$ & $8.80e-04$ & $6.22e-01$ & $3.78e-01$ & $4.1170$ & $9.1470$ & $1.53e+06$ & $375$ \\
& $1.0502$ & $1.14e-03$ & $8.07e-04$ & $6.13e-01$ & $3.87e-01$ & $4.1450$ & $9.4060$ & $1.60e+06$ & $400$ \\
& $1.0585$ & $1.06e-03$ & $7.20e-04$ & $6.08e-01$ & $3.92e-01$ & $4.1520$ & $9.6470$ & $1.66e+06$ & $425$ \\
& $1.0670$ & $9.75e-04$ & $6.35e-04$ & $6.01e-01$ & $3.99e-01$ & $4.1530$ & $9.8740$ & $1.72e+06$ & $450$ \\
& $1.0750$ & $9.90e-04$ & $7.41e-04$ & $5.95e-01$ & $4.05e-01$ & $4.2410$ & $10.0800$ & $1.77e+06$ & $475$ \\
& $1.0823$ & $9.24e-04$ & $6.68e-04$ & $5.96e-01$ & $4.04e-01$ & $4.2400$ & $10.2400$ & $1.82e+06$ & $500$ \\
& $1.0899$ & $8.51e-04$ & $4.91e-04$ & $6.34e-01$ & $3.66e-01$ & $4.1080$ & $10.4200$ & $1.87e+06$ & $525$ \\
& $1.0968$ & $8.30e-04$ & $5.70e-04$ & $6.31e-01$ & $3.69e-01$ & $4.1560$ & $10.5600$ & $1.92e+06$ & $550$ \\
& $1.1033$ & $7.83e-04$ & $5.17e-04$ & $5.86e-01$ & $4.14e-01$ & $4.2890$ & $10.7000$ & $1.96e+06$ & $575$ \\
& $1.1100$ & $7.23e-04$ & $4.55e-04$ & $5.83e-01$ & $4.17e-01$ & $4.2750$ & $10.8300$ & $2.00e+06$ & $600$ \\
& $1.1166$ & $6.81e-04$ & $4.14e-04$ & $5.67e-01$ & $4.33e-01$ & $4.2900$ & $10.9800$ & $2.04e+06$ & $625$ \\
& $1.1227$ & $6.66e-04$ & $4.98e-04$ & $5.63e-01$ & $4.37e-01$ & $4.3330$ & $11.1000$ & $2.08e+06$ & $650$ \\
& $1.1278$ & $6.27e-04$ & $4.62e-04$ & $5.59e-01$ & $4.41e-01$ & $4.3390$ & $11.1900$ & $2.12e+06$ & $675$ \\
& $1.1332$ & $6.02e-04$ & $4.36e-04$ & $5.63e-01$ & $4.37e-01$ & $4.3450$ & $11.2900$ & $2.16e+06$ & $700$ \\
& $1.1384$ & $6.01e-04$ & $4.43e-04$ & $5.57e-01$ & $4.43e-01$ & $4.3950$ & $11.4000$ & $2.19e+06$ & $725$ \\
& $1.1435$ & $5.78e-04$ & $3.20e-04$ & $6.00e-01$ & $4.00e-01$ & $4.2710$ & $11.5000$ & $2.23e+06$ & $750$ \\
& $1.1485$ & $5.74e-04$ & $3.18e-04$ & $5.93e-01$ & $4.07e-01$ & $4.3230$ & $11.6100$ & $2.26e+06$ & $775$ \\
& $1.1529$ & $5.57e-04$ & $4.02e-04$ & $5.45e-01$ & $4.55e-01$ & $4.4930$ & $11.7000$ & $2.29e+06$ & $800$ \\
& $1.1566$ & $5.43e-04$ & $3.90e-04$ & $5.77e-01$ & $4.23e-01$ & $4.3730$ & $11.7500$ & $2.32e+06$ & $825$ \\
& $1.1603$ & $5.33e-04$ & $3.98e-04$ & $5.81e-01$ & $4.19e-01$ & $4.3990$ & $11.8000$ & $2.35e+06$ & $850$ \\
& $1.1638$ & $5.14e-04$ & $3.80e-04$ & $5.77e-01$ & $4.23e-01$ & $4.4050$ & $11.8500$ & $2.38e+06$ & $875$ \\
& $1.1672$ & $4.84e-04$ & $3.56e-04$ & $5.72e-01$ & $4.28e-01$ & $4.3860$ & $11.9100$ & $2.41e+06$ & $900$ \\
& $1.1706$ & $4.92e-04$ & $3.53e-04$ & $5.64e-01$ & $4.36e-01$ & $4.4460$ & $11.9600$ & $2.44e+06$ & $925$ \\
& $1.1737$ & $4.54e-04$ & $3.28e-04$ & $5.73e-01$ & $4.27e-01$ & $4.3980$ & $12.0100$ & $2.46e+06$ & $950$ \\
& $1.1768$ & $4.47e-04$ & $3.26e-04$ & $5.66e-01$ & $4.34e-01$ & $4.4250$ & $12.0600$ & $2.49e+06$ & $975$ \\
& $1.1798$ & $4.35e-04$ & $3.13e-04$ & $5.64e-01$ & $4.36e-01$ & $4.4320$ & $12.1100$ & $2.52e+06$ & $1000$ \\
& $1.1827$ & $4.33e-04$ & $3.12e-04$ & $5.62e-01$ & $4.38e-01$ & $4.4630$ & $12.1700$ & $2.54e+06$ & $1025$ \\
& $1.1856$ & $4.23e-04$ & $3.14e-04$ & $5.65e-01$ & $4.35e-01$ & $4.4760$ & $12.2200$ & $2.57e+06$ & $1050$ \\
& $1.1884$ & $4.18e-04$ & $3.11e-04$ & $5.60e-01$ & $4.40e-01$ & $4.5000$ & $12.2700$ & $2.59e+06$ & $1075$ \\

\hline\hline

\multirow{53}{*}{\begin{tabular}{@{}c@{}} \vspace{-15.0cm}\\18\\\\$0.70[M_\odot]$\\\\$3\times10^{-7}[M_\odot\rm yr^{-1}]$\end{tabular}}
& $0.8639$ & $3.52e-03$ & $2.26e-03$ & $7.09e-01$ & $2.91e-01$ & $3.7870$ & $4.7500$ & $1.14e+06$ & $75$ \\
& $0.9293$ & $3.35e-03$ & $2.65e-03$ & $6.82e-01$ & $3.18e-01$ & $4.2410$ & $6.6330$ & $2.06e+06$ & $150$ \\
& $0.9742$ & $2.83e-03$ & $2.29e-03$ & $6.59e-01$ & $3.41e-01$ & $4.4720$ & $8.0810$ & $2.86e+06$ & $225$ \\
& $1.0109$ & $2.37e-03$ & $1.91e-03$ & $6.44e-01$ & $3.56e-01$ & $4.6080$ & $8.6120$ & $3.53e+06$ & $300$ \\
& $1.0417$ & $2.05e-03$ & $1.67e-03$ & $6.31e-01$ & $3.69e-01$ & $4.7360$ & $8.9390$ & $4.10e+06$ & $375$ \\
& $1.0692$ & $1.73e-03$ & $1.37e-03$ & $6.18e-01$ & $3.82e-01$ & $4.8160$ & $9.2570$ & $4.58e+06$ & $450$ \\
& $1.0937$ & $1.48e-03$ & $1.19e-03$ & $6.04e-01$ & $3.96e-01$ & $4.9260$ & $9.5770$ & $4.99e+06$ & $525$ \\
& $1.1151$ & $1.30e-03$ & $1.05e-03$ & $5.91e-01$ & $4.09e-01$ & $5.0180$ & $9.8760$ & $5.34e+06$ & $600$ \\
& $1.1339$ & $1.14e-03$ & $9.07e-04$ & $5.81e-01$ & $4.19e-01$ & $5.0850$ & $10.1600$ & $5.66e+06$ & $675$ \\
& $1.1508$ & $1.00e-03$ & $7.90e-04$ & $5.74e-01$ & $4.26e-01$ & $5.1350$ & $10.4400$ & $5.93e+06$ & $750$ \\
& $1.1663$ & $9.08e-04$ & $7.15e-04$ & $5.70e-01$ & $4.30e-01$ & $5.2020$ & $10.7300$ & $6.18e+06$ & $825$ \\
& $1.1803$ & $8.00e-04$ & $6.20e-04$ & $5.66e-01$ & $4.34e-01$ & $5.2260$ & $11.0200$ & $6.39e+06$ & $900$ \\
& $1.1931$ & $7.14e-04$ & $5.44e-04$ & $5.60e-01$ & $4.40e-01$ & $5.2600$ & $11.3000$ & $6.59e+06$ & $975$ \\
& $1.2049$ & $6.44e-04$ & $4.94e-04$ & $5.48e-01$ & $4.52e-01$ & $5.3180$ & $11.5900$ & $6.76e+06$ & $1050$ \\
& $1.2159$ & $5.75e-04$ & $4.35e-04$ & $5.47e-01$ & $4.53e-01$ & $5.3530$ & $11.8900$ & $6.91e+06$ & $1125$ \\
& $1.2260$ & $5.18e-04$ & $3.88e-04$ & $5.36e-01$ & $4.64e-01$ & $5.3920$ & $12.1900$ & $7.05e+06$ & $1200$ \\
& $1.2356$ & $4.64e-04$ & $3.45e-04$ & $5.36e-01$ & $4.64e-01$ & $5.3960$ & $12.4900$ & $7.18e+06$ & $1275$ \\
& $1.2444$ & $4.23e-04$ & $3.13e-04$ & $5.27e-01$ & $4.73e-01$ & $5.4400$ & $12.8000$ & $7.29e+06$ & $1350$ \\
& $1.2527$ & $3.77e-04$ & $2.77e-04$ & $5.18e-01$ & $4.82e-01$ & $5.4480$ & $13.1100$ & $7.39e+06$ & $1425$ \\
& $1.2605$ & $3.42e-04$ & $2.42e-04$ & $5.22e-01$ & $4.78e-01$ & $5.4350$ & $13.4200$ & $7.49e+06$ & $1500$ \\
& $1.2678$ & $3.05e-04$ & $2.05e-04$ & $5.15e-01$ & $4.85e-01$ & $5.4370$ & $13.7500$ & $7.57e+06$ & $1575$ \\
& $1.2747$ & $2.76e-04$ & $1.86e-04$ & $5.05e-01$ & $4.95e-01$ & $5.4500$ & $14.0700$ & $7.65e+06$ & $1650$ \\
& $1.2812$ & $2.48e-04$ & $1.68e-04$ & $4.98e-01$ & $5.02e-01$ & $5.4530$ & $14.4000$ & $7.72e+06$ & $1725$ \\
& $1.2874$ & $2.28e-04$ & $1.48e-04$ & $5.10e-01$ & $4.90e-01$ & $5.4240$ & $14.7300$ & $7.78e+06$ & $1800$ \\
& $1.2934$ & $2.04e-04$ & $1.24e-04$ & $5.05e-01$ & $4.95e-01$ & $5.4090$ & $15.0900$ & $7.83e+06$ & $1875$ \\
& $1.2988$ & $1.89e-04$ & $1.20e-04$ & $4.95e-01$ & $5.05e-01$ & $5.4410$ & $15.4100$ & $7.89e+06$ & $1950$ \\
& $1.3040$ & $1.67e-04$ & $9.66e-05$ & $4.93e-01$ & $5.07e-01$ & $5.3980$ & $15.7600$ & $7.93e+06$ & $2025$ \\
& $1.3087$ & $1.61e-04$ & $1.01e-04$ & $4.83e-01$ & $5.17e-01$ & $5.4750$ & $16.0700$ & $7.98e+06$ & $2100$ \\
& $1.3132$ & $1.43e-04$ & $8.32e-05$ & $4.79e-01$ & $5.21e-01$ & $5.4340$ & $16.3900$ & $8.02e+06$ & $2175$ \\
& $1.3177$ & $1.32e-04$ & $7.17e-05$ & $4.73e-01$ & $5.27e-01$ & $5.4390$ & $16.7400$ & $8.05e+06$ & $2250$ \\
& $1.3215$ & $1.25e-04$ & $7.46e-05$ & $4.65e-01$ & $5.35e-01$ & $5.4730$ & $17.0200$ & $8.09e+06$ & $2325$ \\
& $1.3252$ & $1.12e-04$ & $6.17e-05$ & $4.62e-01$ & $5.38e-01$ & $5.4340$ & $17.3300$ & $8.12e+06$ & $2400$ \\
& $1.3290$ & $9.97e-05$ & $4.97e-05$ & $4.55e-01$ & $5.45e-01$ & $5.3880$ & $17.6700$ & $8.15e+06$ & $2475$ \\
& $1.3324$ & $9.87e-05$ & $5.86e-05$ & $4.45e-01$ & $5.55e-01$ & $5.4840$ & $17.9900$ & $8.18e+06$ & $2550$ \\
& $1.3354$ & $8.98e-05$ & $4.98e-05$ & $4.41e-01$ & $5.59e-01$ & $5.4510$ & $18.2600$ & $8.20e+06$ & $2625$ \\
& $1.3384$ & $8.01e-05$ & $4.01e-05$ & $4.40e-01$ & $5.60e-01$ & $5.3940$ & $18.5500$ & $8.23e+06$ & $2700$ \\
& $1.3414$ & $7.15e-05$ & $3.17e-05$ & $4.37e-01$ & $5.63e-01$ & $5.3400$ & $18.8800$ & $8.25e+06$ & $2775$ \\
& $1.3444$ & $6.25e-05$ & $2.24e-05$ & $4.36e-01$ & $5.64e-01$ & $5.2600$ & $19.2500$ & $8.27e+06$ & $2850$ \\
& $1.3471$ & $6.55e-05$ & $3.55e-05$ & $4.17e-01$ & $5.83e-01$ & $5.4140$ & $19.5700$ & $8.28e+06$ & $2925$ \\
& $1.3493$ & $5.97e-05$ & $2.99e-05$ & $4.12e-01$ & $5.88e-01$ & $5.3750$ & $19.8100$ & $8.30e+06$ & $3000$ \\
& $1.3516$ & $5.43e-05$ & $2.43e-05$ & $4.09e-01$ & $5.91e-01$ & $5.3340$ & $20.0800$ & $8.32e+06$ & $3075$ \\
& $1.3538$ & $4.93e-05$ & $1.90e-05$ & $4.07e-01$ & $5.93e-01$ & $5.2730$ & $20.3700$ & $8.34e+06$ & $3150$ \\
& $1.3561$ & $4.41e-05$ & $1.41e-05$ & $4.06e-01$ & $5.94e-01$ & $5.2160$ & $20.7000$ & $8.35e+06$ & $3225$ \\
& $1.3583$ & $4.05e-05$ & $1.04e-05$ & $4.02e-01$ & $5.98e-01$ & $5.1800$ & $21.0600$ & $8.36e+06$ & $3300$ \\
& $1.3606$ & $3.66e-05$ & $6.50e-06$ & $3.88e-01$ & $6.12e-01$ & $5.1400$ & $21.4400$ & $8.38e+06$ & $3375$ \\
& $1.3628$ & $3.14e-05$ & $1.57e-06$ & $4.04e-01$ & $5.96e-01$ & $5.0310$ & $21.8600$ & $8.38e+06$ & $3450$ \\
& $1.3650$ & $3.04e-05$ & $6.35e-07$ & $3.95e-01$ & $6.05e-01$ & $5.0790$ & $22.3000$ & $8.40e+06$ & $3525$ \\
& $1.3670$ & $3.04e-05$ & $3.38e-07$ & $3.84e-01$ & $6.16e-01$ & $5.1560$ & $22.6700$ & $8.40e+06$ & $3600$ \\
& $1.3685$ & $2.99e-05$ & $1.01e-05$ & $3.50e-01$ & $6.50e-01$ & $5.2120$ & $22.9300$ & $8.42e+06$ & $3675$ \\
& $1.3700$ & $2.91e-05$ & $9.12e-06$ & $3.44e-01$ & $6.56e-01$ & $5.2410$ & $23.1900$ & $8.42e+06$ & $3750$ \\
& $1.3715$ & $2.62e-05$ & $6.23e-06$ & $3.30e-01$ & $6.70e-01$ & $5.1810$ & $23.4600$ & $8.43e+06$ & $3825$ \\
& $1.3730$ & $2.21e-05$ & $2.03e-06$ & $3.39e-01$ & $6.61e-01$ & $5.1930$ & $23.7700$ & $8.44e+06$ & $3900$ \\
& $1.3745$ & $2.04e-05$ & $6.58e-07$ & $3.35e-01$ & $6.65e-01$ & $5.2140$ & $24.1000$ & $8.45e+06$ & $3975$ \\

\hline\hline

\multirow{44}{*}{\begin{tabular}{@{}c@{}} \vspace{-12.0cm}\\19\\\\$0.70[M_\odot]$\\\\$1\times10^{-7}[M_\odot\rm yr^{-1}]$\end{tabular}}
& $0.7460$ & $2.18e-02$ & $2.08e-02$ & $7.70e-01$ & $2.30e-01$ & $4.5110$ & $5.9460$ & $1.19e+07$ & $50$ \\
& $0.7745$ & $2.04e-02$ & $1.99e-02$ & $7.61e-01$ & $2.39e-01$ & $4.6850$ & $6.0980$ & $2.20e+07$ & $100$ \\
& $0.8020$ & $1.90e-02$ & $1.85e-02$ & $7.47e-01$ & $2.54e-01$ & $4.8710$ & $6.0800$ & $3.20e+07$ & $150$ \\
& $0.8270$ & $1.69e-02$ & $1.64e-02$ & $7.35e-01$ & $2.65e-01$ & $4.9800$ & $6.1710$ & $4.11e+07$ & $200$ \\
& $0.8520$ & $1.54e-02$ & $1.49e-02$ & $7.28e-01$ & $2.72e-01$ & $5.0720$ & $6.2730$ & $4.93e+07$ & $250$ \\
& $0.8770$ & $1.36e-02$ & $1.31e-02$ & $7.17e-01$ & $2.83e-01$ & $5.1730$ & $6.3650$ & $5.67e+07$ & $300$ \\
& $0.9010$ & $1.27e-02$ & $1.22e-02$ & $7.05e-01$ & $2.95e-01$ & $5.3260$ & $6.4750$ & $6.33e+07$ & $350$ \\
& $0.9145$ & $1.23e-02$ & $1.23e-02$ & $6.86e-01$ & $3.14e-01$ & $5.4400$ & $6.4410$ & $6.97e+07$ & $400$ \\
& $0.9240$ & $1.22e-02$ & $1.22e-02$ & $6.77e-01$ & $3.23e-01$ & $5.4790$ & $6.4240$ & $7.59e+07$ & $450$ \\
& $0.9310$ & $1.16e-02$ & $1.11e-02$ & $6.86e-01$ & $3.14e-01$ & $5.4860$ & $6.4160$ & $8.21e+07$ & $500$ \\
& $0.9425$ & $1.03e-02$ & $9.76e-03$ & $6.79e-01$ & $3.21e-01$ & $5.4430$ & $6.5230$ & $8.76e+07$ & $550$ \\
& $0.9505$ & $1.06e-02$ & $1.06e-02$ & $6.61e-01$ & $3.39e-01$ & $5.5720$ & $6.5420$ & $9.30e+07$ & $600$ \\
& $0.9610$ & $9.62e-03$ & $9.62e-03$ & $6.61e-01$ & $3.39e-01$ & $5.6310$ & $6.6150$ & $9.80e+07$ & $650$ \\
& $0.9700$ & $9.11e-03$ & $8.61e-03$ & $6.64e-01$ & $3.36e-01$ & $5.5800$ & $6.6490$ & $1.03e+08$ & $700$ \\
& $0.9810$ & $7.88e-03$ & $7.88e-03$ & $6.27e-01$ & $3.72e-01$ & $5.5020$ & $6.7390$ & $1.07e+08$ & $750$ \\
& $0.9895$ & $8.23e-03$ & $8.24e-03$ & $6.28e-01$ & $3.72e-01$ & $5.6620$ & $6.7590$ & $1.12e+08$ & $800$ \\
& $0.9975$ & $7.78e-03$ & $7.28e-03$ & $6.52e-01$ & $3.48e-01$ & $5.6670$ & $6.7900$ & $1.16e+08$ & $850$ \\
& $1.0025$ & $7.65e-03$ & $7.64e-03$ & $6.15e-01$ & $3.85e-01$ & $5.7040$ & $6.7810$ & $1.20e+08$ & $900$ \\
& $1.0080$ & $7.34e-03$ & $7.34e-03$ & $6.09e-01$ & $3.91e-01$ & $5.7080$ & $6.7990$ & $1.23e+08$ & $950$ \\
& $1.0135$ & $6.65e-03$ & $6.65e-03$ & $6.06e-01$ & $3.94e-01$ & $5.6290$ & $6.8330$ & $1.27e+08$ & $1000$ \\
& $1.0190$ & $6.78e-03$ & $6.28e-03$ & $6.38e-01$ & $3.62e-01$ & $5.7160$ & $6.8690$ & $1.30e+08$ & $1050$ \\
& $1.0240$ & $6.43e-03$ & $6.43e-03$ & $5.94e-01$ & $4.06e-01$ & $5.7000$ & $6.8990$ & $1.34e+08$ & $1100$ \\
& $1.0295$ & $6.28e-03$ & $6.28e-03$ & $5.87e-01$ & $4.13e-01$ & $5.7290$ & $6.9410$ & $1.37e+08$ & $1150$ \\
& $1.0360$ & $5.90e-03$ & $5.39e-03$ & $6.26e-01$ & $3.73e-01$ & $5.7070$ & $7.0020$ & $1.40e+08$ & $1200$ \\
& $1.0440$ & $5.18e-03$ & $5.19e-03$ & $5.68e-01$ & $4.32e-01$ & $5.6110$ & $7.0970$ & $1.43e+08$ & $1250$ \\
& $1.0520$ & $4.83e-03$ & $4.83e-03$ & $5.57e-01$ & $4.43e-01$ & $5.5980$ & $7.1840$ & $1.46e+08$ & $1300$ \\
& $1.0610$ & $4.73e-03$ & $4.22e-03$ & $6.04e-01$ & $3.96e-01$ & $5.6600$ & $7.2900$ & $1.48e+08$ & $1350$ \\
& $1.0825$ & $2.69e-03$ & $2.19e-03$ & $5.77e-01$ & $4.23e-01$ & $5.2670$ & $7.7930$ & $1.50e+08$ & $1400$ \\
& $1.1075$ & $2.37e-03$ & $1.87e-03$ & $5.62e-01$ & $4.38e-01$ & $5.4290$ & $8.5250$ & $1.52e+08$ & $1450$ \\
& $1.1285$ & $2.02e-03$ & $1.74e-03$ & $5.04e-01$ & $4.96e-01$ & $5.2710$ & $8.7870$ & $1.53e+08$ & $1500$ \\
& $1.1414$ & $1.58e-03$ & $1.23e-03$ & $5.09e-01$ & $4.91e-01$ & $5.3220$ & $8.8120$ & $1.54e+08$ & $1550$ \\
& $1.1580$ & $1.03e-03$ & $5.28e-04$ & $4.56e-01$ & $5.44e-01$ & $4.9250$ & $9.0820$ & $1.55e+08$ & $1600$ \\
& $1.1830$ & $5.31e-04$ & $2.85e-05$ & $3.71e-01$ & $6.29e-01$ & $4.2880$ & $9.8850$ & $1.56e+08$ & $1650$ \\
& $1.2079$ & $4.91e-04$ & $0.00e+00$ & $0.00e+00$ & $0.00e+00$ & $5.0470$ & $10.8200$ & $1.56e+08$ & $1700$ \\
& $1.2239$ & $3.82e-04$ & $0.00e+00$ & $0.00e+00$ & $0.00e+00$ & $3.9190$ & $11.4100$ & $1.56e+08$ & $1750$ \\
& $1.2363$ & $3.45e-04$ & $0.00e+00$ & $0.00e+00$ & $0.00e+00$ & $4.9400$ & $11.8200$ & $1.56e+08$ & $1800$ \\
& $1.2459$ & $1.43e-04$ & $0.00e+00$ & $0.00e+00$ & $0.00e+00$ & $4.3670$ & $12.0900$ & $1.57e+08$ & $1850$ \\
& $1.2542$ & $2.58e-04$ & $0.00e+00$ & $0.00e+00$ & $0.00e+00$ & $4.4980$ & $12.3000$ & $1.57e+08$ & $1900$ \\
& $1.2613$ & $1.20e-04$ & $0.00e+00$ & $0.00e+00$ & $0.00e+00$ & $4.4930$ & $12.4500$ & $1.57e+08$ & $1950$ \\
& $1.2675$ & $8.61e-05$ & $0.00e+00$ & $0.00e+00$ & $0.00e+00$ & $4.8250$ & $12.5500$ & $1.57e+08$ & $2000$ \\
& $1.2729$ & $8.17e-05$ & $0.00e+00$ & $0.00e+00$ & $0.00e+00$ & $4.8820$ & $12.6300$ & $1.57e+08$ & $2050$ \\
& $1.2779$ & $7.46e-05$ & $0.00e+00$ & $0.00e+00$ & $0.00e+00$ & $4.9170$ & $12.7000$ & $1.58e+08$ & $2100$ \\
& $1.2822$ & $9.30e-05$ & $0.00e+00$ & $0.00e+00$ & $0.00e+00$ & $4.6930$ & $12.7400$ & $1.58e+08$ & $2150$ \\
& $1.2864$ & $6.95e-05$ & $0.00e+00$ & $0.00e+00$ & $0.00e+00$ & $5.0200$ & $12.7500$ & $1.58e+08$ & $2200$ \\

\hline\hline

\multirow{48}{*}{\begin{tabular}{@{}c@{}} \vspace{-12.0cm}\\28\\\\$0.80[M_\odot]$\\\\$3\times10^{-5}[M_\odot\rm yr^{-1}]$\end{tabular}}
& $0.8238$ & $9.53e-04$ & $0.00e+00$ & $0.00e+00$ & $0.00e+00$ & $2.7130$ & $3.1490$ & $2.50e+04$ & $25$ \\
& $0.8479$ & $9.54e-04$ & $0.00e+00$ & $0.00e+00$ & $0.00e+00$ & $2.7460$ & $3.3380$ & $5.32e+04$ & $50$ \\
& $0.8710$ & $9.09e-04$ & $0.00e+00$ & $0.00e+00$ & $0.00e+00$ & $2.7990$ & $3.5380$ & $7.97e+04$ & $75$ \\
& $0.8929$ & $8.49e-04$ & $0.00e+00$ & $0.00e+00$ & $0.00e+00$ & $2.9370$ & $3.7470$ & $1.04e+05$ & $100$ \\
& $0.9132$ & $7.93e-04$ & $0.00e+00$ & $0.00e+00$ & $0.00e+00$ & $3.0430$ & $3.9620$ & $1.26e+05$ & $125$ \\
& $0.9320$ & $7.20e-04$ & $0.00e+00$ & $0.00e+00$ & $0.00e+00$ & $3.1260$ & $4.1800$ & $1.45e+05$ & $150$ \\
& $0.9493$ & $6.62e-04$ & $0.00e+00$ & $0.00e+00$ & $0.00e+00$ & $3.1920$ & $4.4020$ & $1.63e+05$ & $175$ \\
& $0.9652$ & $6.13e-04$ & $0.00e+00$ & $0.00e+00$ & $0.00e+00$ & $3.2450$ & $4.6240$ & $1.78e+05$ & $200$ \\
& $0.9798$ & $5.59e-04$ & $0.00e+00$ & $0.00e+00$ & $0.00e+00$ & $3.2830$ & $4.8440$ & $1.92e+05$ & $225$ \\
& $0.9932$ & $5.08e-04$ & $0.00e+00$ & $0.00e+00$ & $0.00e+00$ & $3.3160$ & $5.0640$ & $2.05e+05$ & $250$ \\
& $1.0058$ & $4.96e-04$ & $0.00e+00$ & $0.00e+00$ & $0.00e+00$ & $3.3440$ & $5.2840$ & $2.16e+05$ & $275$ \\
& $1.0175$ & $4.50e-04$ & $0.00e+00$ & $0.00e+00$ & $0.00e+00$ & $3.3670$ & $5.5020$ & $2.27e+05$ & $300$ \\
& $1.0285$ & $4.23e-04$ & $0.00e+00$ & $0.00e+00$ & $0.00e+00$ & $3.3880$ & $5.7160$ & $2.37e+05$ & $325$ \\
& $1.0386$ & $3.97e-04$ & $0.00e+00$ & $0.00e+00$ & $0.00e+00$ & $3.4080$ & $5.9280$ & $2.46e+05$ & $350$ \\
& $1.0482$ & $3.74e-04$ & $0.00e+00$ & $0.00e+00$ & $0.00e+00$ & $3.4290$ & $6.1350$ & $2.54e+05$ & $375$ \\
& $1.0571$ & $3.49e-04$ & $0.00e+00$ & $0.00e+00$ & $0.00e+00$ & $3.4460$ & $6.3420$ & $2.61e+05$ & $400$ \\
& $1.0657$ & $3.32e-04$ & $0.00e+00$ & $0.00e+00$ & $0.00e+00$ & $3.4590$ & $6.5470$ & $2.68e+05$ & $425$ \\
& $1.0737$ & $3.12e-04$ & $0.00e+00$ & $0.00e+00$ & $0.00e+00$ & $3.4730$ & $6.7500$ & $2.75e+05$ & $450$ \\
& $1.0812$ & $2.88e-04$ & $0.00e+00$ & $0.00e+00$ & $0.00e+00$ & $3.4830$ & $6.9470$ & $2.82e+05$ & $475$ \\
& $1.0884$ & $2.86e-04$ & $0.00e+00$ & $0.00e+00$ & $0.00e+00$ & $3.4920$ & $7.1420$ & $2.87e+05$ & $500$ \\
& $1.0951$ & $2.66e-04$ & $0.00e+00$ & $0.00e+00$ & $0.00e+00$ & $3.5030$ & $7.3340$ & $2.93e+05$ & $525$ \\
& $1.1016$ & $2.51e-04$ & $0.00e+00$ & $0.00e+00$ & $0.00e+00$ & $3.5120$ & $7.5260$ & $2.98e+05$ & $550$ \\
& $1.1077$ & $2.36e-04$ & $0.00e+00$ & $0.00e+00$ & $0.00e+00$ & $3.5230$ & $7.7130$ & $3.03e+05$ & $575$ \\
& $1.1135$ & $2.32e-04$ & $0.00e+00$ & $0.00e+00$ & $0.00e+00$ & $3.5340$ & $7.8980$ & $3.08e+05$ & $600$ \\
& $1.1191$ & $2.18e-04$ & $0.00e+00$ & $0.00e+00$ & $0.00e+00$ & $3.5410$ & $8.0810$ & $3.12e+05$ & $625$ \\
& $1.1245$ & $2.08e-04$ & $0.00e+00$ & $0.00e+00$ & $0.00e+00$ & $3.5470$ & $8.2630$ & $3.17e+05$ & $650$ \\
& $1.1296$ & $2.01e-04$ & $0.00e+00$ & $0.00e+00$ & $0.00e+00$ & $3.5560$ & $8.4420$ & $3.21e+05$ & $675$ \\
& $1.1345$ & $1.94e-04$ & $0.00e+00$ & $0.00e+00$ & $0.00e+00$ & $3.5640$ & $8.6210$ & $3.25e+05$ & $700$ \\
& $1.1393$ & $1.90e-04$ & $0.00e+00$ & $0.00e+00$ & $0.00e+00$ & $3.5730$ & $8.8000$ & $3.28e+05$ & $725$ \\
& $1.1439$ & $1.86e-04$ & $0.00e+00$ & $0.00e+00$ & $0.00e+00$ & $3.5790$ & $8.9780$ & $3.32e+05$ & $750$ \\
& $1.1484$ & $1.76e-04$ & $0.00e+00$ & $0.00e+00$ & $0.00e+00$ & $3.5870$ & $9.1550$ & $3.36e+05$ & $775$ \\
& $1.1527$ & $1.68e-04$ & $0.00e+00$ & $0.00e+00$ & $0.00e+00$ & $3.5970$ & $9.3290$ & $3.39e+05$ & $800$ \\
& $1.1568$ & $1.62e-04$ & $0.00e+00$ & $0.00e+00$ & $0.00e+00$ & $3.6040$ & $9.5040$ & $3.42e+05$ & $825$ \\
& $1.1608$ & $1.60e-04$ & $0.00e+00$ & $0.00e+00$ & $0.00e+00$ & $3.6110$ & $9.6780$ & $3.45e+05$ & $850$ \\
& $1.1647$ & $1.55e-04$ & $0.00e+00$ & $0.00e+00$ & $0.00e+00$ & $3.6160$ & $9.8510$ & $3.48e+05$ & $875$ \\
& $1.1685$ & $1.49e-04$ & $0.00e+00$ & $0.00e+00$ & $0.00e+00$ & $3.6230$ & $10.0200$ & $3.51e+05$ & $900$ \\
& $1.1722$ & $1.45e-04$ & $0.00e+00$ & $0.00e+00$ & $0.00e+00$ & $3.6290$ & $10.1900$ & $3.54e+05$ & $925$ \\
& $1.1757$ & $1.40e-04$ & $0.00e+00$ & $0.00e+00$ & $0.00e+00$ & $3.6330$ & $10.3600$ & $3.57e+05$ & $950$ \\
& $1.1792$ & $1.37e-04$ & $0.00e+00$ & $0.00e+00$ & $0.00e+00$ & $3.6410$ & $10.5300$ & $3.59e+05$ & $975$ \\
& $1.1825$ & $1.31e-04$ & $0.00e+00$ & $0.00e+00$ & $0.00e+00$ & $3.6490$ & $10.7000$ & $3.62e+05$ & $1000$ \\
& $1.1858$ & $1.27e-04$ & $0.00e+00$ & $0.00e+00$ & $0.00e+00$ & $3.6580$ & $10.8700$ & $3.64e+05$ & $1025$ \\
& $1.1889$ & $1.23e-04$ & $0.00e+00$ & $0.00e+00$ & $0.00e+00$ & $3.6660$ & $11.0400$ & $3.67e+05$ & $1050$ \\
& $1.1920$ & $1.25e-04$ & $0.00e+00$ & $0.00e+00$ & $0.00e+00$ & $3.6710$ & $11.2000$ & $3.69e+05$ & $1075$ \\
& $1.1950$ & $1.18e-04$ & $0.00e+00$ & $0.00e+00$ & $0.00e+00$ & $3.6750$ & $11.3700$ & $3.71e+05$ & $1100$ \\
& $1.1980$ & $1.14e-04$ & $0.00e+00$ & $0.00e+00$ & $0.00e+00$ & $3.6760$ & $11.5300$ & $3.73e+05$ & $1125$ \\
& $1.2008$ & $1.14e-04$ & $0.00e+00$ & $0.00e+00$ & $0.00e+00$ & $3.6800$ & $11.7000$ & $3.76e+05$ & $1150$ \\
& $1.2036$ & $1.09e-04$ & $0.00e+00$ & $0.00e+00$ & $0.00e+00$ & $3.6860$ & $11.8600$ & $3.78e+05$ & $1175$ \\
& $1.2063$ & $1.07e-04$ & $0.00e+00$ & $0.00e+00$ & $0.00e+00$ & $3.6900$ & $12.0300$ & $3.80e+05$ & $1200$ \\

\hline\hline

\multirow{45}{*}{\begin{tabular}{@{}c@{}} \vspace{-8.0cm}\\29\\\\$0.80[M_\odot]$\\\\$1\times10^{-5}[M_\odot\rm yr^{-1}]$\end{tabular}}
& $0.9465$ & $5.75e-04$ & $0.00e+00$ & $0.00e+00$ & $0.00e+00$ & $3.1100$ & $4.3820$ & $1.76e+05$ & $200$ \\
& $1.0371$ & $3.46e-04$ & $0.00e+00$ & $0.00e+00$ & $0.00e+00$ & $3.3270$ & $5.9230$ & $2.68e+05$ & $400$ \\
& $1.0958$ & $2.44e-04$ & $0.00e+00$ & $0.00e+00$ & $0.00e+00$ & $3.4300$ & $7.3870$ & $3.21e+05$ & $600$ \\
& $1.1375$ & $1.80e-04$ & $0.00e+00$ & $0.00e+00$ & $0.00e+00$ & $3.4990$ & $8.7720$ & $3.57e+05$ & $800$ \\
& $1.1691$ & $1.40e-04$ & $0.00e+00$ & $0.00e+00$ & $0.00e+00$ & $3.5500$ & $10.1000$ & $3.84e+05$ & $1000$ \\
& $1.1942$ & $1.16e-04$ & $0.00e+00$ & $0.00e+00$ & $0.00e+00$ & $3.6040$ & $11.3700$ & $4.04e+05$ & $1200$ \\
& $1.2151$ & $9.45e-05$ & $0.00e+00$ & $0.00e+00$ & $0.00e+00$ & $3.6420$ & $12.6300$ & $4.21e+05$ & $1400$ \\
& $1.2325$ & $7.95e-05$ & $0.00e+00$ & $0.00e+00$ & $0.00e+00$ & $3.6780$ & $13.8300$ & $4.34e+05$ & $1600$ \\
& $1.2473$ & $6.89e-05$ & $0.00e+00$ & $0.00e+00$ & $0.00e+00$ & $3.7070$ & $14.9700$ & $4.46e+05$ & $1800$ \\
& $1.2601$ & $5.99e-05$ & $0.00e+00$ & $0.00e+00$ & $0.00e+00$ & $3.7320$ & $16.0400$ & $4.56e+05$ & $2000$ \\
& $1.2713$ & $5.20e-05$ & $0.00e+00$ & $0.00e+00$ & $0.00e+00$ & $3.7620$ & $17.0200$ & $4.65e+05$ & $2200$ \\
& $1.2811$ & $4.63e-05$ & $0.00e+00$ & $0.00e+00$ & $0.00e+00$ & $3.7880$ & $17.9100$ & $4.72e+05$ & $2400$ \\
& $1.2898$ & $4.00e-05$ & $0.00e+00$ & $0.00e+00$ & $0.00e+00$ & $3.8040$ & $18.7000$ & $4.79e+05$ & $2600$ \\
& $1.2978$ & $3.86e-05$ & $0.00e+00$ & $0.00e+00$ & $0.00e+00$ & $3.8240$ & $19.4300$ & $4.85e+05$ & $2800$ \\
& $1.3050$ & $3.39e-05$ & $0.00e+00$ & $0.00e+00$ & $0.00e+00$ & $3.8400$ & $20.1000$ & $4.90e+05$ & $3000$ \\
& $1.3115$ & $3.04e-05$ & $0.00e+00$ & $0.00e+00$ & $0.00e+00$ & $3.8570$ & $20.7000$ & $4.95e+05$ & $3200$ \\
& $1.3173$ & $2.79e-05$ & $0.00e+00$ & $0.00e+00$ & $0.00e+00$ & $3.8760$ & $21.2600$ & $5.00e+05$ & $3400$ \\
& $1.3227$ & $2.57e-05$ & $0.00e+00$ & $0.00e+00$ & $0.00e+00$ & $3.8890$ & $21.7800$ & $5.04e+05$ & $3600$ \\
& $1.3277$ & $2.40e-05$ & $0.00e+00$ & $0.00e+00$ & $0.00e+00$ & $3.9030$ & $22.2700$ & $5.07e+05$ & $3800$ \\
& $1.3322$ & $2.19e-05$ & $0.00e+00$ & $0.00e+00$ & $0.00e+00$ & $3.9110$ & $22.7500$ & $5.11e+05$ & $4000$ \\
& $1.3365$ & $2.02e-05$ & $0.00e+00$ & $0.00e+00$ & $0.00e+00$ & $3.9240$ & $23.2000$ & $5.14e+05$ & $4200$ \\
& $1.3404$ & $1.85e-05$ & $0.00e+00$ & $0.00e+00$ & $0.00e+00$ & $3.9360$ & $23.6400$ & $5.17e+05$ & $4400$ \\
& $1.3440$ & $1.60e-05$ & $0.00e+00$ & $0.00e+00$ & $0.00e+00$ & $3.9500$ & $24.0700$ & $5.20e+05$ & $4600$ \\
& $1.3473$ & $1.67e-05$ & $0.00e+00$ & $0.00e+00$ & $0.00e+00$ & $3.9610$ & $24.4800$ & $5.22e+05$ & $4800$ \\
& $1.3505$ & $1.50e-05$ & $0.00e+00$ & $0.00e+00$ & $0.00e+00$ & $3.9710$ & $24.8900$ & $5.25e+05$ & $5000$ \\
& $1.3534$ & $1.44e-05$ & $0.00e+00$ & $0.00e+00$ & $0.00e+00$ & $3.9900$ & $25.2900$ & $5.27e+05$ & $5200$ \\
& $1.3562$ & $1.32e-05$ & $0.00e+00$ & $0.00e+00$ & $0.00e+00$ & $3.9940$ & $25.6700$ & $5.29e+05$ & $5400$ \\
& $1.3588$ & $1.26e-05$ & $0.00e+00$ & $0.00e+00$ & $0.00e+00$ & $4.0330$ & $26.0600$ & $5.31e+05$ & $5600$ \\
& $1.3612$ & $1.19e-05$ & $0.00e+00$ & $0.00e+00$ & $0.00e+00$ & $4.0650$ & $26.4300$ & $5.33e+05$ & $5800$ \\
& $1.3635$ & $1.06e-05$ & $0.00e+00$ & $0.00e+00$ & $0.00e+00$ & $4.0280$ & $26.8000$ & $5.35e+05$ & $6000$ \\
& $1.3657$ & $9.90e-06$ & $0.00e+00$ & $0.00e+00$ & $0.00e+00$ & $4.0340$ & $27.1700$ & $5.36e+05$ & $6200$ \\
& $1.3677$ & $1.02e-05$ & $0.00e+00$ & $0.00e+00$ & $0.00e+00$ & $4.1520$ & $27.5300$ & $5.38e+05$ & $6400$ \\
& $1.3697$ & $9.51e-06$ & $0.00e+00$ & $0.00e+00$ & $0.00e+00$ & $4.1550$ & $27.8900$ & $5.39e+05$ & $6600$ \\
& $1.3715$ & $9.18e-06$ & $0.00e+00$ & $0.00e+00$ & $0.00e+00$ & $4.2320$ & $28.2400$ & $5.41e+05$ & $6800$ \\
& $1.3733$ & $8.41e-06$ & $0.00e+00$ & $0.00e+00$ & $0.00e+00$ & $4.1700$ & $28.5900$ & $5.42e+05$ & $7000$ \\
& $1.3749$ & $8.12e-06$ & $0.00e+00$ & $0.00e+00$ & $0.00e+00$ & $4.2020$ & $28.9300$ & $5.43e+05$ & $7200$ \\
& $1.3765$ & $8.10e-06$ & $2.41e-07$ & $9.80e-01$ & $2.00e-02$ & $4.2760$ & $29.2700$ & $5.45e+05$ & $7400$ \\
& $1.3781$ & $7.28e-06$ & $1.62e-07$ & $9.80e-01$ & $2.00e-02$ & $4.2400$ & $29.6100$ & $5.46e+05$ & $7600$ \\
& $1.3795$ & $7.38e-06$ & $2.11e-07$ & $9.80e-01$ & $2.00e-02$ & $4.3110$ & $29.9400$ & $5.47e+05$ & $7800$ \\
& $1.3809$ & $6.95e-06$ & $2.24e-07$ & $9.80e-01$ & $2.00e-02$ & $4.3180$ & $30.2700$ & $5.48e+05$ & $8000$ \\
& $1.3822$ & $6.08e-06$ & $5.14e-08$ & $9.80e-01$ & $2.00e-02$ & $4.2370$ & $30.6000$ & $5.49e+05$ & $8200$ \\
& $1.3834$ & $6.51e-06$ & $2.01e-07$ & $9.80e-01$ & $2.00e-02$ & $4.3640$ & $30.9300$ & $5.50e+05$ & $8400$ \\
& $1.3846$ & $6.00e-06$ & $1.29e-07$ & $9.80e-01$ & $2.00e-02$ & $4.3400$ & $31.2500$ & $5.51e+05$ & $8600$ \\
& $1.3858$ & $5.47e-06$ & $1.10e-07$ & $9.80e-01$ & $2.00e-02$ & $4.3180$ & $31.5700$ & $5.52e+05$ & $8800$ \\
& $1.3869$ & $5.59e-06$ & $0.00e+00$ & $0.00e+00$ & $0.00e+00$ & $4.3740$ & $31.8800$ & $5.53e+05$ & $9000$ \\

\hline\hline

\multirow{51}{*}{\begin{tabular}{@{}c@{}} \vspace{-8.0cm}\\31\\\\$0.80[M_\odot]$\\\\$5\times10^{-6}[M_\odot\rm yr^{-1}]$\end{tabular}}
& $0.9091$ & $6.33e-04$ & $0.00e+00$ & $0.00e+00$ & $0.00e+00$ & $2.9080$ & $3.9350$ & $1.45e+05$ & $150$ \\
& $0.9898$ & $4.33e-04$ & $0.00e+00$ & $0.00e+00$ & $0.00e+00$ & $3.1670$ & $5.0410$ & $2.39e+05$ & $300$ \\
& $1.0471$ & $3.38e-04$ & $0.00e+00$ & $0.00e+00$ & $0.00e+00$ & $3.2830$ & $6.1530$ & $2.99e+05$ & $450$ \\
& $1.0903$ & $2.52e-04$ & $0.00e+00$ & $0.00e+00$ & $0.00e+00$ & $3.3540$ & $7.2420$ & $3.41e+05$ & $600$ \\
& $1.1241$ & $2.00e-04$ & $0.00e+00$ & $0.00e+00$ & $0.00e+00$ & $3.4120$ & $8.3020$ & $3.72e+05$ & $750$ \\
& $1.1515$ & $1.71e-04$ & $0.00e+00$ & $0.00e+00$ & $0.00e+00$ & $3.4590$ & $9.3380$ & $3.97e+05$ & $900$ \\
& $1.1743$ & $1.43e-04$ & $0.00e+00$ & $0.00e+00$ & $0.00e+00$ & $3.4940$ & $10.3600$ & $4.17e+05$ & $1050$ \\
& $1.1932$ & $1.18e-04$ & $0.00e+00$ & $0.00e+00$ & $0.00e+00$ & $3.5260$ & $11.3400$ & $4.34e+05$ & $1200$ \\
& $1.2094$ & $9.79e-05$ & $0.00e+00$ & $0.00e+00$ & $0.00e+00$ & $3.5570$ & $12.2900$ & $4.48e+05$ & $1350$ \\
& $1.2236$ & $8.87e-05$ & $0.00e+00$ & $0.00e+00$ & $0.00e+00$ & $3.5820$ & $13.2200$ & $4.60e+05$ & $1500$ \\
& $1.2361$ & $7.87e-05$ & $0.00e+00$ & $0.00e+00$ & $0.00e+00$ & $3.6070$ & $14.1200$ & $4.71e+05$ & $1650$ \\
& $1.2472$ & $6.98e-05$ & $0.00e+00$ & $0.00e+00$ & $0.00e+00$ & $3.6270$ & $14.9800$ & $4.80e+05$ & $1800$ \\
& $1.2570$ & $6.22e-05$ & $0.00e+00$ & $0.00e+00$ & $0.00e+00$ & $3.6490$ & $15.7800$ & $4.88e+05$ & $1950$ \\
& $1.2658$ & $5.56e-05$ & $0.00e+00$ & $0.00e+00$ & $0.00e+00$ & $3.6680$ & $16.5300$ & $4.95e+05$ & $2100$ \\
& $1.2738$ & $4.98e-05$ & $0.00e+00$ & $0.00e+00$ & $0.00e+00$ & $3.6860$ & $17.2300$ & $5.02e+05$ & $2250$ \\
& $1.2811$ & $4.66e-05$ & $0.00e+00$ & $0.00e+00$ & $0.00e+00$ & $3.6970$ & $17.8800$ & $5.08e+05$ & $2400$ \\
& $1.2878$ & $4.23e-05$ & $0.00e+00$ & $0.00e+00$ & $0.00e+00$ & $3.7090$ & $18.4700$ & $5.14e+05$ & $2550$ \\
& $1.2939$ & $3.93e-05$ & $0.00e+00$ & $0.00e+00$ & $0.00e+00$ & $3.7240$ & $19.0200$ & $5.19e+05$ & $2700$ \\
& $1.2995$ & $3.55e-05$ & $0.00e+00$ & $0.00e+00$ & $0.00e+00$ & $3.7330$ & $19.5200$ & $5.23e+05$ & $2850$ \\
& $1.3047$ & $3.27e-05$ & $0.00e+00$ & $0.00e+00$ & $0.00e+00$ & $3.7440$ & $19.9800$ & $5.28e+05$ & $3000$ \\
& $1.3094$ & $3.06e-05$ & $0.00e+00$ & $0.00e+00$ & $0.00e+00$ & $3.7490$ & $20.4000$ & $5.32e+05$ & $3150$ \\
& $1.3140$ & $2.98e-05$ & $0.00e+00$ & $0.00e+00$ & $0.00e+00$ & $3.7710$ & $20.8200$ & $5.35e+05$ & $3300$ \\
& $1.3183$ & $2.86e-05$ & $0.00e+00$ & $0.00e+00$ & $0.00e+00$ & $3.7780$ & $21.2200$ & $5.39e+05$ & $3450$ \\
& $1.3222$ & $2.53e-05$ & $0.00e+00$ & $0.00e+00$ & $0.00e+00$ & $3.7880$ & $21.6000$ & $5.42e+05$ & $3600$ \\
& $1.3260$ & $2.39e-05$ & $0.00e+00$ & $0.00e+00$ & $0.00e+00$ & $3.8010$ & $21.9600$ & $5.45e+05$ & $3750$ \\
& $1.3295$ & $2.25e-05$ & $0.00e+00$ & $0.00e+00$ & $0.00e+00$ & $3.8110$ & $22.3200$ & $5.48e+05$ & $3900$ \\
& $1.3328$ & $2.19e-05$ & $0.00e+00$ & $0.00e+00$ & $0.00e+00$ & $3.8320$ & $22.6600$ & $5.51e+05$ & $4050$ \\
& $1.3360$ & $2.08e-05$ & $0.00e+00$ & $0.00e+00$ & $0.00e+00$ & $3.8560$ & $23.0000$ & $5.53e+05$ & $4200$ \\
& $1.3389$ & $1.90e-05$ & $0.00e+00$ & $0.00e+00$ & $0.00e+00$ & $3.8450$ & $23.3300$ & $5.56e+05$ & $4350$ \\
& $1.3417$ & $1.85e-05$ & $0.00e+00$ & $0.00e+00$ & $0.00e+00$ & $3.8940$ & $23.6600$ & $5.58e+05$ & $4500$ \\
& $1.3444$ & $1.75e-05$ & $0.00e+00$ & $0.00e+00$ & $0.00e+00$ & $3.9190$ & $23.9700$ & $5.60e+05$ & $4650$ \\
& $1.3469$ & $1.64e-05$ & $0.00e+00$ & $0.00e+00$ & $0.00e+00$ & $3.9170$ & $24.2800$ & $5.62e+05$ & $4800$ \\
& $1.3493$ & $1.59e-05$ & $0.00e+00$ & $0.00e+00$ & $0.00e+00$ & $3.9650$ & $24.5900$ & $5.64e+05$ & $4950$ \\
& $1.3517$ & $1.49e-05$ & $0.00e+00$ & $0.00e+00$ & $0.00e+00$ & $3.9730$ & $24.9000$ & $5.66e+05$ & $5100$ \\
& $1.3539$ & $1.43e-05$ & $0.00e+00$ & $0.00e+00$ & $0.00e+00$ & $3.9930$ & $25.2000$ & $5.68e+05$ & $5250$ \\
& $1.3559$ & $1.35e-05$ & $0.00e+00$ & $0.00e+00$ & $0.00e+00$ & $4.0060$ & $25.4900$ & $5.69e+05$ & $5400$ \\
& $1.3580$ & $1.29e-05$ & $0.00e+00$ & $0.00e+00$ & $0.00e+00$ & $4.0360$ & $25.7800$ & $5.71e+05$ & $5550$ \\
& $1.3599$ & $1.24e-05$ & $0.00e+00$ & $0.00e+00$ & $0.00e+00$ & $4.0590$ & $26.0700$ & $5.73e+05$ & $5700$ \\
& $1.3617$ & $1.19e-05$ & $0.00e+00$ & $0.00e+00$ & $0.00e+00$ & $4.0780$ & $26.3600$ & $5.74e+05$ & $5850$ \\
& $1.3634$ & $1.14e-05$ & $0.00e+00$ & $0.00e+00$ & $0.00e+00$ & $4.0950$ & $26.6400$ & $5.76e+05$ & $6000$ \\
& $1.3651$ & $1.13e-05$ & $1.14e-07$ & $9.80e-01$ & $2.00e-02$ & $4.1440$ & $26.9200$ & $5.77e+05$ & $6150$ \\
& $1.3667$ & $1.04e-05$ & $0.00e+00$ & $0.00e+00$ & $0.00e+00$ & $4.1220$ & $27.1900$ & $5.78e+05$ & $6300$ \\
& $1.3682$ & $1.02e-05$ & $1.06e-07$ & $9.80e-01$ & $2.00e-02$ & $4.1730$ & $27.4600$ & $5.79e+05$ & $6450$ \\
& $1.3697$ & $9.89e-06$ & $5.06e-08$ & $9.80e-01$ & $2.00e-02$ & $4.1950$ & $27.7300$ & $5.81e+05$ & $6600$ \\
& $1.3711$ & $9.49e-06$ & $1.10e-07$ & $9.80e-01$ & $2.00e-02$ & $4.2060$ & $28.0000$ & $5.82e+05$ & $6750$ \\
& $1.3725$ & $9.06e-06$ & $0.00e+00$ & $0.00e+00$ & $0.00e+00$ & $4.2090$ & $28.2700$ & $5.83e+05$ & $6900$ \\
& $1.3738$ & $8.02e-06$ & $0.00e+00$ & $0.00e+00$ & $0.00e+00$ & $4.1470$ & $28.5300$ & $5.84e+05$ & $7050$ \\
& $1.3750$ & $8.63e-06$ & $1.55e-07$ & $9.80e-01$ & $2.00e-02$ & $4.2680$ & $28.7800$ & $5.85e+05$ & $7200$ \\
& $1.3762$ & $7.53e-06$ & $0.00e+00$ & $0.00e+00$ & $0.00e+00$ & $4.1870$ & $29.0300$ & $5.86e+05$ & $7350$ \\
& $1.3774$ & $7.71e-06$ & $0.00e+00$ & $0.00e+00$ & $0.00e+00$ & $4.2790$ & $29.2900$ & $5.87e+05$ & $7500$ \\
& $1.3785$ & $7.53e-06$ & $0.00e+00$ & $0.00e+00$ & $0.00e+00$ & $4.3260$ & $29.5400$ & $5.88e+05$ & $7650$ \\

\hline\hline

\multirow{41}{*}{\begin{tabular}{@{}c@{}} \vspace{-3.0cm}\\32\\\\$0.80[M_\odot]$\\\\$1\times10^{-6}[M_\odot\rm yr^{-1}]$\end{tabular}}
& $0.9391$ & $7.63e-04$ & $0.00e+00$ & $0.00e+00$ & $0.00e+00$ & $3.0690$ & $4.3190$ & $2.60e+05$ & $150$ \\
& $1.0334$ & $5.08e-04$ & $0.00e+00$ & $0.00e+00$ & $0.00e+00$ & $3.2920$ & $5.9010$ & $4.13e+05$ & $300$ \\
& $1.0972$ & $3.62e-04$ & $9.67e-06$ & $6.77e-01$ & $3.23e-01$ & $3.5280$ & $7.5000$ & $5.10e+05$ & $450$ \\
& $1.1431$ & $2.79e-04$ & $2.15e-05$ & $6.64e-01$ & $3.36e-01$ & $3.6440$ & $9.0760$ & $5.78e+05$ & $600$ \\
& $1.1758$ & $2.20e-04$ & $3.47e-05$ & $6.08e-01$ & $3.92e-01$ & $3.8170$ & $10.5300$ & $6.29e+05$ & $750$ \\
& $1.2013$ & $1.83e-04$ & $3.29e-05$ & $5.95e-01$ & $4.05e-01$ & $3.9030$ & $11.9200$ & $6.70e+05$ & $900$ \\
& $1.2221$ & $1.50e-04$ & $2.99e-05$ & $5.78e-01$ & $4.22e-01$ & $3.9550$ & $13.2300$ & $7.04e+05$ & $1050$ \\
& $1.2395$ & $1.30e-04$ & $2.52e-05$ & $5.86e-01$ & $4.14e-01$ & $3.9940$ & $14.4300$ & $7.32e+05$ & $1200$ \\
& $1.2543$ & $1.10e-04$ & $2.04e-05$ & $5.73e-01$ & $4.27e-01$ & $4.0480$ & $15.5000$ & $7.55e+05$ & $1350$ \\
& $1.2670$ & $9.51e-05$ & $1.50e-05$ & $5.91e-01$ & $4.09e-01$ & $4.0220$ & $16.4300$ & $7.76e+05$ & $1500$ \\
& $1.2782$ & $8.45e-05$ & $9.33e-06$ & $6.11e-01$ & $3.89e-01$ & $4.0760$ & $17.2300$ & $7.93e+05$ & $1650$ \\
& $1.2881$ & $7.15e-05$ & $6.52e-06$ & $6.17e-01$ & $3.84e-01$ & $4.0970$ & $17.9400$ & $8.09e+05$ & $1800$ \\
& $1.2970$ & $6.31e-05$ & $8.08e-06$ & $5.86e-01$ & $4.14e-01$ & $4.1450$ & $18.5900$ & $8.22e+05$ & $1950$ \\
& $1.3047$ & $6.06e-05$ & $1.31e-05$ & $5.74e-01$ & $4.26e-01$ & $4.2200$ & $19.1400$ & $8.34e+05$ & $2100$ \\
& $1.3114$ & $5.46e-05$ & $1.26e-05$ & $5.50e-01$ & $4.50e-01$ & $4.3930$ & $19.6300$ & $8.46e+05$ & $2250$ \\
& $1.3175$ & $4.94e-05$ & $1.09e-05$ & $5.46e-01$ & $4.54e-01$ & $4.3130$ & $20.1000$ & $8.56e+05$ & $2400$ \\
& $1.3232$ & $4.52e-05$ & $8.73e-06$ & $5.75e-01$ & $4.25e-01$ & $4.3310$ & $20.5600$ & $8.65e+05$ & $2550$ \\
& $1.3283$ & $4.09e-05$ & $7.95e-06$ & $5.63e-01$ & $4.37e-01$ & $4.4560$ & $21.0100$ & $8.73e+05$ & $2700$ \\
& $1.3330$ & $3.73e-05$ & $6.77e-06$ & $5.67e-01$ & $4.33e-01$ & $4.3860$ & $21.4500$ & $8.81e+05$ & $2850$ \\
& $1.3374$ & $3.47e-05$ & $7.17e-06$ & $5.42e-01$ & $4.58e-01$ & $4.4960$ & $21.8900$ & $8.88e+05$ & $3000$ \\
& $1.3414$ & $3.19e-05$ & $5.88e-06$ & $5.53e-01$ & $4.47e-01$ & $4.5090$ & $22.3100$ & $8.95e+05$ & $3150$ \\
& $1.3452$ & $3.00e-05$ & $5.50e-06$ & $5.45e-01$ & $4.55e-01$ & $4.5430$ & $22.7400$ & $9.01e+05$ & $3300$ \\
& $1.3486$ & $2.73e-05$ & $4.80e-06$ & $5.46e-01$ & $4.54e-01$ & $4.5510$ & $23.1500$ & $9.06e+05$ & $3450$ \\
& $1.3519$ & $2.56e-05$ & $4.13e-06$ & $5.67e-01$ & $4.33e-01$ & $4.5870$ & $23.5700$ & $9.12e+05$ & $3600$ \\
& $1.3550$ & $2.34e-05$ & $3.87e-06$ & $5.24e-01$ & $4.76e-01$ & $4.5970$ & $23.9800$ & $9.16e+05$ & $3750$ \\
& $1.3578$ & $2.20e-05$ & $3.01e-06$ & $5.47e-01$ & $4.53e-01$ & $4.6290$ & $24.3800$ & $9.21e+05$ & $3900$ \\
& $1.3605$ & $2.02e-05$ & $3.19e-06$ & $5.28e-01$ & $4.72e-01$ & $4.6340$ & $24.7800$ & $9.25e+05$ & $4050$ \\
& $1.3630$ & $1.91e-05$ & $2.63e-06$ & $5.64e-01$ & $4.36e-01$ & $4.6510$ & $25.1800$ & $9.29e+05$ & $4200$ \\
& $1.3654$ & $1.67e-05$ & $2.21e-06$ & $5.38e-01$ & $4.62e-01$ & $4.6180$ & $25.5700$ & $9.33e+05$ & $4350$ \\
& $1.3677$ & $1.68e-05$ & $1.82e-06$ & $5.56e-01$ & $4.44e-01$ & $4.7020$ & $25.9600$ & $9.36e+05$ & $4500$ \\
& $1.3698$ & $1.56e-05$ & $1.63e-06$ & $5.64e-01$ & $4.36e-01$ & $4.7120$ & $26.3500$ & $9.40e+05$ & $4650$ \\
& $1.3718$ & $1.37e-05$ & $1.17e-06$ & $5.54e-01$ & $4.46e-01$ & $4.6780$ & $26.7300$ & $9.42e+05$ & $4800$ \\
& $1.3737$ & $1.30e-05$ & $1.45e-06$ & $5.18e-01$ & $4.82e-01$ & $4.6920$ & $27.1100$ & $9.45e+05$ & $4950$ \\
& $1.3755$ & $1.31e-05$ & $1.08e-06$ & $5.69e-01$ & $4.31e-01$ & $4.7810$ & $27.4800$ & $9.48e+05$ & $5100$ \\
& $1.3772$ & $1.23e-05$ & $1.27e-06$ & $5.25e-01$ & $4.75e-01$ & $4.8010$ & $27.8600$ & $9.51e+05$ & $5250$ \\
& $1.3788$ & $1.17e-05$ & $1.21e-06$ & $5.21e-01$ & $4.79e-01$ & $4.8030$ & $28.2200$ & $9.53e+05$ & $5400$ \\
& $1.3804$ & $1.08e-05$ & $7.80e-07$ & $5.13e-01$ & $4.87e-01$ & $4.8000$ & $28.5900$ & $9.56e+05$ & $5550$ \\
& $1.3819$ & $1.03e-05$ & $7.80e-07$ & $5.02e-01$ & $4.98e-01$ & $4.8130$ & $28.9500$ & $9.58e+05$ & $5700$ \\
& $1.3833$ & $9.96e-06$ & $4.51e-07$ & $6.17e-01$ & $3.83e-01$ & $4.8560$ & $29.3100$ & $9.60e+05$ & $5850$ \\
& $1.3846$ & $9.31e-06$ & $3.16e-07$ & $5.00e-01$ & $5.00e-01$ & $4.8560$ & $29.6700$ & $9.62e+05$ & $6000$ \\
& $1.3859$ & $8.70e-06$ & $1.95e-07$ & $5.63e-01$ & $4.37e-01$ & $4.8740$ & $30.0300$ & $9.64e+05$ & $6150$ \\

\hline\pagebreak\hline

\multirow{55}{*}{\begin{tabular}{@{}c@{}} 33\\\\$0.80[M_\odot]$\\\\$1\times10^{-7}[M_\odot\rm yr^{-1}]$\end{tabular}}
& $0.8429$ & $1.48e-02$ & $1.40e-02$ & $7.28e-01$ & $2.72e-01$ & $5.0100$ & $5.8520$ & $8.50e+06$ & $50$ \\
& $0.8767$ & $1.37e-02$ & $1.31e-02$ & $7.14e-01$ & $2.86e-01$ & $5.2410$ & $6.4060$ & $1.57e+07$ & $100$ \\
& $0.9067$ & $1.18e-02$ & $1.12e-02$ & $6.99e-01$ & $3.01e-01$ & $5.3730$ & $6.5290$ & $2.22e+07$ & $150$ \\
& $0.9347$ & $1.11e-02$ & $1.07e-02$ & $6.92e-01$ & $3.08e-01$ & $5.5740$ & $6.6610$ & $2.79e+07$ & $200$ \\
& $0.9547$ & $1.02e-02$ & $9.80e-03$ & $6.83e-01$ & $3.17e-01$ & $5.6900$ & $6.6860$ & $3.33e+07$ & $250$ \\
& $0.9747$ & $9.14e-03$ & $8.74e-03$ & $6.74e-01$ & $3.26e-01$ & $5.7770$ & $6.7940$ & $3.82e+07$ & $300$ \\
& $0.9947$ & $8.21e-03$ & $7.81e-03$ & $6.71e-01$ & $3.29e-01$ & $5.8290$ & $6.9120$ & $4.26e+07$ & $350$ \\
& $1.0147$ & $7.19e-03$ & $6.79e-03$ & $6.62e-01$ & $3.38e-01$ & $5.8870$ & $7.0500$ & $4.66e+07$ & $400$ \\
& $1.0347$ & $6.26e-03$ & $5.86e-03$ & $6.53e-01$ & $3.47e-01$ & $5.9360$ & $7.2110$ & $5.00e+07$ & $450$ \\
& $1.0507$ & $5.86e-03$ & $5.67e-03$ & $6.45e-01$ & $3.55e-01$ & $6.0520$ & $7.2990$ & $5.31e+07$ & $500$ \\
& $1.0607$ & $5.96e-03$ & $5.76e-03$ & $6.42e-01$ & $3.58e-01$ & $6.2100$ & $7.2590$ & $5.63e+07$ & $550$ \\
& $1.0707$ & $5.49e-03$ & $5.28e-03$ & $6.36e-01$ & $3.64e-01$ & $6.2210$ & $7.2860$ & $5.92e+07$ & $600$ \\
& $1.0807$ & $5.03e-03$ & $4.83e-03$ & $6.31e-01$ & $3.69e-01$ & $6.2290$ & $7.3580$ & $6.19e+07$ & $650$ \\
& $1.0907$ & $4.61e-03$ & $4.41e-03$ & $6.25e-01$ & $3.76e-01$ & $6.2340$ & $7.4480$ & $6.43e+07$ & $700$ \\
& $1.1007$ & $4.26e-03$ & $4.06e-03$ & $6.18e-01$ & $3.82e-01$ & $6.2610$ & $7.5510$ & $6.66e+07$ & $750$ \\
& $1.1107$ & $3.90e-03$ & $3.70e-03$ & $6.10e-01$ & $3.90e-01$ & $6.2800$ & $7.6620$ & $6.87e+07$ & $800$ \\
& $1.1207$ & $3.56e-03$ & $3.36e-03$ & $6.00e-01$ & $4.00e-01$ & $6.2040$ & $7.7760$ & $7.06e+07$ & $850$ \\
& $1.1305$ & $3.25e-03$ & $3.25e-03$ & $5.69e-01$ & $4.31e-01$ & $6.2050$ & $7.8920$ & $7.24e+07$ & $900$ \\
& $1.1359$ & $3.43e-03$ & $3.23e-03$ & $5.89e-01$ & $4.11e-01$ & $6.3750$ & $7.8880$ & $7.41e+07$ & $950$ \\
& $1.1417$ & $2.98e-03$ & $2.97e-03$ & $5.57e-01$ & $4.43e-01$ & $6.2380$ & $7.9020$ & $7.58e+07$ & $1000$ \\
& $1.1485$ & $2.99e-03$ & $2.78e-03$ & $5.80e-01$ & $4.20e-01$ & $6.3450$ & $7.9800$ & $7.74e+07$ & $1050$ \\
& $1.1535$ & $2.86e-03$ & $2.66e-03$ & $5.76e-01$ & $4.24e-01$ & $6.3480$ & $8.0330$ & $7.88e+07$ & $1100$ \\
& $1.1587$ & $2.57e-03$ & $2.57e-03$ & $5.37e-01$ & $4.63e-01$ & $6.2630$ & $8.0990$ & $8.02e+07$ & $1150$ \\
& $1.1639$ & $2.51e-03$ & $2.31e-03$ & $5.67e-01$ & $4.33e-01$ & $6.3030$ & $8.1850$ & $8.16e+07$ & $1200$ \\
& $1.1699$ & $2.05e-03$ & $1.89e-03$ & $5.62e-01$ & $4.38e-01$ & $6.2530$ & $8.3230$ & $8.28e+07$ & $1250$ \\
& $1.1777$ & $2.01e-03$ & $1.81e-03$ & $5.51e-01$ & $4.49e-01$ & $6.1530$ & $8.6470$ & $8.38e+07$ & $1300$ \\
& $1.1859$ & $1.73e-03$ & $1.53e-03$ & $5.44e-01$ & $4.56e-01$ & $6.2330$ & $8.9330$ & $8.48e+07$ & $1350$ \\
& $1.1919$ & $1.72e-03$ & $1.63e-03$ & $5.51e-01$ & $4.49e-01$ & $6.3570$ & $8.9390$ & $8.57e+07$ & $1400$ \\
& $1.1962$ & $1.53e-03$ & $1.50e-03$ & $5.39e-01$ & $4.61e-01$ & $6.2380$ & $8.8680$ & $8.66e+07$ & $1450$ \\
& $1.1995$ & $1.59e-03$ & $1.43e-03$ & $5.54e-01$ & $4.46e-01$ & $6.2010$ & $8.7850$ & $8.74e+07$ & $1500$ \\
& $1.2023$ & $1.45e-03$ & $1.25e-03$ & $5.47e-01$ & $4.53e-01$ & $6.0270$ & $8.7050$ & $8.82e+07$ & $1550$ \\
& $1.2041$ & $1.52e-03$ & $1.52e-03$ & $5.47e-01$ & $4.53e-01$ & $6.3920$ & $8.6170$ & $8.90e+07$ & $1600$ \\
& $1.2057$ & $1.46e-03$ & $1.46e-03$ & $5.42e-01$ & $4.58e-01$ & $6.3600$ & $8.5390$ & $8.98e+07$ & $1650$ \\
& $1.2067$ & $1.47e-03$ & $1.47e-03$ & $5.42e-01$ & $4.58e-01$ & $6.3910$ & $8.4550$ & $9.06e+07$ & $1700$ \\
& $1.2085$ & $1.39e-03$ & $1.39e-03$ & $5.31e-01$ & $4.69e-01$ & $6.3430$ & $8.4180$ & $9.14e+07$ & $1750$ \\
& $1.2111$ & $1.36e-03$ & $1.19e-03$ & $5.47e-01$ & $4.53e-01$ & $6.1130$ & $8.4280$ & $9.21e+07$ & $1800$ \\
& $1.2131$ & $1.24e-03$ & $1.09e-03$ & $5.32e-01$ & $4.68e-01$ & $6.0080$ & $8.4300$ & $9.28e+07$ & $1850$ \\
& $1.2154$ & $1.22e-03$ & $1.22e-03$ & $5.17e-01$ & $4.83e-01$ & $6.2870$ & $8.4550$ & $9.34e+07$ & $1900$ \\
& $1.2167$ & $1.20e-03$ & $1.04e-03$ & $5.28e-01$ & $4.72e-01$ & $6.0030$ & $8.4460$ & $9.41e+07$ & $1950$ \\
& $1.2200$ & $1.03e-03$ & $1.00e-03$ & $5.00e-01$ & $5.00e-01$ & $6.0790$ & $8.5200$ & $9.47e+07$ & $2000$ \\
& $1.2224$ & $1.00e-03$ & $9.45e-04$ & $5.10e-01$ & $4.90e-01$ & $6.0730$ & $8.5640$ & $9.53e+07$ & $2050$ \\
& $1.2273$ & $9.69e-04$ & $8.20e-04$ & $4.98e-01$ & $5.02e-01$ & $5.8470$ & $8.7230$ & $9.58e+07$ & $2100$ \\
& $1.2316$ & $8.35e-04$ & $8.27e-04$ & $4.63e-01$ & $5.37e-01$ & $6.0870$ & $8.8640$ & $9.63e+07$ & $2150$ \\
& $1.2357$ & $7.31e-04$ & $6.01e-04$ & $4.73e-01$ & $5.27e-01$ & $5.8510$ & $9.0000$ & $9.68e+07$ & $2200$ \\
& $1.2423$ & $6.20e-04$ & $4.92e-04$ & $4.43e-01$ & $5.57e-01$ & $5.7280$ & $9.2750$ & $9.71e+07$ & $2250$ \\
& $1.2511$ & $4.21e-04$ & $2.22e-04$ & $4.05e-01$ & $5.95e-01$ & $5.3090$ & $9.6960$ & $9.75e+07$ & $2300$ \\
& $1.2611$ & $2.16e-04$ & $1.42e-05$ & $3.30e-01$ & $6.70e-01$ & $4.5340$ & $10.2300$ & $9.77e+07$ & $2350$ \\
& $1.2711$ & $2.11e-04$ & $7.01e-06$ & $3.23e-01$ & $6.77e-01$ & $4.6550$ & $10.8500$ & $9.79e+07$ & $2400$ \\
& $1.2811$ & $2.05e-04$ & $7.22e-06$ & $3.12e-01$ & $6.88e-01$ & $4.8030$ & $11.5200$ & $9.80e+07$ & $2450$ \\
& $1.2903$ & $5.06e-05$ & $0.00e+00$ & $0.00e+00$ & $0.00e+00$ & $5.1390$ & $12.2100$ & $9.82e+07$ & $2500$ \\
& $1.2967$ & $5.02e-05$ & $0.00e+00$ & $0.00e+00$ & $0.00e+00$ & $5.2890$ & $12.7000$ & $9.83e+07$ & $2550$ \\
& $1.3017$ & $6.38e-05$ & $0.00e+00$ & $0.00e+00$ & $0.00e+00$ & $5.3660$ & $13.0900$ & $9.84e+07$ & $2600$ \\
& $1.3067$ & $6.43e-05$ & $0.00e+00$ & $0.00e+00$ & $0.00e+00$ & $4.6510$ & $13.4900$ & $9.85e+07$ & $2650$ \\
& $1.3111$ & $6.05e-05$ & $0.00e+00$ & $0.00e+00$ & $0.00e+00$ & $4.7260$ & $13.8400$ & $9.86e+07$ & $2700$ \\
& $1.3145$ & $3.87e-05$ & $0.00e+00$ & $0.00e+00$ & $0.00e+00$ & $5.7190$ & $14.0900$ & $9.86e+07$ & $2750$ \\

\hline\hline

\multirow{36}{*}{\begin{tabular}{@{}c@{}} 35\\\\$0.80[M_\odot]$\\\\$5\times10^{-8}[M_\odot\rm yr^{-1}]$\end{tabular}}
& $0.8285$ & $3.07e-02$ & $3.07e-02$ & $7.30e-01$ & $2.70e-01$ & $5.6670$ & $5.8460$ & $3.41e+07$ & $50$ \\
& $0.8535$ & $2.73e-02$ & $2.73e-02$ & $7.18e-01$ & $2.82e-01$ & $5.7750$ & $5.9480$ & $6.37e+07$ & $100$ \\
& $0.8215$ & $3.21e-02$ & $3.21e-02$ & $7.31e-01$ & $2.69e-01$ & $5.6620$ & $5.7350$ & $9.22e+07$ & $150$ \\
& $0.8465$ & $2.82e-02$ & $2.82e-02$ & $7.21e-01$ & $2.79e-01$ & $5.7400$ & $5.9210$ & $1.23e+08$ & $200$ \\
& $0.8735$ & $2.49e-02$ & $2.39e-02$ & $7.16e-01$ & $2.84e-01$ & $5.7900$ & $6.0310$ & $1.50e+08$ & $250$ \\
& $0.8315$ & $3.08e-02$ & $2.98e-02$ & $7.35e-01$ & $2.65e-01$ & $5.6640$ & $5.8660$ & $1.83e+08$ & $300$ \\
& $0.8565$ & $2.70e-02$ & $2.70e-02$ & $7.16e-01$ & $2.84e-01$ & $5.7960$ & $5.9610$ & $2.13e+08$ & $350$ \\
& $0.8325$ & $3.06e-02$ & $2.96e-02$ & $7.34e-01$ & $2.66e-01$ & $5.6650$ & $5.8590$ & $2.42e+08$ & $400$ \\
& $0.8575$ & $2.74e-02$ & $2.64e-02$ & $7.24e-01$ & $2.76e-01$ & $5.7630$ & $5.9700$ & $2.71e+08$ & $450$ \\
& $0.8235$ & $3.21e-02$ & $3.11e-02$ & $7.37e-01$ & $2.63e-01$ & $5.6350$ & $5.8310$ & $3.01e+08$ & $500$ \\
& $0.8485$ & $2.85e-02$ & $2.75e-02$ & $7.28e-01$ & $2.72e-01$ & $5.7260$ & $5.9340$ & $3.32e+08$ & $550$ \\
& $0.8009$ & $2.45e-02$ & $9.81e-02$ & $2.98e-01$ & $7.02e-01$ & $5.7880$ & $4.8990$ & $3.59e+08$ & $600$ \\
& $0.8325$ & $3.07e-02$ & $2.97e-02$ & $7.34e-01$ & $2.66e-01$ & $5.6610$ & $5.8760$ & $3.92e+08$ & $650$ \\
& $0.8585$ & $2.71e-02$ & $2.61e-02$ & $7.25e-01$ & $2.75e-01$ & $5.7560$ & $5.9780$ & $4.22e+08$ & $700$ \\
& $0.8185$ & $3.29e-02$ & $3.19e-02$ & $7.39e-01$ & $2.61e-01$ & $5.6290$ & $5.8230$ & $4.52e+08$ & $750$ \\
& $0.8435$ & $2.93e-02$ & $2.83e-02$ & $7.30e-01$ & $2.70e-01$ & $5.7110$ & $5.9160$ & $4.83e+08$ & $800$ \\
& $0.8715$ & $2.51e-02$ & $2.41e-02$ & $7.18e-01$ & $2.82e-01$ & $5.7830$ & $6.0320$ & $5.11e+08$ & $850$ \\
& $0.8675$ & $2.57e-02$ & $2.57e-02$ & $7.11e-01$ & $2.89e-01$ & $5.8390$ & $5.9970$ & $5.38e+08$ & $900$ \\
& $0.8425$ & $2.88e-02$ & $2.88e-02$ & $7.23e-01$ & $2.77e-01$ & $5.7330$ & $5.9080$ & $5.68e+08$ & $950$ \\
& $0.8705$ & $2.55e-02$ & $2.45e-02$ & $7.18e-01$ & $2.82e-01$ & $5.7970$ & $6.0280$ & $5.96e+08$ & $1000$ \\
& $0.8265$ & $3.13e-02$ & $3.13e-02$ & $7.31e-01$ & $2.70e-01$ & $5.6810$ & $5.8480$ & $6.28e+08$ & $1050$ \\
& $0.8515$ & $2.79e-02$ & $2.79e-02$ & $7.19e-01$ & $2.81e-01$ & $5.7880$ & $5.9470$ & $6.58e+08$ & $1100$ \\
& $0.8135$ & $3.39e-02$ & $3.28e-02$ & $7.40e-01$ & $2.60e-01$ & $5.6130$ & $5.7230$ & $6.87e+08$ & $1150$ \\
& $0.8385$ & $2.96e-02$ & $2.96e-02$ & $7.25e-01$ & $2.75e-01$ & $5.7300$ & $5.8960$ & $7.19e+08$ & $1200$ \\
& $0.8665$ & $2.62e-02$ & $2.52e-02$ & $7.20e-01$ & $2.80e-01$ & $5.7910$ & $6.0110$ & $7.48e+08$ & $1250$ \\
& $0.8305$ & $3.11e-02$ & $3.01e-02$ & $7.35e-01$ & $2.65e-01$ & $5.6660$ & $5.8650$ & $7.81e+08$ & $1300$ \\
& $0.8565$ & $2.76e-02$ & $2.66e-02$ & $7.25e-01$ & $2.75e-01$ & $5.7630$ & $5.9760$ & $8.11e+08$ & $1350$ \\
& $0.8175$ & $3.32e-02$ & $3.22e-02$ & $7.38e-01$ & $2.62e-01$ & $5.6330$ & $5.8190$ & $8.41e+08$ & $1400$ \\
& $0.8425$ & $2.92e-02$ & $2.92e-02$ & $7.23e-01$ & $2.77e-01$ & $5.7550$ & $5.9060$ & $8.72e+08$ & $1450$ \\
& $0.8715$ & $2.54e-02$ & $2.44e-02$ & $7.18e-01$ & $2.82e-01$ & $5.7990$ & $6.0320$ & $9.00e+08$ & $1500$ \\
& $0.8315$ & $3.06e-02$ & $3.06e-02$ & $7.28e-01$ & $2.72e-01$ & $5.7050$ & $5.8680$ & $9.33e+08$ & $1550$ \\
& $0.8585$ & $2.74e-02$ & $2.64e-02$ & $7.24e-01$ & $2.77e-01$ & $5.7720$ & $5.9820$ & $9.63e+08$ & $1600$ \\
& $0.8165$ & $3.33e-02$ & $3.23e-02$ & $7.39e-01$ & $2.62e-01$ & $5.6220$ & $5.8190$ & $9.93e+08$ & $1650$ \\
& $0.8425$ & $2.95e-02$ & $2.84e-02$ & $7.31e-01$ & $2.70e-01$ & $5.7110$ & $5.9160$ & $1.02e+09$ & $1700$ \\
& $0.8705$ & $2.51e-02$ & $2.51e-02$ & $7.09e-01$ & $2.91e-01$ & $5.8360$ & $6.0290$ & $1.05e+09$ & $1750$ \\
& $0.8325$ & $3.05e-02$ & $3.05e-02$ & $7.28e-01$ & $2.72e-01$ & $5.7100$ & $5.8760$ & $1.09e+09$ & $1800$ \\

\hline\hline

\multirow{26}{*}{\begin{tabular}{@{}c@{}} \vspace{-5.0cm}\\36\\\\$0.80[M_\odot]$\\\\$4\times10^{-8}[M_\odot\rm yr^{-1}]$\end{tabular}}
& $0.8176$ & $3.84e-02$ & $3.74e-02$ & $6.74e-01$ & $3.26e-01$ & $5.7080$ & $5.8900$ & $2.05e+07$ & $20$ \\
& $0.8369$ & $3.63e-02$ & $3.53e-02$ & $6.61e-01$ & $3.39e-01$ & $5.8040$ & $5.9480$ & $3.94e+07$ & $40$ \\
& $0.8550$ & $3.41e-02$ & $3.32e-02$ & $6.51e-01$ & $3.48e-01$ & $5.9140$ & $6.0120$ & $5.73e+07$ & $60$ \\
& $0.8721$ & $3.19e-02$ & $3.10e-02$ & $6.41e-01$ & $3.59e-01$ & $6.0010$ & $6.0550$ & $7.41e+07$ & $80$ \\
& $0.8883$ & $3.04e-02$ & $2.96e-02$ & $6.42e-01$ & $3.58e-01$ & $6.1590$ & $6.1200$ & $9.00e+07$ & $100$ \\
& $0.9035$ & $2.84e-02$ & $2.77e-02$ & $6.33e-01$ & $3.67e-01$ & $6.2240$ & $6.1870$ & $1.05e+08$ & $120$ \\
& $0.9178$ & $2.65e-02$ & $2.58e-02$ & $6.24e-01$ & $3.76e-01$ & $6.2830$ & $6.2430$ & $1.19e+08$ & $140$ \\
& $0.9312$ & $2.52e-02$ & $2.45e-02$ & $6.28e-01$ & $3.72e-01$ & $6.4080$ & $6.3040$ & $1.32e+08$ & $160$ \\
& $0.9438$ & $2.36e-02$ & $2.29e-02$ & $6.20e-01$ & $3.80e-01$ & $6.4570$ & $6.3620$ & $1.45e+08$ & $180$ \\
& $0.9557$ & $2.21e-02$ & $2.15e-02$ & $6.23e-01$ & $3.77e-01$ & $6.5370$ & $6.4160$ & $1.56e+08$ & $200$ \\
& $0.9669$ & $2.09e-02$ & $2.03e-02$ & $6.17e-01$ & $3.83e-01$ & $6.5900$ & $6.4710$ & $1.68e+08$ & $220$ \\
& $0.9774$ & $1.95e-02$ & $1.89e-02$ & $6.08e-01$ & $3.92e-01$ & $6.6060$ & $6.5170$ & $1.78e+08$ & $240$ \\
& $0.9872$ & $1.83e-02$ & $1.79e-02$ & $6.10e-01$ & $3.90e-01$ & $6.6520$ & $6.5670$ & $1.88e+08$ & $260$ \\
& $0.9964$ & $1.72e-02$ & $1.68e-02$ & $6.03e-01$ & $3.97e-01$ & $6.6660$ & $6.6100$ & $1.97e+08$ & $280$ \\
& $1.0052$ & $1.62e-02$ & $1.58e-02$ & $5.96e-01$ & $4.04e-01$ & $6.6700$ & $6.6520$ & $2.06e+08$ & $300$ \\
& $1.0133$ & $1.51e-02$ & $1.48e-02$ & $5.94e-01$ & $4.06e-01$ & $6.6670$ & $6.6900$ & $2.14e+08$ & $320$ \\
& $1.0209$ & $1.42e-02$ & $1.38e-02$ & $5.85e-01$ & $4.15e-01$ & $6.6450$ & $6.7240$ & $2.21e+08$ & $340$ \\
& $1.0282$ & $1.36e-02$ & $1.32e-02$ & $5.80e-01$ & $4.20e-01$ & $6.6590$ & $6.7590$ & $2.28e+08$ & $360$ \\
& $1.0350$ & $1.28e-02$ & $1.25e-02$ & $5.71e-01$ & $4.29e-01$ & $6.6320$ & $6.7940$ & $2.35e+08$ & $380$ \\
& $1.0415$ & $1.19e-02$ & $1.16e-02$ & $5.88e-01$ & $4.12e-01$ & $6.7450$ & $6.8600$ & $2.42e+08$ & $400$ \\
& $1.0478$ & $1.12e-02$ & $1.09e-02$ & $5.82e-01$ & $4.18e-01$ & $6.7010$ & $6.9030$ & $2.48e+08$ & $420$ \\
& $1.0549$ & $1.01e-02$ & $9.78e-03$ & $5.68e-01$ & $4.32e-01$ & $6.6090$ & $6.9570$ & $2.53e+08$ & $440$ \\
& $1.0646$ & $9.07e-03$ & $8.51e-03$ & $5.53e-01$ & $4.47e-01$ & $6.4830$ & $7.0620$ & $2.58e+08$ & $460$ \\
& $1.0771$ & $7.23e-03$ & $6.55e-03$ & $5.70e-01$ & $4.30e-01$ & $6.4410$ & $7.2420$ & $2.63e+08$ & $480$ \\
& $1.0898$ & $6.50e-03$ & $5.93e-03$ & $5.61e-01$ & $4.39e-01$ & $6.4350$ & $7.4160$ & $2.67e+08$ & $500$ \\
& $1.1000$ & $5.59e-03$ & $5.09e-03$ & $5.52e-01$ & $4.48e-01$ & $6.3500$ & $7.5250$ & $2.70e+08$ & $520$ \\

\hline\hline

\multirow{19}{*}{\begin{tabular}{@{}c@{}} 43\\\\$1.0[M_\odot]$\\\\$3\times10^{-5}[M_\odot\rm yr^{-1}]$\end{tabular}}
& $1.0162$ & $3.29e-04$ & $0.00e+00$ & $0.00e+00$ & $0.00e+00$ & $3.0320$ & $3.1480$ & $1.24e+04$ & $50$ \\
& $1.0326$ & $3.16e-04$ & $0.00e+00$ & $0.00e+00$ & $0.00e+00$ & $3.0410$ & $3.3050$ & $2.61e+04$ & $100$ \\
& $1.0483$ & $3.06e-04$ & $0.00e+00$ & $0.00e+00$ & $0.00e+00$ & $3.1530$ & $3.4690$ & $3.91e+04$ & $150$ \\
& $1.0632$ & $2.88e-04$ & $0.00e+00$ & $0.00e+00$ & $0.00e+00$ & $3.2770$ & $3.6390$ & $5.16e+04$ & $200$ \\
& $1.0772$ & $2.72e-04$ & $0.00e+00$ & $0.00e+00$ & $0.00e+00$ & $3.3600$ & $3.8150$ & $6.34e+04$ & $250$ \\
& $1.0902$ & $2.49e-04$ & $0.00e+00$ & $0.00e+00$ & $0.00e+00$ & $3.4260$ & $3.9920$ & $7.40e+04$ & $300$ \\
& $1.1022$ & $2.37e-04$ & $0.00e+00$ & $0.00e+00$ & $0.00e+00$ & $3.4750$ & $4.1700$ & $8.37e+04$ & $350$ \\
& $1.1132$ & $2.12e-04$ & $0.00e+00$ & $0.00e+00$ & $0.00e+00$ & $3.5060$ & $4.3490$ & $9.27e+04$ & $400$ \\
& $1.1235$ & $1.97e-04$ & $0.00e+00$ & $0.00e+00$ & $0.00e+00$ & $3.5300$ & $4.5280$ & $1.01e+05$ & $450$ \\
& $1.1332$ & $1.87e-04$ & $0.00e+00$ & $0.00e+00$ & $0.00e+00$ & $3.5550$ & $4.7060$ & $1.09e+05$ & $500$ \\
& $1.1422$ & $1.77e-04$ & $0.00e+00$ & $0.00e+00$ & $0.00e+00$ & $3.5730$ & $4.8850$ & $1.16e+05$ & $550$ \\
& $1.1506$ & $1.62e-04$ & $0.00e+00$ & $0.00e+00$ & $0.00e+00$ & $3.5910$ & $5.0620$ & $1.22e+05$ & $600$ \\
& $1.1585$ & $1.54e-04$ & $0.00e+00$ & $0.00e+00$ & $0.00e+00$ & $3.6080$ & $5.2380$ & $1.28e+05$ & $650$ \\
& $1.1660$ & $1.43e-04$ & $0.00e+00$ & $0.00e+00$ & $0.00e+00$ & $3.6240$ & $5.4140$ & $1.34e+05$ & $700$ \\
& $1.1730$ & $1.38e-04$ & $0.00e+00$ & $0.00e+00$ & $0.00e+00$ & $3.6360$ & $5.5890$ & $1.40e+05$ & $750$ \\
& $1.1797$ & $1.31e-04$ & $0.00e+00$ & $0.00e+00$ & $0.00e+00$ & $3.6490$ & $5.7640$ & $1.45e+05$ & $800$ \\
& $1.1860$ & $1.23e-04$ & $0.00e+00$ & $0.00e+00$ & $0.00e+00$ & $3.6610$ & $5.9360$ & $1.49e+05$ & $850$ \\
& $1.1920$ & $1.18e-04$ & $0.00e+00$ & $0.00e+00$ & $0.00e+00$ & $3.6710$ & $6.1080$ & $1.54e+05$ & $900$ \\
& $1.1977$ & $1.13e-04$ & $0.00e+00$ & $0.00e+00$ & $0.00e+00$ & $3.6810$ & $6.2800$ & $1.58e+05$ & $950$ \\

\hline\hline

\multirow{36}{*}{\begin{tabular}{@{}c@{}} \vspace{-8.0cm}\\44\\\\$1.0[M_\odot]$\\\\$1\times10^{-5}[M_\odot\rm yr^{-1}]$\end{tabular}}
& $1.0224$ & $2.95e-04$ & $0.00e+00$ & $0.00e+00$ & $0.00e+00$ & $3.0240$ & $3.2070$ & $1.92e+04$ & $75$ \\
& $1.0440$ & $2.83e-04$ & $0.00e+00$ & $0.00e+00$ & $0.00e+00$ & $3.0410$ & $3.4240$ & $3.89e+04$ & $150$ \\
& $1.0641$ & $2.58e-04$ & $0.00e+00$ & $0.00e+00$ & $0.00e+00$ & $3.2150$ & $3.6520$ & $5.71e+04$ & $225$ \\
& $1.0825$ & $2.37e-04$ & $0.00e+00$ & $0.00e+00$ & $0.00e+00$ & $3.3240$ & $3.8880$ & $7.35e+04$ & $300$ \\
& $1.0993$ & $2.10e-04$ & $0.00e+00$ & $0.00e+00$ & $0.00e+00$ & $3.3940$ & $4.1300$ & $8.83e+04$ & $375$ \\
& $1.1147$ & $1.96e-04$ & $0.00e+00$ & $0.00e+00$ & $0.00e+00$ & $3.4360$ & $4.3780$ & $1.02e+05$ & $450$ \\
& $1.1287$ & $1.79e-04$ & $0.00e+00$ & $0.00e+00$ & $0.00e+00$ & $3.4700$ & $4.6270$ & $1.14e+05$ & $525$ \\
& $1.1415$ & $1.65e-04$ & $0.00e+00$ & $0.00e+00$ & $0.00e+00$ & $3.5020$ & $4.8750$ & $1.25e+05$ & $600$ \\
& $1.1532$ & $1.49e-04$ & $0.00e+00$ & $0.00e+00$ & $0.00e+00$ & $3.5280$ & $5.1230$ & $1.34e+05$ & $675$ \\
& $1.1640$ & $1.36e-04$ & $0.00e+00$ & $0.00e+00$ & $0.00e+00$ & $3.5510$ & $5.3710$ & $1.43e+05$ & $750$ \\
& $1.1740$ & $1.31e-04$ & $0.00e+00$ & $0.00e+00$ & $0.00e+00$ & $3.5690$ & $5.6180$ & $1.51e+05$ & $825$ \\
& $1.1832$ & $1.19e-04$ & $0.00e+00$ & $0.00e+00$ & $0.00e+00$ & $3.5890$ & $5.8640$ & $1.59e+05$ & $900$ \\
& $1.1918$ & $1.08e-04$ & $0.00e+00$ & $0.00e+00$ & $0.00e+00$ & $3.6060$ & $6.1090$ & $1.66e+05$ & $975$ \\
& $1.1998$ & $1.00e-04$ & $0.00e+00$ & $0.00e+00$ & $0.00e+00$ & $3.6210$ & $6.3520$ & $1.72e+05$ & $1050$ \\
& $1.2073$ & $9.68e-05$ & $0.00e+00$ & $0.00e+00$ & $0.00e+00$ & $3.6380$ & $6.5930$ & $1.78e+05$ & $1125$ \\
& $1.2143$ & $9.01e-05$ & $0.00e+00$ & $0.00e+00$ & $0.00e+00$ & $3.6510$ & $6.8350$ & $1.84e+05$ & $1200$ \\
& $1.2210$ & $8.81e-05$ & $0.00e+00$ & $0.00e+00$ & $0.00e+00$ & $3.6600$ & $7.0790$ & $1.89e+05$ & $1275$ \\
& $1.2273$ & $8.12e-05$ & $0.00e+00$ & $0.00e+00$ & $0.00e+00$ & $3.6690$ & $7.3230$ & $1.94e+05$ & $1350$ \\
& $1.2332$ & $7.82e-05$ & $0.00e+00$ & $0.00e+00$ & $0.00e+00$ & $3.6820$ & $7.5650$ & $1.99e+05$ & $1425$ \\
& $1.2388$ & $7.20e-05$ & $0.00e+00$ & $0.00e+00$ & $0.00e+00$ & $3.6950$ & $7.8060$ & $2.03e+05$ & $1500$ \\
& $1.2440$ & $6.77e-05$ & $0.00e+00$ & $0.00e+00$ & $0.00e+00$ & $3.7080$ & $8.0470$ & $2.07e+05$ & $1575$ \\
& $1.2491$ & $6.42e-05$ & $0.00e+00$ & $0.00e+00$ & $0.00e+00$ & $3.7180$ & $8.2870$ & $2.11e+05$ & $1650$ \\
& $1.2538$ & $6.39e-05$ & $0.00e+00$ & $0.00e+00$ & $0.00e+00$ & $3.7290$ & $8.5270$ & $2.15e+05$ & $1725$ \\
& $1.2584$ & $5.77e-05$ & $0.00e+00$ & $0.00e+00$ & $0.00e+00$ & $3.7390$ & $8.7670$ & $2.18e+05$ & $1800$ \\
& $1.2627$ & $5.70e-05$ & $0.00e+00$ & $0.00e+00$ & $0.00e+00$ & $3.7470$ & $9.0080$ & $2.22e+05$ & $1875$ \\
& $1.2668$ & $5.37e-05$ & $0.00e+00$ & $0.00e+00$ & $0.00e+00$ & $3.7570$ & $9.2480$ & $2.25e+05$ & $1950$ \\
& $1.2707$ & $4.88e-05$ & $0.00e+00$ & $0.00e+00$ & $0.00e+00$ & $3.7660$ & $9.4860$ & $2.28e+05$ & $2025$ \\
& $1.2745$ & $4.97e-05$ & $0.00e+00$ & $0.00e+00$ & $0.00e+00$ & $3.7750$ & $9.7250$ & $2.31e+05$ & $2100$ \\
& $1.2781$ & $4.57e-05$ & $0.00e+00$ & $0.00e+00$ & $0.00e+00$ & $3.7860$ & $9.9630$ & $2.33e+05$ & $2175$ \\
& $1.2815$ & $4.61e-05$ & $0.00e+00$ & $0.00e+00$ & $0.00e+00$ & $3.7980$ & $10.2000$ & $2.36e+05$ & $2250$ \\
& $1.2849$ & $4.36e-05$ & $0.00e+00$ & $0.00e+00$ & $0.00e+00$ & $3.8030$ & $10.4400$ & $2.39e+05$ & $2325$ \\
& $1.2881$ & $4.17e-05$ & $0.00e+00$ & $0.00e+00$ & $0.00e+00$ & $3.8110$ & $10.6800$ & $2.41e+05$ & $2400$ \\
& $1.2911$ & $3.88e-05$ & $0.00e+00$ & $0.00e+00$ & $0.00e+00$ & $3.8150$ & $10.9200$ & $2.43e+05$ & $2475$ \\
& $1.2941$ & $3.83e-05$ & $0.00e+00$ & $0.00e+00$ & $0.00e+00$ & $3.8180$ & $11.1600$ & $2.46e+05$ & $2550$ \\
& $1.2969$ & $3.85e-05$ & $0.00e+00$ & $0.00e+00$ & $0.00e+00$ & $3.8230$ & $11.4000$ & $2.48e+05$ & $2625$ \\
& $1.2997$ & $3.48e-05$ & $0.00e+00$ & $0.00e+00$ & $0.00e+00$ & $3.8290$ & $11.6500$ & $2.50e+05$ & $2700$ \\

\hline\hline

\multirow{13}{*}{\begin{tabular}{@{}c@{}} 45\\\\$1.0[M_\odot]$\\\\$1\times10^{-6}[M_\odot\rm yr^{-1}]$\end{tabular}}
& $1.0103$ & $5.86e-04$ & $1.86e-04$ & $6.54e-01$ & $3.46e-01$ & $3.4480$ & $3.1020$ & $2.45e+04$ & $25$ \\
& $1.0207$ & $5.46e-04$ & $1.47e-04$ & $6.43e-01$ & $3.57e-01$ & $3.4280$ & $3.2020$ & $4.40e+04$ & $50$ \\
& $1.0307$ & $5.13e-04$ & $1.14e-04$ & $6.32e-01$ & $3.68e-01$ & $3.4360$ & $3.3030$ & $6.27e+04$ & $75$ \\
& $1.0407$ & $4.95e-04$ & $9.57e-05$ & $6.20e-01$ & $3.80e-01$ & $3.4640$ & $3.4100$ & $8.07e+04$ & $100$ \\
& $1.0501$ & $4.77e-04$ & $1.58e-04$ & $5.92e-01$ & $4.08e-01$ & $3.5050$ & $3.5180$ & $9.80e+04$ & $125$ \\
& $1.0589$ & $4.81e-04$ & $1.63e-04$ & $5.79e-01$ & $4.21e-01$ & $3.5800$ & $3.6230$ & $1.15e+05$ & $150$ \\
& $1.0669$ & $4.48e-04$ & $1.29e-04$ & $6.49e-01$ & $3.51e-01$ & $3.4200$ & $3.7260$ & $1.31e+05$ & $175$ \\
& $1.0749$ & $4.20e-04$ & $1.02e-04$ & $6.27e-01$ & $3.72e-01$ & $3.4780$ & $3.8320$ & $1.46e+05$ & $200$ \\
& $1.0829$ & $3.97e-04$ & $7.87e-05$ & $6.00e-01$ & $4.00e-01$ & $3.5020$ & $3.9430$ & $1.60e+05$ & $225$ \\
& $1.0909$ & $3.80e-04$ & $6.04e-05$ & $5.89e-01$ & $4.11e-01$ & $3.5360$ & $4.0600$ & $1.74e+05$ & $250$ \\
& $1.0989$ & $3.52e-04$ & $3.34e-05$ & $5.75e-01$ & $4.25e-01$ & $3.5510$ & $4.1820$ & $1.87e+05$ & $275$ \\
& $1.1068$ & $3.37e-04$ & $9.66e-05$ & $5.55e-01$ & $4.45e-01$ & $3.5860$ & $4.3100$ & $1.99e+05$ & $300$ \\
& $1.1128$ & $3.56e-04$ & $1.18e-04$ & $5.32e-01$ & $4.68e-01$ & $3.7040$ & $4.4150$ & $2.11e+05$ & $325$ \\

\hline\hline

\multirow{19}{*}{\begin{tabular}{@{}c@{}} 46\\\\$1.0[M_\odot]$\\\\$1\times10^{-7}[M_\odot\rm yr^{-1}]$\end{tabular}}
& $1.1906$ & $2.79e-03$ & $2.63e-03$ & $5.89e-01$ & $4.11e-01$ & $7.1650$ & $8.3840$ & $3.78e+07$ & $750$ \\
& $1.2640$ & $1.48e-03$ & $1.42e-03$ & $5.39e-01$ & $4.61e-01$ & $7.7430$ & $9.2760$ & $5.27e+07$ & $1500$ \\
& $1.2920$ & $9.61e-04$ & $9.40e-04$ & $5.10e-01$ & $4.90e-01$ & $7.8070$ & $9.5500$ & $6.14e+07$ & $2250$ \\
& $1.3058$ & $7.05e-04$ & $7.32e-04$ & $4.57e-01$ & $5.43e-01$ & $7.7090$ & $9.7900$ & $6.76e+07$ & $3000$ \\
& $1.3109$ & $6.20e-04$ & $5.94e-04$ & $4.72e-01$ & $5.28e-01$ & $7.6290$ & $9.9530$ & $7.26e+07$ & $3750$ \\
& $1.3141$ & $5.90e-04$ & $5.90e-04$ & $4.64e-01$ & $5.36e-01$ & $7.6850$ & $9.9900$ & $7.72e+07$ & $4500$ \\
& $1.3163$ & $5.50e-04$ & $5.74e-04$ & $4.44e-01$ & $5.56e-01$ & $7.6440$ & $10.0200$ & $8.15e+07$ & $5250$ \\
& $1.3186$ & $5.16e-04$ & $5.16e-04$ & $4.56e-01$ & $5.44e-01$ & $7.6020$ & $10.0800$ & $8.56e+07$ & $6000$ \\
& $1.3218$ & $4.73e-04$ & $4.73e-04$ & $4.49e-01$ & $5.51e-01$ & $7.5680$ & $10.1900$ & $8.94e+07$ & $6750$ \\
& $1.3241$ & $4.20e-04$ & $4.30e-04$ & $4.37e-01$ & $5.63e-01$ & $7.4410$ & $10.3600$ & $9.28e+07$ & $7500$ \\
& $1.3255$ & $3.98e-04$ & $4.18e-04$ & $4.20e-01$ & $5.80e-01$ & $7.3870$ & $10.4500$ & $9.61e+07$ & $8250$ \\
& $1.3269$ & $3.97e-04$ & $3.78e-04$ & $4.53e-01$ & $5.47e-01$ & $7.4510$ & $10.5800$ & $9.92e+07$ & $9000$ \\
& $1.3256$ & $3.86e-04$ & $3.77e-04$ & $4.51e-01$ & $5.49e-01$ & $7.4000$ & $10.5600$ & $1.02e+08$ & $9750$ \\
& $1.3262$ & $3.98e-04$ & $3.78e-04$ & $4.59e-01$ & $5.41e-01$ & $7.4210$ & $10.6100$ & $1.05e+08$ & $10500$ \\
& $1.3264$ & $3.98e-04$ & $3.78e-04$ & $4.58e-01$ & $5.42e-01$ & $7.4340$ & $10.6000$ & $1.08e+08$ & $11250$ \\
& $1.3272$ & $3.79e-04$ & $3.51e-04$ & $4.56e-01$ & $5.44e-01$ & $7.4240$ & $10.6100$ & $1.11e+08$ & $12000$ \\
& $1.3273$ & $3.82e-04$ & $3.57e-04$ & $4.58e-01$ & $5.42e-01$ & $7.3680$ & $10.6100$ & $1.14e+08$ & $12750$ \\
& $1.3259$ & $3.82e-04$ & $3.94e-04$ & $4.11e-01$ & $5.89e-01$ & $7.2520$ & $10.5600$ & $1.17e+08$ & $13500$ \\
& $1.3259$ & $3.83e-04$ & $3.94e-04$ & $4.18e-01$ & $5.82e-01$ & $7.3180$ & $10.6600$ & $1.20e+08$ & $14250$ \\

\hline\hline

\multirow{43}{*}{\begin{tabular}{@{}c@{}} \vspace{-15.0cm}\\47\\\\$1.0[M_\odot]$\\\\$5\times10^{-8}[M_\odot\rm yr^{-1}]$\end{tabular}}
& $1.0064$ & $1.24e-02$ & $1.24e-02$ & $5.68e-01$ & $4.32e-01$ & $6.2290$ & $6.2900$ & $2.72e+07$ & $100$ \\
& $1.0064$ & $1.24e-02$ & $1.24e-02$ & $5.68e-01$ & $4.32e-01$ & $6.2290$ & $6.2980$ & $5.35e+07$ & $200$ \\
& $1.0064$ & $1.24e-02$ & $1.24e-02$ & $5.68e-01$ & $4.32e-01$ & $6.2270$ & $6.2990$ & $7.98e+07$ & $300$ \\
& $1.0064$ & $1.24e-02$ & $1.24e-02$ & $5.68e-01$ & $4.32e-01$ & $6.2270$ & $6.2970$ & $1.06e+08$ & $400$ \\
& $1.0064$ & $1.24e-02$ & $1.24e-02$ & $5.68e-01$ & $4.32e-01$ & $6.2300$ & $6.2980$ & $1.32e+08$ & $500$ \\
& $1.0064$ & $1.24e-02$ & $1.24e-02$ & $5.68e-01$ & $4.32e-01$ & $6.2260$ & $6.2980$ & $1.58e+08$ & $600$ \\
& $1.0064$ & $1.24e-02$ & $1.24e-02$ & $5.68e-01$ & $4.32e-01$ & $6.2280$ & $6.2970$ & $1.84e+08$ & $700$ \\
& $1.0064$ & $1.24e-02$ & $1.24e-02$ & $5.68e-01$ & $4.32e-01$ & $6.2260$ & $6.2970$ & $2.10e+08$ & $800$ \\
& $1.0064$ & $1.24e-02$ & $1.24e-02$ & $5.68e-01$ & $4.32e-01$ & $6.2280$ & $6.2970$ & $2.37e+08$ & $900$ \\
& $1.0064$ & $1.24e-02$ & $1.24e-02$ & $5.68e-01$ & $4.32e-01$ & $6.2300$ & $6.2960$ & $2.63e+08$ & $1000$ \\
& $1.0064$ & $1.24e-02$ & $1.24e-02$ & $5.68e-01$ & $4.32e-01$ & $6.2280$ & $6.2970$ & $2.89e+08$ & $1100$ \\
& $1.0064$ & $1.24e-02$ & $1.24e-02$ & $5.69e-01$ & $4.31e-01$ & $6.2290$ & $6.2970$ & $3.15e+08$ & $1200$ \\
& $1.0064$ & $1.24e-02$ & $1.24e-02$ & $5.68e-01$ & $4.32e-01$ & $6.2230$ & $6.2980$ & $3.41e+08$ & $1300$ \\
& $1.0064$ & $1.24e-02$ & $1.24e-02$ & $5.68e-01$ & $4.32e-01$ & $6.2270$ & $6.2980$ & $3.67e+08$ & $1400$ \\
& $1.0064$ & $1.24e-02$ & $1.24e-02$ & $5.68e-01$ & $4.32e-01$ & $6.2310$ & $6.2980$ & $3.93e+08$ & $1500$ \\
& $1.0064$ & $1.24e-02$ & $1.24e-02$ & $5.68e-01$ & $4.32e-01$ & $6.2240$ & $6.2990$ & $4.20e+08$ & $1600$ \\
& $1.0064$ & $1.24e-02$ & $1.24e-02$ & $5.69e-01$ & $4.31e-01$ & $6.2290$ & $6.2980$ & $4.46e+08$ & $1700$ \\
& $1.0064$ & $1.24e-02$ & $1.24e-02$ & $5.68e-01$ & $4.32e-01$ & $6.2240$ & $6.2990$ & $4.72e+08$ & $1800$ \\
& $1.0064$ & $1.24e-02$ & $1.24e-02$ & $5.68e-01$ & $4.32e-01$ & $6.2280$ & $6.2970$ & $4.98e+08$ & $1900$ \\
& $1.0064$ & $1.24e-02$ & $1.24e-02$ & $5.68e-01$ & $4.32e-01$ & $6.2310$ & $6.2970$ & $5.24e+08$ & $2000$ \\
& $1.0064$ & $1.24e-02$ & $1.24e-02$ & $5.68e-01$ & $4.32e-01$ & $6.2310$ & $6.2960$ & $5.50e+08$ & $2100$ \\
& $1.0064$ & $1.24e-02$ & $1.24e-02$ & $5.68e-01$ & $4.32e-01$ & $6.2290$ & $6.2980$ & $5.76e+08$ & $2200$ \\
& $1.0064$ & $1.24e-02$ & $1.24e-02$ & $5.68e-01$ & $4.32e-01$ & $6.2280$ & $6.2970$ & $6.02e+08$ & $2300$ \\
& $1.0064$ & $1.24e-02$ & $1.24e-02$ & $5.68e-01$ & $4.32e-01$ & $6.2290$ & $6.2980$ & $6.28e+08$ & $2400$ \\
& $1.0064$ & $1.24e-02$ & $1.24e-02$ & $5.68e-01$ & $4.32e-01$ & $6.2240$ & $6.2970$ & $6.54e+08$ & $2500$ \\
& $1.0064$ & $1.24e-02$ & $1.24e-02$ & $5.68e-01$ & $4.32e-01$ & $6.2280$ & $6.2980$ & $6.80e+08$ & $2600$ \\
& $1.0064$ & $1.24e-02$ & $1.24e-02$ & $5.69e-01$ & $4.31e-01$ & $6.2290$ & $6.2980$ & $7.06e+08$ & $2700$ \\
& $1.0064$ & $1.24e-02$ & $1.24e-02$ & $5.68e-01$ & $4.32e-01$ & $6.2310$ & $6.2970$ & $7.32e+08$ & $2800$ \\
& $1.0064$ & $1.24e-02$ & $1.24e-02$ & $5.69e-01$ & $4.31e-01$ & $6.2290$ & $6.2970$ & $7.58e+08$ & $2900$ \\
& $1.0064$ & $1.24e-02$ & $1.24e-02$ & $5.68e-01$ & $4.32e-01$ & $6.2290$ & $6.2970$ & $7.85e+08$ & $3000$ \\
& $1.0064$ & $1.24e-02$ & $1.24e-02$ & $5.69e-01$ & $4.31e-01$ & $6.2290$ & $6.2970$ & $8.11e+08$ & $3100$ \\
& $1.0064$ & $1.24e-02$ & $1.24e-02$ & $5.68e-01$ & $4.32e-01$ & $6.2250$ & $6.2980$ & $8.37e+08$ & $3200$ \\
& $1.0064$ & $1.24e-02$ & $1.24e-02$ & $5.68e-01$ & $4.32e-01$ & $6.2310$ & $6.2960$ & $8.63e+08$ & $3300$ \\
& $1.0064$ & $1.24e-02$ & $1.24e-02$ & $5.68e-01$ & $4.32e-01$ & $6.2250$ & $6.2980$ & $8.89e+08$ & $3400$ \\
& $1.0064$ & $1.24e-02$ & $1.24e-02$ & $5.69e-01$ & $4.31e-01$ & $6.2290$ & $6.2960$ & $9.15e+08$ & $3500$ \\
& $1.0064$ & $1.24e-02$ & $1.24e-02$ & $5.68e-01$ & $4.32e-01$ & $6.2330$ & $6.2960$ & $9.41e+08$ & $3600$ \\
& $1.0064$ & $1.24e-02$ & $1.24e-02$ & $5.68e-01$ & $4.32e-01$ & $6.2310$ & $6.2970$ & $9.67e+08$ & $3700$ \\
& $1.0064$ & $1.24e-02$ & $1.24e-02$ & $5.68e-01$ & $4.32e-01$ & $6.2300$ & $6.2960$ & $9.93e+08$ & $3800$ \\
& $1.0064$ & $1.24e-02$ & $1.24e-02$ & $5.69e-01$ & $4.31e-01$ & $6.2320$ & $6.2960$ & $1.02e+09$ & $3900$ \\
& $1.0064$ & $1.24e-02$ & $1.24e-02$ & $5.69e-01$ & $4.31e-01$ & $6.2290$ & $6.2960$ & $1.04e+09$ & $4000$ \\
& $1.0064$ & $1.24e-02$ & $1.24e-02$ & $5.68e-01$ & $4.32e-01$ & $6.2320$ & $6.2960$ & $1.07e+09$ & $4100$ \\
& $1.0064$ & $1.24e-02$ & $1.24e-02$ & $5.68e-01$ & $4.32e-01$ & $6.2250$ & $6.2970$ & $1.10e+09$ & $4200$ \\
& $1.0064$ & $1.24e-02$ & $1.24e-02$ & $5.68e-01$ & $4.32e-01$ & $6.2330$ & $6.2970$ & $1.12e+09$ & $4300$ \\

\hline
  \end{longtable}
\twocolumn


\bibliographystyle{elsarticle-harv} 
\bibliography{rfrncs}






\end{document}